\documentclass[bibyear]{aa}  

\usepackage{graphicx}
\usepackage{txfonts}
\usepackage{natbib}
\bibpunct[]{(}{)}{;}{a}{}{,}
\begin{document}

   \title{Exploring Molecular Complexity with ALMA (EMoCA):\\
     Deuterated complex organic molecules in Sagittarius~B2(N2)}

   %\subtitle{I. Overviewing the $\kappa$-mechanism}
   \titlerunning{Deuterated complex organic molecules in Sgr~B2(N2)}
   \author{A. Belloche\inst{1}
          \and
          H.~S.~P. M\"uller\inst{2}
          \and
          R.~T. Garrod\inst{3}
          \and
          K.~M. Menten\inst{1}
          }

   \institute{Max-Planck-Institut f\"ur Radioastronomie, Auf dem H\"ugel 69,
              53121 Bonn, Germany\\
              \email{belloche@mpifr-bonn.mpg.de}
         \and
             I. Physikalisches Institut, Universit{\"a}t zu K{\"o}ln, 
             Z{\"u}lpicher Str. 77, 50937 K{\"o}ln, Germany
         \and
             Depts. of Astronomy and Chemistry, University of Virginia, 
             Charlottesville, VA 22904, USA
             }

   \date{Received 28 August 2015; accepted 16 November 2015}

% \abstract{}{}{}{}{} 
% 5 {} token are mandatory
 
  \abstract
   {Deuteration is a powerful tracer of the history of the cold prestellar
   phase in star forming regions. Apart from methanol, little is known about 
   deuterium fractionation of complex organic molecules in the interstellar 
   medium, especially in high mass star forming regions.}
   {Our goal is to detect deuterated complex organic molecules toward the high
   mass star forming region Sagittarius~B2 (Sgr~B2) and derive their level of 
   deuteration.}
   {We use a complete 3~mm spectral line survey performed with the Atacama 
   Large Millimeter/submillimeter Array (ALMA) to search for deuterated 
   complex organic molecules toward the hot molecular core Sgr~B2(N2).
   Population diagrams and integrated intensity maps are constructed to fit 
   rotational temperatures and emission sizes for each molecule. Column 
   densities are derived by modelling the full spectrum under the assumption 
   of local thermodynamic equilibrium. The results are compared to predictions 
   of two astrochemical models that treat the deuteration process.}
   {We report the detection of CH$_2$DCN toward Sgr~B2(N2) with a deuteration
   level of 0.4\%, and tentative detections of CH$_2$DOH, 
   CH$_2$DCH$_2$CN, the chiral molecule CH$_3$CHDCN, and DC$_3$N with 
   levels in the range 
   0.05\%--0.12\%. A stringent deuteration upper limit is obtained for CH$_3$OD 
   ($<0.07\%$). Upper limits in the range 0.5--1.8\% are derived for the 
   three deuterated isotopologues of vinyl cyanide, the four deuterated 
   species of ethanol, and CH$_2$DOCHO. Ethyl cyanide is less deuterated than 
   methyl cyanide by at least a factor five. The [CH$_2$DOH]/[CH$_3$OD] 
   abundance ratio is higher than 1.8. It may still be consistent with the 
   value obtained in Orion~KL. Except for methyl cyanide, the measured 
   deuteration levels lie at least a factor four below the predictions of 
   current astrochemical models. The deuteration levels in Sgr~B2(N2) are also 
   lower than in Orion~KL by a factor of a few up to a factor ten.}
   {The discrepancy between the deuteration levels of Sgr~B2(N2) and the 
   predictions of chemical models, and the difference between Sgr~B2(N2) and 
   Orion~KL may both be due to the higher kinetic temperatures that 
   characterize the Galactic Center region compared to nearby clouds. 
   Alternatively, they may result from a lower overall abundance of deuterium 
   itself in the Galactic Center region by up to a factor ten.}


   \maketitle
%
%________________________________________________________________

\section{Introduction}
\label{s:intro}

Deuteration is a powerful tracer of the history of the cold prestellar phase
in star forming regions \citep[see, e.g.,][]{Caselli12,Ceccarelli14}. At low
temperatures and high densities that prevail in prestellar cores, the 
gas-phase abundance of H$_3^+$ gets enhanced thanks to the depletion of its 
main destroyer, CO, onto the surface of dust grains. The reaction of 
H$_3^+$ with HD, the reservoir of deuterium in molecular clouds, produces 
H$_2$D$^+$. The exothermicity of this reaction leads to an enhancement of the
[H$_2$D$^+$]/[H$_3^+$] abundance ratio. The reaction of H$_2$D$^+$ with other 
molecules then transfers the deuterium to other species, producing a 
significant increase in their deuterium fractionation that can reach several 
orders of magnitude compared to the low cosmic abundance of deuterium with 
respect to hydrogen \citep[$\sim 1.5 \times 10^{-5}$,][]{Linsky03}. 
Deuteration levels -- i.e. the abundance ratio [XD]/[XH] of a deuterated 
molecule XD to the corresponding hydrogen-bearing one XH -- above 10\% are 
commonly found in low-mass star forming regions \citep[e.g.,][]{Parise06}. 
Even molecules containing several deuterium atoms have been detected 
\citep[e.g.,][]{Parise04}. Levels of deuteration above 10\% have also been
found for small molecules such as ammonia in high mass star forming regions 
\citep[][]{Pillai07,Fontani15}. Deuteration at somewhat higher temperatures
($\sim$ 30--50~K), at which the H$_2$D$^+$ channel is not efficient anymore, 
can also occur in the gas phase via CH$_2$D$^+$ which is, like H$_2$D$^+$, 
produced by an exothermic reaction of CH$_3^+$ with HD 
\citep[e.g.,][]{Parise09}.

The overall increase in the deuterium fractionation of molecules in cold, 
prestellar cores leads to an abundance enhancement of atomic deuterium in the 
gas phase. These atoms can stick to the surface of dust grains and in turn 
participate in the deuterium fractionation of molecules already or 
subsequently formed on these grains, in particular of the complex organic 
ones\footnote{We adopt the definition of \citet{Herbst09}: a complex organic 
molecule is an organic molecule containing six atoms or more.}. Once a 
protostar is formed, the energy it radiates heats up its 
circumstellar envelope and complex organic molecules desorb from the grain 
surfaces \citep[e.g.,][]{Garrod08}. During this hot core phase, the relative 
abundances of these freshly desorbed molecules is expected to provide clues 
about their formation paths during the past evolution of the core. In 
particular, their degree of deuterium fractionation is a relic of the 
conditions that were prevailing in earlier, colder stages 
\citep[e.g.][]{Taquet12,Taquet14,Aikawa12}.

Apart from methanol, relatively little is known about deuteration of complex 
organic molecules in the interstellar medium, in 
particular in high mass star forming regions. A few detections or tentative
detections of deuterated (complex) organic molecules have been reported toward
Orion~KL \citep[][]{Gerin92,Daly13,Esplugues13,Neill13,Coudert13}, but no 
systematic study of the deuteration of complex organic molecules in high mass 
star forming regions has been reported so far.

Sagittarius~B2, hereafter Sgr~B2, is one of the most massive star forming 
regions in the Galaxy. It is located in the Central Molecular Zone, close to 
the Galactic Center. Its current star formation rate qualifies Sgr~B2 as a 
mini-starburst region \citep[see, e.g., Appendix A of][]{Belloche13}. The cloud
contains several sites of on-going high-mass star formation. One of them, 
Sgr~B2(N), contains two hot molecular cores that we name here Sgr~B2(N1) and 
(N2)\footnote{They were named P1 and P2 in \citet{Belloche08} and SMA1 and 
SMA2 in \citet{Qin11}.}. Their angular separation is $5\arcsec$ in the
north-south direction, corresponding to $0.2$~pc in projection at a distance 
of 8.3~kpc \citep[][]{Reid14}. Their velocities projected along the line
of sight differ by 9--10 km~s$^{-1}$. Both are characterized by extremely high 
H$_2$ column densities \citep[$> 10^{25}$~cm$^{-2}$ over few arcsec, 
see][]{Belloche08,Belloche14,Qin11}. They are both in an early stage of star
formation when a (massive) protostar has already formed and started to heat up 
its circumstellar envelope. The high kinetic temperatures of the hot cores
($\sim 150$--200~K) lead to the sublimation of molecules that formed in the ice 
mantles of dust grains during the prestellar phase and the warming-up period of
the protostellar phase. As a result of both the high temperatures and column
densities, numerous complex organic molecules have been detected toward 
Sgr~B2(N) -- many of them for the first time in the interstellar medium -- 
since the beginning of radioastronomy nearly five decades ago.

Following up a molecular line survey of Sgr~B2(N) performed with the IRAM 30~m
telescope that led to the first detection in space of a few new complex organic
molecules \citep[][]{Belloche08,Belloche09,Belloche13}, we performed a 
systematic line survey of Sgr~B2(N) in the 3~mm atmospheric window at high 
angular resolution and sensitivity with the Atacama Large 
Millimeter/submillimeter Array (ALMA) in its Cycles 0 and 1. This survey is 
called EMoCA, which stands for ``Exploring Molecular Complexity with ALMA''. It 
aims at investigating molecular complexity in the interstellar 
medium. One of the initial results of EMoCA was the first interstellar 
detection of a branched alkyl molecule \citep[][]{Belloche14}. Here, we take
advantage of this sensitive survey to explore, for the first time in a 
systematic way, the deuterium fractionation of complex organic molecules in 
Sgr~B2(N2). We focus on Sgr~B2(N2) rather than the main hot core Sgr~B2(N1)
because the former has relatively narrow linewidths ($\sim 5$~km~s$^{-1}$) at 
the angular resolution of EMoCA ($\sim 1.8\arcsec$) while the latter still has 
prominent linewings like in our previous single-dish survey.
A companion paper reports on the detection of alkanols and 
alkanethiols based on EMoCA \citep[][]{Mueller15c}.

The article is structured as follows. The observational setup and the process 
of data reduction are described in Sect.~\ref{s:observations}. 
Section~\ref{s:ltemodeling} explains the method employed to model the observed 
spectra in the approximation of local thermodynamic equilibrium and 
Sect.~\ref{s:spectro} gives some details about the spectroscopic predictions
used to generate the synthetic spectra. The results of the analysis are
reported in Sect.~\ref{s:results} and discussed in Sect.~\ref{s:discussion}.
Section~\ref{s:conclusions} gives our conclusions about deuterium fractionation
of complex organic molecules in Sgr~B2(N2).

%__________________________________________________________________

\section{Observations and data reduction}
\label{s:observations}

\begin{table*}
\caption{Observational setups of the EMoCA survey.}
\label{t:setup}
\centering
\begin{tabular}{cccccccccccccc}
\hline
\hline
\noalign{\smallskip}
Setup & \multicolumn{2}{c}{Frequency range} & Date of & $t_{\rm start}$\tablefootmark{a} & $N_{\rm a}$\tablefootmark{b} & Baseline  & $t_{\rm int}$\tablefootmark{d} & \multicolumn{3}{c}{Calibrators\tablefootmark{e}} & & \multicolumn{2}{c}{Peak\tablefootmark{f}} \\
%\noalign{\smallskip}
\cline{2-3} \cline{9-11} \cline{13-14}
\noalign{\smallskip}
      & LSB & USB  & observation & (UTC) &  & range\tablefootmark{c} & & B & A & P & & $\Delta\alpha$ & $\Delta\delta$ \\
      & GHz & GHz & yyyy-mm-dd & hh:mm & & m & min & & & & & $''$ & $''$ \\
\hline
\noalign{\smallskip}
S1 & 84.1--87.8  & 96.2--99.8   & 2012-08-27 & 01:05 & 26 & 17--400 & 54.7 & 1 & 3 & 2 & & 0.2        & $-2.4$ \\
S2 & 87.7--91.4  & 99.7--103.4  & 2012-06-05 & 09:03 & 20 & 14--392 & 24.2 & 1 & 3 & 2 & & 1.2$^\star$ & $-2.4$ \\
   &             &              & 2012-07-01 & 07:04 & 13 & 25--393 & 40.0 & 1 & 3 & 2 & & 1.2$^\star$ & $-2.6$ \\
   &             &              & 2012-07-03 & 06:36 & 21 & 14--395 & 48.4 & 1 & 3 & 2 & & 0.6$^\star$ & $-2.3$ \\
   &             &              & 2012-09-28 & 21:57 & 25 & 20--387 & 44.1 & 1 & 4 & 2 & & 0.2        & $-2.5$ \\
S3 & 91.4--95.1  & 103.4--107.1 & 2012-06-06 & 08:20 & 18 & 15--395 & 40.2 & 1 & 3 & 2 & & 0.7$^\star$ & $-2.4$ \\
   &             &              & 2012-06-18 & 07:29 & 22 & 15--395 & 40.4 & 1 & 3 & 2 & & 0.6$^\star$ & $-2.2$ \\
S4 & 95.0--98.7  & 107.0--110.7 & 2012-07-04 & 05:38 & 21 & 17--398 &  8.1 & 1 & 3 & 2 & & 0.1        & $-2.3$ \\
   &             &              & 2012-08-01 & 02:32 & 24 & 19--442 & 34.9 & 1 & 3 & 2 & & 0.1        & $-2.4$ \\
   &             &              & 2012-08-10 & 00:45 & 26 & 21--400 & 35.0 & 1 & 3 & 2 & & 0.2        & $-2.4$ \\
S5 & 98.7--102.4 & 110.7--114.4 & 2014-04-05 & 06:22 & 38 & 15--413 & 24.4 & 2 & 4 & 5 & & 0.2        & $-2.4$ \\
\hline
\end{tabular}
\tablefoot{
\tablefoottext{a}{Start time of observation.}
\tablefoottext{b}{Number of ALMA 12\,m antennas.}
\tablefoottext{c}{Minimum and maximum projected baseline separations.}
\tablefoottext{d}{On-source integration time.}
\tablefoottext{e}{Bandpass (B), amplitude (A), and phase (P) calibrators. The 
calibrators are: 1: B1730-130, 2: J1700-2610, 3: Neptune, 4: Titan, 
5: J1744-3116.}
\tablefoottext{f}{Offset of the continuum peak position of Sgr~B2(N1) with 
respect to the phase center, in equatorial coordinate system (J2000). 
Measurement sets with $\Delta\alpha$ differing from $0.1''$ by more than 
$0.2''$ (marked with a $\star$) are believed to be affected by an astrometric 
problem.}
}
\end{table*}

\subsection{Observations}
\label{ss:observations}

We used the Atacama Large Millimeter/submillimeter Array (ALMA) to perform a 
complete spectral line survey toward Sgr~B2(N) between 84.1 and 114.4~GHz. 
The field was centered at ($\alpha, \delta$)$_{\rm J2000}$=
($17^{\rm h}47^{\rm m}19.87^{\rm s}, -28^\circ22'16''$),
half way between Sgr~B2(N1) and (N2) that are separated by $4.9''$ in the 
north-south direction. The size (HPBW) of the primary beam of the 12\,m 
antennas varies between $69''$ at 84~GHz and $51''$ at 114~GHz 
\citep{Remijan15}.
% 1.13 lambda/D from Section 3.2 page 19 of ALMA Cycle 3 Technical Guide

The spectral line survey was divided into five spectral setups. Each setup was 
observed in one polarisation and delivered four spectral windows, two per 
sideband. The separation between the centers of the lower and upper sidebands 
is 12 GHz. Each spectral window has a bandwidth of 1875 MHz and a channel 
spacing of 244.141~kHz, but the spectra were smoothed to a spectral resolution 
of 488.3~kHz (1.7 to 1.3~km~s$^{-1}$). Each pair of adjacent spectral windows 
has an overlap of about 50~MHz. Details about the frequency coverage, the date 
of observation, the number of antennas, the range of baselines, the on-source 
integration time, and the bandpass, amplitude, and phase calibrators are given 
in Table~\ref{t:setup}. Setups S1 to S4 were observed in Cycle 0 while setup S5 
was observed in Cycle 1. As reported in Table~\ref{t:setup}, setups S1 and S5 
were observed only once, but setups S2, S3, and S4 were observed on several
days, between two and four times each.

\subsection{Data reduction}
\label{ss:reduction}

The data was calibrated and imaged with CASA. We used version 4.2.0 (r28322) 
for setups S1 to S4 and version 4.2.1 (r29047) for setup S5. We used the 
standard procedures provided by the Joint ALMA Observatory to apply the 
bandpass, amplitude, and phase calibrations. The deconvolution was performed 
with the \textit{csclean} imager mode and a \textit{Briggs} weighting scheme 
with a \textit{robust} parameter of 0.5. The cell size was set to $0.3''$. In 
addition, three or four iterations of self-calibration were performed using a 
strong spectral line detected toward Sgr~B2(N1) in each setup. This improved 
the dynamical range in the resulting images significantly.

The spectra toward Sgr~B2(N1) and (N2) are full of lines and close 
to the confusion limit. It is thus difficult to separate the line emission 
from the continuum emission in a systematic way for the full data cubes, but 
it is a necessary step to produce separate line and continuum maps. For each
spectral window of each setup, we selected six groups of few channels that 
seemed to be free of strong line emission. A first-order baseline was fitted 
to these selected channels and the result of the fit was used to split each 
data cube into two cubes, one for the line emission and one for the continuum 
emission. Given the difference in systemic velocity between the two hot cores 
\citep[$\sim 10$~km~s$^{-1}$, see][]{Belloche13}, we selected different sets 
of channels for the northern and southern parts of the field. This process of
baseline subtraction was performed with the CLASS 
software\footnote{\label{fn:gildas}See http://www.iram.fr/IRAMFR/GILDAS.}.

We checked the accuracy of the relative astrometry between the ten measurement
sets by fitting the peak position of the continuum emission toward Sgr~B2(N1) 
using selected channels of the line+continuum data cubes that appeared to be 
free of line emission. It turns out that the first three measurement sets of 
setup S2 and both measurement sets of setup S3 are affected by an astrometric 
problem: the continuum peak of Sgr~B2(N1) is shifted by 0.6$''$ to 1.2$''$ in 
right ascension with respect to all other measurement sets (see 
Table~\ref{t:setup}). The dispersion of the peak position in declination is 
also a bit higher for the former compared to the latter. The average peak 
position of Sgr~B2(N1) in all non-affected measurement sets is at 
($\Delta\alpha$, $\Delta\delta$) = 
($0.15''$, $-2.40''$), i.e. ($\alpha, \delta$)$_{\rm J2000}$=
($17^{\rm h}47^{\rm m}19.881^{\rm s}, -28^\circ22'18.40''$).
The five affected measurement sets were obtained after transit when the source 
was setting and the phase calibrator was at low elevation, which lets us think
that the shift of the affected measurement sets may be due to an inacurrate 
calibration of the atmospheric phase fluctuations. As a result, we ignored the 
affected measurement sets of setup S2 and used only its fourth one. For setup 
S3, both measurement sets were used but the offset was approximately 
compensated for by modifying the visibilities of the phase calibrator with the 
CASA task \textit{fixvis} before the phase calibration. After this correction, 
the relative positional accuracy of all measurement sets used in this work is 
on the order of $\pm 0.1''$ in both right ascension and declination.

\begin{table*}
\caption{Beam sizes and noise levels.}
\label{t:beam_noise}
\centering
\begin{tabular}{cccccccc}
\hline
\hline
\noalign{\smallskip}
Setup & SPW\tablefootmark{a} & Frequency & \multicolumn{2}{c}{Synthesized beam} & & \multicolumn{2}{c}{rms\tablefootmark{c}} \\
%\noalign{\smallskip}
\cline{4-5} \cline{7-8}
\noalign{\smallskip}
      & & range & $HPBW$ & $P.A.$\tablefootmark{b} & & mJy & \\
\noalign{\smallskip}
      & & MHz & $''\times''$ & $^\circ$ & & beam$^{-1}$ & K\\
\hline
\noalign{\smallskip}
S1 & 0 &  84091 -- 85966  & $2.1 \times 1.5$ & $-85$ & & 3.0 & 0.16 \\
   & 1 &  85904 -- 87779  & $2.0 \times 1.5$ & $-83$ & & 2.7 & 0.14 \\
   & 2 &  96154 -- 98029  & $1.8 \times 1.4$ & $-85$ & & 3.0 & 0.16 \\
   & 3 &  97904 -- 99779  & $1.8 \times 1.3$ & $-85$ & & 3.1 & 0.16 \\
S2 & 0 &  87729 -- 89604  & $1.9 \times 1.7$ &  $86$ & & 3.1 & 0.15 \\
   & 1 &  89554 -- 91429  & $1.8 \times 1.6$ &  $52$ & & 2.8 & 0.15 \\
   & 2 &  99728 -- 101602 & $1.6 \times 1.4$ &  $48$ & & 2.7 & 0.14 \\
   & 3 & 101552 -- 103427 & $1.6 \times 1.4$ &  $49$ & & 2.7 & 0.14 \\
S3 & 0 &  91368 -- 93242  & $2.9 \times 1.5$ &  $84$ & & 3.4 & 0.12 \\
   & 1 &  93193 -- 95067  & $2.8 \times 1.5$ &  $83$ & & 3.1 & 0.10 \\
   & 2 & 103365 -- 105239 & $2.5 \times 1.3$ &  $82$ & & 3.4 & 0.11 \\
   & 3 & 105189 -- 107064 & $2.5 \times 1.3$ &  $82$ & & 3.6 & 0.12 \\
S4 & 0 &  95021 -- 96896  & $1.9 \times 1.4$ & $-82$ & & 1.9 & 0.10 \\
   & 1 &  96846 -- 98720  & $1.8 \times 1.3$ & $-82$ & & 1.9 & 0.10 \\
   & 2 & 107019 -- 108893 & $1.7 \times 1.2$ & $-83$ & & 2.2 & 0.11 \\
   & 3 & 108843 -- 110718 & $1.6 \times 1.2$ & $-82$ & & 2.3 & 0.12 \\
S5 & 0 &  98672 -- 100546 & $1.8 \times 1.4$ & $-76$ & & 2.8 & 0.14 \\
   & 1 & 100496 -- 102370 & $1.7 \times 1.4$ & $-76$ & & 2.7 & 0.13 \\
   & 2 & 110669 -- 112543 & $1.6 \times 1.3$ & $-72$ & & 3.5 & 0.17 \\
   & 3 & 112494 -- 114368 & $1.6 \times 1.2$ & $-77$ & & 4.9 & 0.24 \\
\hline
\end{tabular}
\tablefoot{
\tablefoottext{a}{Spectral window.}
\tablefoottext{b}{Position angle east from north.}
\tablefoottext{c}{Median rms noise level measured in the channel maps of the 
continuum-removed data cubes.}
}
\end{table*}

The measurement sets of setup S3 were merged into one single measurement set 
with the CASA task \textit{concat} before imaging. The same was done for
setup S4.
Only one measurement set was used for the other three setups. The size (HPBW) 
of the synthesized beam and the rms noise level in the final cubes are given in 
Table~\ref{t:beam_noise}. The noise level of each spectral window corresponds 
to the median of the noise levels measured in all channel maps using the 
procedure \textit{go noise} in GREG$^{\rm \ref{fn:gildas}}$. The noise levels 
reported in Table~\ref{t:beam_noise} were measured in the continuum-subtracted 
datacubes. They are typically a factor $\sim 2$ higher in the line+continuum 
datacubes.

Based on the redundancies of the measurement sets and the spectral overlap 
between the setups, we estimate the relative calibration uncertainty on the 
flux density to be on the order of 15\%. 

%__________________________________________________________________

\section{Radiative transfer modeling of the line survey}
\label{s:ltemodeling}

We used the input parameters of our LTE model of the IRAM 30\,m spectrum of 
Sgr~B2(N) \citep[][]{Belloche13} as a starting point to assign the lines 
detected in the ALMA spectra and model the emission of the detected molecules. 
Given the high H$_2$ densities of Sgr~B2(N1) and (N2) 
\citep[$\sim 10^8$~cm$^{-3}$ at arcsecond 
scale,][]{Belloche08,Belloche14,Qin11}, the LTE approximation is appropriate.
We used the software Weeds \citep[][]{Maret11} to produce synthetic LTE 
spectra that take into account the radiative transfer and the (spectral-window- 
and measurement-set-dependent) angular resolution of the observations. 
The modeling was performed for each species
separately, and the contributions of all detected species to the emitted 
spectra were then added linearly to obtain the final synthetic spectrum
(hereafter called the full LTE model). This approximation is valid for 
optically thin lines that overlap in frequency space or for (optically thick 
or thin) lines of species that are emitted from separated regions within the 
beam, but it is not correct anymore for frequency-overlapping optically thick 
lines of species that are cospatial or aligned along the line of sight. In such 
cases, the synthetic spectrum overestimates the actual line flux density.

The model of each species is defined by five parameters: angular size of the 
emitting region assumed to be Gaussian, column density, rotational 
temperature, velocity offset with respect to the assumed systemic velocity of 
the source, and linewidth ($FWHM$). For a given 
species, the source size was derived from two-dimensional Gaussian fits to the 
integrated intensity maps of all transitions that were well detected and found 
to be free of contamination (based on the full LTE model). The source size was 
set to the median deconvolved size of all such transitions. The other four 
parameters were optimized by eye. Population diagrams were constructed based 
on the transitions that are well detected and not severely contaminated by 
transitions of other species. In the case where a transition was partially 
contaminated, the contributions of the contaminating species was removed from 
the measured integrated intensities, on the basis of the full LTE model. Each 
population diagram was also corrected for optical depth following the method
described in
\citet{Goldsmith99}, using the opacities delivered by Weeds. The population 
diagrams were used to verify that the rotational temperature derived in the
course of the (manual) modeling with Weeds made sense.

In the population diagrams corrected for optical depth and contamination, the 
residual dispersion of the synthetic datapoints (red crosses) results in part
from the frequency boundaries set to integrate the intensity. These boundaries
are a compromise between covering the line as much as possible and limiting as 
much as possible the contamination from other species emitting at nearby 
frequencies. In addition, the correction for optical depth is an 
approximation, and may also introduce some bias. Finally, another limitation 
of this fit is that it can be biased by residual contamination remaining even
after removal of the contribution of the identified contaminating species.  
Therefore, we believe that the formal errors on the rotational temperature 
derived from the fit to the population diagrams do not necessarily represent 
the true errors on this temperature and should be viewed with caution.

The emission of vibrationally excited states of a given 
molecule were modeled independentely of the vibrational ground state. The 
emission of isotopologues of a given molecule were also modeled separately.

The physical structure of the source assumed for the modeling is uniform. 
This may sound simplistic given that temperature and density gradients are 
expected in the envelope of Sgr~B2(N2) \citep[e.g.,][]{Rolffs11}. However, it 
turns out that, even with such a simple assumption, the spectra of most 
complex organic molecules detected toward Sgr B2(N2) can be well reproduced so 
we are confident that the rotational temperatures and column densities derived 
from our analysis are reliable.

In the following, we count a line of a given species as a ``detected'' line 
if its frequency, peak intensity, and width are well reproduced by our model 
and it is not 
(or barely) contaminated by emission from other species. As a counter example,
a synthetic line that is consistent with the observed spectrum -- i.e. that has
a peak intensity simply below the intensity of the detected signal -- but 
would still remain consistent if it were shifted by a frequency offset 
typically equal to its linewidth is not counted as ``detected''. We emphasize
that our definition of a ``detected'' line is very conservative but we believe
that it is required to avoid unsecure molecule detections.

The complete list of transitions identified in our survey is presently 
not available but the interested reader can refer to the list of lines 
identified in our previous single-dish survey of Sgr~B2(N)
\citep[][]{Belloche13}.

\section{Spectroscopic predictions}
\label{s:spectro}

The origin of the spectroscopic predictions used to model the emission of the 
species reported in Sect.~\ref{s:results} is provided here. 

Predictions for the three singly-deuterated species of ethyl cyanide were 
taken from the catalog of the Cologne Database for Molecular Spectroscopy 
\citep[CDMS\footnote{See http://www.cdms.de.},][]{Mueller01,Mueller05} 
(tags 56509, 56510, and 56511, all version 1). They are based on 
\citet{Margules09}. All other ethyl cyanide data were also taken from the 
CDMS. The main species predictions are based on \citet{EtCN_rot_2009} with 
published data in the range of our survey from \citet{EtCN_rot_1996}. 
Transition frequencies of the isotopologues containing one $^{13}$C were taken 
from \citet{13C-EtCN_rot_2012}, those for the $^{15}$N isotopologue from
\citet{Margules09}. Vibrational corrections to the rotational partition 
function, and thus to the column density, were derived for the main 
isotopologue from \citet{EtCN_IR_1981} and applied to all species. Only 
limited isotopic data are available. It is safe to assume that differences 
among the isotopologues are small, most likely not exceeding a few percent, 
because of the large number of heavy atoms in the molecule.

Predictions for singly-deuterated methyl cyanide were taken from the CDMS 
catalog (tag 42511, version 2). This entry is based on \citet{D-MeCN_rot_2013}.
Predictions for CH$_3$CN in its $\varv _8 = 1$ and 2 excited states are based 
on preliminary results from \citet{MeCN_rot-16_2015}, those in $\varv _4 = 1$ 
are preliminary data from a subsequent study (M{\"u}ller et al., in 
preparation). Transition frequencies in the range of our study are from 
\citet{MeCN_rot-8_1969} and \citet{MeCN_rot-16_1971} for $\varv _8 = 1$ and 2, 
respectively. 
Predictions for isotopic species with $^{15}$N or one or two $^{13}$C in their 
ground vibrational states are from the CDMS. They are based on 
\citet{MeCN-isos_rot_2009} with transition frequencies in the range of our 
survey from \citet{MeCN-isos_rot_1979}. Predictions for $^{13}$C isotopologues 
in their $\varv _8 = 1$ states are based on preliminary data from 
\citet{MeCN_isos-v8_2015}.
Vibrational corrections to the partition function were included in the 
private entries for the main isotopic species. They are now available in 
\citet{MeCN_rot-16_2015}. As the corrections are small, the error by using the 
values from the main isotopic species also for the other isotopologues 
is very small, even for CH$_2$DCN.

Predictions for deuterated vinyl cyanide CH$_2$CDCN were taken from the
molecular spectroscopic database of the Jet Propulsion Laboratory 
\citep[JPL\footnote{See http://spec.jpl.nasa.gov.},][]{Pickett98}
(tag 54004, version 2). Predictions for \textit{cis}-CHDCHCN and 
\textit{trans}-CHDCHCN were prepared by us. All predictions are based on 
\citet{VyCN-isos_rot_1997}. Predictions for C$_2$H$_3$CN and isotopic species 
with one $^{13}$C or with $^{15}$N were taken from the CDMS and are based on 
\citet{VyCN-isos_rot_2008}. Transition frequencies in the range of our survey 
are mostly from that study. For the main species, they are, to a large 
extent, also from \citet{VyCN_rot_1996}. Predictions for excited states of 
vinyl cyanide used in the present work are based on 
\citet{VyCN_rot-vib_1988} and unpublished data from one of us (HSPM).
These data included vibrational corrections which are essentially complete at 
200~K. The vibrational energies were gathered from several sources. A recent 
compilation of low-lying vibrational states is avalaible in 
\citet{VyCN_IR_2015}. It is safe to assume that differences among the 
isotopologues are small, most likely not exceeding a few percent, because of 
the large number of heavy atoms in the molecule.

Predictions for deuterated cyanoacetylene were taken from the CDMS catalog
(tag 52508, version 2). This entry is based on \citet{DC3N_rot_2008}. 
All other cyanoacetylene predictions were also taken from the CDMS. 
The $\varv _7 = 1$ predictions of the main species are based on 
\citet{HC3N_rot-vib_2000} with data in the range of our survey from 
\citet{HC3N_rot-vib_1986}. All predictions of iotopologues containing one or 
two $^{13}$C are based on \citet{HC3N-isos_rot-vib_2001}, those for 
HC$_3^{15}$N on \citet{HC3N-15_rot-vib_2004}. Ground state transition 
frequencies for singly substituted species in the range of our survey were 
taken from \citet{HC3N-isos_rot_1977}.
Vibrational contributions to the partition functions of HC$_3$N and 
DC$_3$N can be evaluated from the study of their low-lying vibrational states 
by \citet{HC3N_DC3N_IR_1982}. Isotopic shifts, in particular of the lowest 
$\nu _7$ mode, are much smaller for $^{13}$C or $^{15}$N species. Therefore, 
using vibrational corrections of the main isotoplogue introduces small 
errors for these species.

Predictions for deuterated methanol CH$_2$DOH were taken from the JPL 
catalog (tag 33004, version 1). They are based on \citet{CH2DOH_rot_2012} with 
rest frequencies almost entirely from that study. Using torsional energies from 
\citet{CH2DOH_FIR_2009}, we estimate a vibrational correction factor to the 
partition function of 1.15 at 160~K.
For CH$_3$OD, we prepared a 
catalog entry based on \citet{Anderson88}, with frequencies updated to the 
values published in \citet{Duan03}. We estimated the partition function to be 
11770 at 150~K and 25550 at 225~K, taking into account torsional energies of 
CH$_3$OD in \citet{Anderson88}.
%11217 at 150~K and 24156 at 225~K.
Details on other methanol isotopologues are given in \citet{Mueller15c}.

Predictions for all singly-deuterated species of ethanol were taken from the 
CDMS (tags 47515 to 47518, all version 1). They are based on 
\citet{D-EtOH_rot_2015} with rest frequencies almost entirely from that study. 
All other ethanol analyses were taken from \citet{Mueller15c}, and further 
details can be found there.
Conformational and vibrational corrections to the partition function 
were taken from the main isotopologue for which only data were available. 
This assumption is reasonable, though errors may not be completely negligible. 
They are, however, difficult to evaluate.

Predictions for both conformers of CH$_2$DOCHO were extracted from Table 7
of \citet{Coudert13} and split into two separate entries. The partition 
function was taken from their Table 6. It is identical for both entries.
Contrary to the CDMS entries for CH$_2$DCH$_2$CN and CH$_2$DCH$_2$OH,
this treatment means that the two entries represent a single species with a 
statistical distribution (the out-of-plane conformer being twice as abundant 
as the in-plane one). This means that the column density derived for each 
conformer represents the total column density of the molecule. However, we 
assumed a statistical distribution (2:1) to compute and report individual 
column densities in Sect.~\ref{s:results}. Predictions for the main 
isotopologue were taken from the JPL catalog. The entry is based on 
\citet{Ilyushin09}. Vibrational corrections to the partition function were 
derived from the $^{13}$C species \citep{Favre14}. These authors provide 
rotational partition function values at different temperatures as well as 
detailed vibrational corrections that are complete up to 300~K. The correction 
factors are 1.59 and 1.23 at 150~K for the deuterated and main isotopic 
species, respectively. These values differ because values for the deuterated 
species refer to the ground state only whereas for the main isotopologue 
contributions of the first excited states were already included.

%__________________________________________________________________

\section{Results}
\label{s:results}

In this section, we report the detection or tentative detection of
deuterated complex organic molecules toward Sgr~B2(N2). Column density upper 
limits are also reported for several non-detections. Each subsection first 
presents the LTE model derived for the main isotopologue and its $^{13}$C 
and/or $^{15}$N isotopologues. This model is then used to obtain constraints
on the column density of the deuterated species. The rotational temperatures 
derived from fits to the population diagrams are reported in 
Table~\ref{t:popfit} and the parameters of the LTE model used to fit the 
spectra are listed in Table~\ref{t:coldens}. The analysis toward Sgr~B2(N2) 
was performed at the offset position 
($\Delta\alpha$, $\Delta\delta$) = ($-0.1\arcsec, 2.6\arcsec$), i.e. 
($\alpha, \delta$)$_{\rm J2000}$=
($17^{\rm h}47^{\rm m}19.86^{\rm s}, -28^\circ22'13.4''$).

\subsection{Deuterated ethyl cyanide CH$_2$DCH$_2$CN and CH$_3$CHDCN}
\label{ss:c2h5cn}

\begin{table}
 {\centering
 \caption{
 Rotational temperatures derived from population diagrams of selected (complex) organic molecules toward Sgr~B2(N2).
}
 \label{t:popfit}
 \vspace*{0.0ex}
 \begin{tabular}{lll}
 \hline\hline
 \multicolumn{1}{c}{Molecule} & \multicolumn{1}{c}{States\tablefootmark{a}} & \multicolumn{1}{c}{$T_{\rm fit}$\tablefootmark{b}} \\ 
  & & \multicolumn{1}{c}{\scriptsize (K)} \\ 
 \hline
CH$_3$CN & $\varv_8=1$, $\varv_8=2$, $\varv_4=1$ &   253 (15) \\ 
$^{13}$CH$_3$CN & $\varv=0$, $\varv_8=1$ &   168 (13) \\ 
CH$_3$$^{13}$CN & $\varv=0$, $\varv_8=1$ & 165.5 (3.3) \\ 
CH$_2$DCN & $\varv=0$ &   136 (14) \\ 
\hline 
C$_2$H$_5$CN & $\varv=0$ & 137.3 (1.6) \\ 
$^{13}$CH$_3$CH$_2$CN & $\varv=0$ & 138.3 (7.5) \\ 
CH$_3$$^{13}$CH$_2$CN & $\varv=0$ &   112 (11) \\ 
CH$_3$CH$_2$$^{13}$CN & $\varv=0$ &   150 (40) \\ 
\hline 
C$_2$H$_3$CN & $\varv=0$, $\varv_{11}=1$, $\varv_{15}=1$, $\varv_{11}=2$ & 199.5 (3.4) \\ 
$^{13}$CH$_2$CHCN & $\varv=0$ &   255 (101) \\ 
CH$_2$$^{13}$CHCN & $\varv=0$ &   140 (31) \\ 
CH$_2$CH$^{13}$CN & $\varv=0$ &   278 (126) \\ 
\hline 
H$^{13}$CCCN & $\varv=0$, $\varv_7=1$ & 171.1 (3.2) \\ 
HC$^{13}$CCN & $\varv=0$, $\varv_7=1$ & 167.7 (5.5) \\ 
HCC$^{13}$CN & $\varv=0$, $\varv_7=1$ &   177 (18) \\ 
\hline 
CH$_3$OCHO & $\varv_{\rm t}$ = 0, $\varv_{\rm t}$ = 1 & 142.4 (4.4) \\ 
\hline 
 \end{tabular}
 }\\[1ex] 
 \tablefoot{
 \tablefoottext{a}{Vibrational states that were taken into account to fit the population diagram.}
 \tablefoottext{b}{The standard deviation of the fit is given in parentheses. As explained in Sect.~\ref{s:ltemodeling}, these uncertainties should be viewed with caution. They may be underestimated.}
 }
 \end{table}

About 154, 54, 38, and 37 lines of ethyl cyanide and its 
singly-substituted $^{13}$C isotopologues, $^{13}$CH$_3$CH$_2$CN, 
CH$_3$$^{13}$CH$_2$CN, and CH$_3$CH$_2$$^{13}$CN, respectively, are detected 
toward Sgr~B2(N2) (Figs.~\ref{f:spec_c2h5cn_ve0}, 
\ref{f:spec_c2h5cn_13c1_ve0}, \ref{f:spec_c2h5cn_13c2_ve0}, and 
\ref{f:spec_c2h5cn_13c3_ve0}). The $^{15}$N isotopologue is also detected 
unambiguously, with 9 detected lines (Fig.~\ref{f:spec_c2h5cn_15n_ve0}). The 
best-fit LTE model fits very well all 
detected transitions, except the very optically thick lines of the main 
isotopologue which it underestimates significantly. We ignored the lines with 
optical depth higher than $\sim 2.5$ to construct the population diagram of 
this species (Fig.~\ref{f:popdiag_c2h5cn}), while all lines of the $^{13}$C 
isotopologues that matched the criteria defined in Sect.~\ref{s:ltemodeling} 
were used for their population diagrams (Figs.~\ref{f:popdiag_c2h5cn_13c1}, 
\ref{f:popdiag_c2h5cn_13c2}, and \ref{f:popdiag_c2h5cn_13c3}). The results of 
the linear fit to the population diagrams of all four isotopologues are given 
in Table~\ref{t:popfit}. The rotational temperature is well
constrained to $\sim 140$~K for both C$_2$H$_5$CN and $^{13}$CH$_3$CH$_2$CN.
This value is consistent with the result of the fit for the two other 
isotopologues within the uncertainties. Note that the temperature derived in
this way depends on the model used to make the opacity correction. With an
earlier best-fit model that assumed a temperature of 170~K (instead of 150~K 
here), the fit to the population diagrams of both C$_2$H$_5$CN and 
$^{13}$CH$_3$CH$_2$CN yielded a temperature of $\sim 150$~K. This is the reason
why we decided to assume a rotational temperature of 150~K for ethyl cyanide and
all its isotopologues.

\begin{table*}[!ht]
 {\centering
 \caption{
 Parameters of our best-fit LTE model (or upper limit) of selected (complex) organic molecules toward Sgr~B2(N2).
}
 \label{t:coldens}
 \vspace*{-1.2ex}
 \begin{tabular}{lcrccccccrc}
 \hline\hline
 \multicolumn{1}{c}{Molecule} & \multicolumn{1}{c}{Status\tablefootmark{a}} & \multicolumn{1}{c}{$N_{\rm det}$\tablefootmark{b}} & \multicolumn{1}{c}{Size\tablefootmark{c}} & \multicolumn{1}{c}{$T_{\mathrm{rot}}$\tablefootmark{d}} & \multicolumn{1}{c}{$N$\tablefootmark{e}} & \multicolumn{1}{c}{$C$\tablefootmark{f}} & \multicolumn{1}{c}{$\Delta V$\tablefootmark{g}} & \multicolumn{1}{c}{$V_{\mathrm{off}}$\tablefootmark{h}} & \multicolumn{1}{c}{$\frac{N_{\rm ref}}{N}$\tablefootmark{i}} & \multicolumn{1}{c}{30\,m\tablefootmark{j}} \\ 
  & & & \multicolumn{1}{c}{\scriptsize ($''$)} & \multicolumn{1}{c}{\scriptsize (K)} & \multicolumn{1}{c}{\scriptsize (cm$^{-2}$)} & & \multicolumn{1}{c}{\scriptsize (km~s$^{-1}$)} & \multicolumn{1}{c}{\scriptsize (km~s$^{-1}$)} & & \\ 
 \hline
 CH$_3$CN, $\varv_8=1$$^\star$ & d & 20 & 1.40 &  170 &  2.2 (18) & 1.00 & 5.4 & -0.5 &       1 & y \\ 
 \hspace*{8ex} $\varv_8=2$ & d & 8 & 1.40 &  170 &  7.5 (18) & 1.00 & 6.5 & -0.5 &    0.29 & y \\ 
 \hspace*{8ex} $\varv_4=1$ & t & 1 & 1.40 &  170 &  2.0 (19) & 1.00 & 6.5 & -0.5 &    0.11 & y \\ 
 $^{13}$CH$_3$CN, $\varv=0$ & d & 8 & 1.40 &  170 &  1.1 (17) & 1.18 & 5.4 & -0.5 &      21 & y \\ 
 \hspace*{9ex} $\varv_8=1$ & d & 3 & 1.40 &  170 &  1.1 (17) & 1.18 & 5.4 & -0.5 &      21 & y \\ 
 CH$_3$$^{13}$CN, $\varv=0$ & d & 7 & 1.40 &  170 &  1.1 (17) & 1.18 & 5.4 & -0.5 &      21 & y \\ 
 \hspace*{9ex} $\varv_8=1$ & d & 9 & 1.40 &  170 &  1.1 (17) & 1.18 & 5.4 & -0.5 &      21 & y \\ 
 $^{13}$CH$_3$$^{13}$CN & d & 1 & 1.40 &  170 &  4.7 (15) & 1.18 & 5.4 & -0.5 &     466 & n \\ 
 CH$_3$C$^{15}$N & t & 0 & 1.40 &  170 &  8.3 (15) & 1.18 & 5.4 & -0.5 &     266 & n \\ 
 CH$_2$DCN & d & 6 & 1.40 &  170 &  8.3 (15) & 1.18 & 5.4 & -0.6 &     266 & n \\ 
\hline 
 C$_2$H$_5$CN$^\star$ & d & 154 & 1.20 &  150 &  6.9 (18) & 1.53 & 5.0 & -0.8 &       1 & y \\ 
 $^{13}$CH$_3$CH$_2$CN & d & 54 & 1.20 &  150 &  2.1 (17) & 1.53 & 5.0 & -0.8 &      32 & y \\ 
 CH$_3$$^{13}$CH$_2$CN & d & 38 & 1.20 &  150 &  2.1 (17) & 1.53 & 5.0 & -0.8 &      32 & y \\ 
 CH$_3$CH$_2$$^{13}$CN & d & 37 & 1.20 &  150 &  2.1 (17) & 1.53 & 5.0 & -0.8 &      32 & y \\ 
 C$_2$H$_5$C$^{15}$N & d & 9 & 1.20 &  150 &  1.4 (16) & 1.53 & 5.0 & -0.8 &     500 & n \\ 
 CH$_2$DCH$_2$CN (out of plane) & t & 2 & 1.20 &  150 &  3.4 (15) & 1.53 & 5.0 & -0.8 &    2045 & n \\ 
 CH$_2$DCH$_2$CN (in plane) & n & 0 & 1.20 &  150 & $<$  1.7 (15) & 1.53 & 5.0 & -0.8 & $>$    4091 & n \\ 
 CH$_3$CHDCN & t & 1 & 1.20 &  150 &  3.4 (15) & 1.53 & 5.0 & -0.8 &    2045 & n \\ 
\hline 
 C$_2$H$_3$CN, $\varv=0$$^\star$ & d & 44 & 1.10 &  200 &  4.2 (17) & 1.00 & 6.0 & -0.6 &       1 & y \\ 
 \hspace*{8.5ex} $\varv_{11}=1$ & d & 30 & 1.10 &  200 &  4.2 (17) & 1.00 & 6.0 & -0.5 &       1 & y \\ 
 \hspace*{8.5ex} $\varv_{15}=1$ & d & 20 & 1.10 &  200 &  4.2 (17) & 1.00 & 6.0 & -0.5 &       1 & y \\ 
 \hspace*{8.5ex} $\varv_{11}=2$ & d & 6 & 1.10 &  200 &  4.2 (17) & 1.00 & 6.0 & -0.5 &       1 & y \\ 
 $^{13}$CH$_2$CHCN & d & 10 & 1.10 &  200 &  2.1 (16) & 1.38 & 6.0 & -0.6 &      20 & y \\ 
 CH$_2$$^{13}$CHCN & d & 9 & 1.10 &  200 &  2.1 (16) & 1.38 & 6.0 & -0.6 &      20 & y \\ 
 CH$_2$CH$^{13}$CN & d & 8 & 1.10 &  200 &  2.1 (16) & 1.38 & 6.0 & -0.6 &      20 & y \\ 
 C$_2$H$_3$C$^{15}$N & n & 0 & 1.10 &  200 & $<$  3.4 (15) & 1.38 & 6.0 & -0.6 & $>$     122 & n \\ 
 \textit{cis-}CHDCHCN & n & 0 & 1.10 &  200 & $<$  3.4 (15) & 1.38 & 6.0 & -0.6 & $>$     122 & n \\ 
 \textit{trans-}CHDCHCN & n & 0 & 1.10 &  200 & $<$  3.4 (15) & 1.38 & 6.0 & -0.6 & $>$     122 & n \\ 
 CH$_2$CDCN & n & 0 & 1.10 &  200 & $<$  2.1 (15) & 1.38 & 6.0 & -0.6 & $>$     203 & n \\ 
\hline 
 HC$_3$N, $\varv_7=1$$^\star$ & d & 6 & 1.30 &  170 &  3.5 (17) & 1.44 & 5.0 & -0.7 &       1 & y \\ 
 H$^{13}$CCCN, $\varv=0$ & d & 2 & 1.30 &  170 &  1.7 (16) & 1.44 & 5.0 & -0.7 &      20 & y \\ 
 \hspace*{10ex} $\varv_7=1$ & d & 4 & 1.30 &  170 &  1.7 (16) & 1.44 & 5.0 & -1.0 &      20 & y \\ 
 HC$^{13}$CCN, $\varv=0$ & d & 3 & 1.30 &  170 &  1.7 (16) & 1.44 & 5.0 & -0.7 &      20 & y \\ 
 \hspace*{10ex} $\varv_7=1$ & d & 3 & 1.30 &  170 &  1.7 (16) & 1.44 & 5.0 & -1.0 &      20 & y \\ 
 HCC$^{13}$CN, $\varv=0$ & d & 3 & 1.30 &  170 &  1.7 (16) & 1.44 & 5.0 & -0.7 &      20 & y \\ 
 \hspace*{10ex} $\varv_7=1$ & d & 3 & 1.30 &  170 &  1.7 (16) & 1.44 & 5.0 & -1.0 &      20 & y \\ 
 H$^{13}$C$^{13}$CCN & t & 1 & 1.30 &  170 &  7.2 (14) & 1.44 & 5.0 & -0.7 &     480 & n \\ 
 H$^{13}$CC$^{13}$CN & t & 0 & 1.30 &  170 &  7.2 (14) & 1.44 & 5.0 & -0.7 &     480 & y \\ 
 HC$^{13}$C$^{13}$CN & t & 1 & 1.30 &  170 &  7.2 (14) & 1.44 & 5.0 & -0.7 &     480 & n \\ 
 HC$_3$$^{15}$N & t & 0 & 1.30 &  170 &  1.2 (15) & 1.44 & 5.0 & -0.7 &     300 & y \\ 
 DC$_3$N & t & 0 & 1.30 &  170 &  3.0 (14) & 1.51 & 5.0 & -0.5 &    1144 & n \\ 
\hline 
 CH$_3$OH, $\varv_{\rm t}=1$$^\star$ & d & 16 & 1.40 &  160 &  4.0 (19) & 1.00 & 5.4 & -0.2 &       1 & y \\ 
 CH$_2$DOH & t & 2 & 1.40 &  160 &  4.8 (16) & 1.15 & 5.4 & -0.5 &     828 & n \\ 
 CH$_3$OD & n & 0 & 1.40 &  160 & $<$  2.6 (16) & 1.05 & 5.4 & -0.5 & $>$    1524 & n \\ 
\hline 
 C$_2$H$_5$OH$^\star$ & d & 168 & 1.25 &  150 &  2.0 (18) & 1.24 & 5.4 & 0.0 &       1 & y \\ 
 CH$_3$CH$_2$OD & n & 0 & 1.25 &  150 & $<$  3.0 (16) & 2.96 & 5.4 & 0.0 & $>$      67 & n \\ 
 CH$_3$CHDOH & n & 0 & 1.25 &  150 & $<$  3.0 (16) & 2.96 & 5.4 & 0.0 & $>$      67 & n \\ 
 CH$_2$DCH$_2$OH (out of plane) & n & 0 & 1.25 &  150 & $<$  3.0 (16) & 2.96 & 5.4 & 0.0 & $>$      67 & n \\ 
 CH$_2$DCH$_2$OH (in plane) & n & 0 & 1.25 &  150 & $<$  2.1 (16) & 2.96 & 5.4 & 0.0 & $>$      96 & n \\ 
\hline 
 CH$_3$OCHO, $\varv_{\rm t}=0$$^\star$ & d & 90 & 1.50 &  150 &  1.2 (18) & 1.23 & 4.7 & -0.4 &       1 & y \\ 
 \hspace*{11ex} $\varv_{\rm t}=1$ & d & 35 & 1.50 &  150 &  1.2 (18) & 1.23 & 4.7 & -0.4 &       1 & y \\ 
 CH$_2$DOCHO (out of plane) & n & 0 & 1.50 &  150 & $<$  2.5 (16) & 1.07 & 4.7 & -0.4 & $>$      50 & n \\ 
 CH$_2$DOCHO (in plane) & n & 0 & 1.50 &  150 & $<$  7.3 (15) & 0.52 & 4.7 & -0.4 & $>$     167 & n \\ 
\hline 
 \end{tabular}
 }\\[-0.5ex] 
 \tablefoot{
 \tablefoottext{a}{d: detection, t: tentative detection, n: non-detection.}
 \tablefoottext{b}{Number of detected lines (conservative estimate, see Sect.~\ref{s:ltemodeling}). One line of a given species may mean a group of transitions of that species that are blended together.}
 \tablefoottext{c}{Source diameter (\textit{FWHM}).}
 \tablefoottext{d}{Rotational temperature.}
 \tablefoottext{e}{Total column density of the molecule. $X$ ($Y$) means $X \times 10^Y$.}
 \tablefoottext{f}{Correction factor that was applied to the column density to account for the contribution of vibrationally or torsionally excited states or other conformers (e.g., \textit{gauche} for ethanol), in the cases where this contribution was not included in the partition function of the spectroscopic predictions. For deuterated methyl formate, it is the scaling factor used to compute the column density of each conformer as if it were an independent species.}
 \tablefoottext{g}{Linewidth (\textit{FWHM}).}
 \tablefoottext{h}{Velocity offset with respect to the assumed systemic velocity of Sgr~B2(N2), V$_{\mathrm{lsr}} = 74$ km~s$^{-1}$.}
 \tablefoottext{i}{Column density ratio, with $N_{\rm ref}$ the column density of the previous reference species marked with a $\star$.}
 \tablefoottext{j}{Detected (y) or not detected (n) toward Sgr~B2(N) (N1 and/or N2) with the IRAM 30\,m telescope \citep[][]{Belloche13}.}
 }
 \end{table*}

\onlfig{
\begin{figure*}
%\centerline{\resizebox{0.9\hsize}{!}{\includegraphics[angle=0]{/homes/belloche/SgrB2/ALMA/Analysis/Weeds/R-NC/Figs/c2h5cn_ve0_p2_1.eps}}}
\centerline{\resizebox{0.9\hsize}{!}{\includegraphics[angle=0]{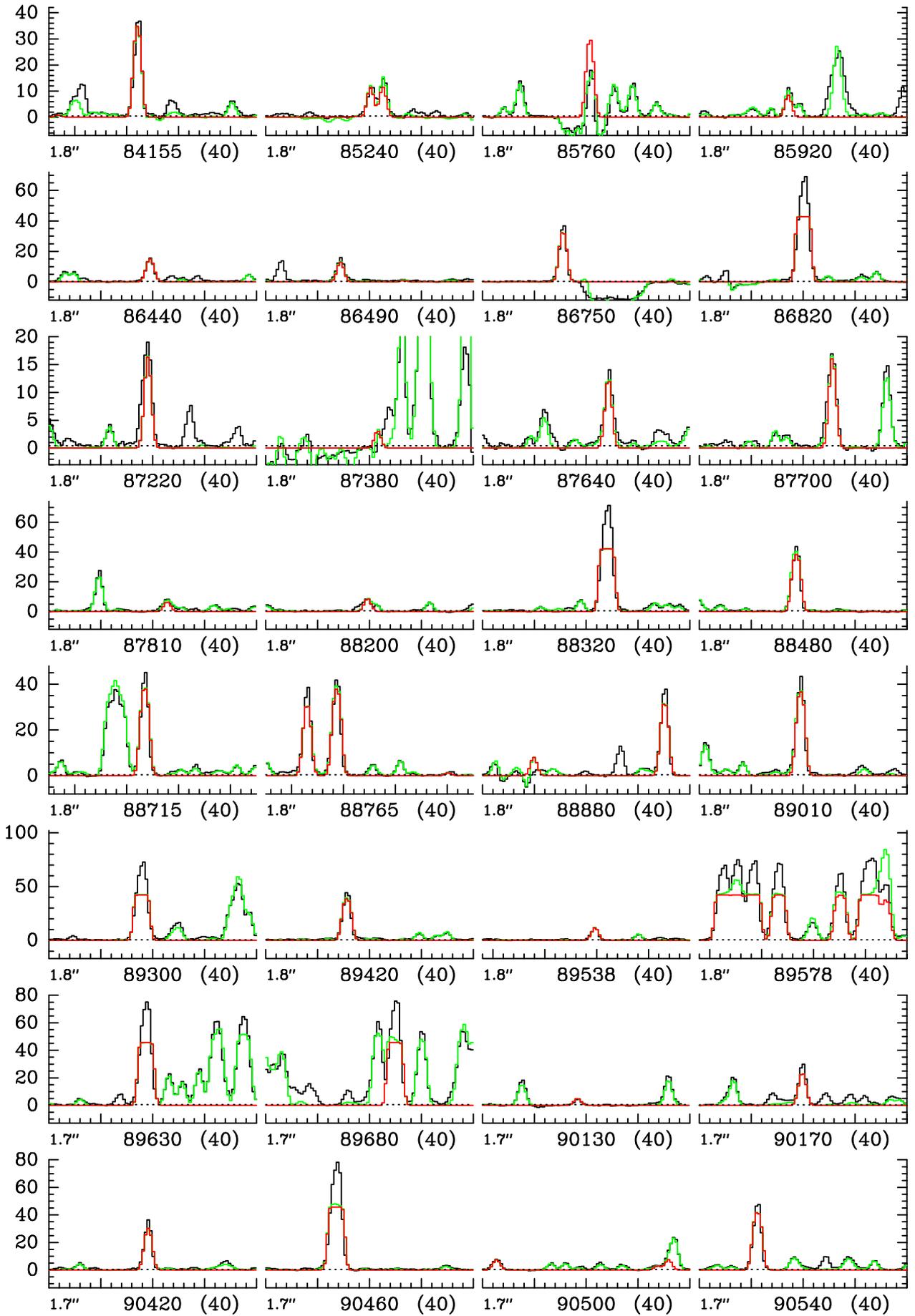}}}
\caption{Transitions of C$_2$H$_5$CN, $\varv = 0$ covered by our ALMA survey. 
The best-fit LTE synthetic spectrum of C$_2$H$_5$CN is displayed in red and 
overlaid on the observed spectrum of Sgr~B2(N2) shown in black. The green 
synthetic spectrum contains the contributions of all molecules identified in 
our survey so far, including the one shown in red. The central 
frequency and width are indicated in MHz below each panel. The angular 
resolution (HPBW) is also indicated. The y-axis is labeled in brightness 
temperature units (K). The dotted line indicates the $3\sigma$ noise level.
}
\label{f:spec_c2h5cn_ve0}
\end{figure*}
}

\onlfig{
\clearpage
\begin{figure*}
\addtocounter{figure}{-1}
%\centerline{\resizebox{0.9\hsize}{!}{\includegraphics[angle=0]{/homes/belloche/SgrB2/ALMA/Analysis/Weeds/R-NC/Figs/c2h5cn_ve0_p2_2.eps}}}
\centerline{\resizebox{0.9\hsize}{!}{\includegraphics[angle=0]{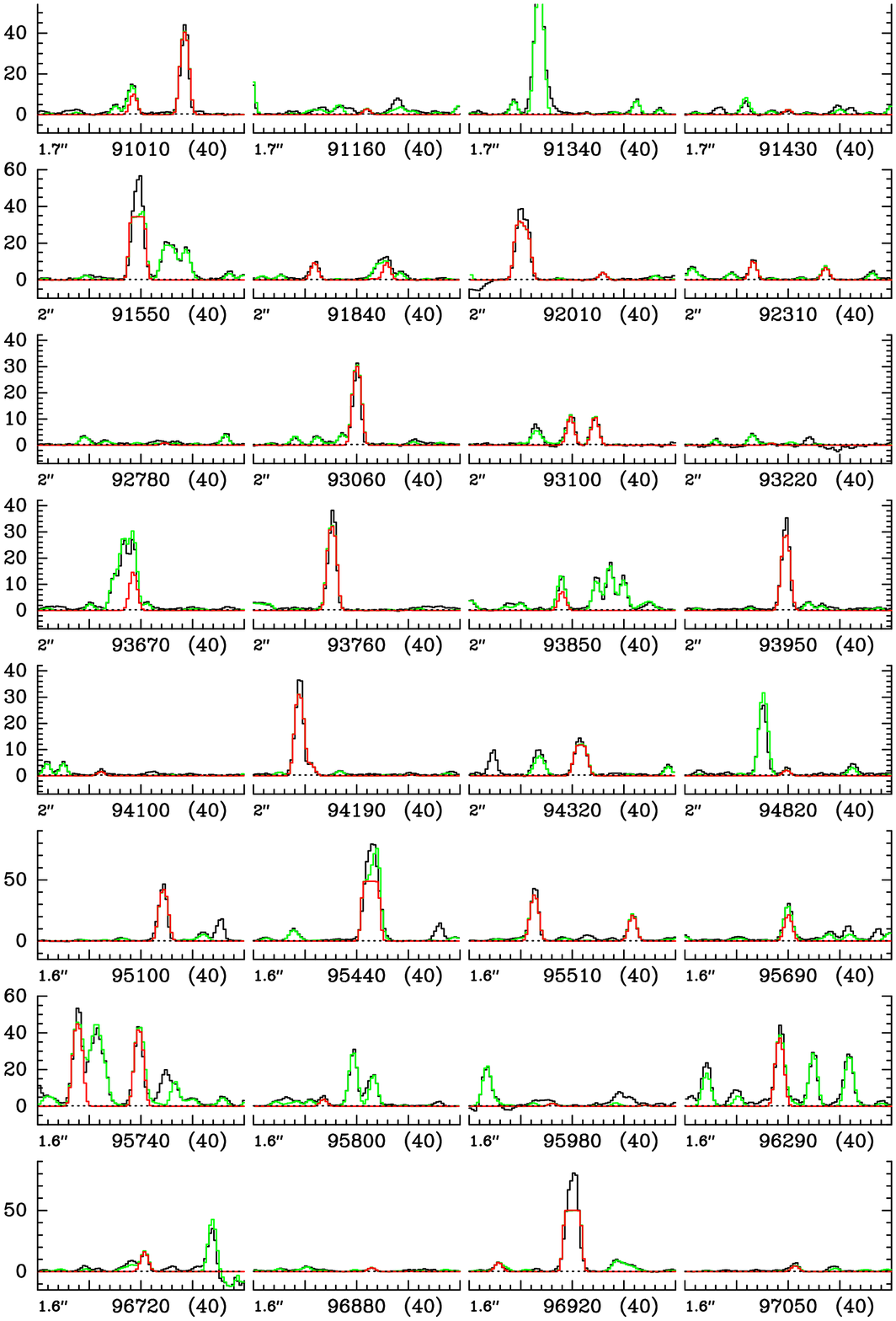}}}
\caption{continued.}
\end{figure*}
}
\addtocounter{figure}{-1}

\onlfig{
\clearpage
\begin{figure*}
\addtocounter{figure}{-1}
%\centerline{\resizebox{0.9\hsize}{!}{\includegraphics[angle=0]{/homes/belloche/SgrB2/ALMA/Analysis/Weeds/R-NC/Figs/c2h5cn_ve0_p2_3.eps}}}
\centerline{\resizebox{0.9\hsize}{!}{\includegraphics[angle=0]{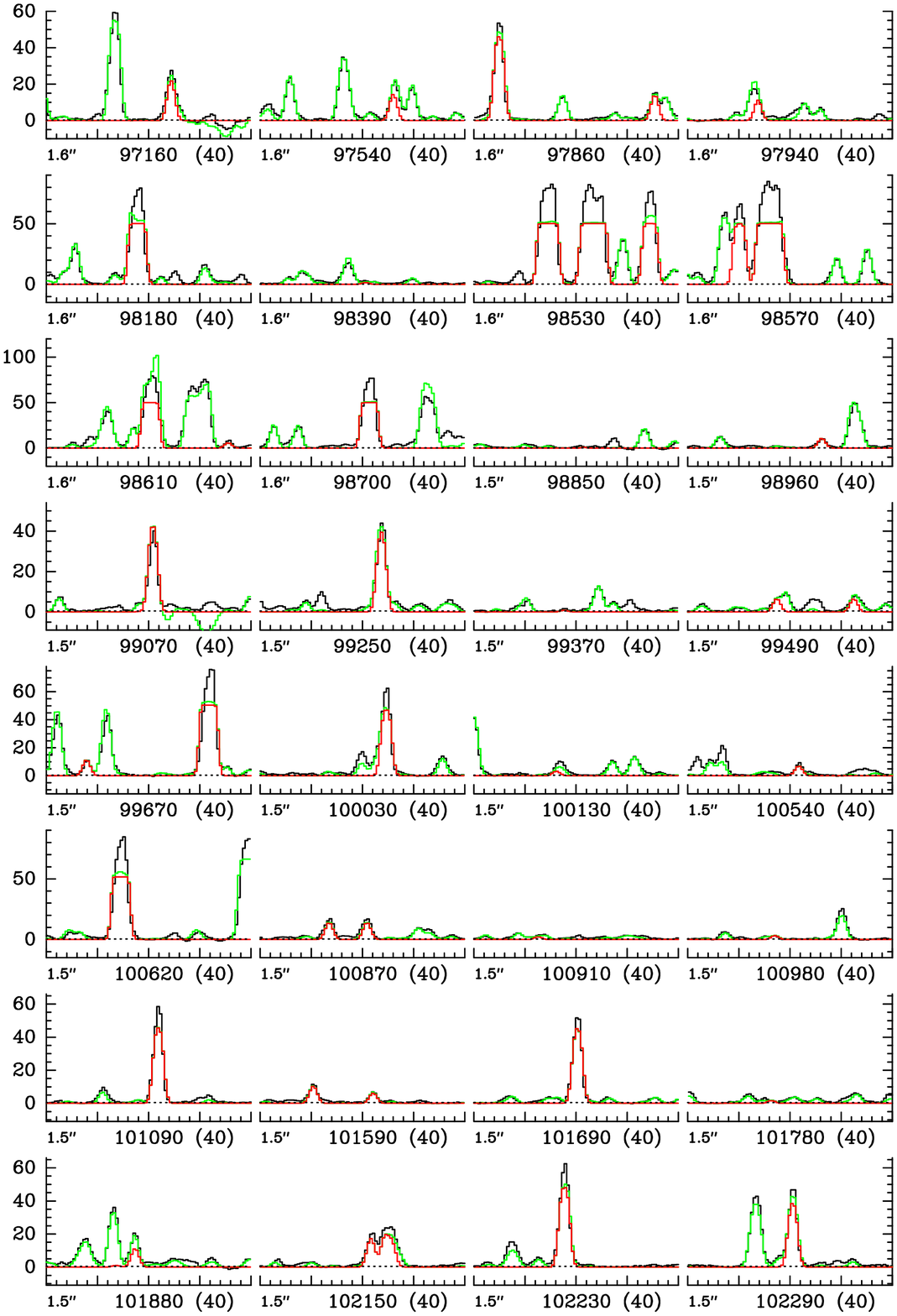}}}
\caption{continued.}
\end{figure*}
}
\addtocounter{figure}{-1}

\onlfig{
\clearpage
\begin{figure*}
\addtocounter{figure}{-1}
%\centerline{\resizebox{0.9\hsize}{!}{\includegraphics[angle=0]{/homes/belloche/SgrB2/ALMA/Analysis/Weeds/R-NC/Figs/c2h5cn_ve0_p2_4.eps}}}
\centerline{\resizebox{0.9\hsize}{!}{\includegraphics[angle=0]{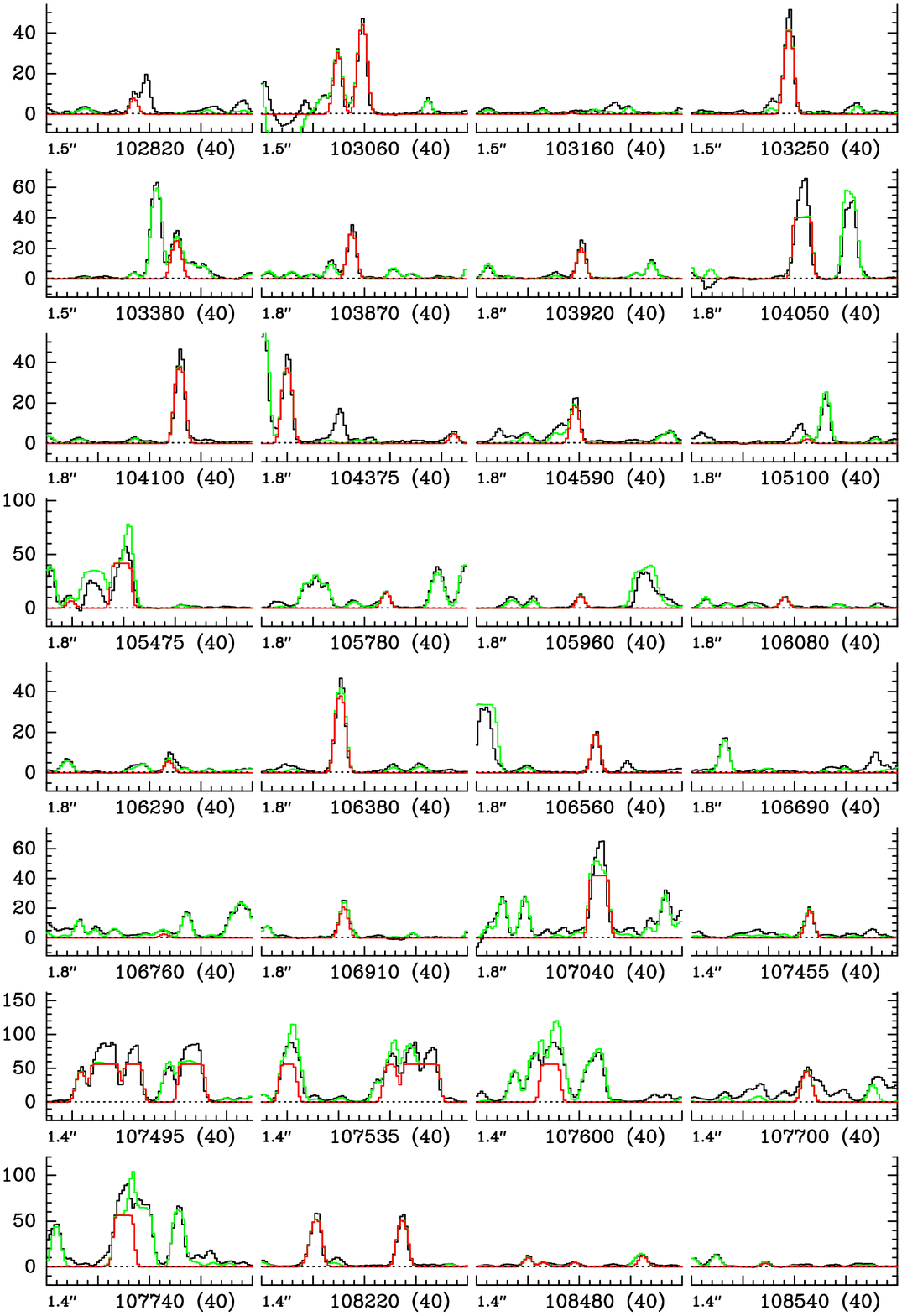}}}
\caption{continued.}
\end{figure*}
}
\addtocounter{figure}{-1}

\onlfig{
\clearpage
\begin{figure*}
\addtocounter{figure}{-1}
%\centerline{\resizebox{0.9\hsize}{!}{\includegraphics[angle=0]{/homes/belloche/SgrB2/ALMA/Analysis/Weeds/R-NC/Figs/c2h5cn_ve0_p2_5.eps}}}
\centerline{\resizebox{0.9\hsize}{!}{\includegraphics[angle=0]{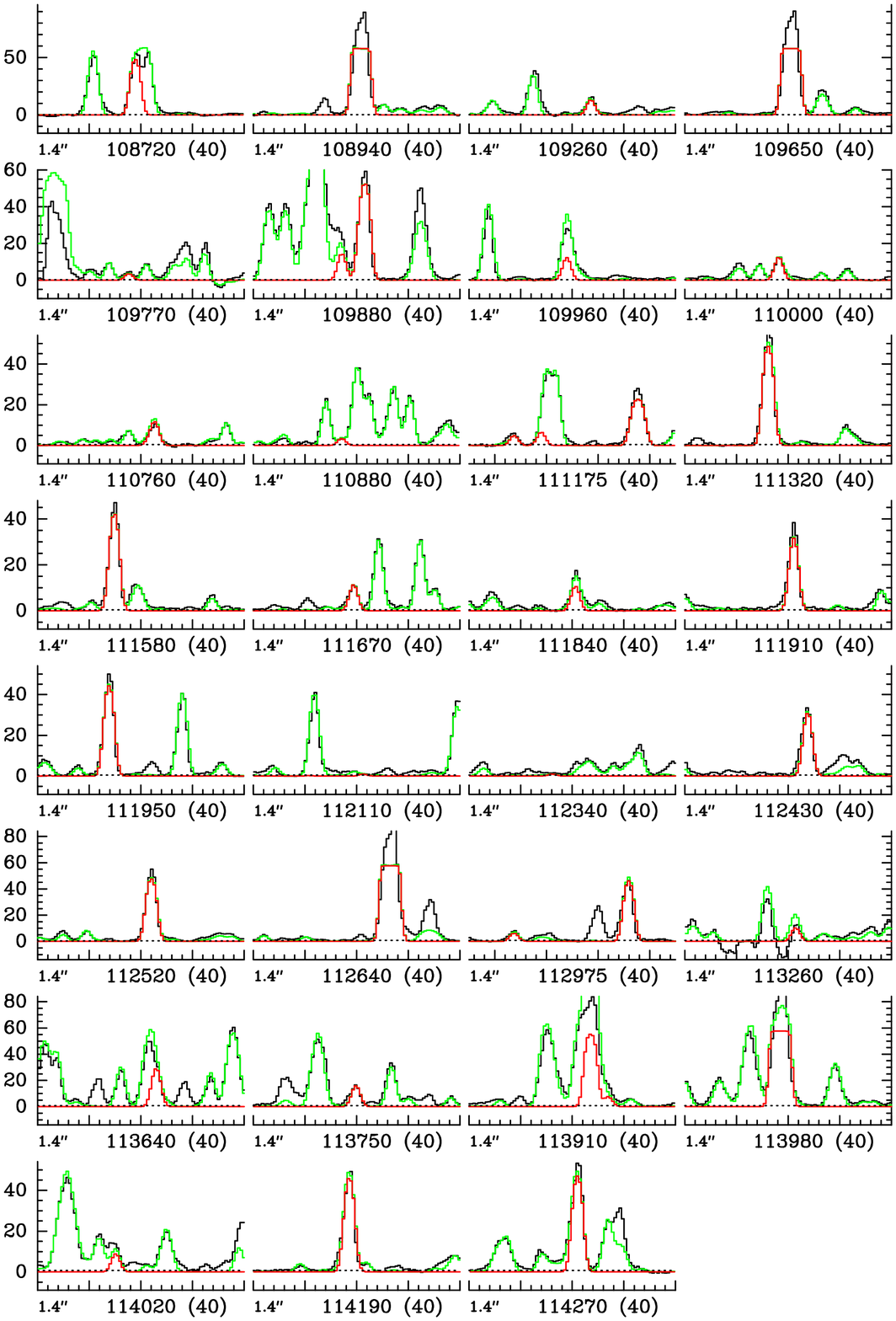}}}
\caption{continued.}
\end{figure*}
}
\addtocounter{figure}{-1}

\onlfig{
\clearpage
\begin{figure*}
%\centerline{\resizebox{0.9\hsize}{!}{\includegraphics[angle=0]{/homes/belloche/SgrB2/ALMA/Analysis/Weeds/R-NC/Figs/c2h5cn_13c1_ve0_p2_1.eps}}}
\centerline{\resizebox{0.9\hsize}{!}{\includegraphics[angle=0]{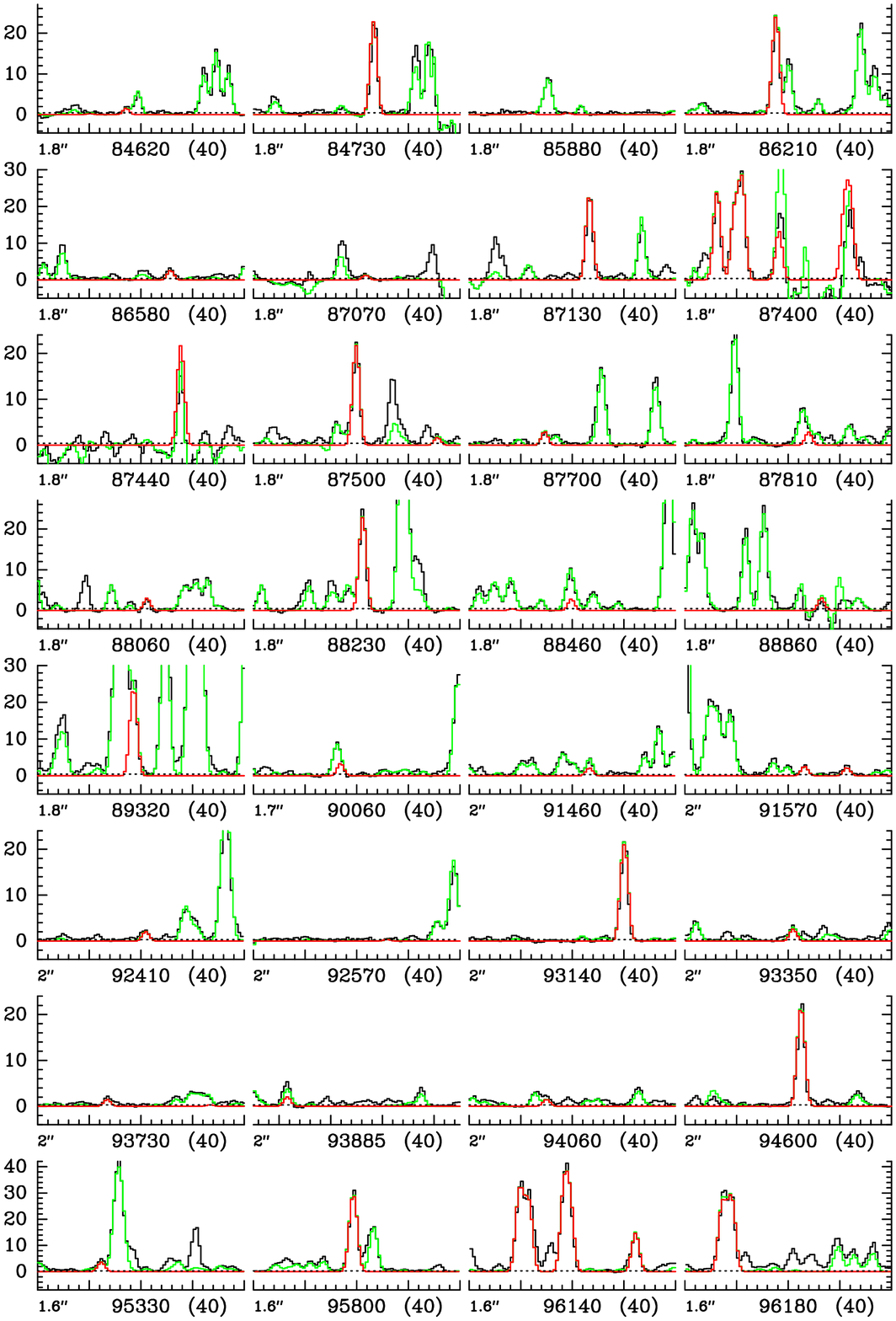}}}
\caption{Same as Fig.~\ref{f:spec_c2h5cn_ve0} for $^{13}$CH$_3$CH$_2$CN, 
$\varv=0$.
}
\label{f:spec_c2h5cn_13c1_ve0}
\end{figure*}
}

\onlfig{
\clearpage
\begin{figure*}
\addtocounter{figure}{-1}
%\centerline{\resizebox{0.9\hsize}{!}{\includegraphics[angle=0]{/homes/belloche/SgrB2/ALMA/Analysis/Weeds/R-NC/Figs/c2h5cn_13c1_ve0_p2_2.eps}}}
\centerline{\resizebox{0.9\hsize}{!}{\includegraphics[angle=0]{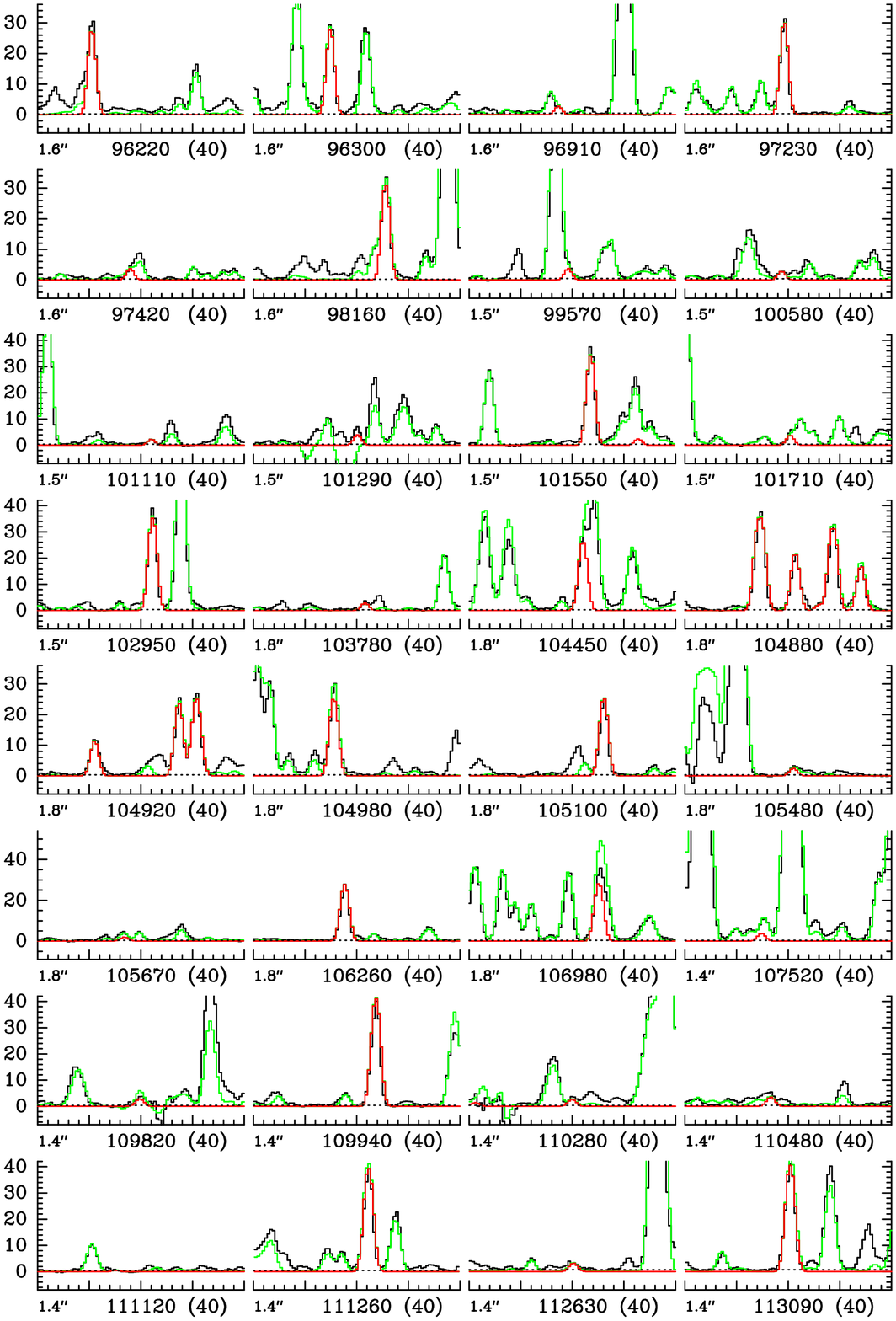}}}
\caption{continued.}
\end{figure*}
}
\addtocounter{figure}{-1}

\onlfig{
\clearpage
\begin{figure*}
\addtocounter{figure}{-1}
%\centerline{\resizebox{0.9\hsize}{!}{\includegraphics[angle=0]{/homes/belloche/SgrB2/ALMA/Analysis/Weeds/R-NC/Figs/c2h5cn_13c1_ve0_p2_3.eps}}}
\centerline{\resizebox{0.9\hsize}{!}{\includegraphics[angle=0]{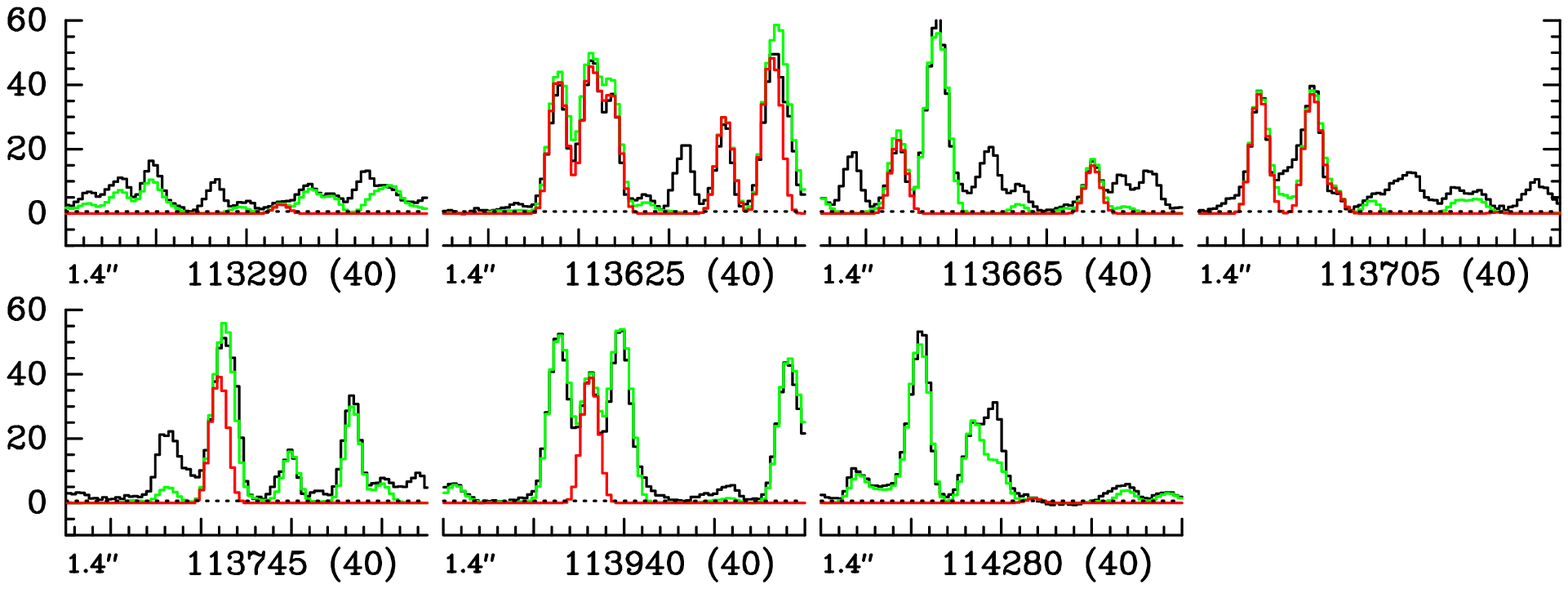}}}
\caption{continued.}
\end{figure*}
}
\addtocounter{figure}{-1}

\onlfig{
\clearpage
\begin{figure*}
%\centerline{\resizebox{0.9\hsize}{!}{\includegraphics[angle=0]{/homes/belloche/SgrB2/ALMA/Analysis/Weeds/R-NC/Figs/c2h5cn_13c2_ve0_p2_1.eps}}}
\centerline{\resizebox{0.9\hsize}{!}{\includegraphics[angle=0]{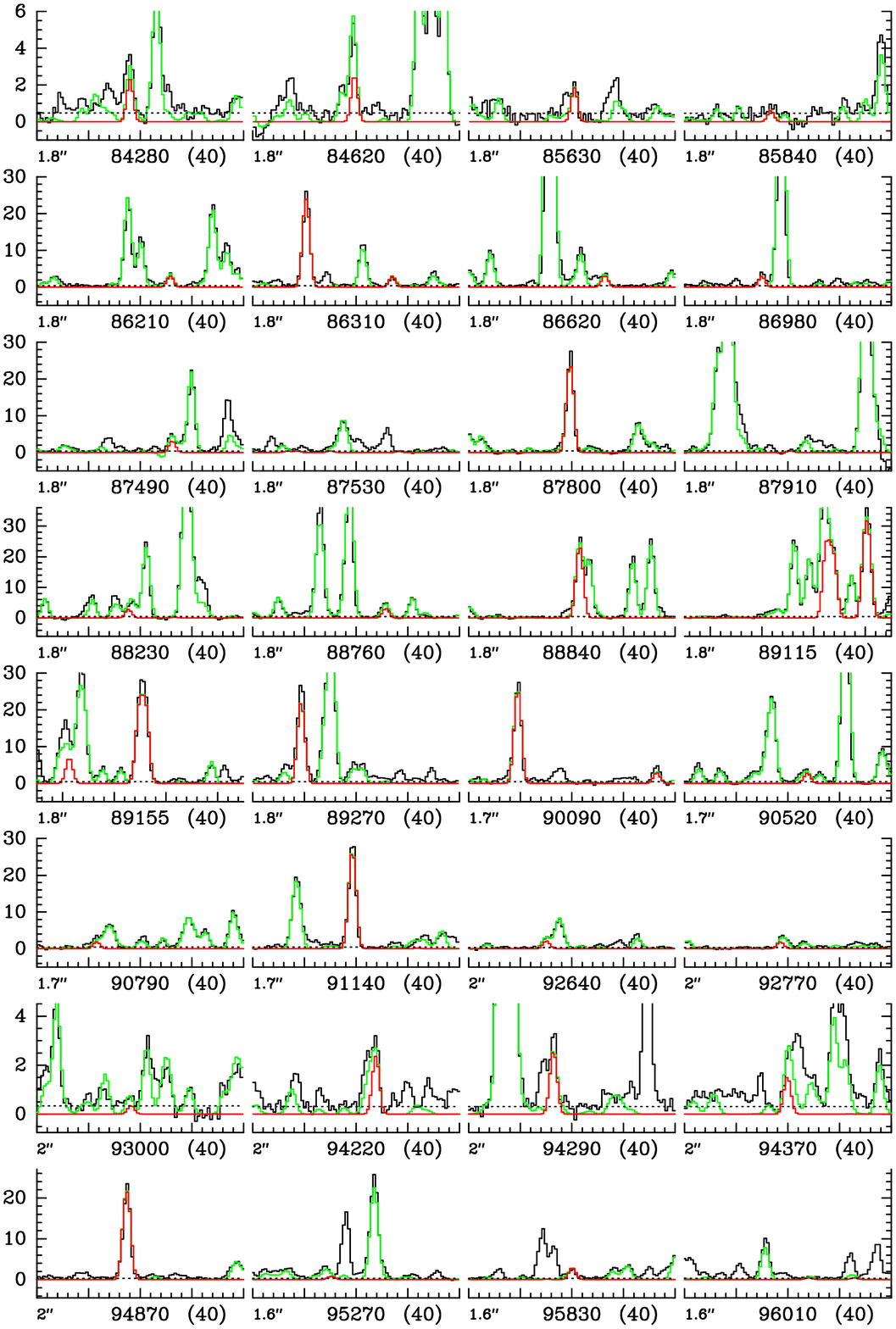}}}
\caption{Same as Fig.~\ref{f:spec_c2h5cn_ve0} for CH$_3$$^{13}$CH$_2$CN, 
$\varv=0$.
}
\label{f:spec_c2h5cn_13c2_ve0}
\end{figure*}
}

\onlfig{
\clearpage
\begin{figure*}
\addtocounter{figure}{-1}
%\centerline{\resizebox{0.9\hsize}{!}{\includegraphics[angle=0]{/homes/belloche/SgrB2/ALMA/Analysis/Weeds/R-NC/Figs/c2h5cn_13c2_ve0_p2_2.eps}}}
\centerline{\resizebox{0.9\hsize}{!}{\includegraphics[angle=0]{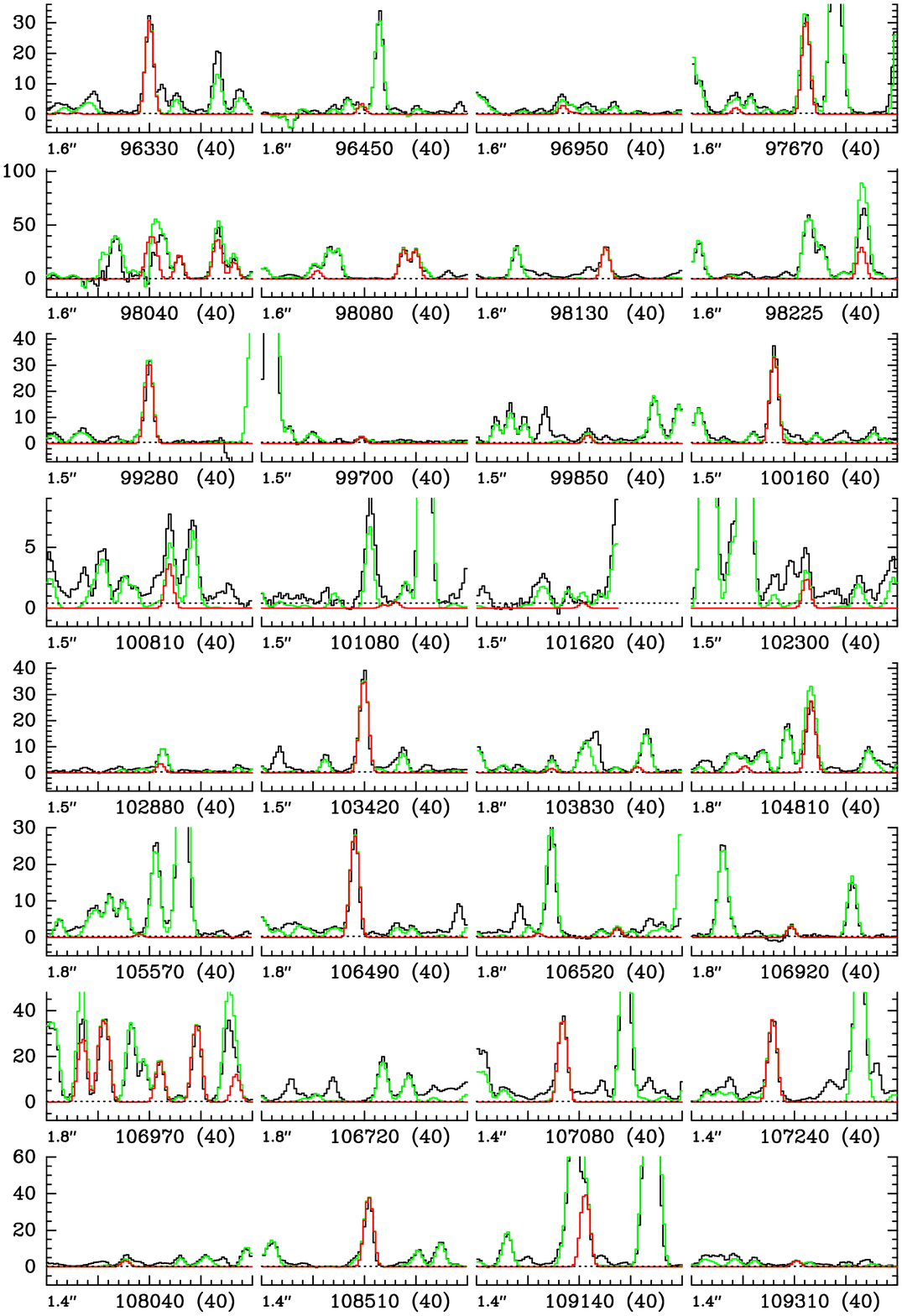}}}
\caption{continued.}
\end{figure*}
}
\addtocounter{figure}{-1}

\onlfig{
\clearpage
\begin{figure*}
\addtocounter{figure}{-1}
%\centerline{\resizebox{0.9\hsize}{!}{\includegraphics[angle=0]{/homes/belloche/SgrB2/ALMA/Analysis/Weeds/R-NC/Figs/c2h5cn_13c2_ve0_p2_3.eps}}}
\centerline{\resizebox{0.9\hsize}{!}{\includegraphics[angle=0]{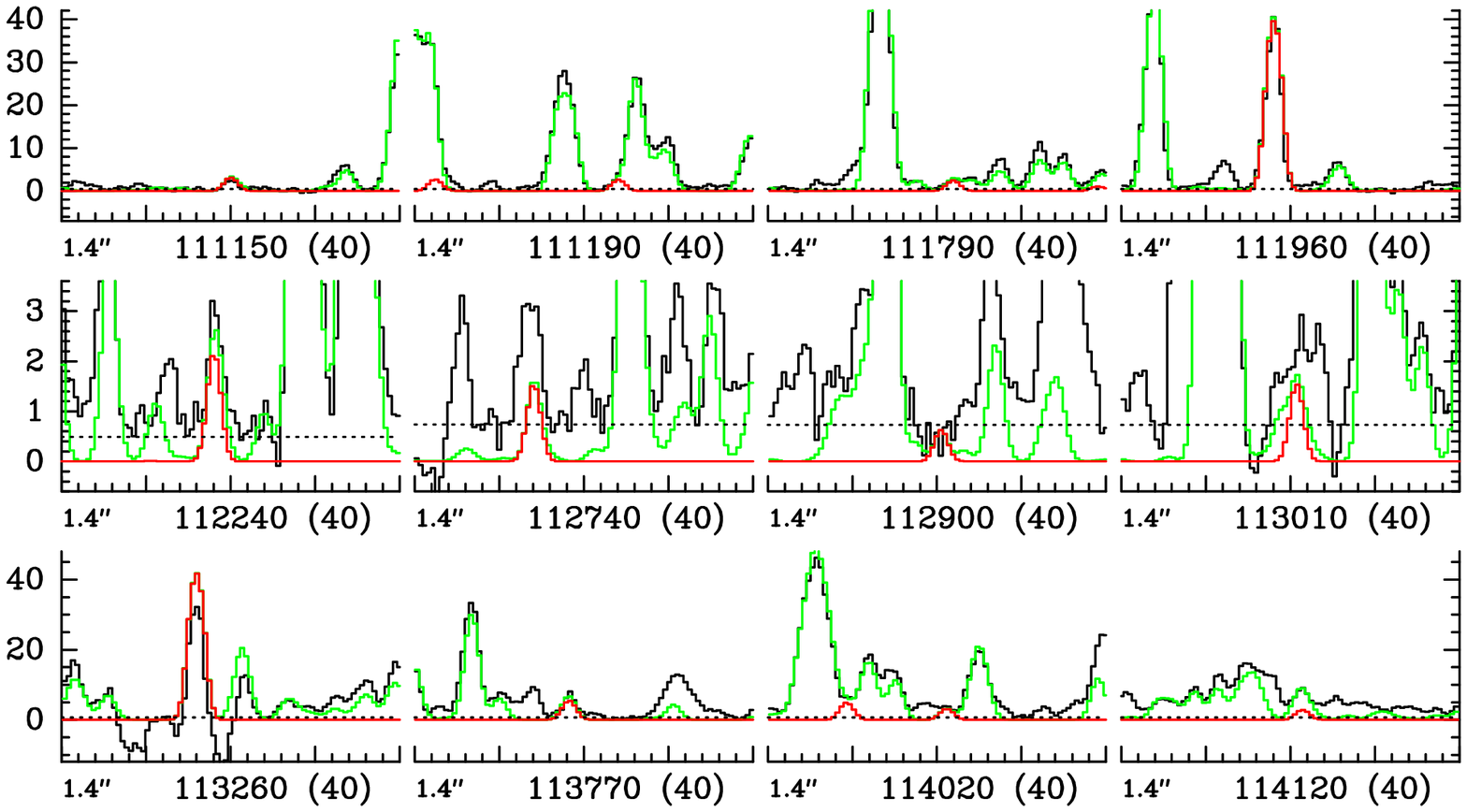}}}
\caption{continued.}
\end{figure*}
}
\addtocounter{figure}{-1}

\onlfig{
\clearpage
\begin{figure*}
%\centerline{\resizebox{0.9\hsize}{!}{\includegraphics[angle=0]{/homes/belloche/SgrB2/ALMA/Analysis/Weeds/R-NC/Figs/c2h5cn_13c3_ve0_p2_1.eps}}}
\centerline{\resizebox{0.9\hsize}{!}{\includegraphics[angle=0]{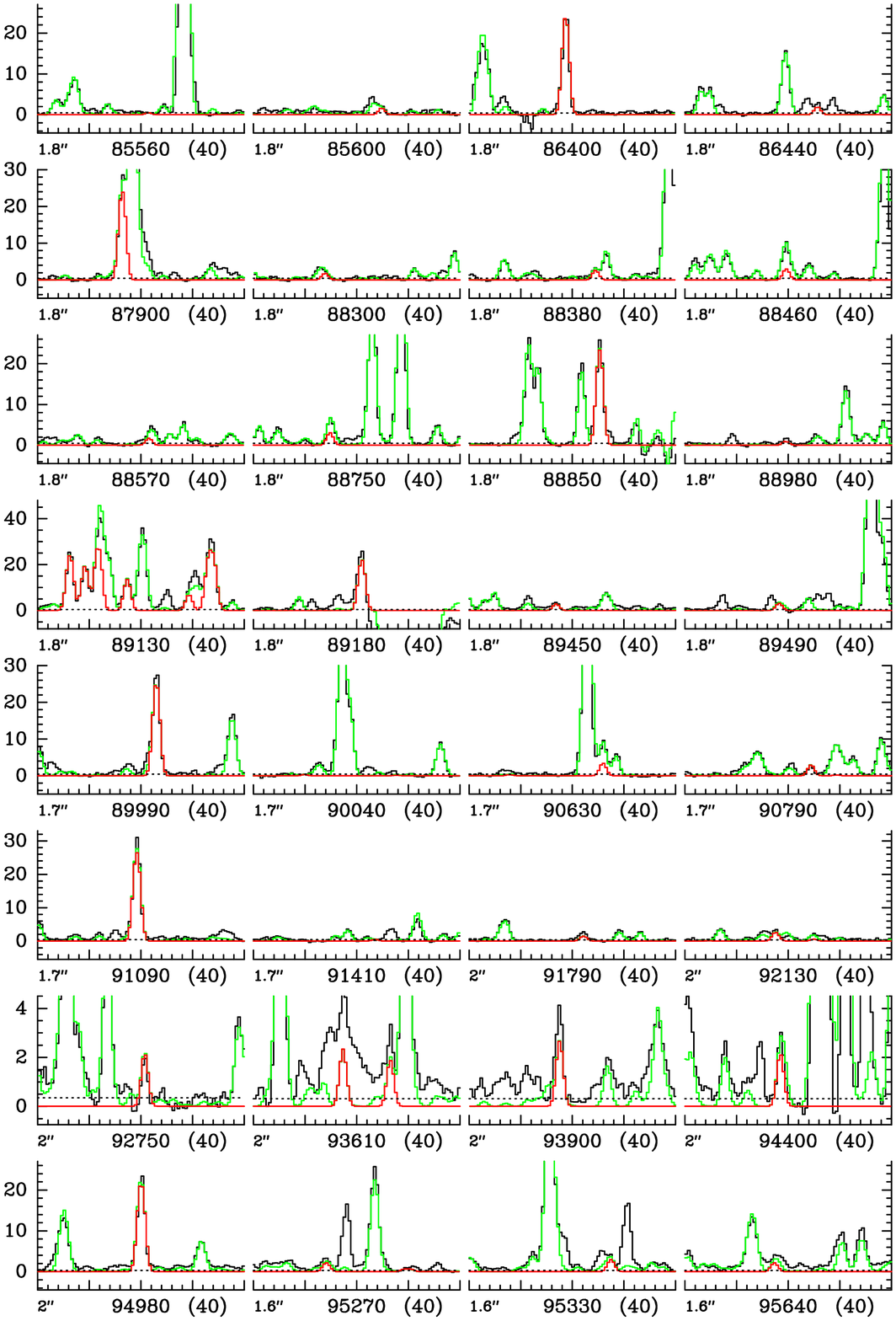}}}
\caption{Same as Fig.~\ref{f:spec_c2h5cn_ve0} for CH$_3$CH$_2$$^{13}$CN, 
$\varv=0$.
}
\label{f:spec_c2h5cn_13c3_ve0}
\end{figure*}
}

\onlfig{
\clearpage
\begin{figure*}
\addtocounter{figure}{-1}
%\centerline{\resizebox{0.9\hsize}{!}{\includegraphics[angle=0]{/homes/belloche/SgrB2/ALMA/Analysis/Weeds/R-NC/Figs/c2h5cn_13c3_ve0_p2_2.eps}}}
\centerline{\resizebox{0.9\hsize}{!}{\includegraphics[angle=0]{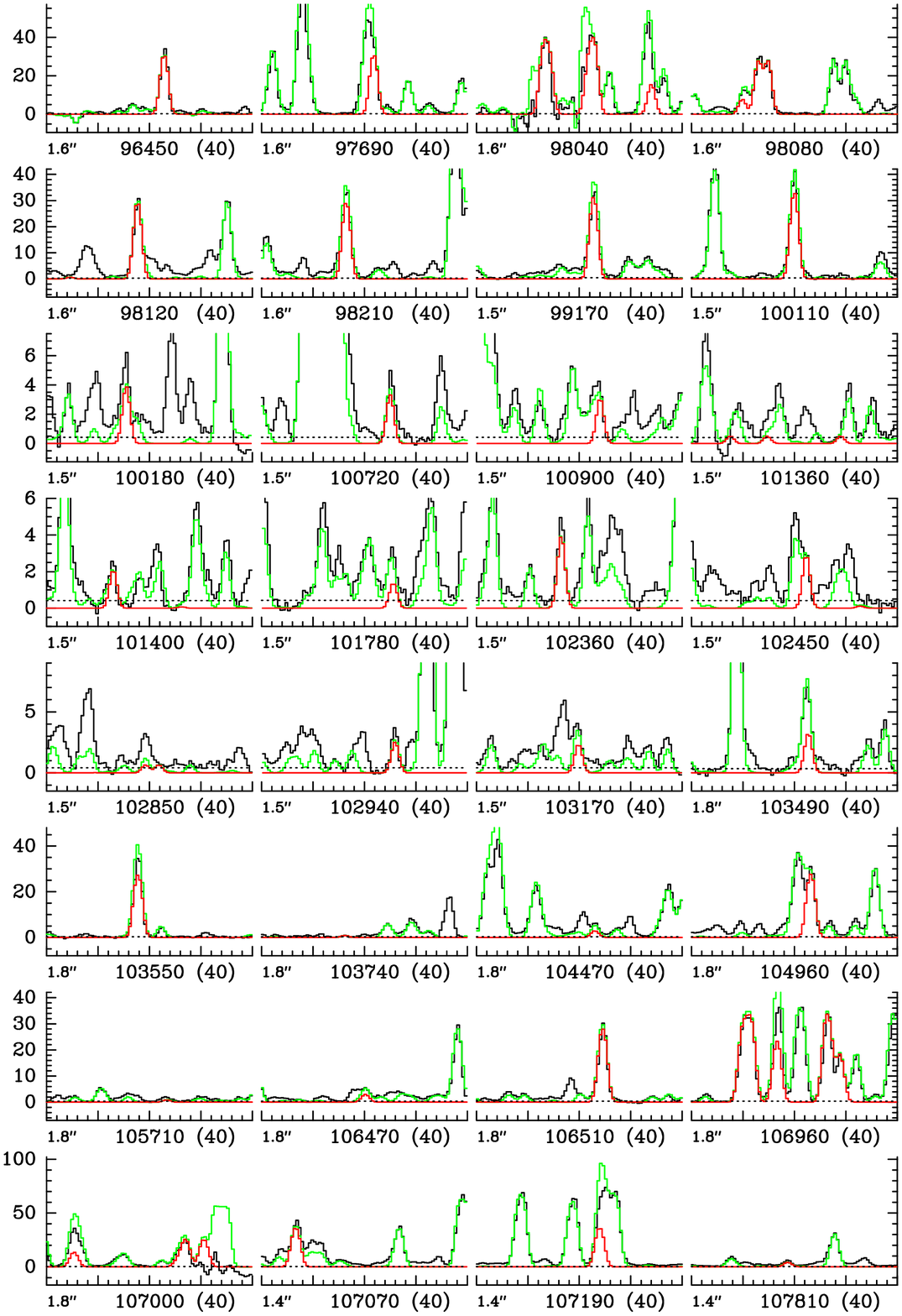}}}
\caption{continued.}
\end{figure*}
}
\addtocounter{figure}{-1}

\onlfig{
\clearpage
\begin{figure*}
\addtocounter{figure}{-1}
%\centerline{\resizebox{0.9\hsize}{!}{\includegraphics[angle=0]{/homes/belloche/SgrB2/ALMA/Analysis/Weeds/R-NC/Figs/c2h5cn_13c3_ve0_p2_3.eps}}}
\centerline{\resizebox{0.9\hsize}{!}{\includegraphics[angle=0]{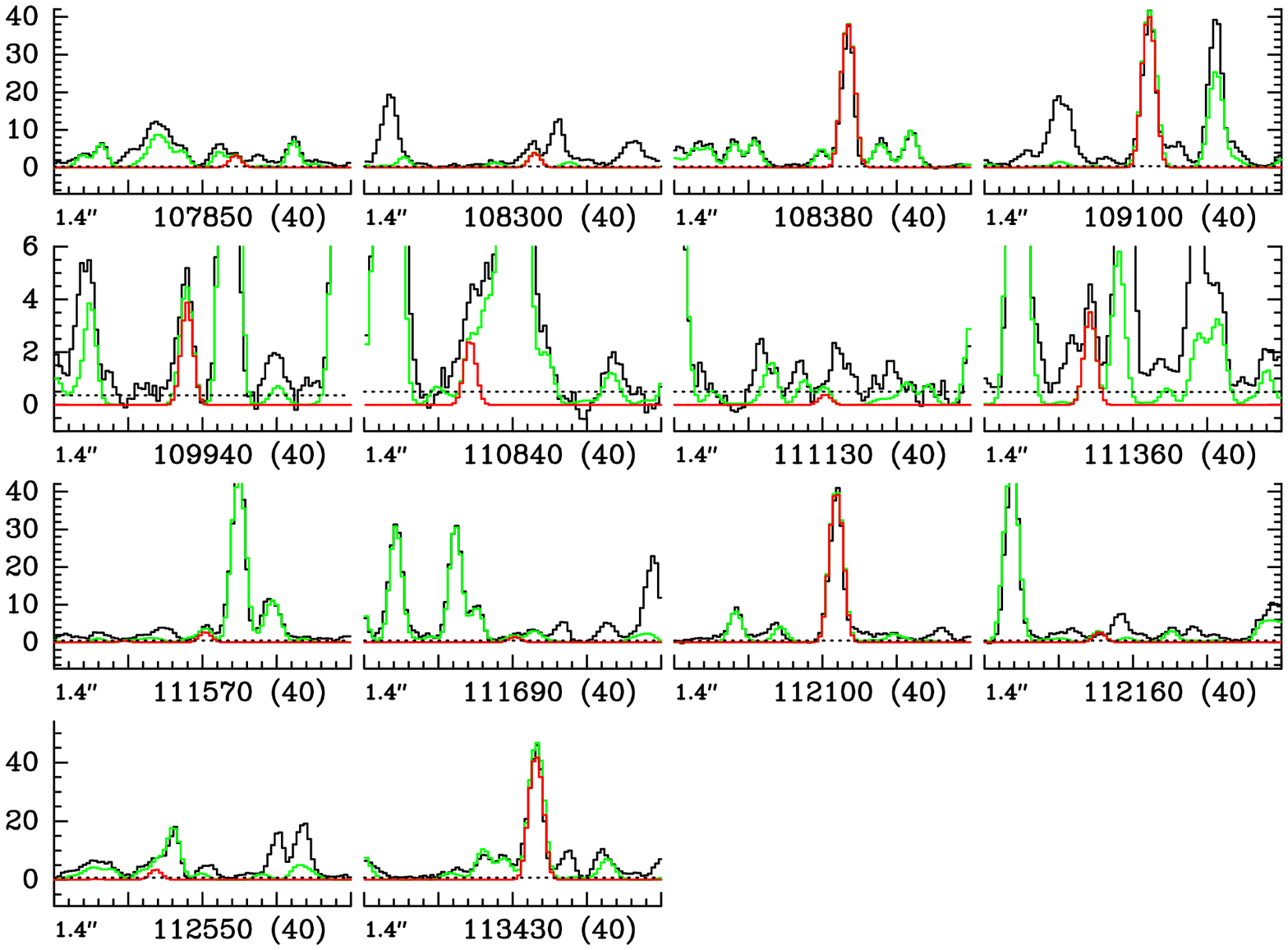}}}
\caption{continued.}
\end{figure*}
}
\addtocounter{figure}{-1}

\onlfig{
\clearpage
\begin{figure*}
%\centerline{\resizebox{0.9\hsize}{!}{\includegraphics[angle=0]{/homes/belloche/SgrB2/ALMA/Analysis/Weeds/R-NC/Figs/c2h5cn_15n_ve0_p2_1.eps}}}
\centerline{\resizebox{0.9\hsize}{!}{\includegraphics[angle=0]{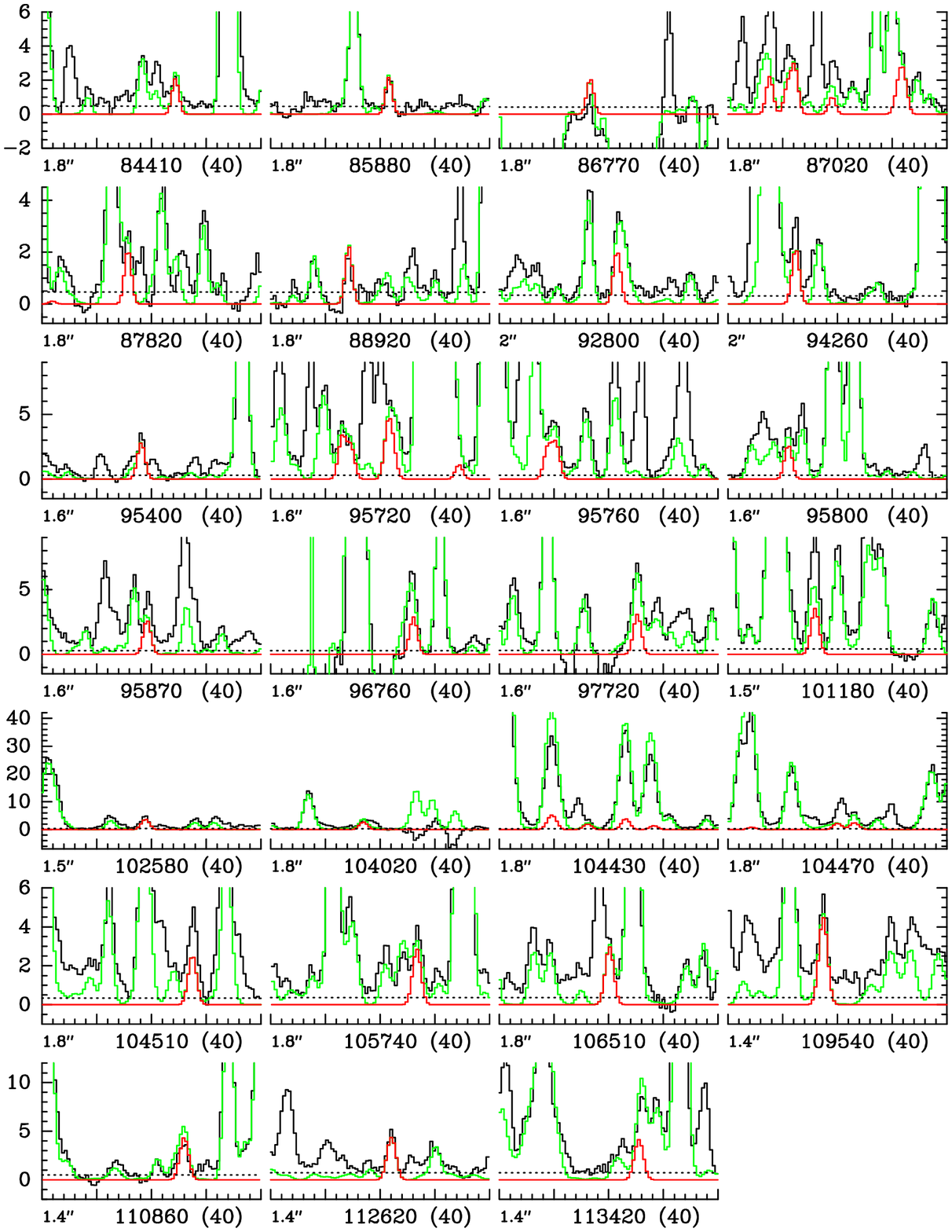}}}
\caption{Same as Fig.~\ref{f:spec_c2h5cn_ve0} for C$_2$H$_5$C$^{15}$N, 
$\varv=0$.
}
\label{f:spec_c2h5cn_15n_ve0}
\end{figure*}
}

\onlfig{
\clearpage
\begin{figure}
%\centerline{\resizebox{1.0\hsize}{!}{\includegraphics[angle=0]{/homes/belloche/SgrB2/ALMA/Analysis/Weeds/R-NC/popdiag_c2h5cn_p2_onlyve0.eps}}}
\centerline{\resizebox{1.0\hsize}{!}{\includegraphics[angle=0]{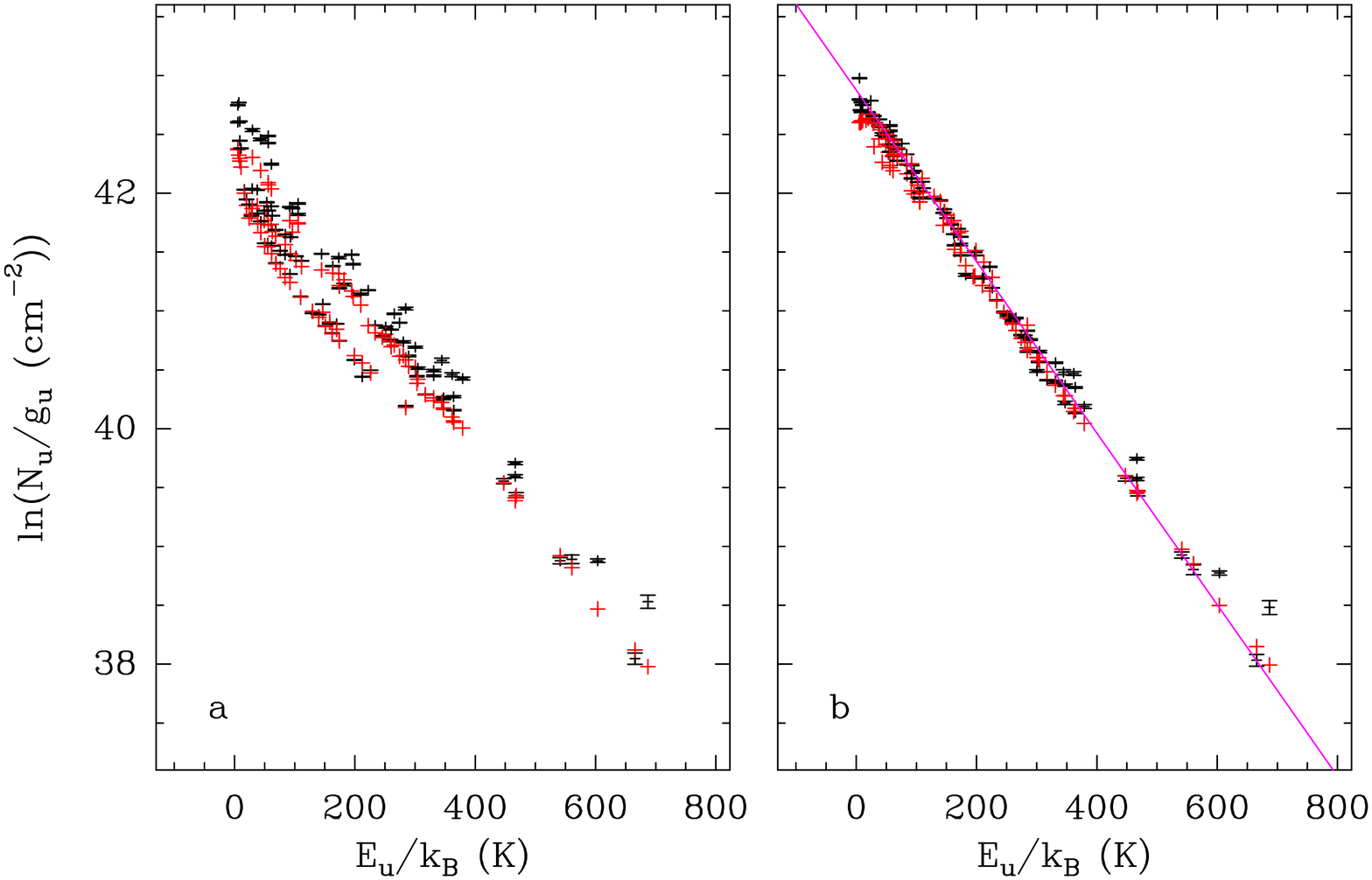}}}
\caption{Population diagram of C$_2$H$_5$CN, $\varv=0$ toward Sgr~B2(N2). Only 
the lines that are clearly detected, do not suffer too much from 
contamination from other species, and have an opacity below 2.5 are displayed. 
The observed datapoints are shown in black while the synthetic populations are 
shown in red. No correction is applied in panel \textbf{a}. In panel 
\textbf{b}, the optical depth correction has been applied to both the observed 
and synthetic populations and the contamination from all other species included
in the full model has been removed from the observed datapoints. The purple 
line is a linear fit to the observed populations (in linear-logarithmic space).
}
\label{f:popdiag_c2h5cn}
\end{figure}
}

\onlfig{
\begin{figure}
%\centerline{\resizebox{1.0\hsize}{!}{\includegraphics[angle=0]{/homes/belloche/SgrB2/ALMA/Analysis/Weeds/R-NC/popdiag_c2h5cn_13c1_p2.eps}}}
\centerline{\resizebox{1.0\hsize}{!}{\includegraphics[angle=0]{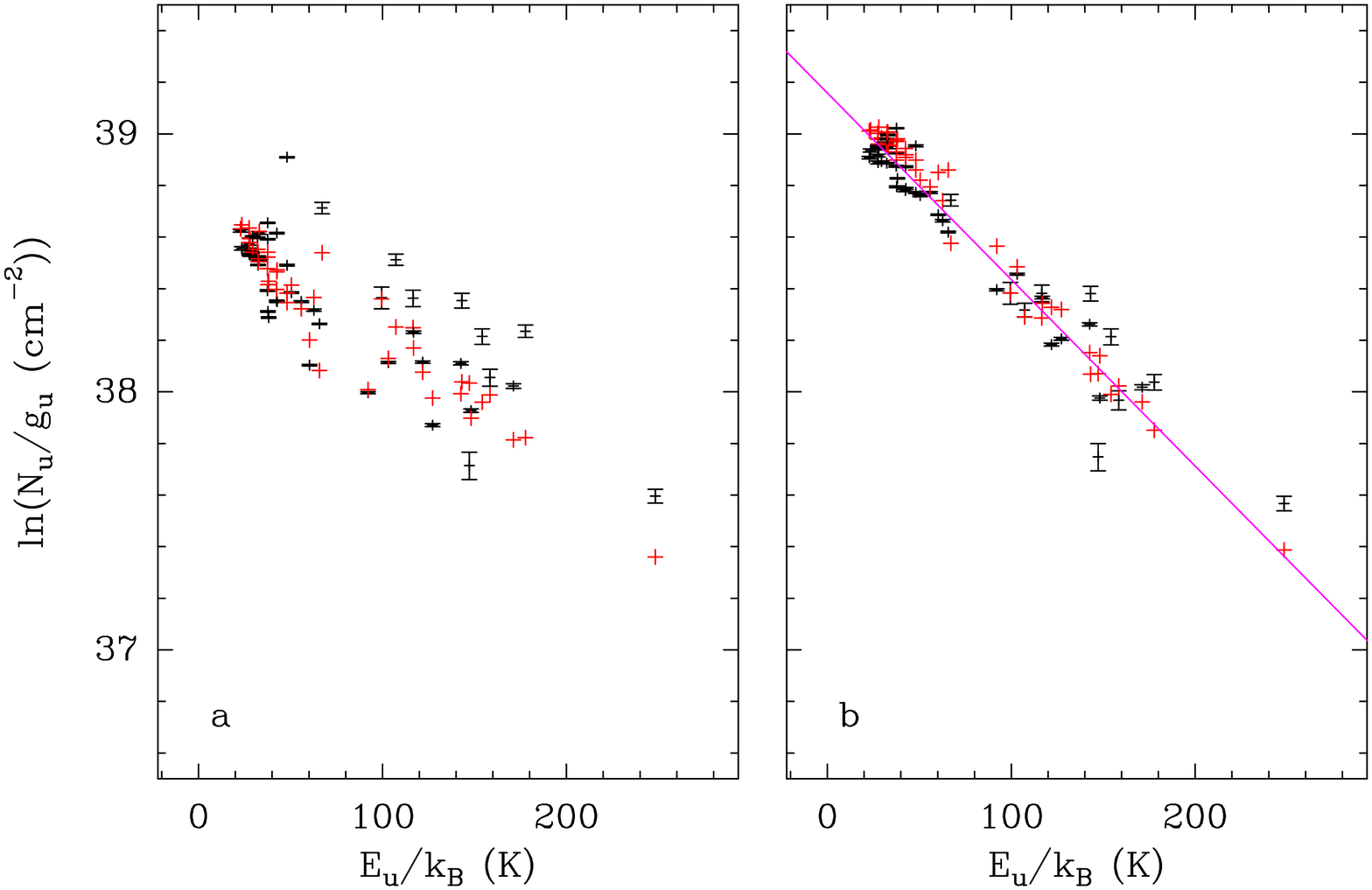}}}
\caption{Same as Fig.~\ref{f:popdiag_c2h5cn} for $^{13}$CH$_3$CH$_2$CN, 
$\varv=0$.}
\label{f:popdiag_c2h5cn_13c1}
\end{figure}
}

\onlfig{
\begin{figure}
%\centerline{\resizebox{1.0\hsize}{!}{\includegraphics[angle=0]{/homes/belloche/SgrB2/ALMA/Analysis/Weeds/R-NC/popdiag_c2h5cn_13c2_p2.eps}}}
\centerline{\resizebox{1.0\hsize}{!}{\includegraphics[angle=0]{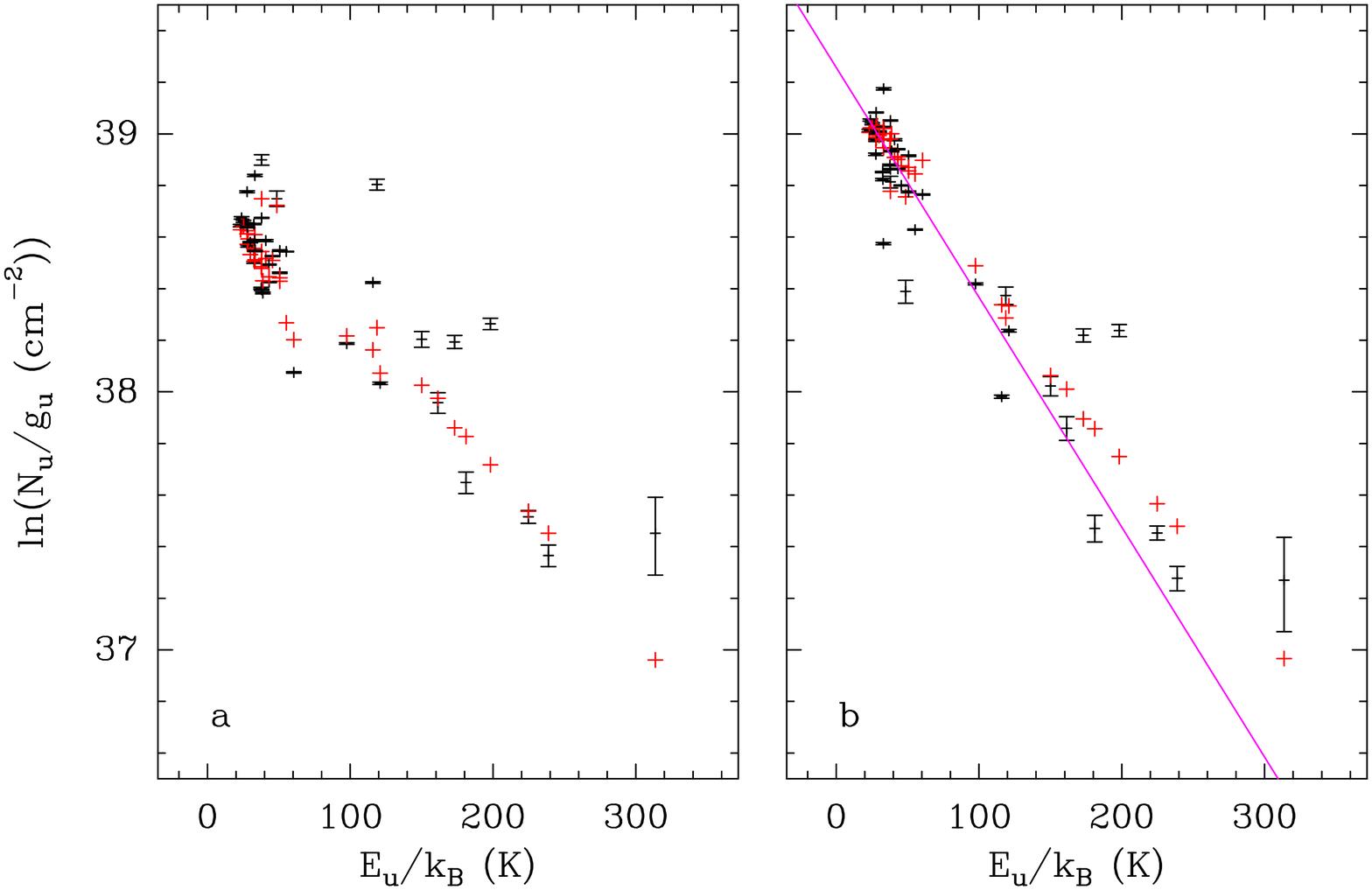}}}
\caption{Same as Fig.~\ref{f:popdiag_c2h5cn} for CH$_3$$^{13}$CH$_2$CN, 
$\varv=0$.}
\label{f:popdiag_c2h5cn_13c2}
\end{figure}
}

\onlfig{
\begin{figure}
%\centerline{\resizebox{1.0\hsize}{!}{\includegraphics[angle=0]{/homes/belloche/SgrB2/ALMA/Analysis/Weeds/R-NC/popdiag_c2h5cn_13c3_p2.eps}}}
\centerline{\resizebox{1.0\hsize}{!}{\includegraphics[angle=0]{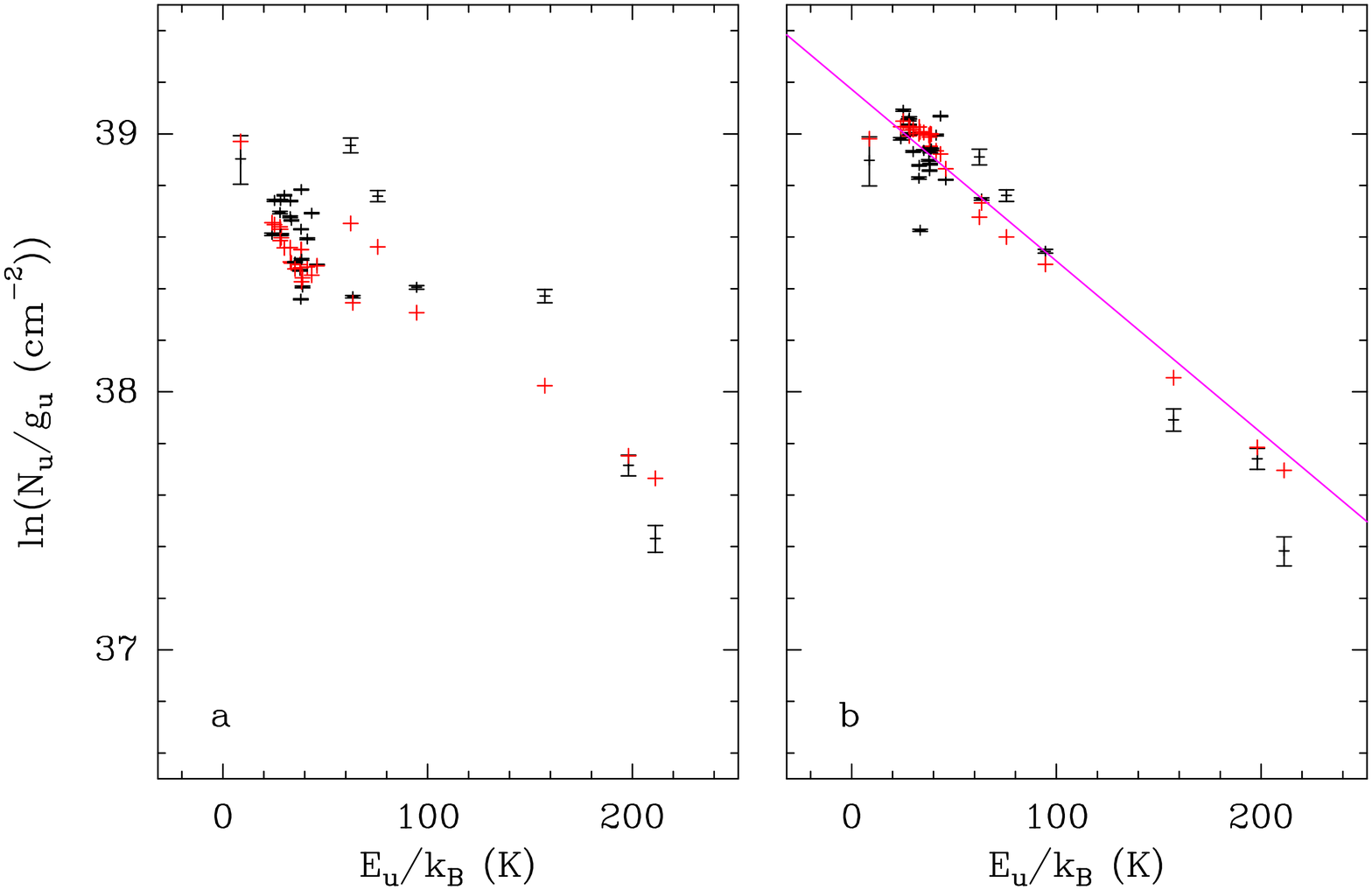}}}
\caption{Same as Fig.~\ref{f:popdiag_c2h5cn} for CH$_3$CH$_2$$^{13}$CN, 
$\varv=0$.}
\label{f:popdiag_c2h5cn_13c3}
\end{figure}
}

\begin{figure*}
%\centerline{\resizebox{0.9\hsize}{!}{\includegraphics[angle=0]{/homes/belloche/SgrB2/ALMA/Analysis/Maps/R-NC/plot_gauss2ddeconv_c2h5cn_art_r-d_p2.eps}}}
\centerline{\resizebox{0.9\hsize}{!}{\includegraphics[angle=0]{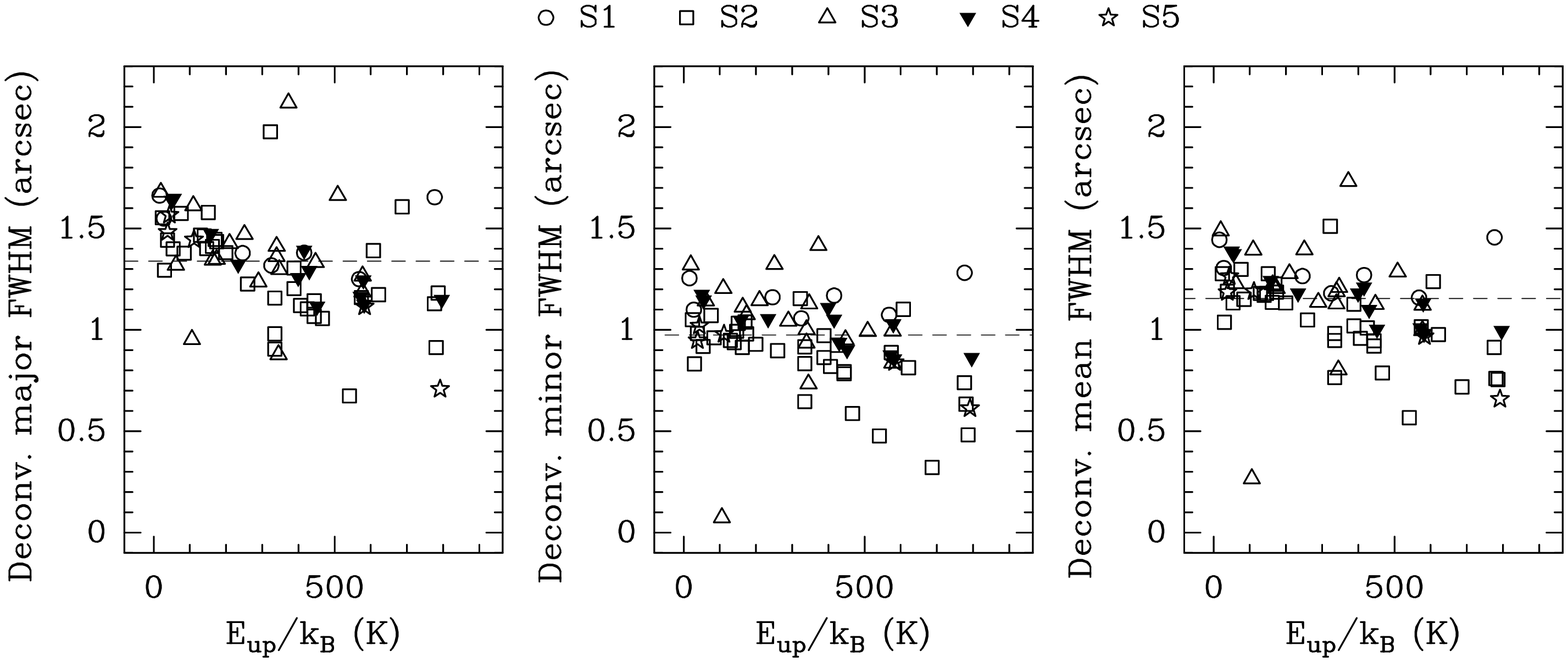}}}
\caption{Deconvolved major, minor, and mean sizes (FWHM) derived for 
uncontaminated C$_2$H$_5$CN transitions detected toward Sgr~B2(N2), plotted as 
a function of upper level energy. The symbols code for the spectral setup 
(S1 to S5, like in Table~\ref{t:beam_noise}). In each panel, the dashed line 
indicates the median value.}
\label{f:size_c2h5cn}
\end{figure*}

The median source size derived for the selected lines of C$_2$H$_5$CN is about 
$1.15\arcsec$, but there seems to be a trend of decreasing size with increasing
upper level energy, from $\sim 1.3\arcsec$ at low energy to $\sim 0.8\arcsec$
at $E_{\rm u} \sim 700$~K (Fig.~\ref{f:size_c2h5cn}). Similar results are
obtained for $^{13}$CH$_3$CH$_2$CN, with a median size of $\sim 1.25\arcsec$
and a hint of a decrease down to $\sim 1.0\arcsec$ at $E_{\rm u} \sim 120$~K.
Our model does not treat such gradients. As a compromise, we used a source
size of $1.2\arcsec$. 

With this source size and a rotational temperature of 150~K, we obtain an 
excellent fit to all emission lines of the four isotopologues, except for the 
very optically thick lines of C$_2$H$_5$CN ($\tau_{\rm max} \sim 60$), as 
mentioned above. A better fit to these lines would be obtained by increasing 
the temperature and/or assuming a larger source size. Increasing the size to 
$1.4\arcsec$ turns out to be insufficient. A larger size would be inconsistent 
with the measured sizes. Increasing the temperature to 200~K and the size to 
$1.3\arcsec$ yields peak temperatures of the optically thick lines similar to 
those observed, but the synthetic lines look too saturated compared to the 
observed ones, and the fit to the optically thin lines becomes worse, lines 
with high upper level energies becoming overpredicted. A more complex model 
with non-uniform physical parameters would probably be needed to reproduce the 
intensity and shape of the very optically thick lines.

Using the source size, rotational temperature, linewidth, and velocity offset
derived for C$_2$H$_5$CN and its $^{13}$C and $^{15}$N isotopologues, we looked 
for emission of the singly-deuterated isotopologues, CH$_3$CHDCN and 
CH$_2$DCH$_2$CN. The former is a chiral molecule because the carbon 
atom in the middle of the chain is linked to four different atoms or functional 
groups. Both isotopologues
are tentatively detected toward Sgr~B2(N2) with 1 and 2 line(s), respectively 
(Figs.~\ref{f:spec_ch3chdcn} and \ref{f:spec_ch2dch2cn-oop}), the latter in its 
out-of-plane conformation only. For the in-plane confomer of CH$_2$DCH$_2$CN, 
we derive an upper limit only. This upper limit is uncertain because the 
apparent inconsistency between the synthetic spectrum and the observed one 
around $\sim 101190$~MHz may result from a slight overestimate of the 
baseline, at the $3\sigma$ level (Fig.\ref{f:spec_ch2dch2cn-ip}). Due to the
limited signal-to-noise ratios, the source size derived from the integrated
intensity maps of the uncontaminated lines assigned to CH$_3$CHDCN and 
CH$_2$DCH$_2$CN is uncertain, varying between unresolved and $\sim 2''$. The
emission looks compact in the maps. Assuming the same source size as the other 
isotopologues looks thus reasonable.

\onlfig{
\clearpage
\begin{figure*}
%\centerline{\resizebox{0.9\hsize}{!}{\includegraphics[angle=0]{/homes/belloche/SgrB2/ALMA/Analysis/Weeds/R-D/Figs/ch3chdcn_ve0_p2.eps}}}
\centerline{\resizebox{0.9\hsize}{!}{\includegraphics[angle=0]{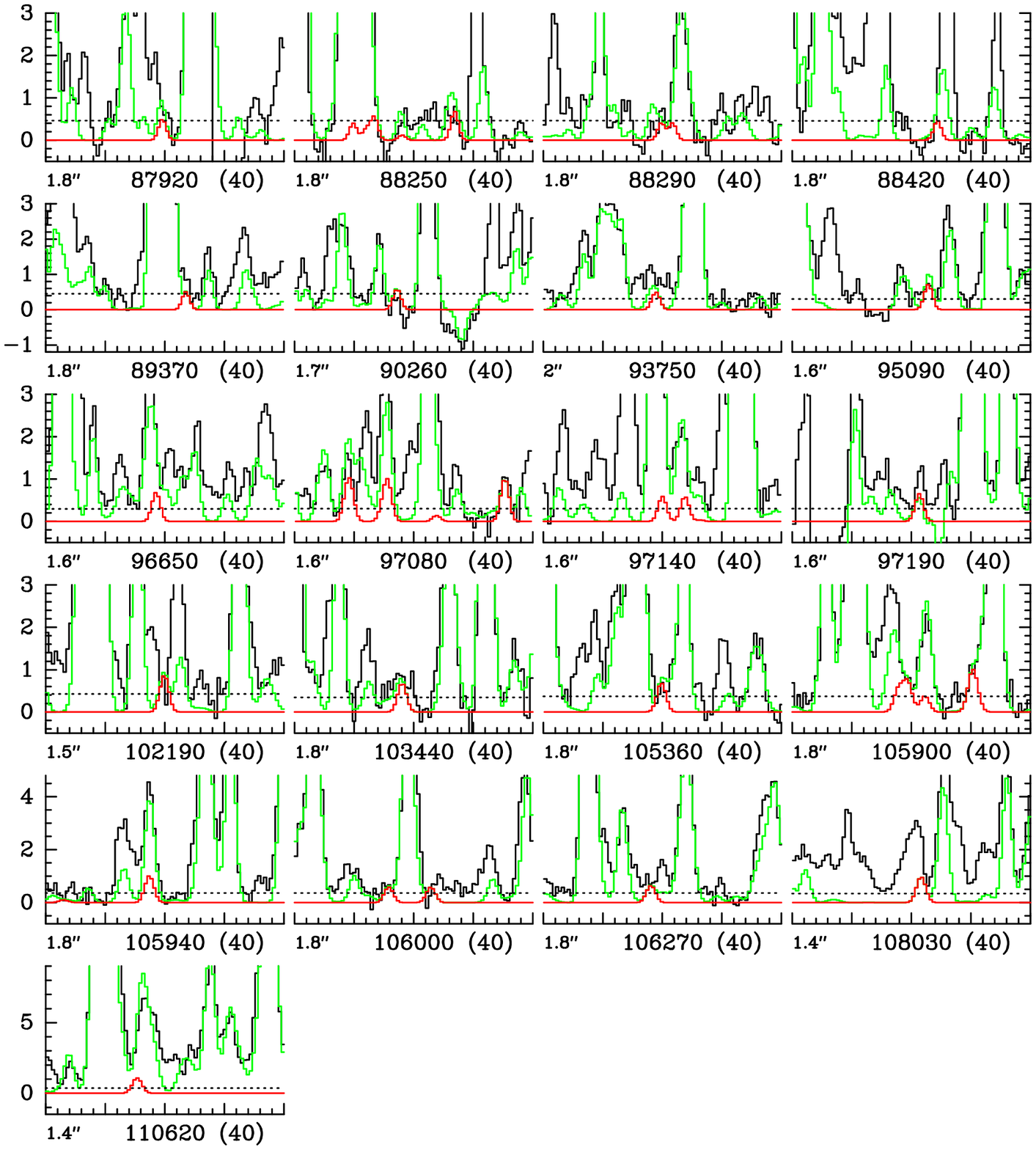}}}
\caption{Same as Fig.~\ref{f:spec_c2h5cn_ve0} for CH$_3$CHDCN.
}
\label{f:spec_ch3chdcn}
\end{figure*}
}

\onlfig{
\clearpage
\begin{figure*}
%\centerline{\resizebox{0.9\hsize}{!}{\includegraphics[angle=0]{/homes/belloche/SgrB2/ALMA/Analysis/Weeds/R-D/Figs/ch2dch2cn-oop_ve0_p2.eps}}}
\centerline{\resizebox{0.9\hsize}{!}{\includegraphics[angle=0]{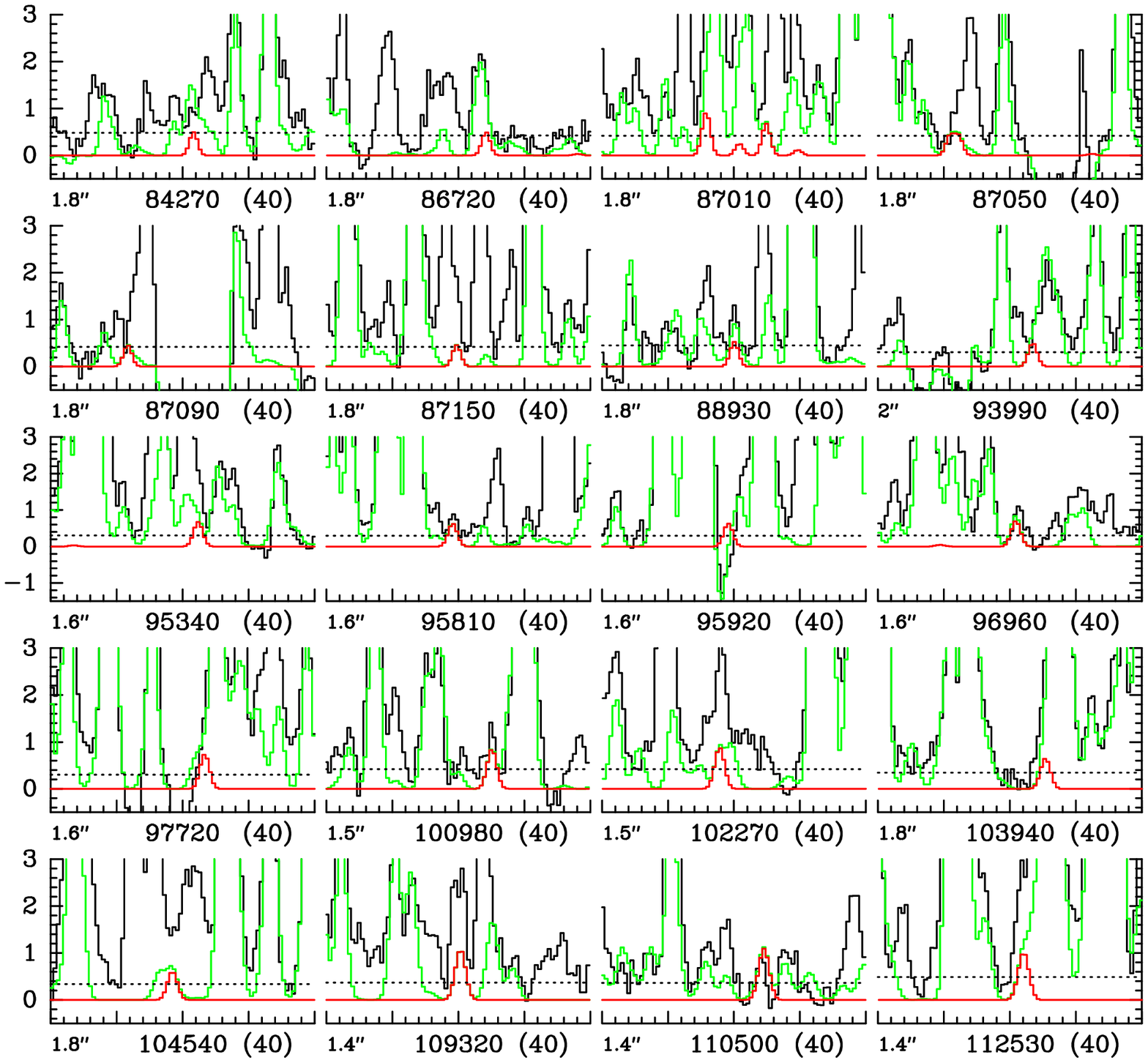}}}
\caption{Same as Fig.~\ref{f:spec_c2h5cn_ve0} for CH$_2$DCH$_2$CN in its out-of-plane conformation.
}
\label{f:spec_ch2dch2cn-oop}
\end{figure*}
}

\onlfig{
\begin{figure*}
%\centerline{\resizebox{0.9\hsize}{!}{\includegraphics[angle=0]{/homes/belloche/SgrB2/ALMA/Analysis/Weeds/R-D/Figs/ch2dch2cn-ip_ve0_p2.eps}}}
\centerline{\resizebox{0.9\hsize}{!}{\includegraphics[angle=0]{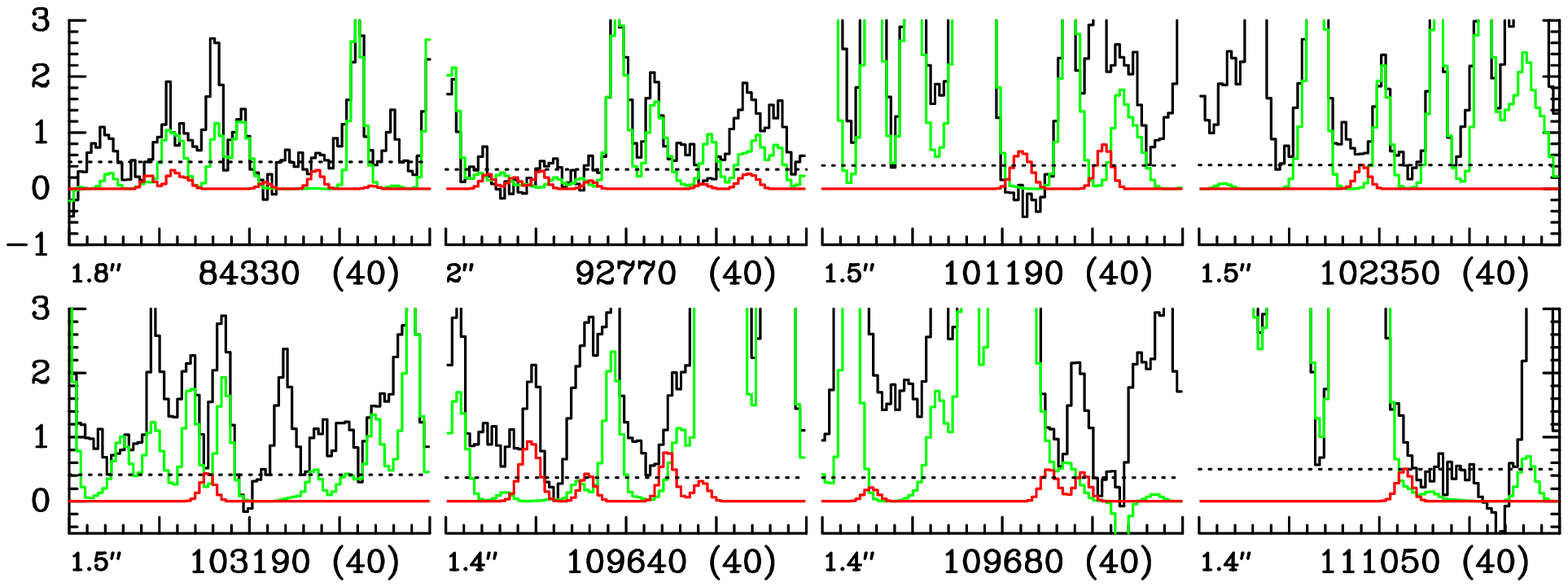}}}
\caption{Same as Fig.~\ref{f:spec_c2h5cn_ve0} for CH$_2$DCH$_2$CN in its 
in-plane conformation. The full synthetic model (in green) does \textit{not} 
contain any contribution of CH$_2$DCH$_2$CN in its in-plane conformation (in 
red).
}
\label{f:spec_ch2dch2cn-ip}
\end{figure*}
}

\subsection{Deuterated methyl cyanide CH$_2$DCN}
\label{ss:ch3cn}

Methyl cyanide is clearly detected in its vibrational ground state toward 
Sgr~B2(N2) but its transitions are very optically thick 
($\tau_{\rm max} \sim 50$) and cannot be properly
fitted in the framework of our simple model (Fig.~\ref{f:spec_ch3cn_ve0}). 
Transitions from within its vibrationally excited states $\varv_8=1$ and
$\varv_8=2$ are also clearly detected, with $\tau_{\rm max} \sim 2.7$ and 0.3, 
respectively (Figs.~\ref{f:spec_ch3cn_v8e1} and \ref{f:spec_ch3cn_v8e2}). We 
also find four transitions from within $\varv_4=1$ around 91520~MHz and 
109820~MHz ($\tau_{\rm max} \sim 0.06$), but they partially suffer from
blends with other species (Fig.~\ref{f:spec_ch3cn_v4e1}). The assignment looks 
reasonable, but the detection should be considered as tentative. 

\onlfig{
\begin{figure*}
%\centerline{\resizebox{0.9\hsize}{!}{\includegraphics[angle=0]{/homes/belloche/SgrB2/ALMA/Analysis/Weeds/R-NC/Figs/ch3cn_ve0_p2.eps}}}
\centerline{\resizebox{0.9\hsize}{!}{\includegraphics[angle=0]{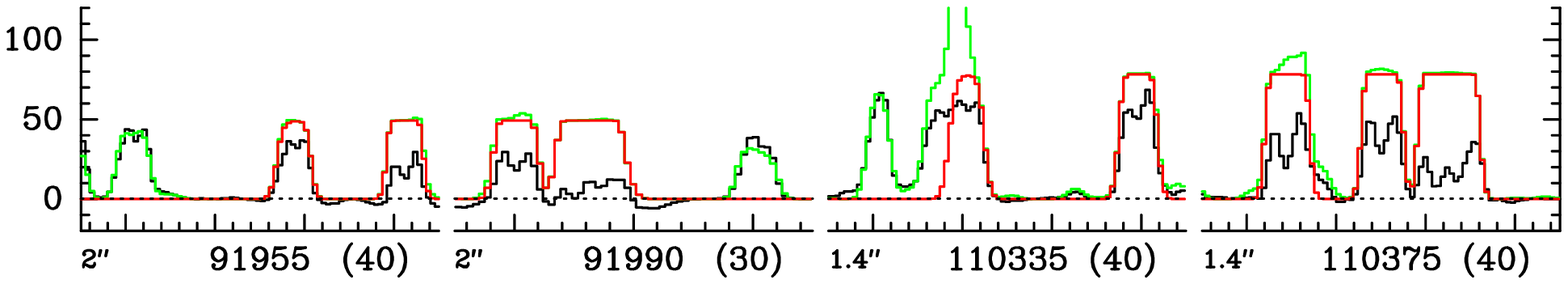}}}
\caption{Same as Fig.~\ref{f:spec_c2h5cn_ve0} for CH$_3$CN, $\varv=0$.
}
\label{f:spec_ch3cn_ve0}
\end{figure*}
}

\onlfig{
\begin{figure*}
%\centerline{\resizebox{0.9\hsize}{!}{\includegraphics[angle=0]{/homes/belloche/SgrB2/ALMA/Analysis/Weeds/R-NC/Figs/ch3cn_v8e1_p2.eps}}}
\centerline{\resizebox{0.9\hsize}{!}{\includegraphics[angle=0]{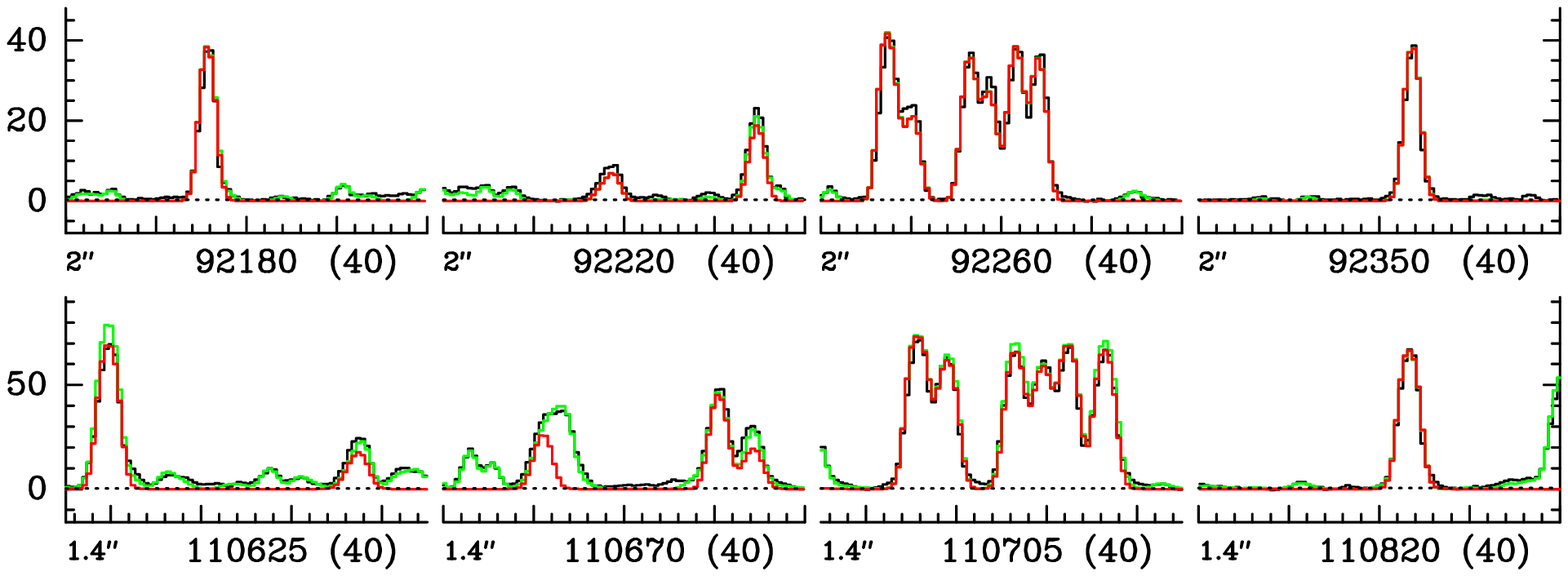}}}
\caption{Same as Fig.~\ref{f:spec_c2h5cn_ve0} for CH$_3$CN, $\varv_8=1$.
}
\label{f:spec_ch3cn_v8e1}
\end{figure*}
}

\onlfig{
\begin{figure*}
%\centerline{\resizebox{0.9\hsize}{!}{\includegraphics[angle=0]{/homes/belloche/SgrB2/ALMA/Analysis/Weeds/R-NC/Figs/ch3cn_v8e2_p2.eps}}}
\centerline{\resizebox{0.9\hsize}{!}{\includegraphics[angle=0]{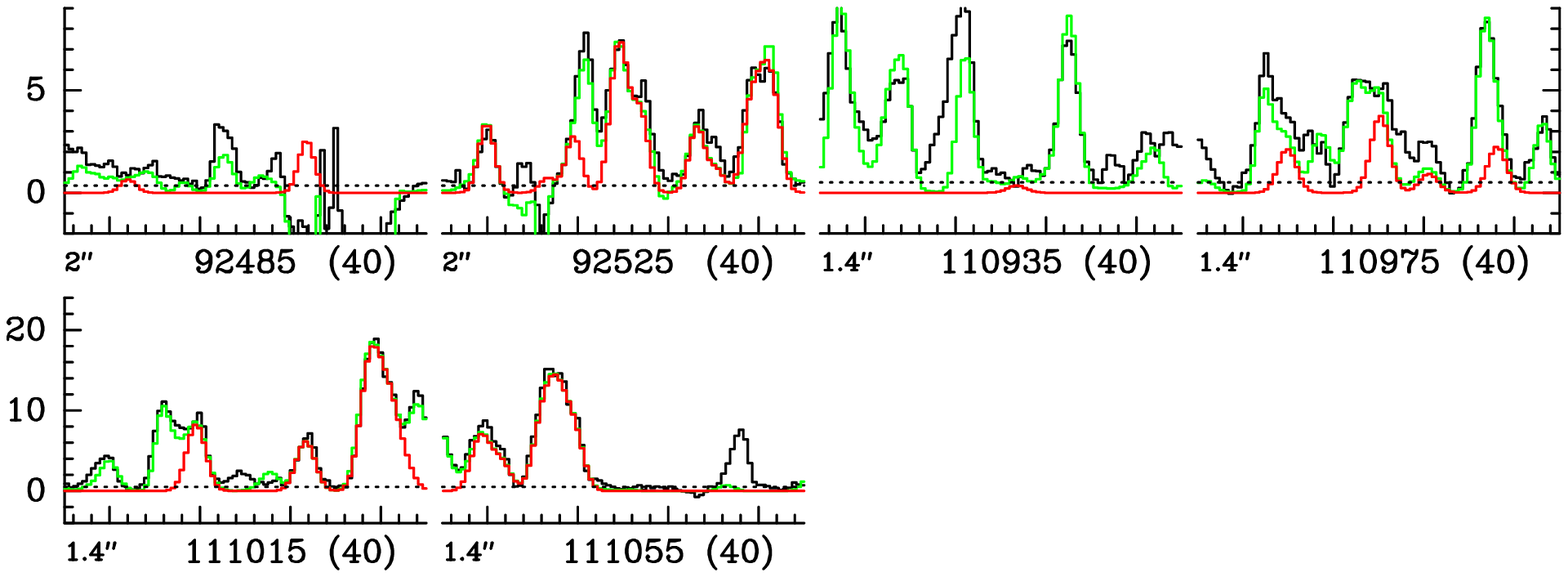}}}
\caption{Same as Fig.~\ref{f:spec_c2h5cn_ve0} for CH$_3$CN, $\varv_8=2$.
}
\label{f:spec_ch3cn_v8e2}
\end{figure*}
}

\onlfig{
\begin{figure*}
%\centerline{\resizebox{0.675\hsize}{!}{\includegraphics[angle=0]{/homes/belloche/SgrB2/ALMA/Analysis/Weeds/R-NC/Figs/ch3cn_v4e1_p2.eps}}}
\centerline{\resizebox{0.675\hsize}{!}{\includegraphics[angle=0]{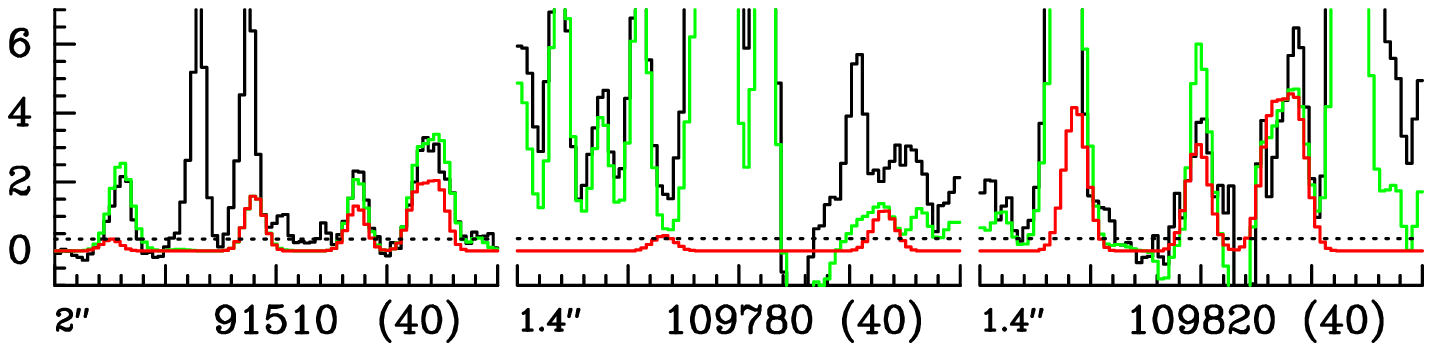}}}
\caption{Same as Fig.~\ref{f:spec_c2h5cn_ve0} for CH$_3$CN, $\varv_4=1$.
}
\label{f:spec_ch3cn_v4e1}
\end{figure*}
}

The singly-substituted $^{13}$C isotopologues are very well detected, both in
their vibrational ground state and in their first vibrationally excited state 
$\varv_8=1$ (Figs.~\ref{f:spec_ch3cn_13c1_ve0}, \ref{f:spec_ch3cn_13c2_ve0}, 
\ref{f:spec_ch3cn_13c1_v8e1}, and \ref{f:spec_ch3cn_13c2_v8e1}). The fit to 
their population diagrams yields rotational temperatures of about 170~K 
(see Table~\ref{t:popfit} and Figs.~\ref{f:popdiag_ch3cn_13c1} and 
\ref{f:popdiag_ch3cn_13c2}).
The analysis of the integrated intensity maps of the $^{13}$C isotopologues
delivers a source size of $\sim 1.4\arcsec$. For the main isotopologue, it
seems that the source size decreases with the vibrational energy 
($\sim 1.2\arcsec$ for $\varv_8=1$ and $\sim 0.8\arcsec$ for $\varv_8=2$).

\onlfig{
\begin{figure*}
%\centerline{\resizebox{0.675\hsize}{!}{\includegraphics[angle=0]{/homes/belloche/SgrB2/ALMA/Analysis/Weeds/R-NC/Figs/ch3cn_13c1_ve0_p2.eps}}}
\centerline{\resizebox{0.675\hsize}{!}{\includegraphics[angle=0]{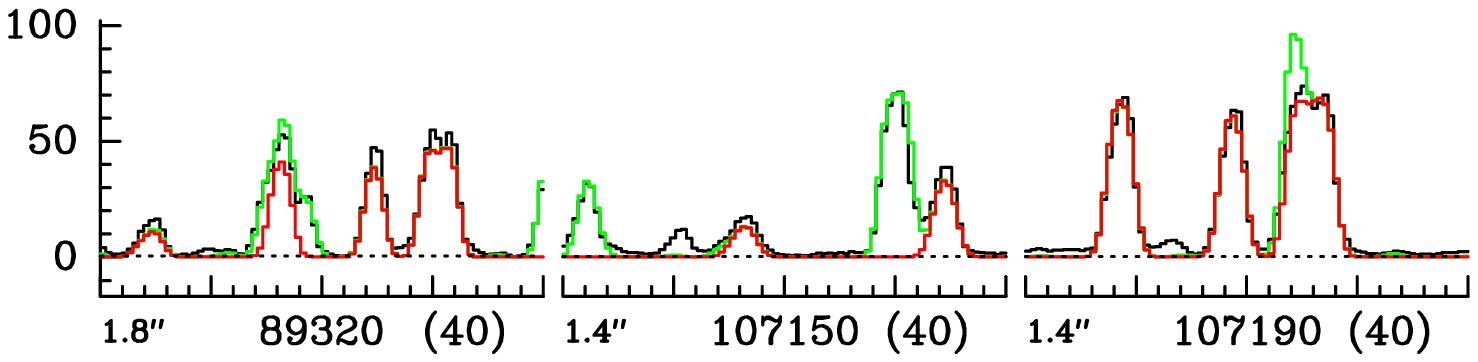}}}
\caption{Same as Fig.~\ref{f:spec_c2h5cn_ve0} for $^{13}$CH$_3$CN, $\varv=0$.
}
\label{f:spec_ch3cn_13c1_ve0}
\end{figure*}
}

\onlfig{
\begin{figure*}
%\centerline{\resizebox{0.675\hsize}{!}{\includegraphics[angle=0]{/homes/belloche/SgrB2/ALMA/Analysis/Weeds/R-NC/Figs/ch3cn_13c2_ve0_p2.eps}}}
\centerline{\resizebox{0.675\hsize}{!}{\includegraphics[angle=0]{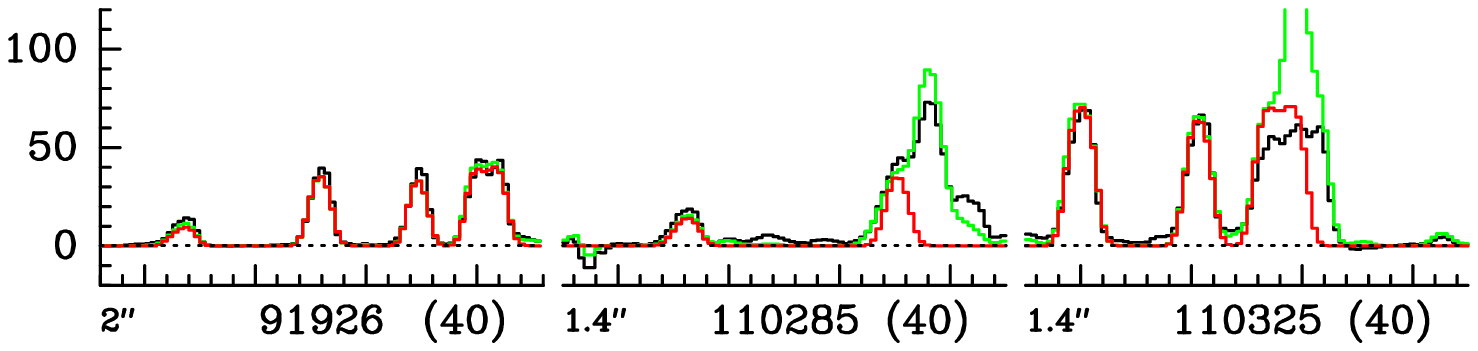}}}
\caption{Same as Fig.~\ref{f:spec_c2h5cn_ve0} for CH$_3$$^{13}$CN, $\varv=0$.
}
\label{f:spec_ch3cn_13c2_ve0}
\end{figure*}
}

\onlfig{
\begin{figure*}
%\centerline{\resizebox{0.9\hsize}{!}{\includegraphics[angle=0]{/homes/belloche/SgrB2/ALMA/Analysis/Weeds/R-NC/Figs/ch3cn_13c1_v8e1_p2.eps}}}
\centerline{\resizebox{0.9\hsize}{!}{\includegraphics[angle=0]{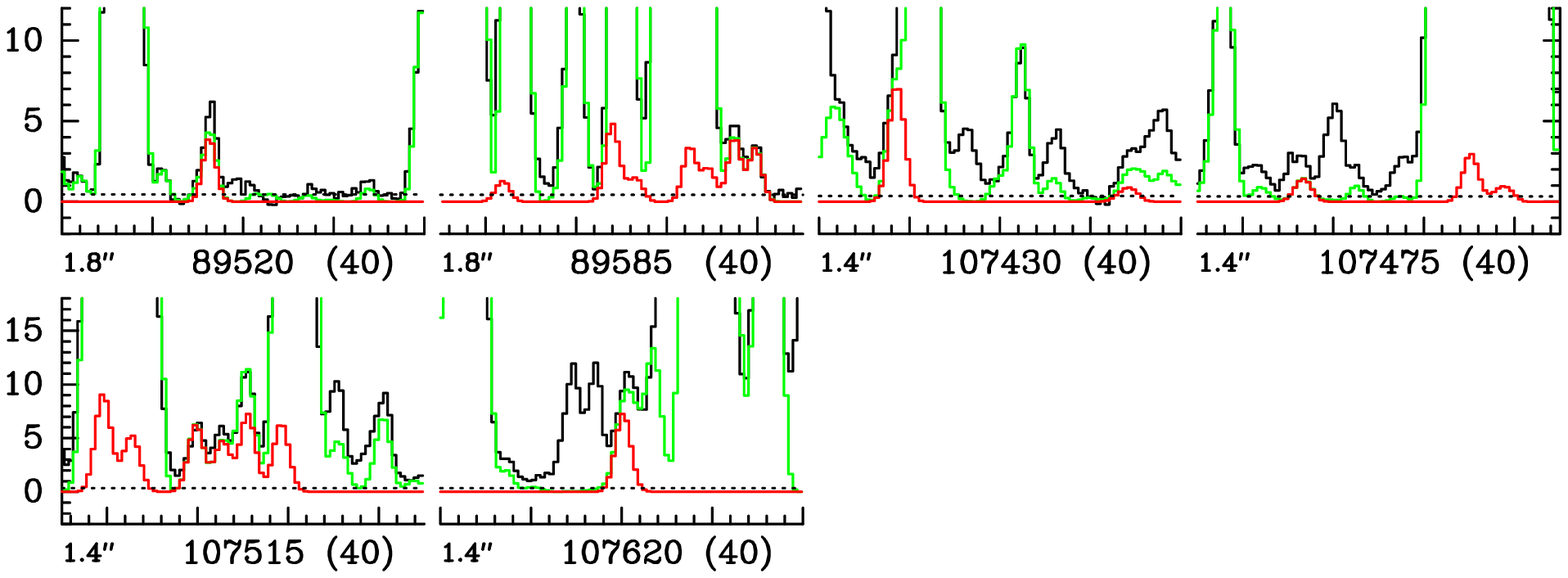}}}
\caption{Same as Fig.~\ref{f:spec_c2h5cn_ve0} for $^{13}$CH$_3$CN, $\varv_8=1$.
}
\label{f:spec_ch3cn_13c1_v8e1}
\end{figure*}
}

\onlfig{
\begin{figure*}
%\centerline{\resizebox{0.9\hsize}{!}{\includegraphics[angle=0]{/homes/belloche/SgrB2/ALMA/Analysis/Weeds/R-NC/Figs/ch3cn_13c2_v8e1_p2.eps}}}
\centerline{\resizebox{0.9\hsize}{!}{\includegraphics[angle=0]{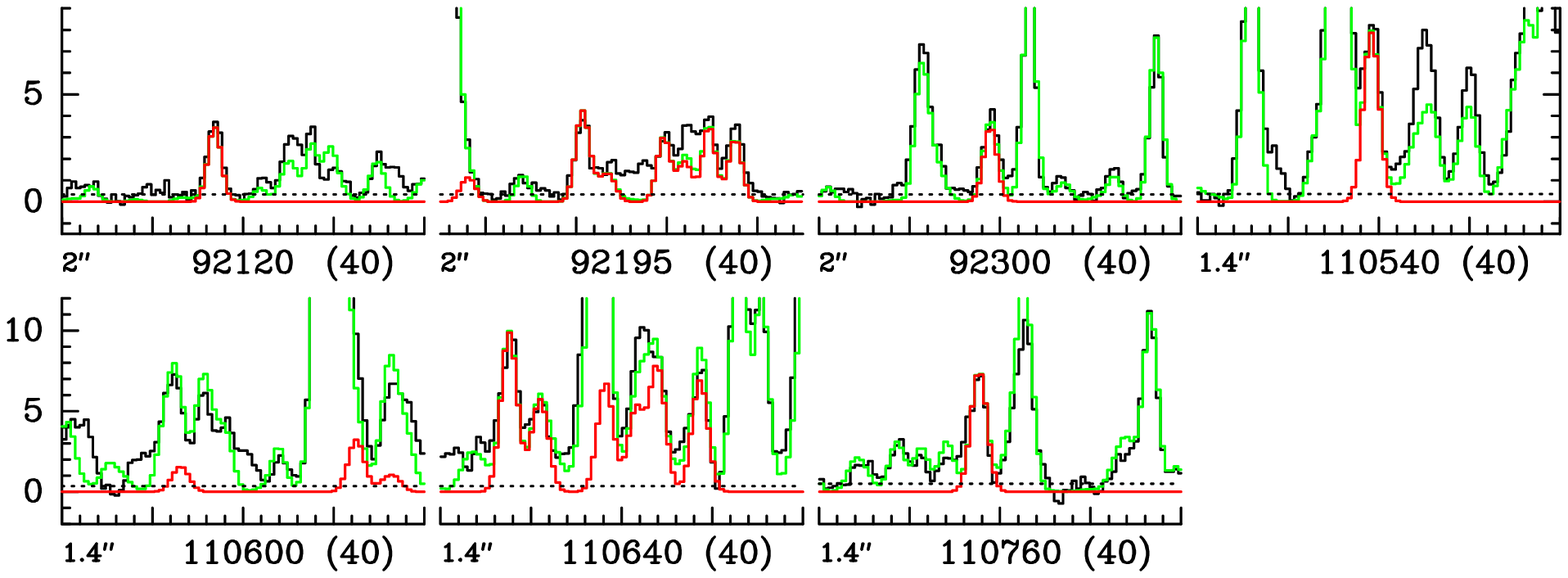}}}
\caption{Same as Fig.~\ref{f:spec_c2h5cn_ve0} for CH$_3$$^{13}$CN, $\varv_8=1$.
}
\label{f:spec_ch3cn_13c2_v8e1}
\end{figure*}
}

\onlfig{
\clearpage
\begin{figure}
%\centerline{\resizebox{1.0\hsize}{!}{\includegraphics[angle=0]{/homes/belloche/SgrB2/ALMA/Analysis/Weeds/R-NC/popdiag_ch3cn_13c1_p2.eps}}}
\centerline{\resizebox{1.0\hsize}{!}{\includegraphics[angle=0]{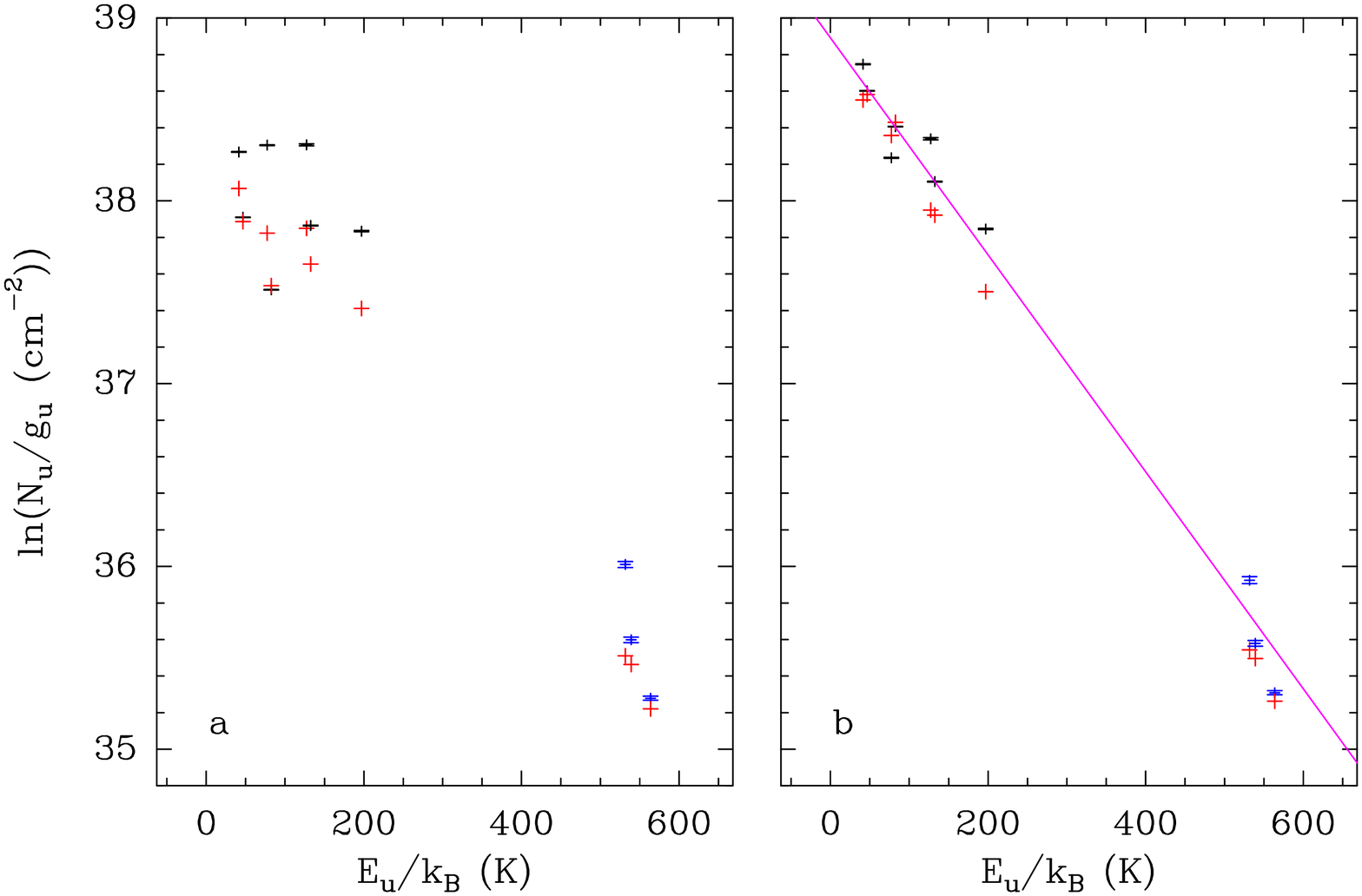}}}
\caption{Same as Fig.~\ref{f:popdiag_c2h5cn} for $^{13}$CH$_3$CN, $\varv=0$ and $\varv_8=1$.}
\label{f:popdiag_ch3cn_13c1}
\end{figure}
}

\onlfig{
\begin{figure}
%\centerline{\resizebox{1.0\hsize}{!}{\includegraphics[angle=0]{/homes/belloche/SgrB2/ALMA/Analysis/Weeds/R-NC/popdiag_ch3cn_13c2_p2.eps}}}
\centerline{\resizebox{1.0\hsize}{!}{\includegraphics[angle=0]{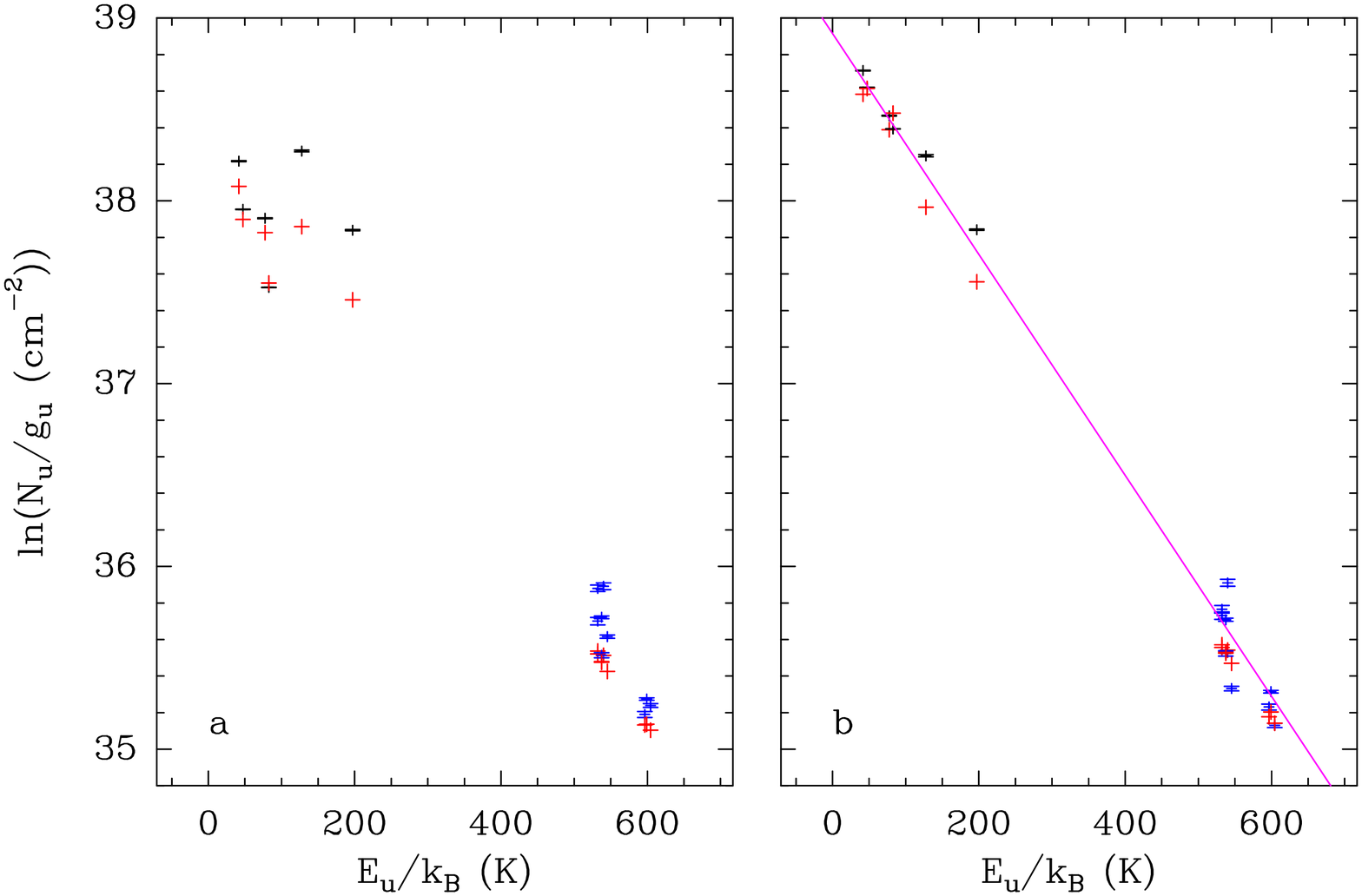}}}
\caption{Same as Fig.~\ref{f:popdiag_c2h5cn} for CH$_3$$^{13}$CN, $\varv=0$ and $\varv_8=1$.}
\label{f:popdiag_ch3cn_13c2}
\end{figure}
}

\onlfig{
\begin{figure}
%\centerline{\resizebox{1.0\hsize}{!}{\includegraphics[angle=0]{/homes/belloche/SgrB2/ALMA/Analysis/Weeds/R-NC/popdiag_ch3cn_p2.eps}}}
\centerline{\resizebox{1.0\hsize}{!}{\includegraphics[angle=0]{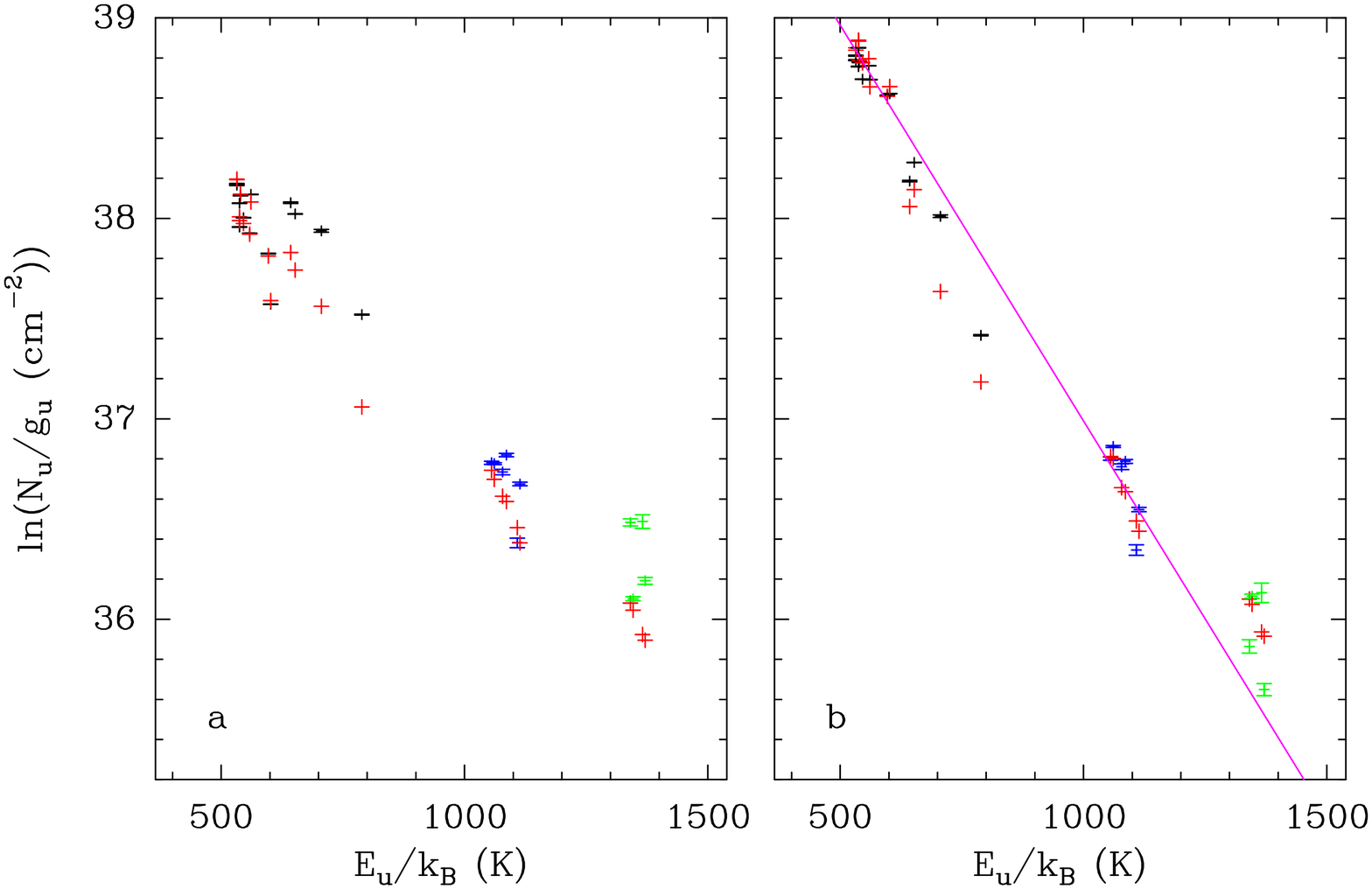}}}
\caption{Same as Fig.~\ref{f:popdiag_c2h5cn} for CH$_3$CN, $\varv_8=1$, 
$\varv_8=2$, and $\varv_4=1$.}
\label{f:popdiag_ch3cn}
\end{figure}
}

Assuming a source size of $1.4\arcsec$ and a temperature of 170~K, our LTE
modeling yields excellent and consistent fits to the $^{13}$C isotopologues 
(both $\varv=0$ and $\varv_8=1$) and to the $\varv_8=1$ transitions of the main 
isotopologue. However, it was necessary to increase the column density and
linewidth in order to fit the transitions of the $\varv_8=2$ and $\varv_4=1$
states. The fit to the population diagram of the main isotopologue including
the three vibrationally excited states suggests a temperature of $\sim 250$~K
(Fig.~\ref{f:popdiag_ch3cn}),
which explains why our 170~K model needs higher column densities to reproduce
the intensities of the $\varv_8=2$ and $\varv_4=1$ transitions. Here again,
a more complex model with a non-uniform physical structure would be necessary 
to fit all transitions in a consistent way. However, given that our simple
LTE model yields a good fit to the $\varv_8=1$ transitions of the main 
isotopologue and all transitions of the $^{13}$C ones with a single set of
parameters, we consider that the derived $^{12}$C/$^{13}$C column density 
ratios are reliable.

On the basis of the LTE model obtained above, we were able to identify emission
from the doubly-substituted $^{13}$C isotopologue of methyl cyanide, 
$^{13}$CH$_3$$^{13}$CN. One 
transition at 107108~MHz is well detected, and a group of transitions at 
89270~MHz is relatively well detected (Fig.~\ref{f:spec_ch3cn_13c13c}). Given
that the column density ratio of the singly- to doubly-substituted 
isotopologues is very close to the ratio between the main isotopologue and the
singly-substituted ones, we consider the identification of $^{13}$CH$_3$$^{13}$CN
as secure.

\onlfig{
\begin{figure}
%\centerline{\resizebox{0.95\hsize}{!}{\includegraphics[angle=0]{/homes/belloche/SgrB2/ALMA/Analysis/Weeds/R-NC/Figs/ch3cn_13c13c_ve0_p2.eps}}}
\centerline{\resizebox{0.95\hsize}{!}{\includegraphics[angle=0]{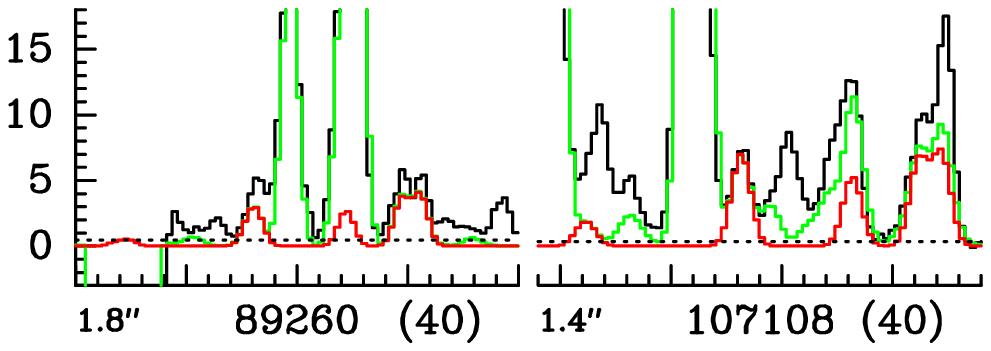}}}
\caption{Same as Fig.~\ref{f:spec_c2h5cn_ve0} for $^{13}$CH$_3$$^{13}$CN, $\varv=0$.
}
\label{f:spec_ch3cn_13c13c}
\end{figure}
}

Our LTE modeling indicates that the $^{15}$N isotopologue contributes 
significantly to the emission detected at 107054~MHz and 107060~MHz
(Fig.~\ref{f:spec_ch3cn_15n}). However,
since there is no clearly detected line, we do not consider this identification
as secure. The derived column density is therefore relatively uncertain and
should rather be considered as an upper limit.

\onlfig{
\clearpage
\begin{figure*}
%\centerline{\resizebox{0.24\hsize}{!}{\includegraphics[angle=0]{/homes/belloche/SgrB2/ALMA/Analysis/Weeds/R-NC/Figs/ch3cn_15n_ve0_p2.eps}}}
\centerline{\resizebox{0.24\hsize}{!}{\includegraphics[angle=0]{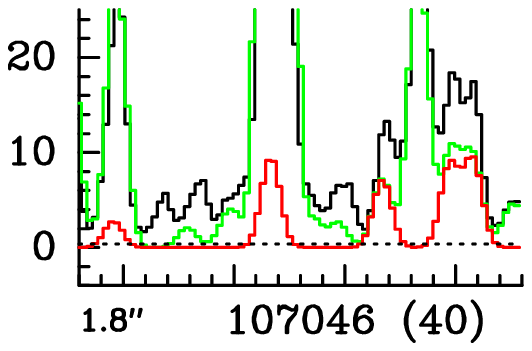}}}
\caption{Same as Fig.~\ref{f:spec_c2h5cn_ve0} for CH$_3$C$^{15}$N, $\varv=0$.
}
\label{f:spec_ch3cn_15n}
\end{figure*}
}

Finally, using the same parameters as for the other isotopologues, we obtain
a secure identification of singly-deuterated methyl cyanide, CH$_2$DCN, with 
about six transitions clearly detected (Fig.~\ref{f:spec_ch2dcn}). The rotation
temperature derived from the population diagram is consistent with the
assumed temperature within $2.4\sigma$ (Table~\ref{t:popfit} and 
Fig.~\ref{f:popdiag_ch2dcn}). The source size can be measured for three of
the detected transitions and is found to be consistent with the size derived 
from the $^{13}$C isotopologues.

\onlfig{
\begin{figure*}
%\centerline{\resizebox{0.95\hsize}{!}{\includegraphics[angle=0]{/homes/belloche/SgrB2/ALMA/Analysis/Weeds/R-D/Figs/ch2dcn_ve0_p2.eps}}}
\centerline{\resizebox{0.95\hsize}{!}{\includegraphics[angle=0]{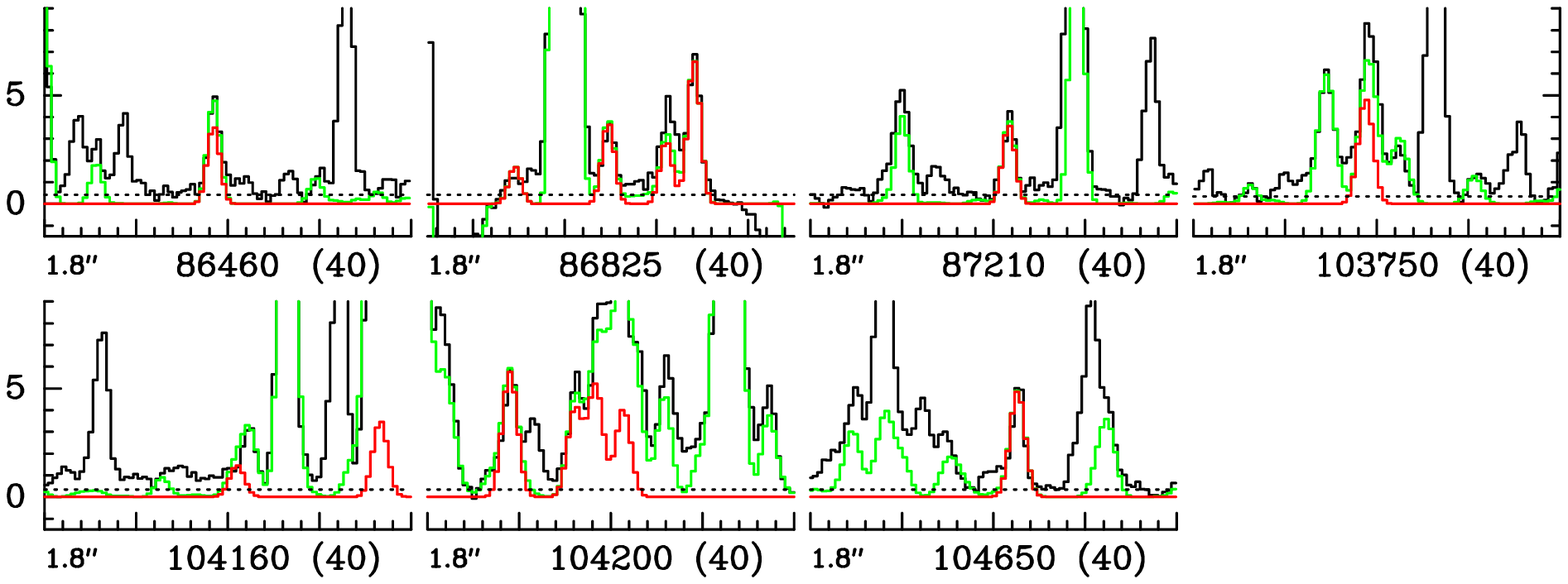}}}
\caption{Same as Fig.~\ref{f:spec_c2h5cn_ve0} for CH$_2$DCN, $\varv=0$.
}
\label{f:spec_ch2dcn}
\end{figure*}
}

\onlfig{
\begin{figure*}
%\centerline{\resizebox{0.5\hsize}{!}{\includegraphics[angle=0]{/homes/belloche/SgrB2/ALMA/Analysis/Weeds/R-D/popdiag_ch2dcn_p2.eps}}}
\centerline{\resizebox{0.5\hsize}{!}{\includegraphics[angle=0]{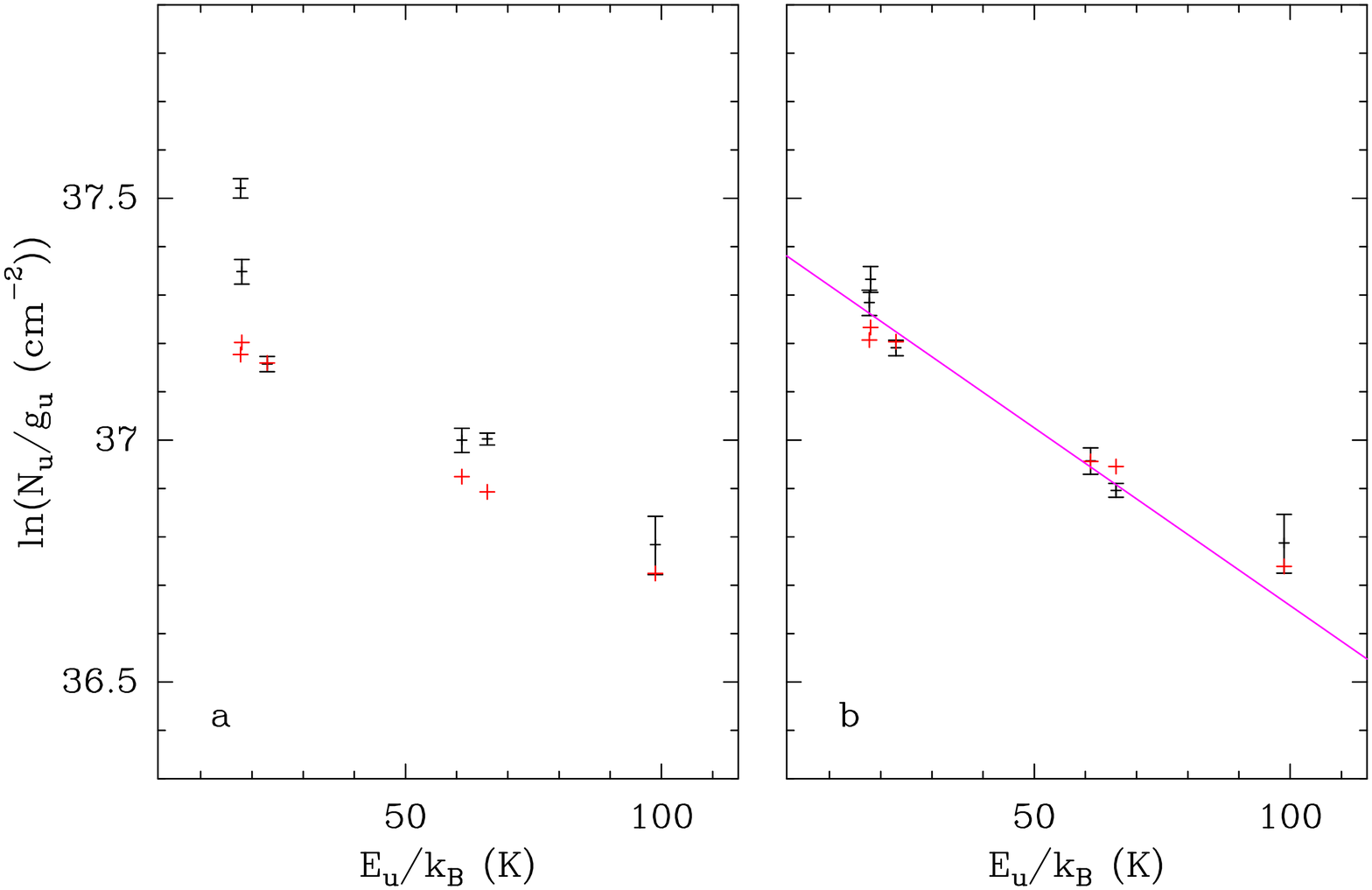}}}
\caption{Same as Fig.~\ref{f:popdiag_c2h5cn} for CH$_2$DCN.}
\label{f:popdiag_ch2dcn}
\end{figure*}
}

\subsection{Deuterated cyanoacetylene DC$_3$N}
\label{ss:hc3n}

Cyanoacetylene is detected toward Sgr~B2(N2) in its vibrational ground state
(Fig.~\ref{f:spec_hc3n_ve0}) but also in many vibrationally excited states: 
$\varv_7=1$ (Fig.~\ref{f:spec_hc3n_v7e1}), $\varv_7=2$ 
(Fig.~\ref{f:spec_hc3n_v7e2}), $\varv_6=1$ (Fig.~\ref{f:spec_hc3n_v6e1}),
$\varv_5=1$ and $\varv_7=3$  (Fig.~\ref{f:spec_hc3n_v5e1v7e3}),
and $\varv_6=\varv_7=1$ (Fig.~\ref{f:spec_hc3n_v6e1v7e1}).
In addition, emission from within the following excited states is tentatively 
detected: $\varv_4=1$ (one detected line, Fig.~\ref{f:spec_hc3n_v4e1}),
$\varv_7=4$ and $\varv_5=\varv_7=1$  (significantly contributes to detected
signal, but no line individually detected, 
Fig.~\ref{f:spec_hc3n_v7e4v5e1v7e1})\footnote{The current model is somewhat 
inconsistent with the observed spectrum at 92129~MHz (blend of 
$\varv_7=4$ $10_{-2}$ -- $9_2$ and $10_4$ -- $9_4$)  and 
100431~MHz ($\varv_5=\varv_7=1$ $l=0^-$ $11_0$  -- $10_0$) but this is most 
likely due to resonant interactions between 
$\varv_7=4$ and $\varv_5=\varv_7=1$ which are not well accounted for in the 
spectroscopic predictions. The frequencies of these transitions may well be
off by a few MHz 
\citep[see CDMS documentation and Sect.~4.4.33 of][]{Belloche13}},
and $\varv_6=2$ (one detected line, Fig.~\ref{f:spec_hc3n_v6e2}).

\onlfig{
\clearpage
\begin{figure*}
%\centerline{\resizebox{0.675\hsize}{!}{\includegraphics[angle=0]{/homes/belloche/SgrB2/ALMA/Analysis/Weeds/R-NC/Figs/hc3n_ve0_p2.eps}}}
\centerline{\resizebox{0.675\hsize}{!}{\includegraphics[angle=0]{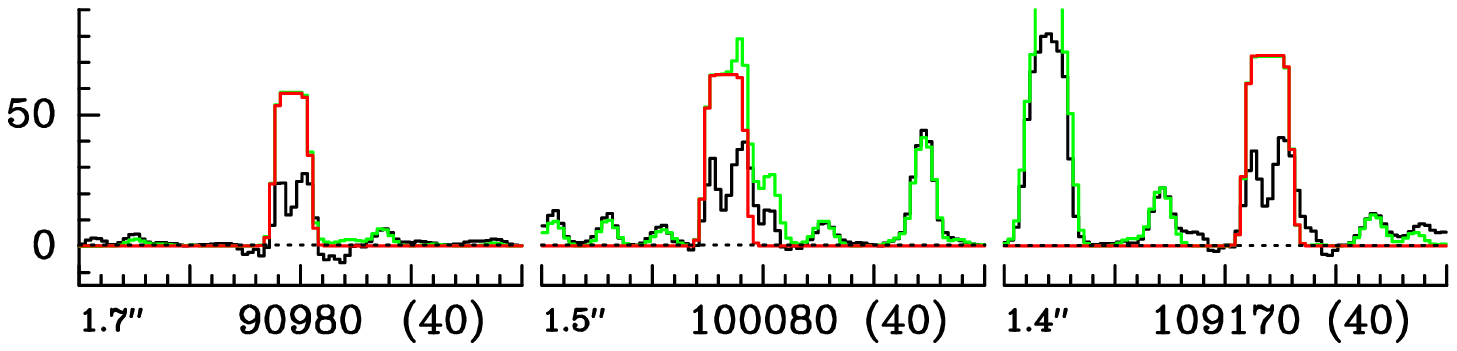}}}
\caption{Same as Fig.~\ref{f:spec_c2h5cn_ve0} for HC$_3$N, $\varv=0$.
}
\label{f:spec_hc3n_ve0}
\end{figure*}
}

\onlfig{
\begin{figure*}
%\centerline{\resizebox{0.9\hsize}{!}{\includegraphics[angle=0]{/homes/belloche/SgrB2/ALMA/Analysis/Weeds/R-NC/Figs/hc3n_v7e1_p2.eps}}}
\centerline{\resizebox{0.9\hsize}{!}{\includegraphics[angle=0]{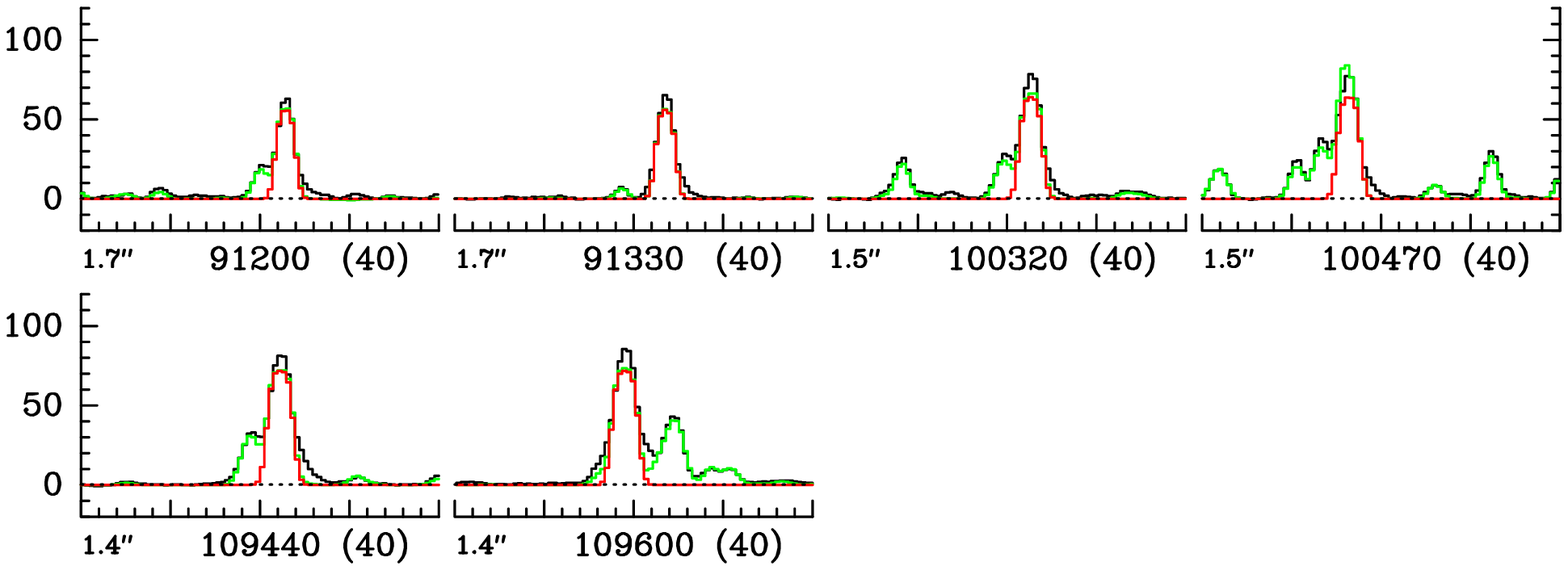}}}
\caption{Same as Fig.~\ref{f:spec_c2h5cn_ve0} for HC$_3$N, $\varv_7=1$.
}
\label{f:spec_hc3n_v7e1}
\end{figure*}
}

\onlfig{
\begin{figure*}
%\centerline{\resizebox{0.675\hsize}{!}{\includegraphics[angle=0]{/homes/belloche/SgrB2/ALMA/Analysis/Weeds/R-NC/Figs/hc3n_v7e2_p2.eps}}}
\centerline{\resizebox{0.675\hsize}{!}{\includegraphics[angle=0]{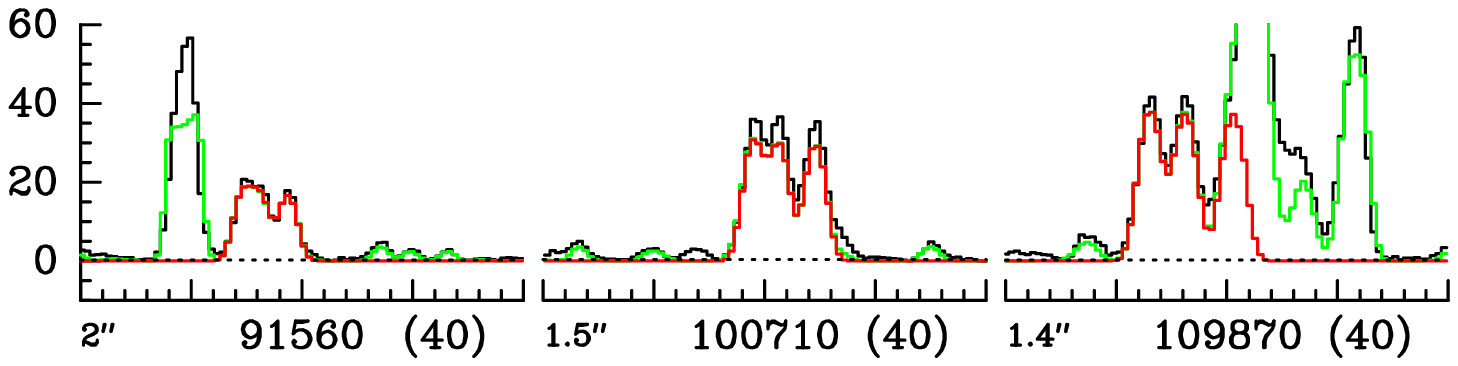}}}
\caption{Same as Fig.~\ref{f:spec_c2h5cn_ve0} for HC$_3$N, $\varv_7=2$.
}
\label{f:spec_hc3n_v7e2}
\end{figure*}
}

\onlfig{
\clearpage
\begin{figure*}
%\centerline{\resizebox{0.9\hsize}{!}{\includegraphics[angle=0]{/homes/belloche/SgrB2/ALMA/Analysis/Weeds/R-NC/Figs/hc3n_v6e1_p2.eps}}}
\centerline{\resizebox{0.9\hsize}{!}{\includegraphics[angle=0]{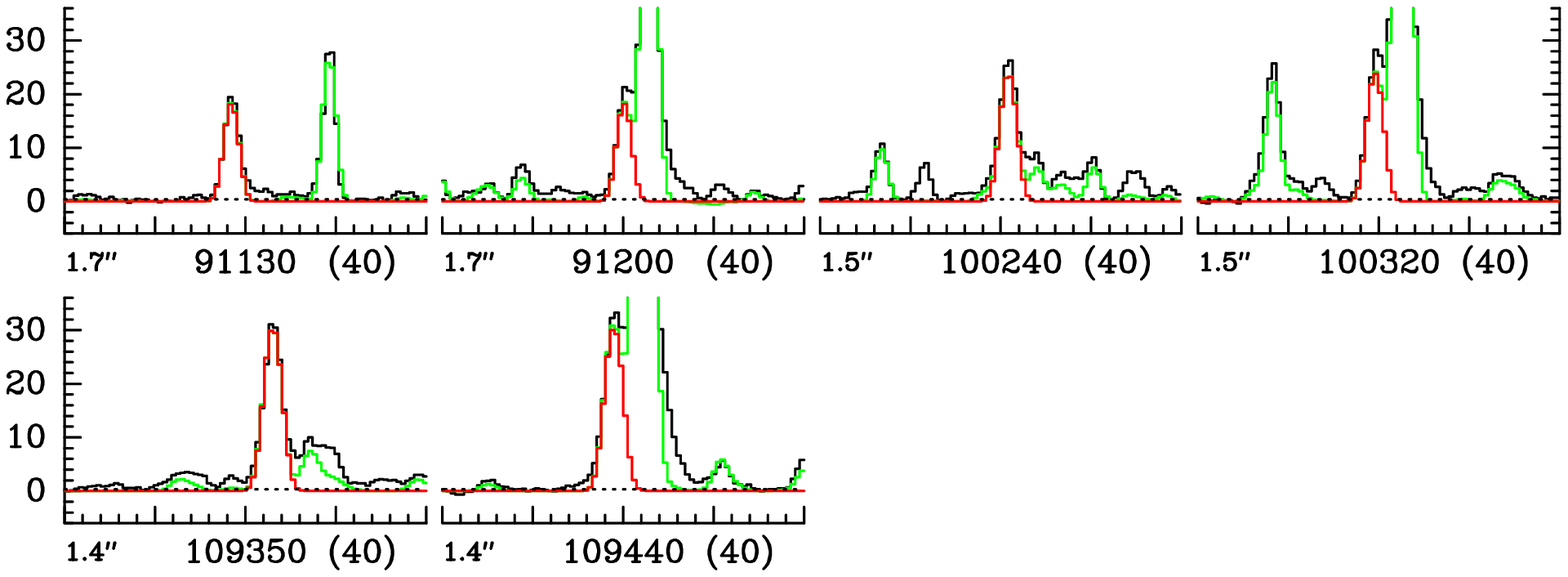}}}
\caption{Same as Fig.~\ref{f:spec_c2h5cn_ve0} for HC$_3$N, $\varv_6=1$.
}
\label{f:spec_hc3n_v6e1}
\end{figure*}
}

\onlfig{
\begin{figure*}
%\centerline{\resizebox{0.9\hsize}{!}{\includegraphics[angle=0]{/homes/belloche/SgrB2/ALMA/Analysis/Weeds/R-NC/Figs/hc3n_v5e1v7e3_p2.eps}}}
\centerline{\resizebox{0.9\hsize}{!}{\includegraphics[angle=0]{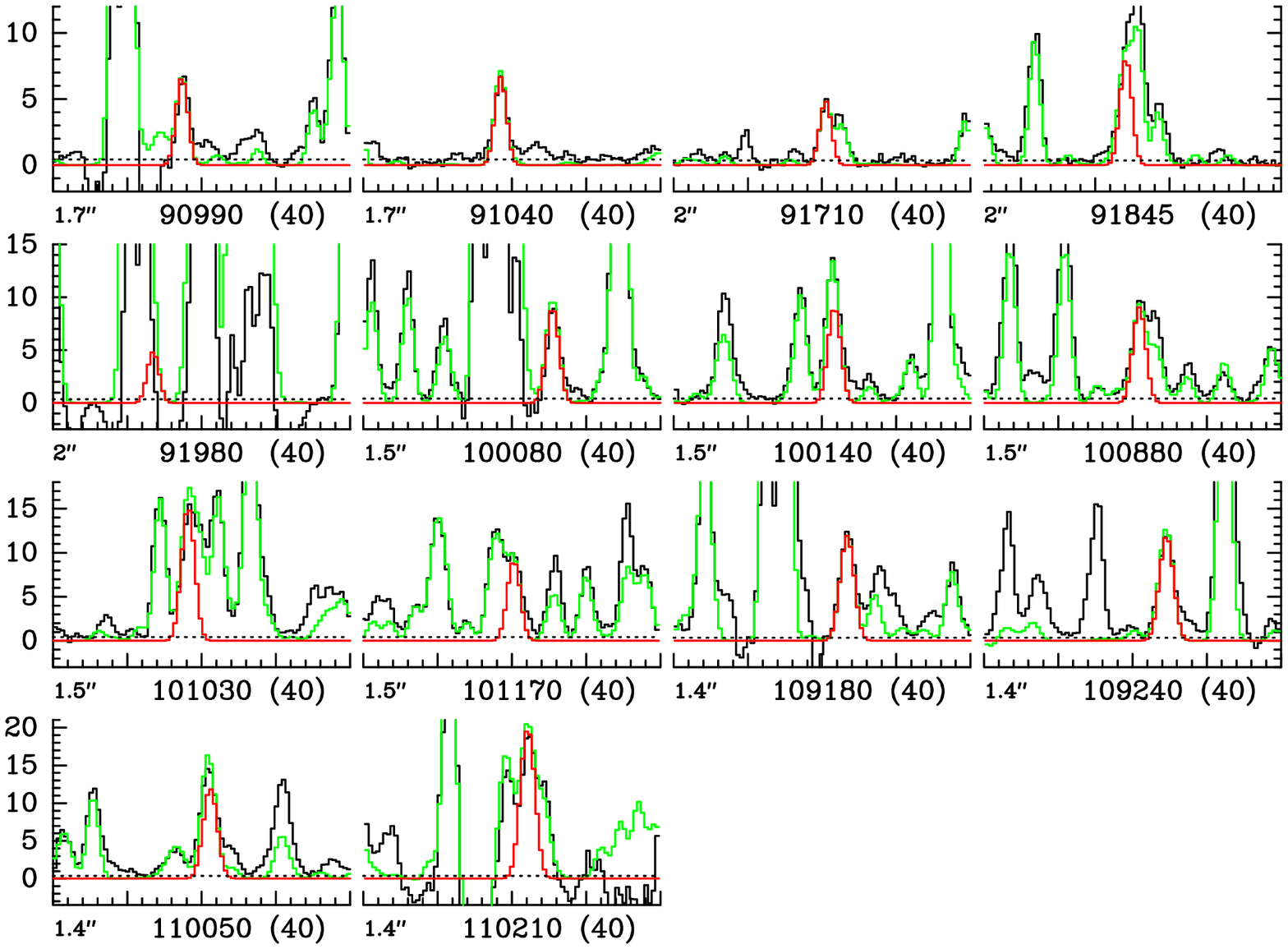}}}
\caption{Same as Fig.~\ref{f:spec_c2h5cn_ve0} for HC$_3$N, $\varv_5=1$ and 
$\varv_7=3$.
}
\label{f:spec_hc3n_v5e1v7e3}
\end{figure*}
}

\onlfig{
\begin{figure*}
%\centerline{\resizebox{0.675\hsize}{!}{\includegraphics[angle=0]{/homes/belloche/SgrB2/ALMA/Analysis/Weeds/R-NC/Figs/hc3n_v6e1v7e1_p2.eps}}}
\centerline{\resizebox{0.675\hsize}{!}{\includegraphics[angle=0]{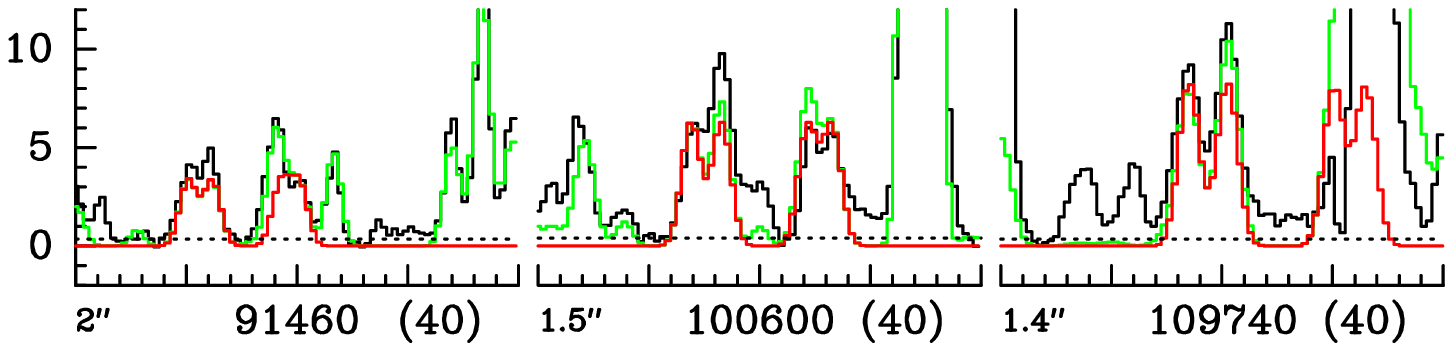}}}
\caption{Same as Fig.~\ref{f:spec_c2h5cn_ve0} for HC$_3$N, $\varv_6=\varv_7=1$.
}
\label{f:spec_hc3n_v6e1v7e1}
\end{figure*}
}

\onlfig{
\clearpage
\begin{figure*}
%\centerline{\resizebox{0.675\hsize}{!}{\includegraphics[angle=0]{/homes/belloche/SgrB2/ALMA/Analysis/Weeds/R-NC/Figs/hc3n_v4e1_p2.eps}}}
\centerline{\resizebox{0.675\hsize}{!}{\includegraphics[angle=0]{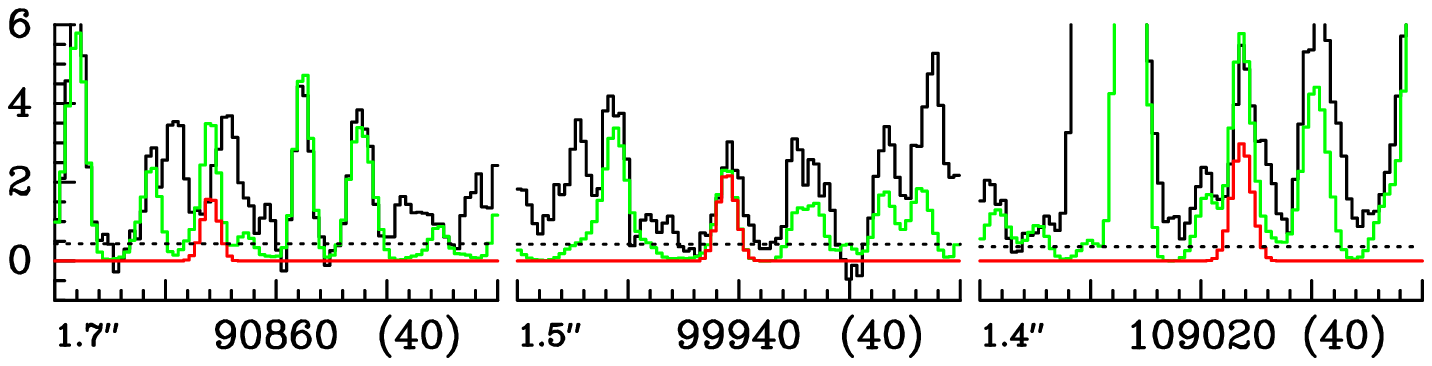}}}
\caption{Same as Fig.~\ref{f:spec_c2h5cn_ve0} for HC$_3$N, $\varv_4=1$.
}
\label{f:spec_hc3n_v4e1}
\end{figure*}
}

\onlfig{
\begin{figure*}
%\centerline{\resizebox{0.9\hsize}{!}{\includegraphics[angle=0]{/homes/belloche/SgrB2/ALMA/Analysis/Weeds/R-NC/Figs/hc3n_v7e4v5e1v7e1_p2.eps}}}
\centerline{\resizebox{0.9\hsize}{!}{\includegraphics[angle=0]{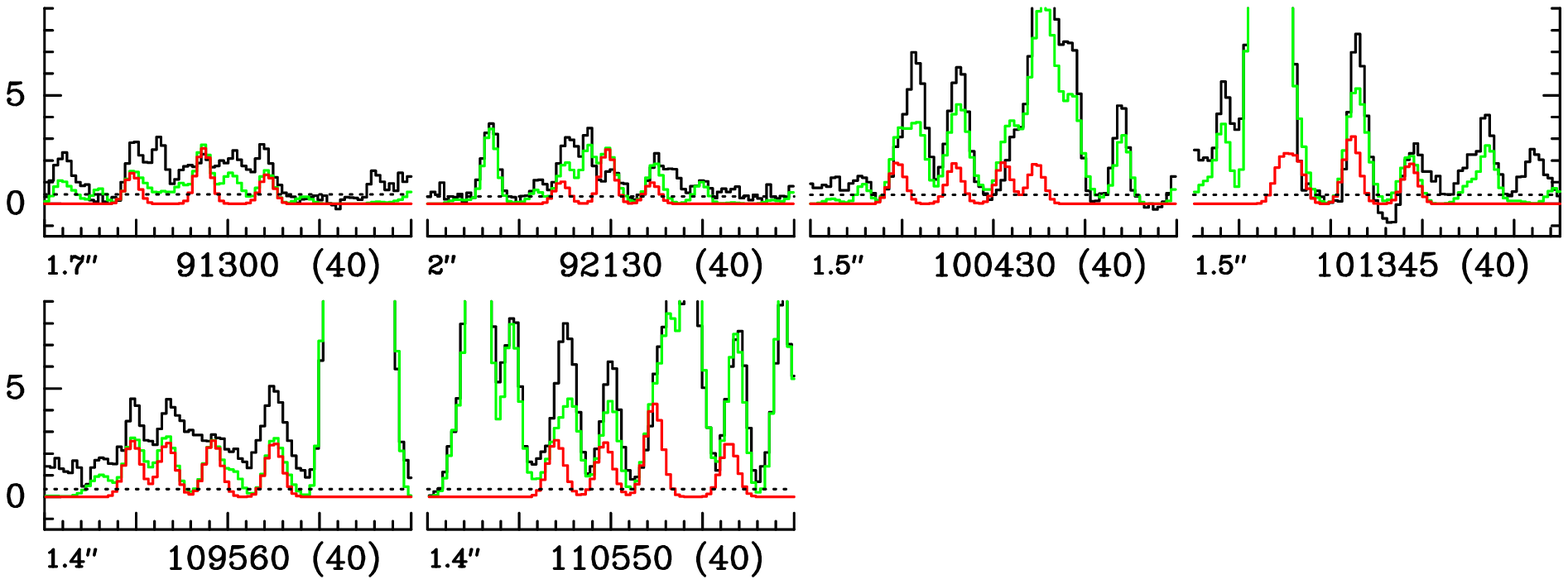}}}
\caption{Same as Fig.~\ref{f:spec_c2h5cn_ve0} for HC$_3$N, $\varv_7=4$ and
$\varv_5=\varv_7=1$.
}
\label{f:spec_hc3n_v7e4v5e1v7e1}
\end{figure*}
}

\onlfig{
\begin{figure*}
%\centerline{\resizebox{0.9\hsize}{!}{\includegraphics[angle=0]{/homes/belloche/SgrB2/ALMA/Analysis/Weeds/R-NC/Figs/hc3n_v6e2_p2.eps}}}
\centerline{\resizebox{0.9\hsize}{!}{\includegraphics[angle=0]{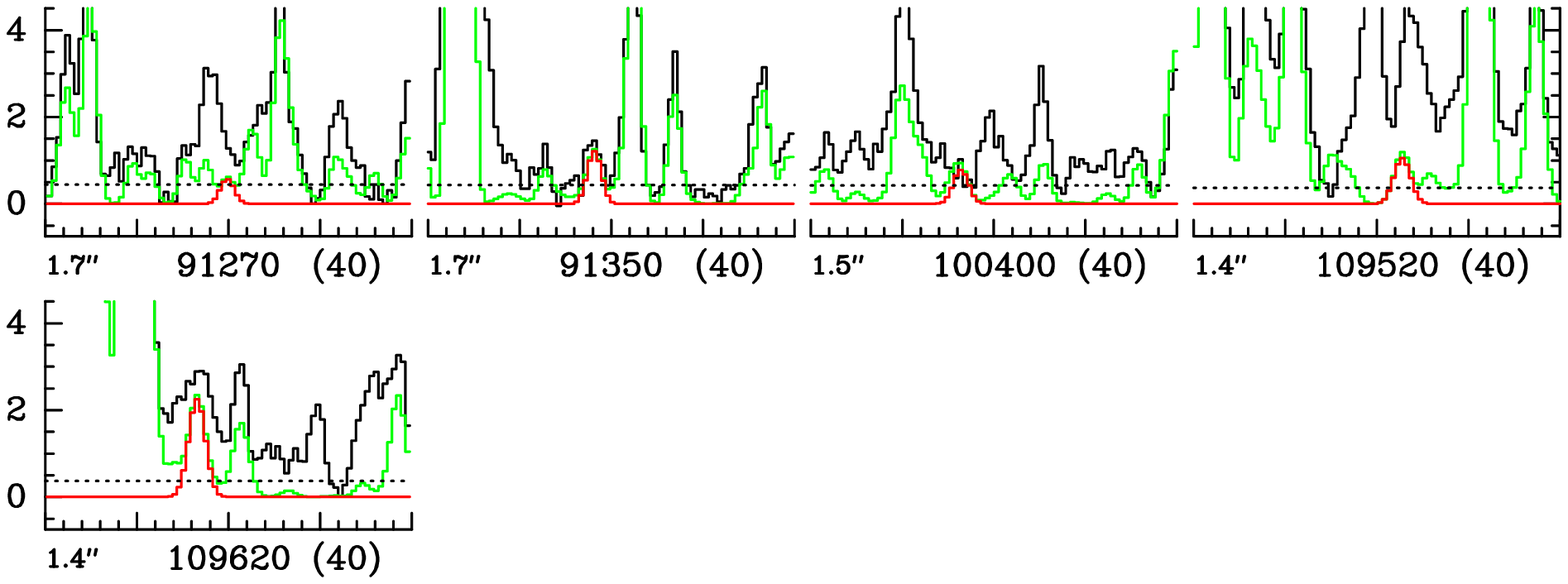}}}
\caption{Same as Fig.~\ref{f:spec_c2h5cn_ve0} for HC$_3$N, $\varv_6=2$.
}
\label{f:spec_hc3n_v6e2}
\end{figure*}
}

The three singly-substituted $^{13}$C isotopologues of cyanoacetylene are also
clearly detected in their vibrational ground state 
(Figs.~\ref{f:spec_hc3n_13c1_ve0}, \ref{f:spec_hc3n_13c2_ve0}, and 
\ref{f:spec_hc3n_13c3_ve0}) and in $\varv_7=1$ 
(Figs.~\ref{f:spec_hc3n_13c1_v7e1}, \ref{f:spec_hc3n_13c2_v7e1}, and 
\ref{f:spec_hc3n_13c3_v7e1}). HC$^{13}$CCN is also detected in $\varv_7=2$ 
(Fig.~\ref{f:spec_hc3n_13c2_v7e2}) while the two other isotopologues are only
tentatively detected in this state (Figs.~\ref{f:spec_hc3n_13c1_v7e2} and
\ref{f:spec_hc3n_13c3_v7e2}). HC$^{13}$CCN and HCC$^{13}$CN are tentatively
detected in $\varv_6=1$ with one detected line each 
(Figs.~\ref{f:spec_hc3n_13c2_v6e1} and \ref{f:spec_hc3n_13c3_v6e1}). Emission
of H$^{13}$CCCN in $\varv_6=1$ significantly contributes to the detected 
signal, but this state cannot be unambiguously identified
(Fig.~\ref{f:spec_hc3n_13c1_v6e1}).

\onlfig{
\begin{figure*}
%\centerline{\resizebox{0.675\hsize}{!}{\includegraphics[angle=0]{/homes/belloche/SgrB2/ALMA/Analysis/Weeds/R-NC/Figs/hc3n_13c1_ve0_p2.eps}}}
\centerline{\resizebox{0.675\hsize}{!}{\includegraphics[angle=0]{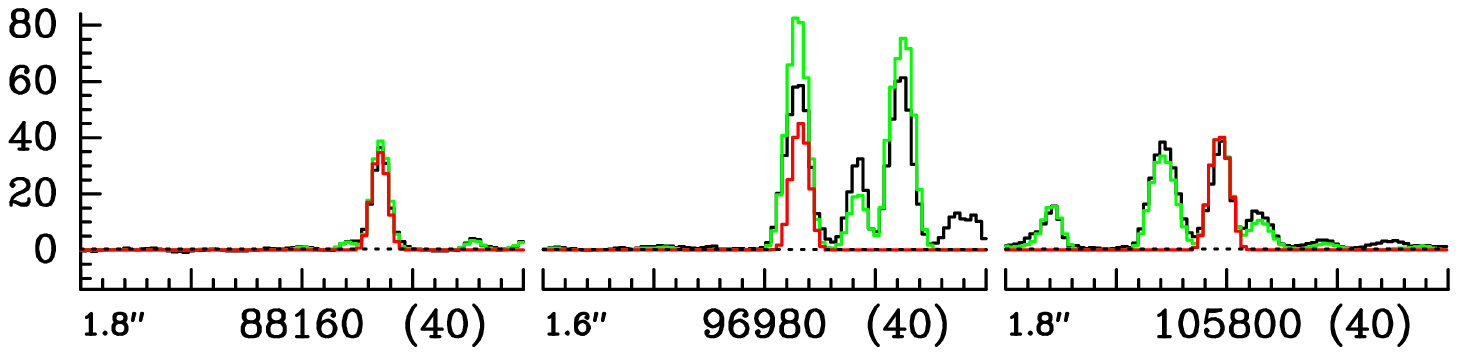}}}
\caption{Same as Fig.~\ref{f:spec_c2h5cn_ve0} for H$^{13}$CCCN, $\varv=0$.
}
\label{f:spec_hc3n_13c1_ve0}
\end{figure*}
}

\onlfig{
\clearpage
\begin{figure*}
%\centerline{\resizebox{0.675\hsize}{!}{\includegraphics[angle=0]{/homes/belloche/SgrB2/ALMA/Analysis/Weeds/R-NC/Figs/hc3n_13c2_ve0_p2.eps}}}
\centerline{\resizebox{0.675\hsize}{!}{\includegraphics[angle=0]{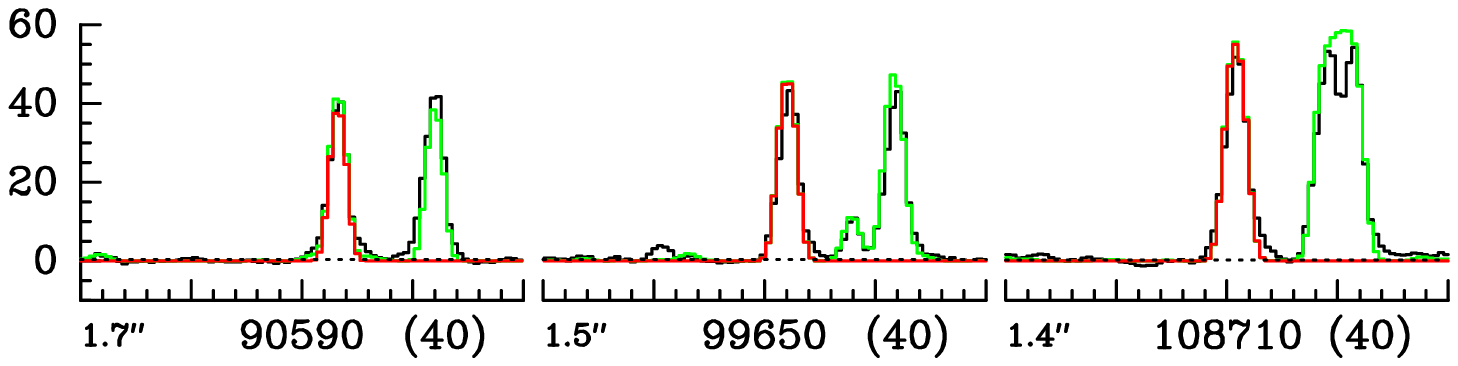}}}
\caption{Same as Fig.~\ref{f:spec_c2h5cn_ve0} for HC$^{13}$CCN, $\varv=0$.
}
\label{f:spec_hc3n_13c2_ve0}
\end{figure*}
}

\onlfig{
\begin{figure*}
%\centerline{\resizebox{0.675\hsize}{!}{\includegraphics[angle=0]{/homes/belloche/SgrB2/ALMA/Analysis/Weeds/R-NC/Figs/hc3n_13c3_ve0_p2.eps}}}
\centerline{\resizebox{0.675\hsize}{!}{\includegraphics[angle=0]{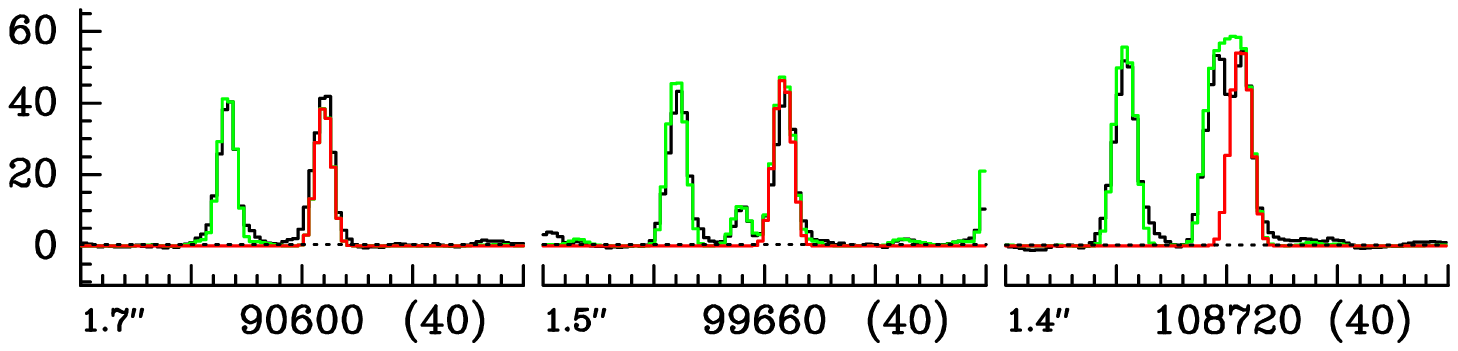}}}
\caption{Same as Fig.~\ref{f:spec_c2h5cn_ve0} for HCC$^{13}$CN, $\varv=0$.
}
\label{f:spec_hc3n_13c3_ve0}
\end{figure*}
}

\onlfig{
\begin{figure*}
%\centerline{\resizebox{0.9\hsize}{!}{\includegraphics[angle=0]{/homes/belloche/SgrB2/ALMA/Analysis/Weeds/R-NC/Figs/hc3n_13c1_v7e1_p2.eps}}}
\centerline{\resizebox{0.9\hsize}{!}{\includegraphics[angle=0]{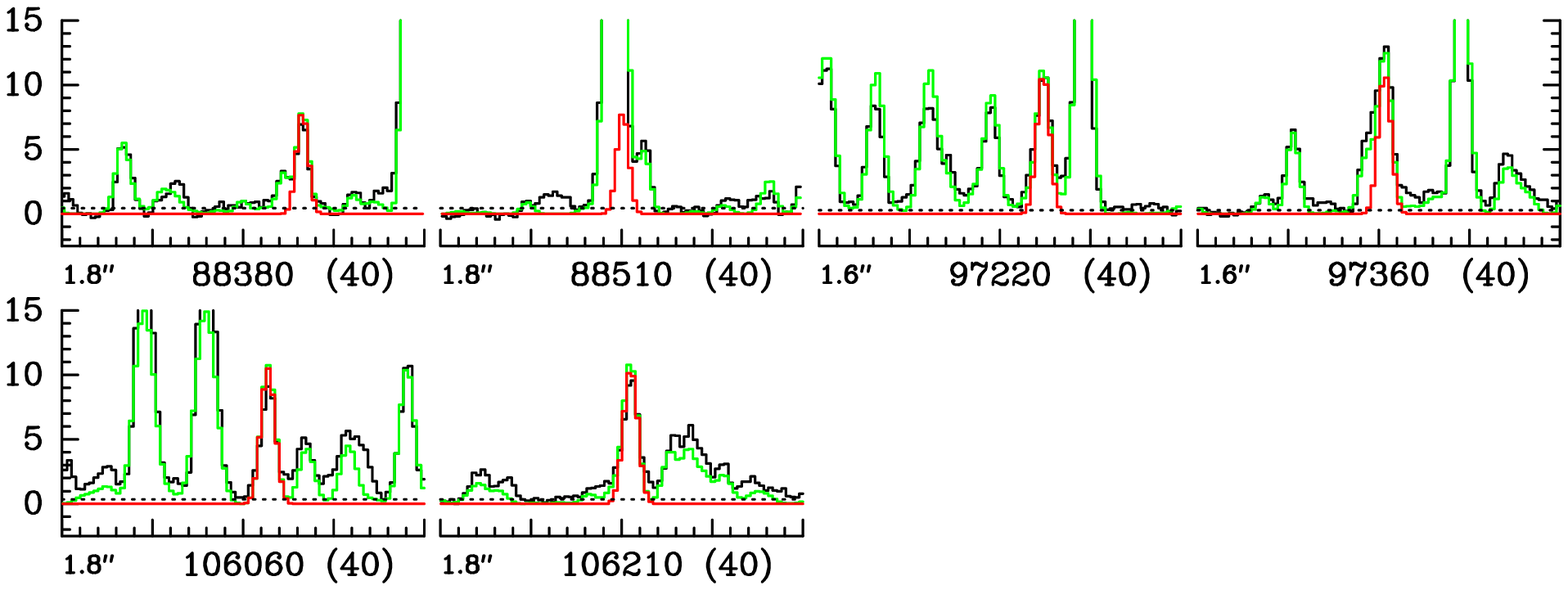}}}
\caption{Same as Fig.~\ref{f:spec_c2h5cn_ve0} for H$^{13}$CCCN, $\varv_7=1$.
}
\label{f:spec_hc3n_13c1_v7e1}
\end{figure*}
}

\onlfig{
\begin{figure*}
%\centerline{\resizebox{0.9\hsize}{!}{\includegraphics[angle=0]{/homes/belloche/SgrB2/ALMA/Analysis/Weeds/R-NC/Figs/hc3n_13c2_v7e1_p2.eps}}}
\centerline{\resizebox{0.9\hsize}{!}{\includegraphics[angle=0]{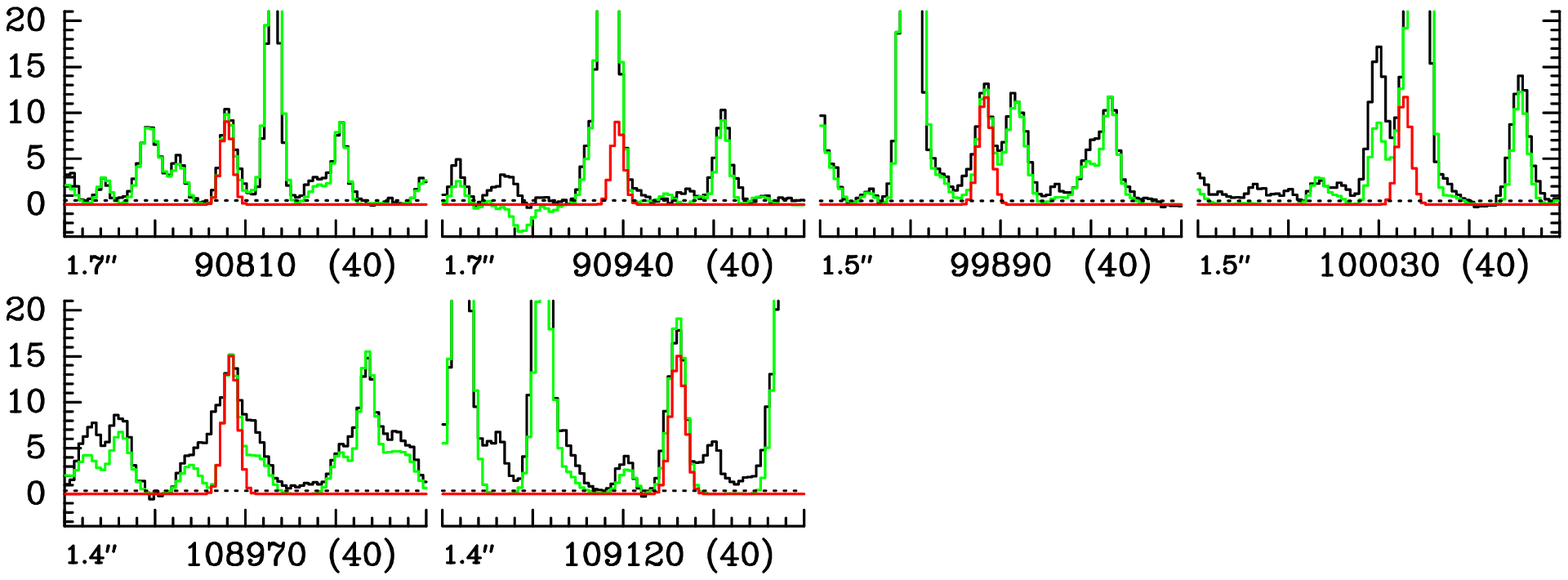}}}
\caption{Same as Fig.~\ref{f:spec_c2h5cn_ve0} for HC$^{13}$CCN, $\varv_7=1$.
}
\label{f:spec_hc3n_13c2_v7e1}
\end{figure*}
}

\onlfig{
\clearpage
\begin{figure*}
%\centerline{\resizebox{0.9\hsize}{!}{\includegraphics[angle=0]{/homes/belloche/SgrB2/ALMA/Analysis/Weeds/R-NC/Figs/hc3n_13c3_v7e1_p2.eps}}}
\centerline{\resizebox{0.9\hsize}{!}{\includegraphics[angle=0]{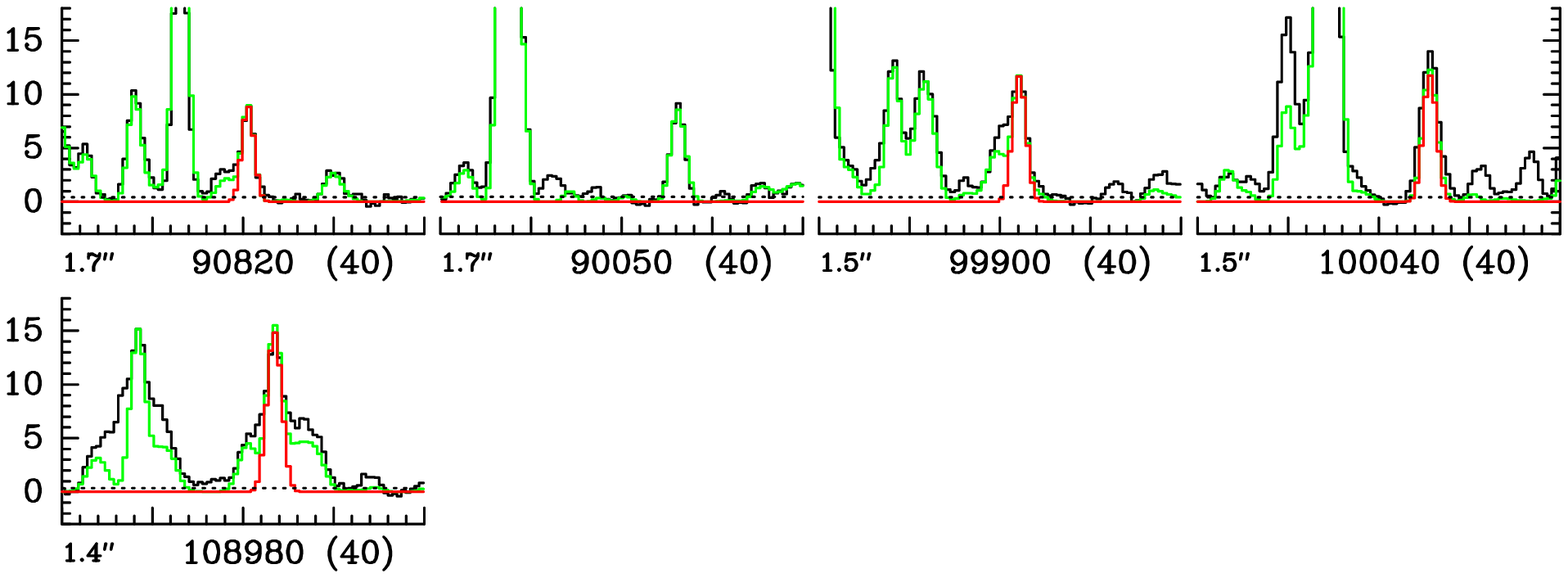}}}
\caption{Same as Fig.~\ref{f:spec_c2h5cn_ve0} for HCC$^{13}$CN, $\varv_7=1$.
}
\label{f:spec_hc3n_13c3_v7e1}
\end{figure*}
}

\onlfig{
\begin{figure*}
%\centerline{\resizebox{0.675\hsize}{!}{\includegraphics[angle=0]{/homes/belloche/SgrB2/ALMA/Analysis/Weeds/R-NC/Figs/hc3n_13c2_v7e2_p2.eps}}}
\centerline{\resizebox{0.675\hsize}{!}{\includegraphics[angle=0]{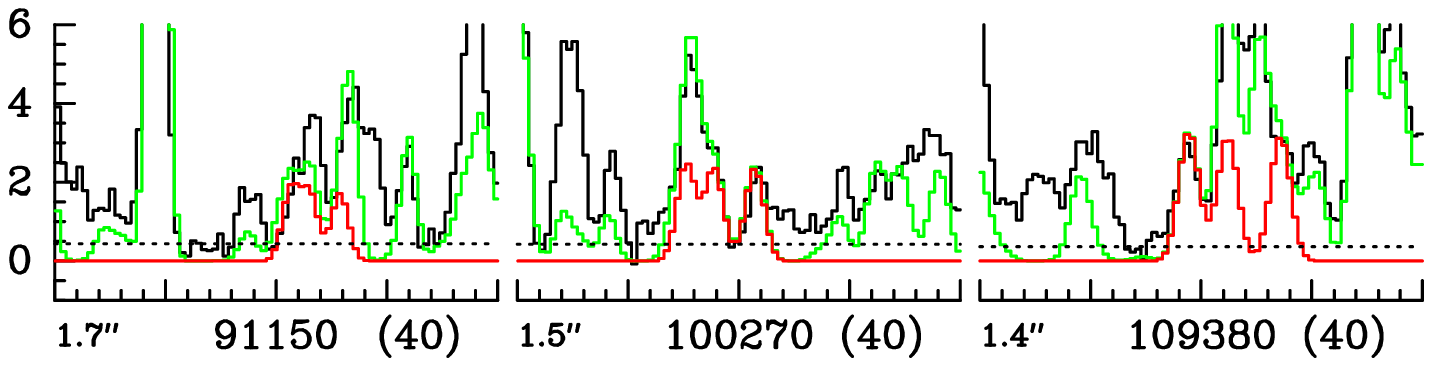}}}
\caption{Same as Fig.~\ref{f:spec_c2h5cn_ve0} for HC$^{13}$CCN, $\varv_7=2$.
}
\label{f:spec_hc3n_13c2_v7e2}
\end{figure*}
}

\onlfig{
\begin{figure*}
%\centerline{\resizebox{0.675\hsize}{!}{\includegraphics[angle=0]{/homes/belloche/SgrB2/ALMA/Analysis/Weeds/R-NC/Figs/hc3n_13c1_v7e2_p2.eps}}}
\centerline{\resizebox{0.675\hsize}{!}{\includegraphics[angle=0]{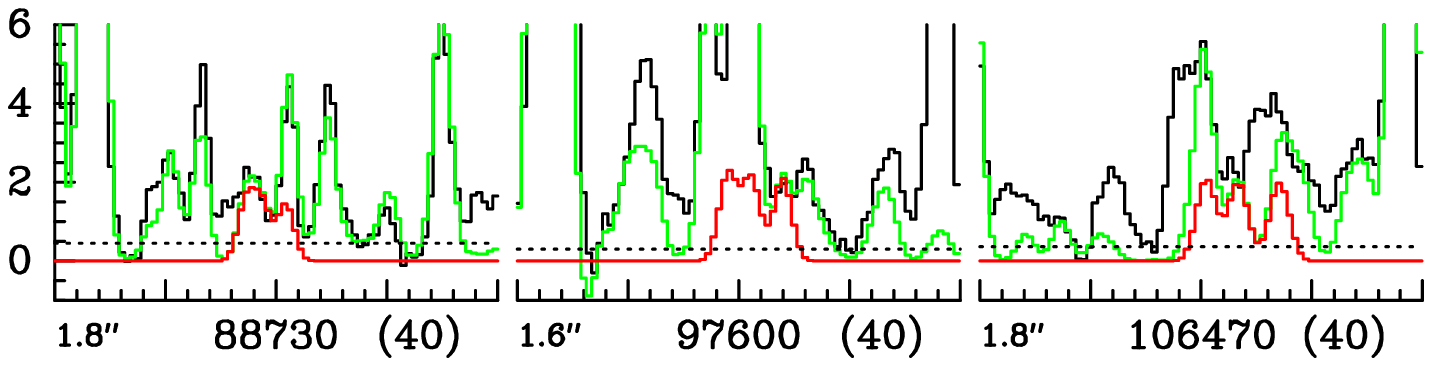}}}
\caption{Same as Fig.~\ref{f:spec_c2h5cn_ve0} for H$^{13}$CCCN, $\varv_7=2$.
}
\label{f:spec_hc3n_13c1_v7e2}
\end{figure*}
}

\onlfig{
\begin{figure*}
%\centerline{\resizebox{0.675\hsize}{!}{\includegraphics[angle=0]{/homes/belloche/SgrB2/ALMA/Analysis/Weeds/R-NC/Figs/hc3n_13c3_v7e2_p2.eps}}}
\centerline{\resizebox{0.675\hsize}{!}{\includegraphics[angle=0]{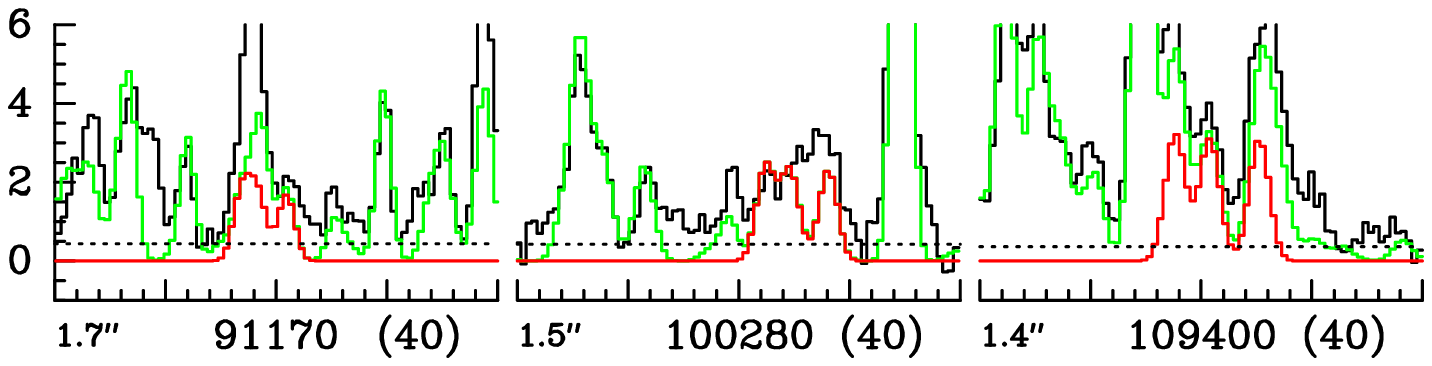}}}
\caption{Same as Fig.~\ref{f:spec_c2h5cn_ve0} for HCC$^{13}$CN, $\varv_7=2$.
}
\label{f:spec_hc3n_13c3_v7e2}
\end{figure*}
}

\onlfig{
\begin{figure*}
%\centerline{\resizebox{0.9\hsize}{!}{\includegraphics[angle=0]{/homes/belloche/SgrB2/ALMA/Analysis/Weeds/R-NC/Figs/hc3n_13c2_v6e1_p2.eps}}}
\centerline{\resizebox{0.9\hsize}{!}{\includegraphics[angle=0]{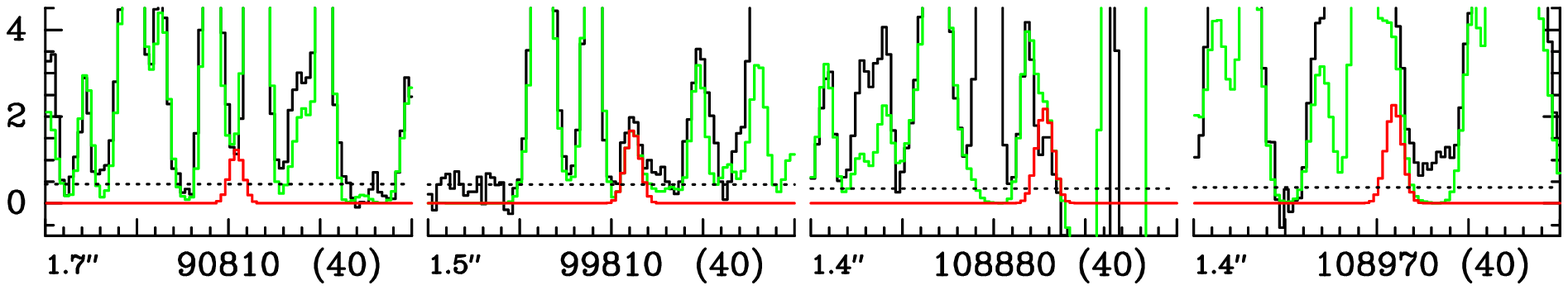}}}
\caption{Same as Fig.~\ref{f:spec_c2h5cn_ve0} for HC$^{13}$CCN, $\varv_6=1$.
}
\label{f:spec_hc3n_13c2_v6e1}
\end{figure*}
}

\onlfig{
\clearpage
\begin{figure*}
%\centerline{\resizebox{0.9\hsize}{!}{\includegraphics[angle=0]{/homes/belloche/SgrB2/ALMA/Analysis/Weeds/R-NC/Figs/hc3n_13c3_v6e1_p2.eps}}}
\centerline{\resizebox{0.9\hsize}{!}{\includegraphics[angle=0]{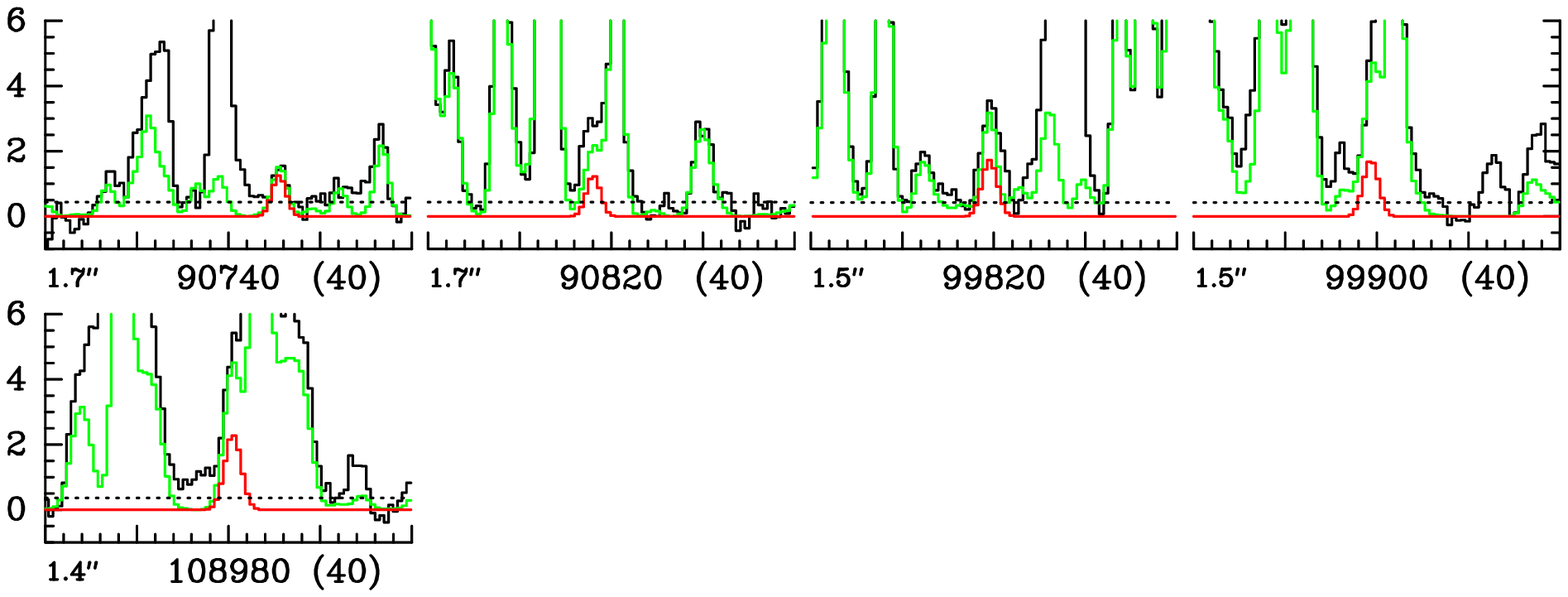}}}
\caption{Same as Fig.~\ref{f:spec_c2h5cn_ve0} for HCC$^{13}$CN, $\varv_6=1$.
}
\label{f:spec_hc3n_13c3_v6e1}
\end{figure*}
}

\onlfig{
\begin{figure*}
%\centerline{\resizebox{0.9\hsize}{!}{\includegraphics[angle=0]{/homes/belloche/SgrB2/ALMA/Analysis/Weeds/R-NC/Figs/hc3n_13c1_v6e1_p2.eps}}}
\centerline{\resizebox{0.9\hsize}{!}{\includegraphics[angle=0]{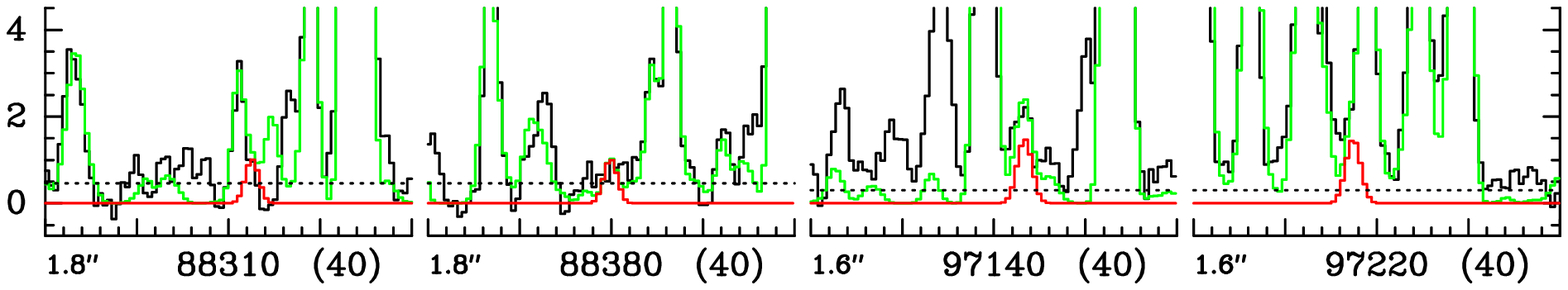}}}
\caption{Same as Fig.~\ref{f:spec_c2h5cn_ve0} for H$^{13}$CCCN, $\varv_6=1$.
}
\label{f:spec_hc3n_13c1_v6e1}
\end{figure*}
}

Two doubly-substituted $^{13}$C isotopologues of cyanoacetylene, 
H$^{13}$C$^{13}$CCN and HC$^{13}$C$^{13}$CN are tentatively detected in their 
vibrational ground state with one line each 
(Figs.~\ref{f:spec_hc3n_13c113c2_ve0} and 
\ref{f:spec_hc3n_13c213c3_ve0}). The third one, H$^{13}$CC$^{13}$CN has no 
clearly detected line, but the model using the same parameters as the former
two isotopologues is fully consistent with the signal detected around 
105328~MHz (Fig.~\ref{f:spec_hc3n_13c113c3_ve0}). Therefore we consider this 
species as tentatively detected too.

\onlfig{
\begin{figure*}
%\centerline{\resizebox{0.9\hsize}{!}{\includegraphics[angle=0]{/homes/belloche/SgrB2/ALMA/Analysis/Weeds/R-NC/Figs/hc3n_13c113c2_ve0_p2.eps}}}
\centerline{\resizebox{0.9\hsize}{!}{\includegraphics[angle=0]{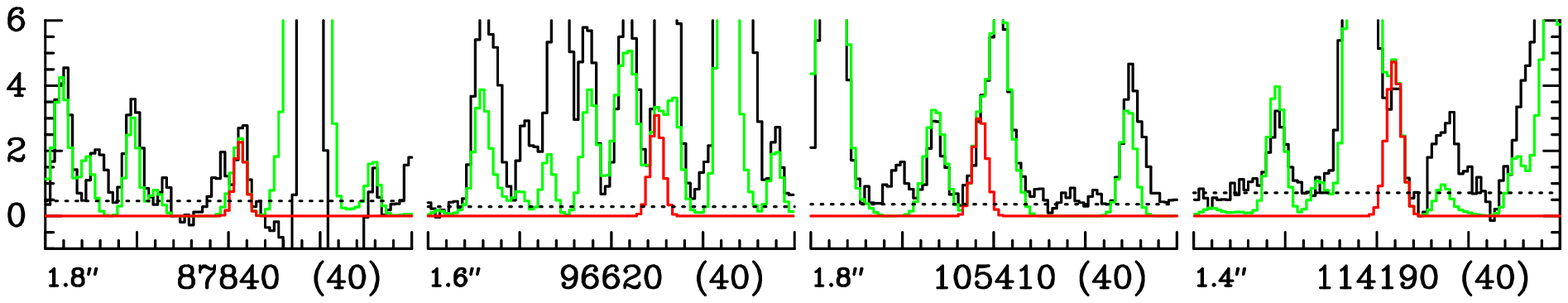}}}
\caption{Same as Fig.~\ref{f:spec_c2h5cn_ve0} for H$^{13}$C$^{13}$CCN, $\varv=0$.
}
\label{f:spec_hc3n_13c113c2_ve0}
\end{figure*}
}

\onlfig{
\begin{figure*}
%\centerline{\resizebox{0.675\hsize}{!}{\includegraphics[angle=0]{/homes/belloche/SgrB2/ALMA/Analysis/Weeds/R-NC/Figs/hc3n_13c213c3_ve0_p2.eps}}}
\centerline{\resizebox{0.675\hsize}{!}{\includegraphics[angle=0]{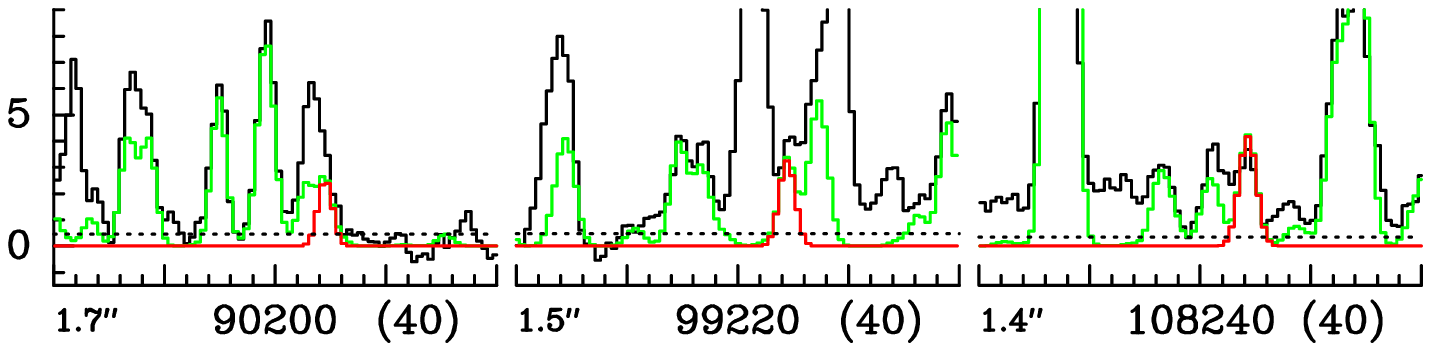}}}
\caption{Same as Fig.~\ref{f:spec_c2h5cn_ve0} for HC$^{13}$C$^{13}$CN, $\varv=0$.
}
\label{f:spec_hc3n_13c213c3_ve0}
\end{figure*}
}

\onlfig{
\begin{figure*}
%\centerline{\resizebox{0.675\hsize}{!}{\includegraphics[angle=0]{/homes/belloche/SgrB2/ALMA/Analysis/Weeds/R-NC/Figs/hc3n_13c113c3_ve0_p2.eps}}}
\centerline{\resizebox{0.675\hsize}{!}{\includegraphics[angle=0]{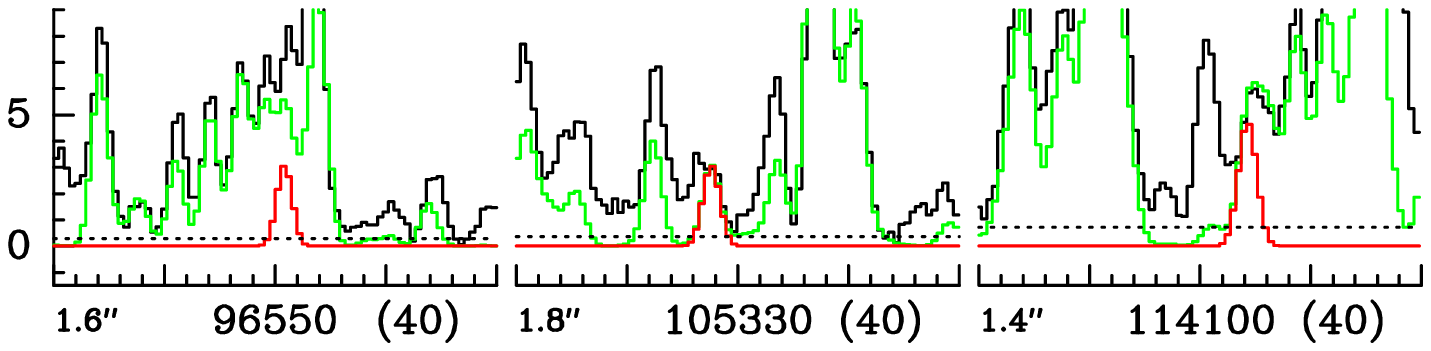}}}
\caption{Same as Fig.~\ref{f:spec_c2h5cn_ve0} for H$^{13}$CC$^{13}$CN, $\varv=0$.
}
\label{f:spec_hc3n_13c113c3_ve0}
\end{figure*}
}

The $^{15}$N isotopologue HC$_3$$^{15}$N is not unambiguously detected in its
vibrational ground state but, if we assume a $^{14}$N/$^{15}$N isotopic ratio
of 300, it contributes significantly to the detected 
signal at 88334~MHz and 105999~MHz and was therefore included in our model
(Fig.~\ref{f:spec_hc3n_15n_ve0}). Its column density should be rather 
considered as an upper limit.

\onlfig{
\clearpage
\begin{figure*}
%\centerline{\resizebox{0.675\hsize}{!}{\includegraphics[angle=0]{/homes/belloche/SgrB2/ALMA/Analysis/Weeds/R-NC/Figs/hc3n_15n_ve0_p2.eps}}}
\centerline{\resizebox{0.675\hsize}{!}{\includegraphics[angle=0]{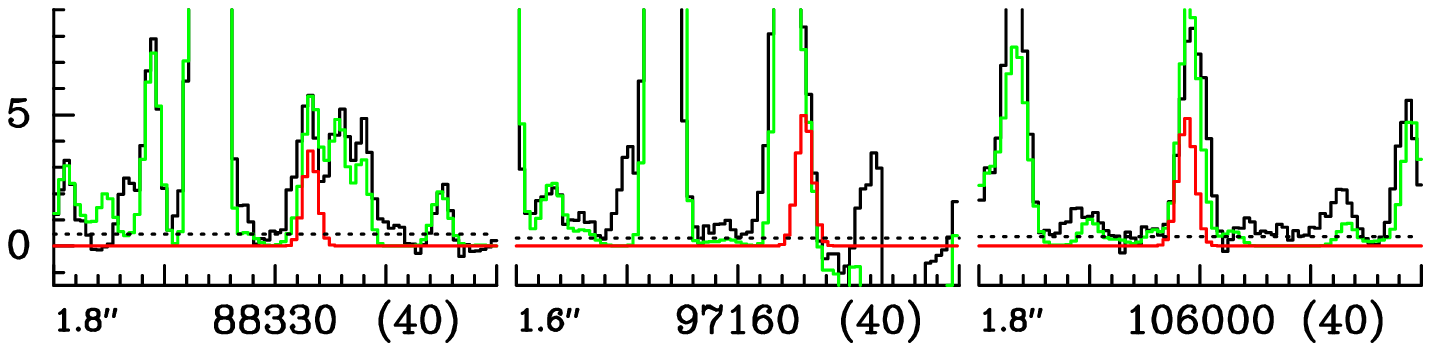}}}
\caption{Same as Fig.~\ref{f:spec_c2h5cn_ve0} for HC$_3$$^{15}$N, $\varv=0$.
}
\label{f:spec_hc3n_15n_ve0}
\end{figure*}
}

The fits to the integrated intensity maps suggest that the size of the 
emission decreases with increasing energy of the vibrational state from within
which the lines are emitted. Since our model cannot 
account for a non-uniform physical structure, we defined two groups of 
vibrational states: $\varv=0$ and $\varv_7=1$ were modeled with a source size 
of $1.3\arcsec$ while the higher excited states were modeled assuming 
$0.9\arcsec$.

The fits to the population diagrams of the singly-substituted $^{13}$C 
isotopologues including both $\varv=0$ and $\varv_7=1$ yield rotational 
temperatures of $\sim 170$--180~K (Table~\ref{t:popfit}, 
Figs.~\ref{f:popdiag_hc3n_13c1}, \ref{f:popdiag_hc3n_13c2}, and 
\ref{f:popdiag_hc3n_13c3}). With a temperature of 170~K and a source size of 
$1.3\arcsec$, the emission of all isotopologues reported above is well fitted 
up to $\varv_7=1$, except for the vibrational ground state of HC$_3$N: its 
transitions are very optically thick ($\tau_{\rm max} \sim 30$) and cannot be 
reproduced with our simple model. For the vibrationally excited states of the 
main and singly-substituted $^{13}$C isotopologues above $\varv_7=1$, we 
assume a source size of $0.9\arcsec$ and obtain a very good fit to the 
observed spectra with a temperature of 200~K and a unique column density 
(divided by 20 for the $^{13}$C isotopologues), 1.5 times higher than for the 
model of the lower states.

\onlfig{
\clearpage
\begin{figure}
%\centerline{\resizebox{1.0\hsize}{!}{\includegraphics[angle=0]{/homes/belloche/SgrB2/ALMA/Analysis/Weeds/R-NC/popdiag_hc3n_13c1_p2.eps}}}
\centerline{\resizebox{1.0\hsize}{!}{\includegraphics[angle=0]{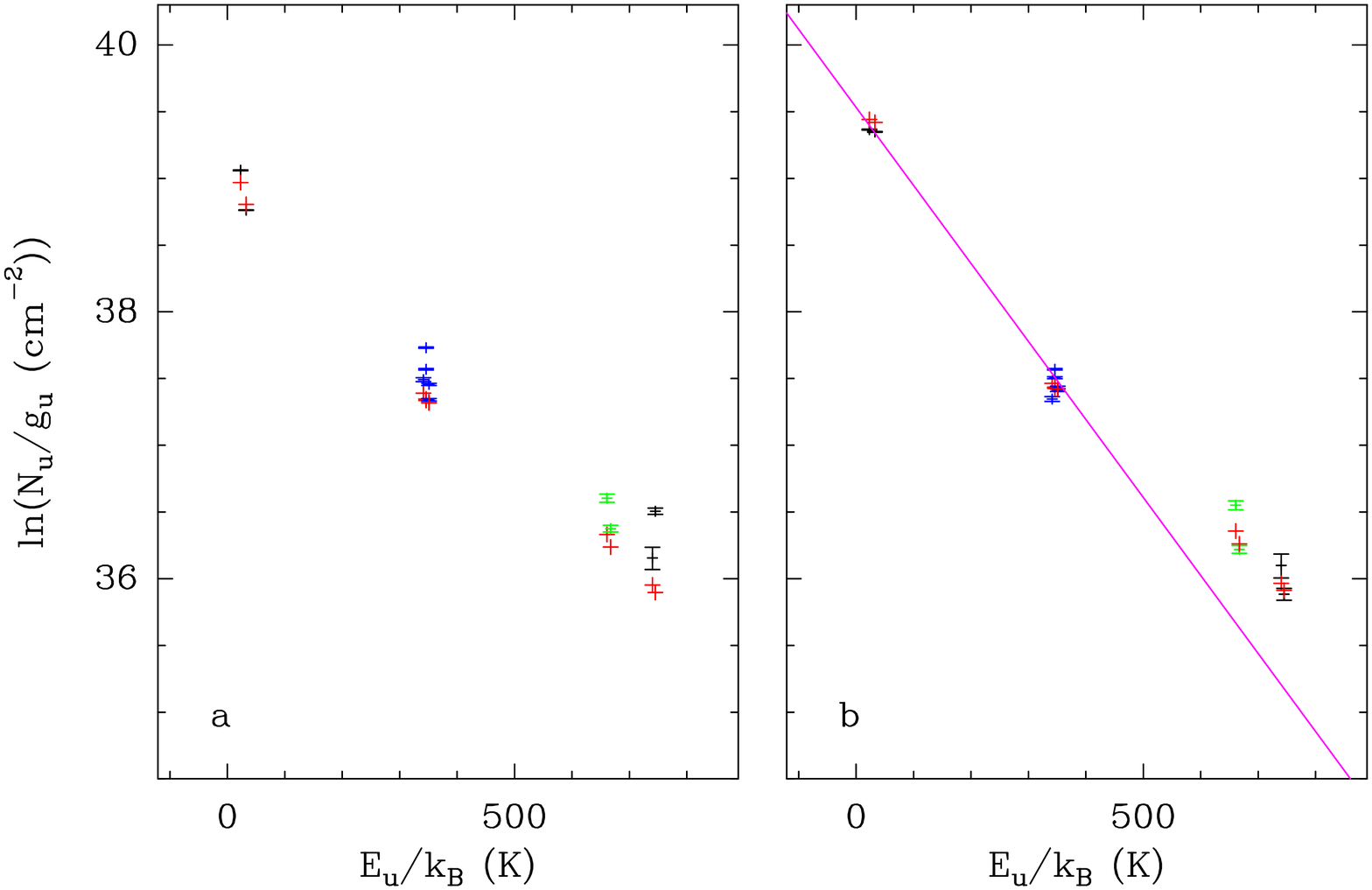}}}
\caption{Same as Fig.~\ref{f:popdiag_c2h5cn} for H$^{13}$CCCN, $\varv=0$, 
$\varv_7=1$, $\varv_7=2$, and $\varv_6=1$.}
\label{f:popdiag_hc3n_13c1}
\end{figure}
}

\onlfig{
\begin{figure}
%\centerline{\resizebox{1.0\hsize}{!}{\includegraphics[angle=0]{/homes/belloche/SgrB2/ALMA/Analysis/Weeds/R-NC/popdiag_hc3n_13c2_p2.eps}}}
\centerline{\resizebox{1.0\hsize}{!}{\includegraphics[angle=0]{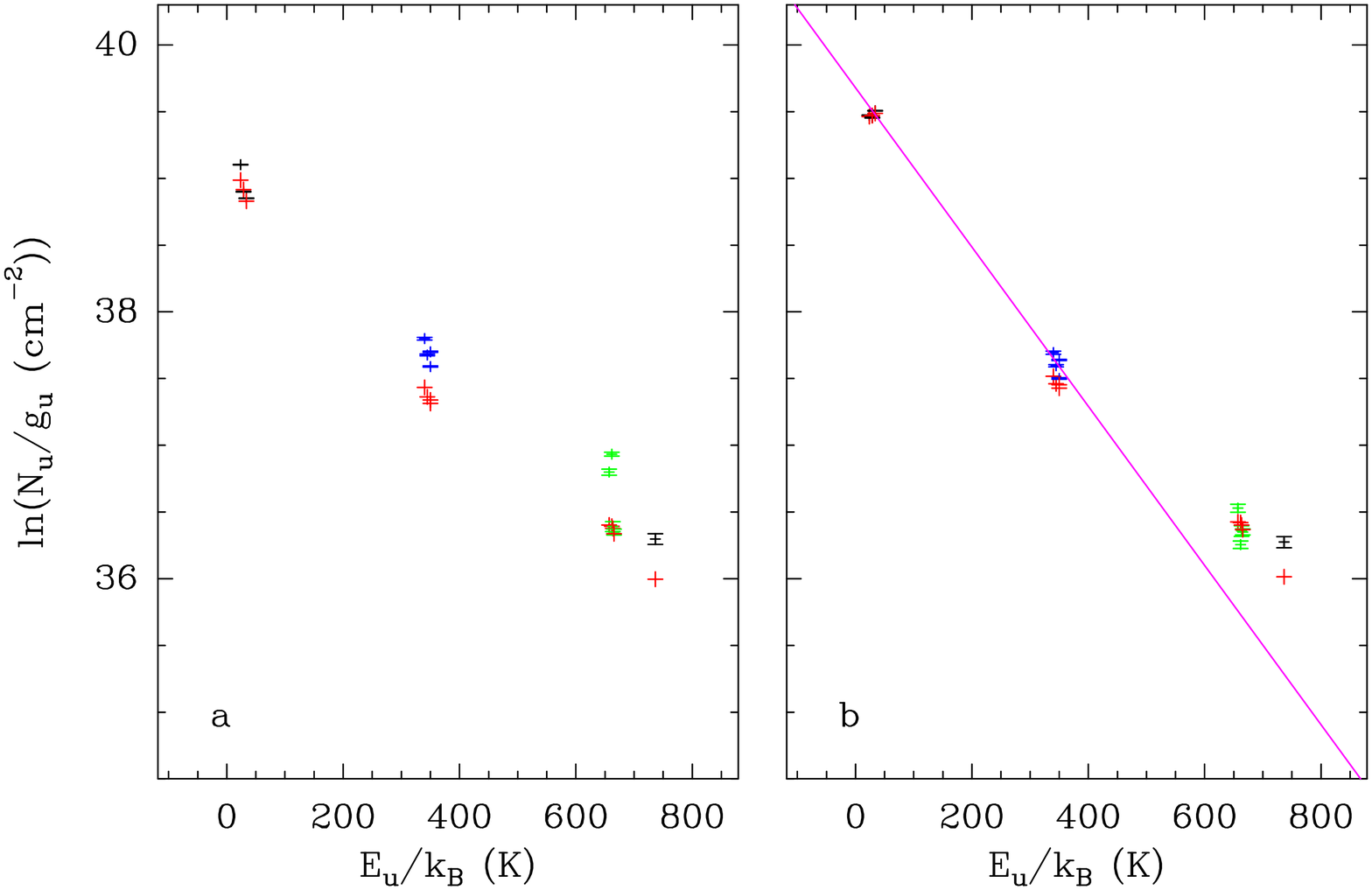}}}
\caption{Same as Fig.~\ref{f:popdiag_c2h5cn} for HC$^{13}$CCN, $\varv=0$, 
$\varv_7=1$, $\varv_7=2$, and $\varv_6=1$.}
\label{f:popdiag_hc3n_13c2}
\end{figure}
}

\onlfig{
\begin{figure}
%\centerline{\resizebox{1.0\hsize}{!}{\includegraphics[angle=0]{/homes/belloche/SgrB2/ALMA/Analysis/Weeds/R-NC/popdiag_hc3n_13c3_p2.eps}}}
\centerline{\resizebox{1.0\hsize}{!}{\includegraphics[angle=0]{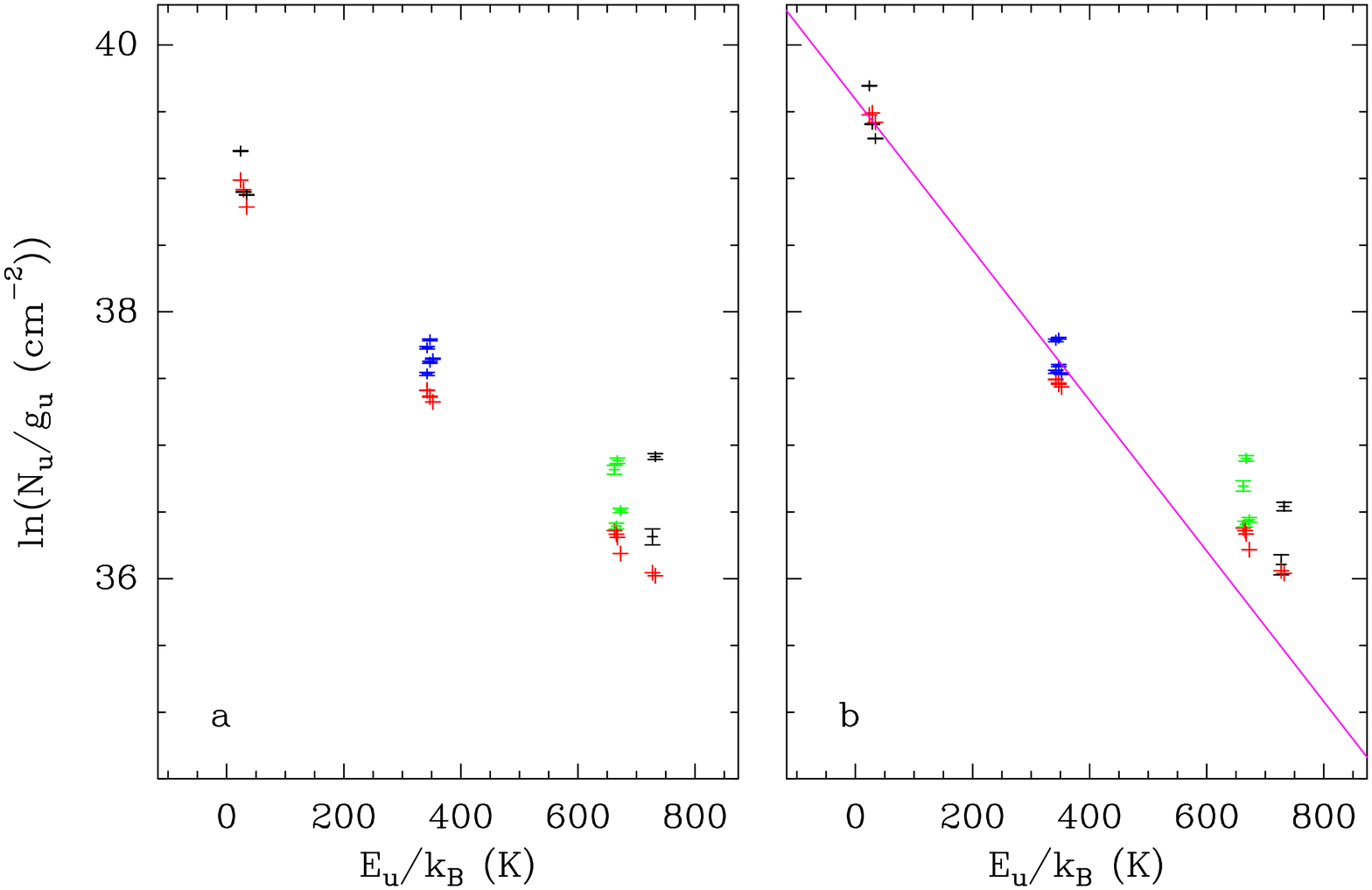}}}
\caption{Same as Fig.~\ref{f:popdiag_c2h5cn} for HCC$^{13}$CN, $\varv=0$, 
$\varv_7=1$, $\varv_7=2$, and $\varv_6=1$.}
\label{f:popdiag_hc3n_13c3}
\end{figure}
}

Using the parameters derived above for the vibrational ground state, we looked
for emission of deuterated cyanoacetylene DC$_3$N. The molecule seems to 
contribute at a level of $\sim 70\%$ to the signal detected at 101315~MHz 
(Fig.~\ref{f:spec_dc3n}). The rest of the emission comes from a transition of 
CH$_2$CO in its vibrationally excited state $\varv_9=1$. The detection of
DC$_3$N is only tentative, and its column density should be rather considered
as an upper limit.

\onlfig{
\clearpage
\begin{figure*}
%\centerline{\resizebox{0.9\hsize}{!}{\includegraphics[angle=0]{/homes/belloche/SgrB2/ALMA/Analysis/Weeds/R-D/Figs/dc3n_ve0_p2.eps}}}
\centerline{\resizebox{0.9\hsize}{!}{\includegraphics[angle=0]{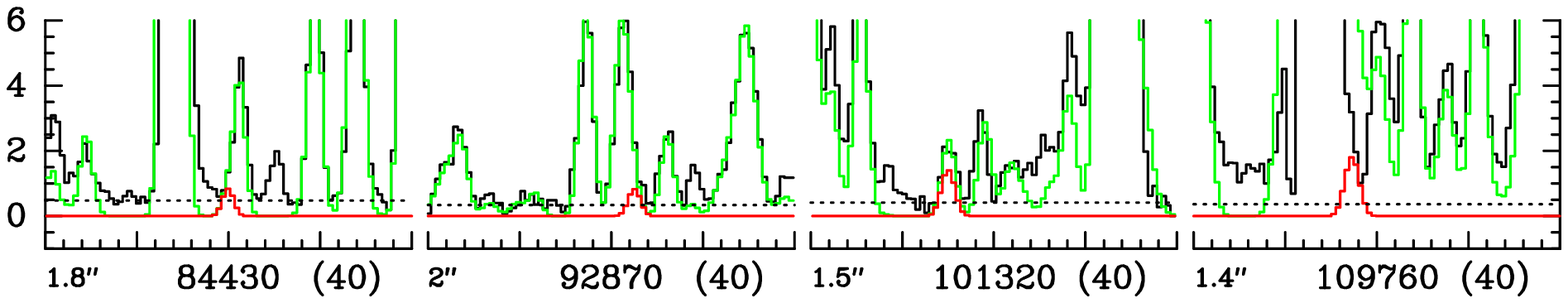}}}
\caption{Same as Fig.~\ref{f:spec_c2h5cn_ve0} for DC$_3$N, $\varv=0$.
}
\label{f:spec_dc3n}
\end{figure*}
}

\subsection{Deuterated methanol CH$_2$DOH}
\label{ss:ch3oh}

Methanol and its $^{13}$C and $^{18}$O isotopologues are well detected toward 
Sgr~B2(N2). The detected lines and detailed modeling of these species is
presented in a companion paper \citep[][]{Mueller15c}. We report in 
Table~\ref{t:coldens} the parameters derived in that paper for the main 
isotopologue based on the analysis of all isotopologues.

Assuming the same source size and rotational temperature, we obtain a tentative
detection of CH$_2$DOH, with two lines detected at 
91587~MHz ($4_{1,3}$ \textit{e}$_0$ -- $4_{0,4}$ \textit{e}$_0$) and 
99672~MHz ($6_{1,5}$ \textit{e}$_0$ -- $6_{0,6}$ \textit{e}$_0$),
two lines tentatively detected at 
85600~MHz ($6_{2,4}$ \textit{e}$_1$ -- $6_{1,6}$ \textit{o}$_1$) and 
94563~MHz ($1_{1,0}$ \textit{o}$_1$ -- $1_{0,1}$ \textit{o}$_1$), and a few other
lines contributing significantly to the detected signal 
(Fig.~\ref{f:spec_ch2doh}). The line appearing at 91589~MHz in the full 
synthetic model with no counterpart in the observed spectrum corresponds to two
transitions of acetone ($23_{18,6}$--$23_{17,7}$ of the EE state and 
$23_{18,6}$--$23_{17,7}$ of the AE state). While acetone is unambiguously 
detected in our ALMA spectrum of Sgr~B2(N2), a significant number of predicted 
lines of acetone do not match the observed spectrum. The 
spectroscopic predictions are not accurate enough for this set of problematic 
lines, the line at 91589~MHz being one of those. The ALMA spectrum suggests
that the true frequency could be 91592~MHz for this acetone line. The source
size derived from the maps of the two detected CH$_2$DOH lines is uncertain 
but the emission looks compact in the integrated intensity maps and is 
consistent with the source size assumed for the modeling.

\onlfig{
\clearpage
\begin{figure*}
%\centerline{\resizebox{0.9\hsize}{!}{\includegraphics[angle=0]{/homes/belloche/SgrB2/ALMA/Analysis/Weeds/R-D/Figs/ch2doh_ve0_p2.eps}}}
\centerline{\resizebox{0.9\hsize}{!}{\includegraphics[angle=0]{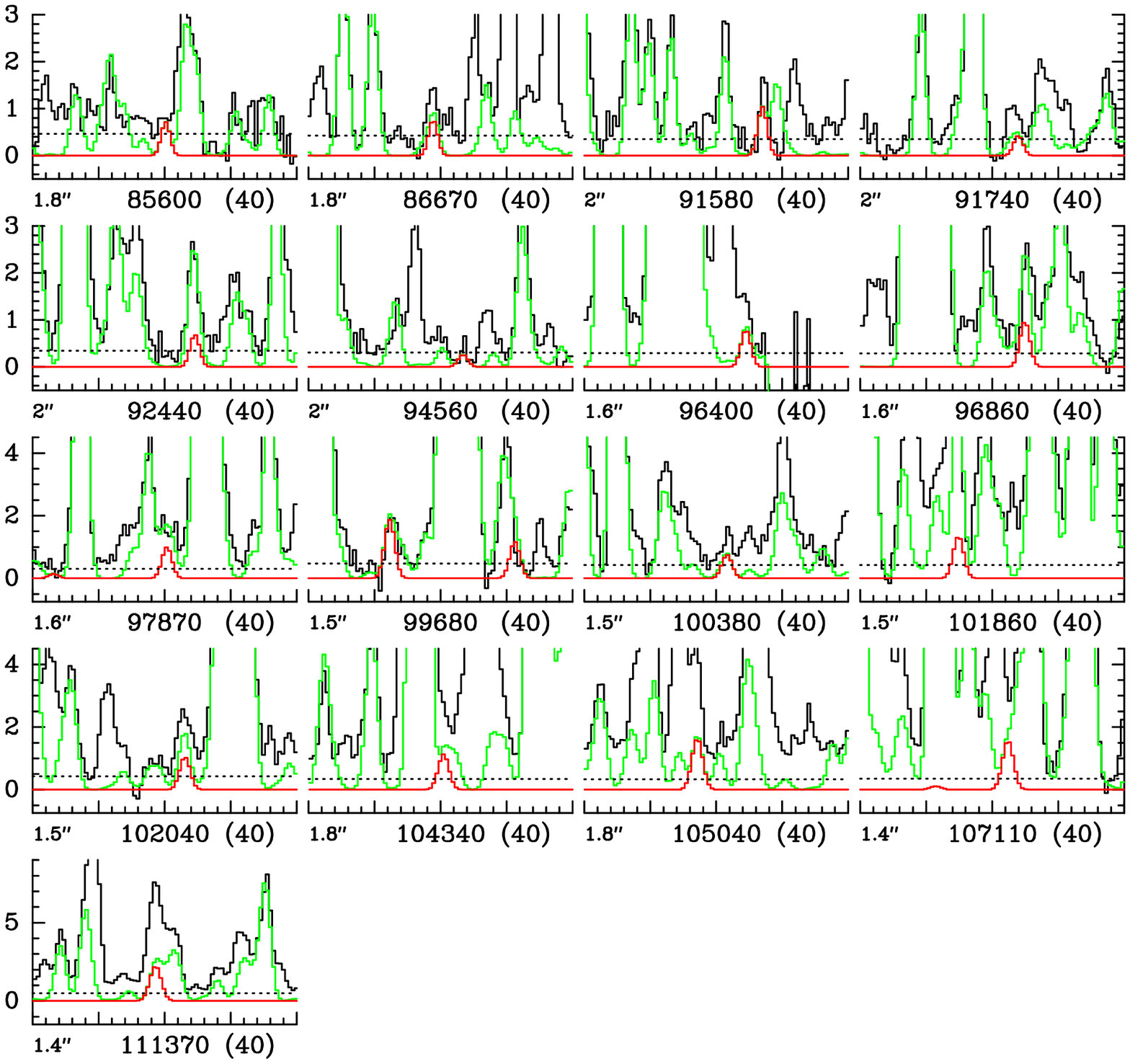}}}
\caption{Same as Fig.~\ref{f:spec_c2h5cn_ve0} for CH$_2$DOH, $\varv=0$.
}
\label{f:spec_ch2doh}
\end{figure*}
}

\subsection{Upper limits}
\label{ss:upper_limits}

\subsubsection{Deuterated methanol CH$_3$OD}
\label{sss:ch3od}

CH$_3$OD is not unambiguously detected toward Sgr~B2(N2). It may significantly 
contribute to the emission detected at 
90743~MHz (blend of $10_{1,1}$ -- $9_{2,1}$ and $2_{1,0}$ -- $1_{1,0}$), 
110951~MHz ($4_{1,0}$ -- $4_{0,0}$),
111846~MHz ($5_{1,0}$ -- $5_{0,0}$), 
and 113352~MHz ($6_{1,0}$ -- $6_{0,0}$), 
but there is no clearly detected line 
(Fig.~\ref{f:spec_ch3od_ve0}). Using the same parameters as for methanol 
(Sect.~\ref{ss:ch3oh}), we derive a column density upper limit a factor 1.8
times lower than the column density tentatively derived for CH$_2$DOH 
(Table~\ref{t:coldens}). This upper limit corresponds to the synthetic spectrum
shown in red in Fig.~\ref{f:spec_ch3od_ve0}.

\onlfig{
\clearpage
\begin{figure*}
%\centerline{\resizebox{0.9\hsize}{!}{\includegraphics[angle=0]{/homes/belloche/SgrB2/ALMA/Analysis/Weeds/R-D/Figs/ch3od_ve0_p2.eps}}}
\centerline{\resizebox{0.9\hsize}{!}{\includegraphics[angle=0]{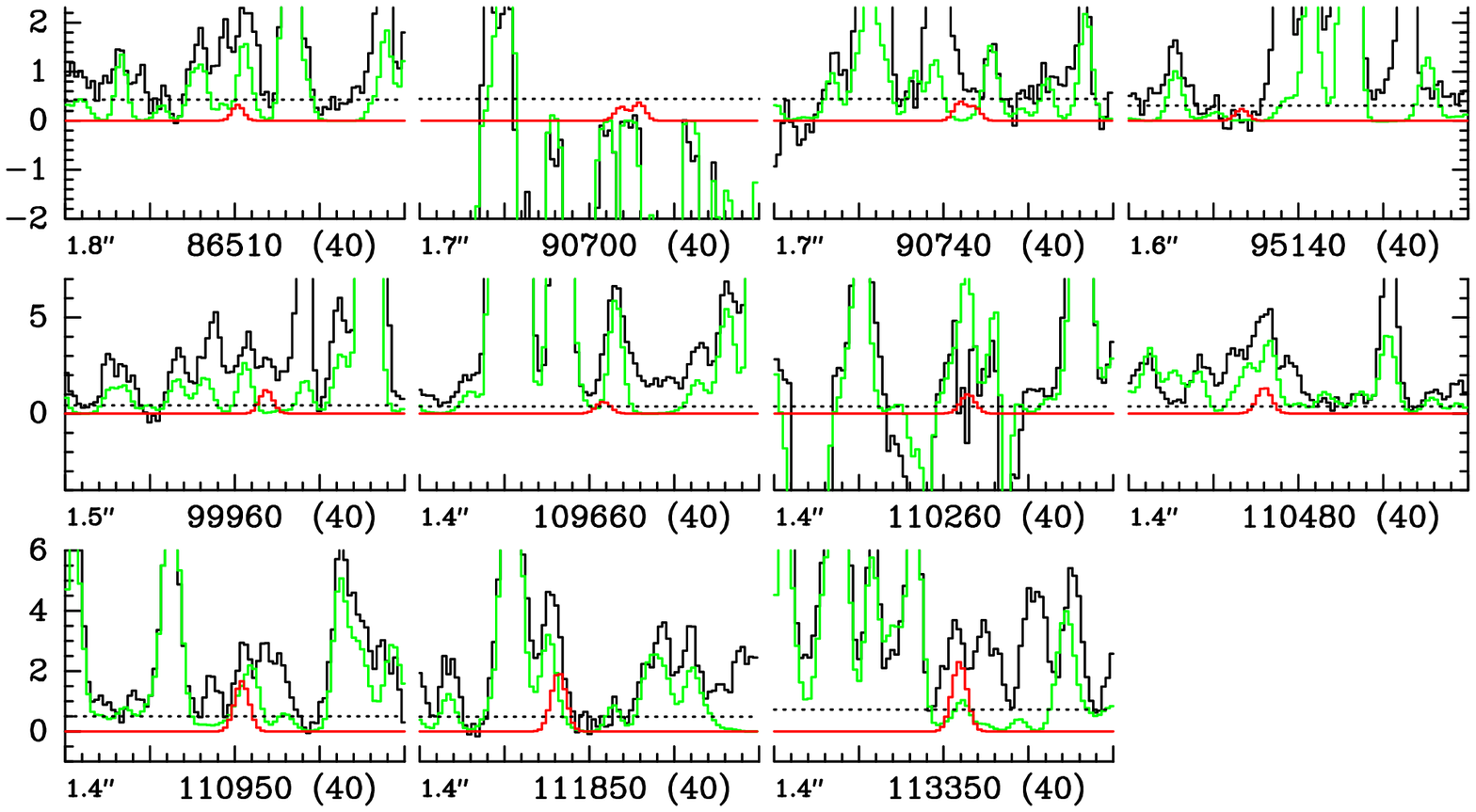}}}
\caption{Same as Fig.~\ref{f:spec_c2h5cn_ve0} for CH$_3$OD, $\varv=0$. The
full synthetic model (in green) does \textit{not} contain any contribution of 
CH$_3$OD (in red).}
\label{f:spec_ch3od_ve0}
\end{figure*}
}

\subsubsection{Deuterated vinyl cyanide CHDCHCN and CH$_2$CDCN}
\label{sss:c2h3cn}

Many lines of vinyl cyanide are detected in its ground state and vibrationally
excited states $\varv_{11}=1$, $\varv_{15}=1$, and $\varv_{11}=2$ 
(Figs.~\ref{f:spec_c2h3cn_ve0}, \ref{f:spec_c2h3cn_v11e1}, 
\ref{f:spec_c2h3cn_v15e1}, \ref{f:spec_c2h3cn_v11e2}). The sizes derived from
the corresponding integrated intensity maps tend to decrease with increasing 
energy, from $\sim 1.2\arcsec$ for $E_{\rm up} < 100$~K to $\sim 0.8\arcsec$ 
for higher-energy transitions. As a compromise we adopt a source size of 
$1.1\arcsec$. With this source size, the analysis of the population diagram 
yields a temperature of $\sim 200$~K (Table~\ref{t:popfit} and 
Fig.~\ref{f:popdiag_c2h3cn}). Transitions from within 
even higher vibrationally excited states are also detected toward Sgr~B2(N2), 
but we do not report about these states in this work.

\onlfig{
\clearpage
\begin{figure*}
%\centerline{\resizebox{0.9\hsize}{!}{\includegraphics[angle=0]{/homes/belloche/SgrB2/ALMA/Analysis/Weeds/R-NC/Figs/c2h3cn_ve0_p2_1.eps}}}
\centerline{\resizebox{0.9\hsize}{!}{\includegraphics[angle=0]{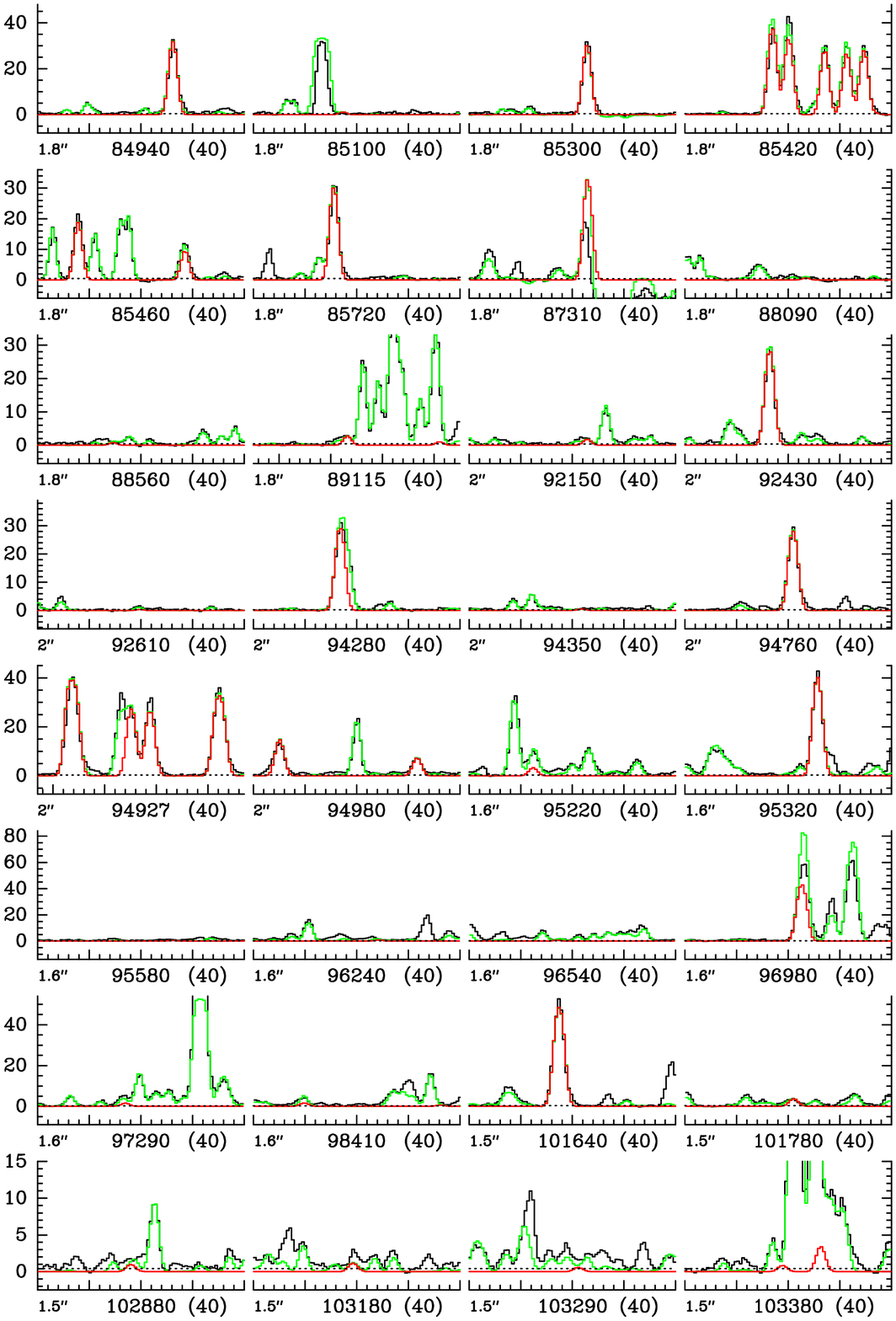}}}
\caption{Same as Fig.~\ref{f:spec_c2h5cn_ve0} for C$_2$H$_3$CN, $\varv=0$.
}
\label{f:spec_c2h3cn_ve0}
\end{figure*}
}

\onlfig{
\clearpage
\begin{figure*}
\addtocounter{figure}{-1}
%\centerline{\resizebox{0.9\hsize}{!}{\includegraphics[angle=0]{/homes/belloche/SgrB2/ALMA/Analysis/Weeds/R-NC/Figs/c2h3cn_ve0_p2_2.eps}}}
\centerline{\resizebox{0.9\hsize}{!}{\includegraphics[angle=0]{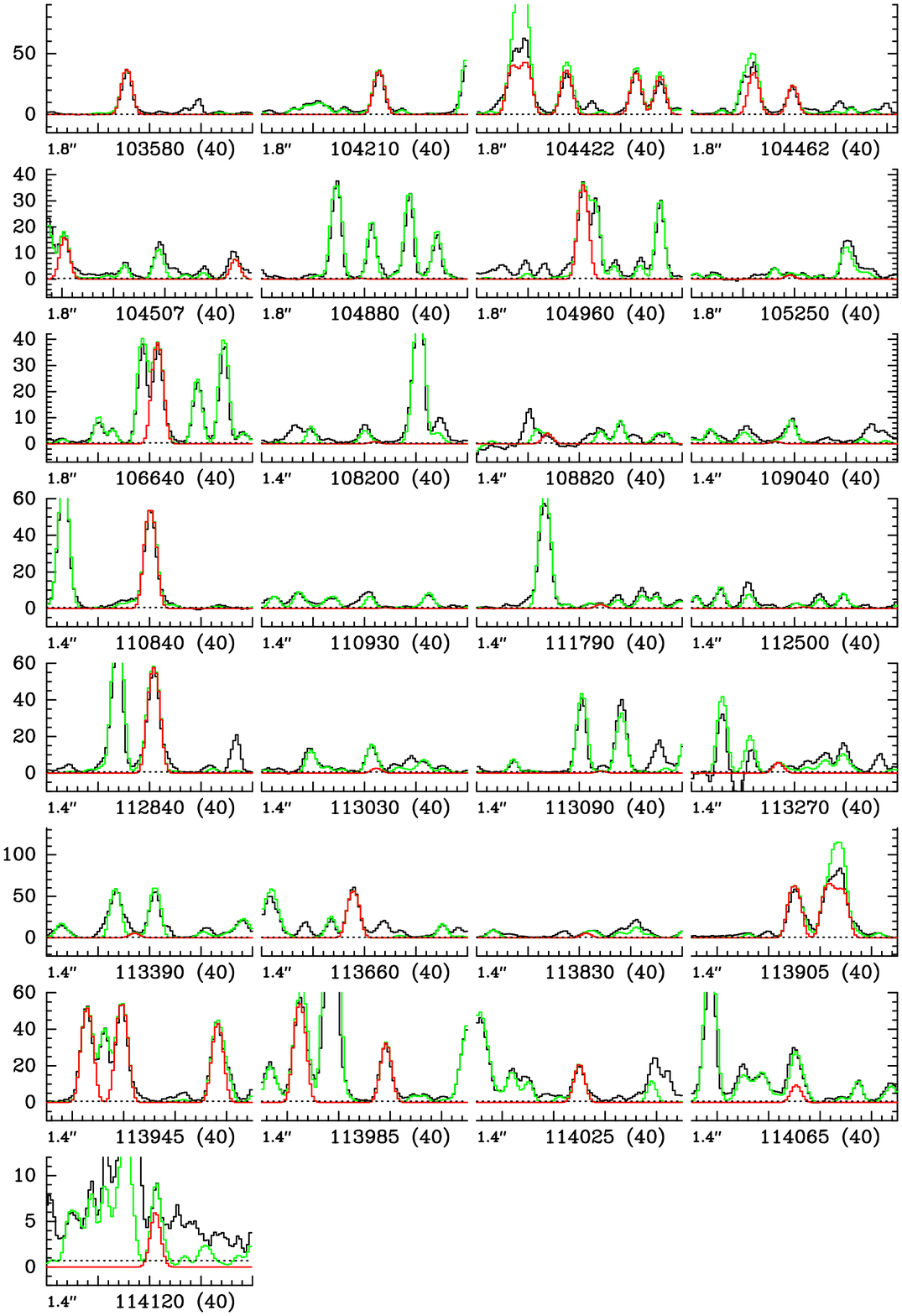}}}
\caption{continued.}
\end{figure*}
}
\addtocounter{figure}{-1}

\onlfig{
\clearpage
\begin{figure*}
%\centerline{\resizebox{0.9\hsize}{!}{\includegraphics[angle=0]{/homes/belloche/SgrB2/ALMA/Analysis/Weeds/R-NC/Figs/c2h3cn_v11e1_p2_1.eps}}}
\centerline{\resizebox{0.9\hsize}{!}{\includegraphics[angle=0]{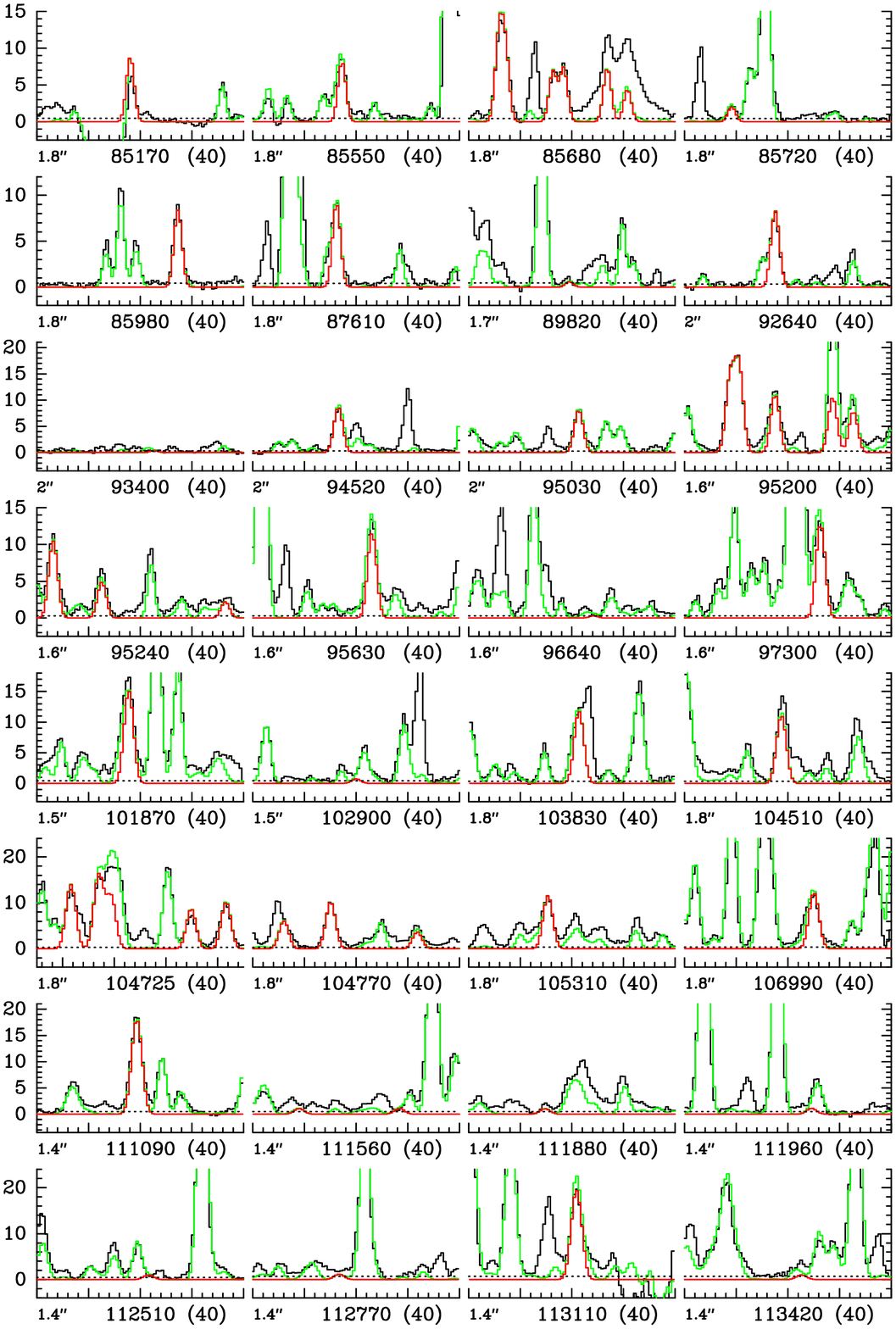}}}
\caption{Same as Fig.~\ref{f:spec_c2h5cn_ve0} for C$_2$H$_3$CN, $\varv_{11}=1$.
}
\label{f:spec_c2h3cn_v11e1}
\end{figure*}
}

\onlfig{
\clearpage
\begin{figure*}
\addtocounter{figure}{-1}
%\centerline{\resizebox{0.9\hsize}{!}{\includegraphics[angle=0]{/homes/belloche/SgrB2/ALMA/Analysis/Weeds/R-NC/Figs/c2h3cn_v11e1_p2_2.eps}}}
\centerline{\resizebox{0.9\hsize}{!}{\includegraphics[angle=0]{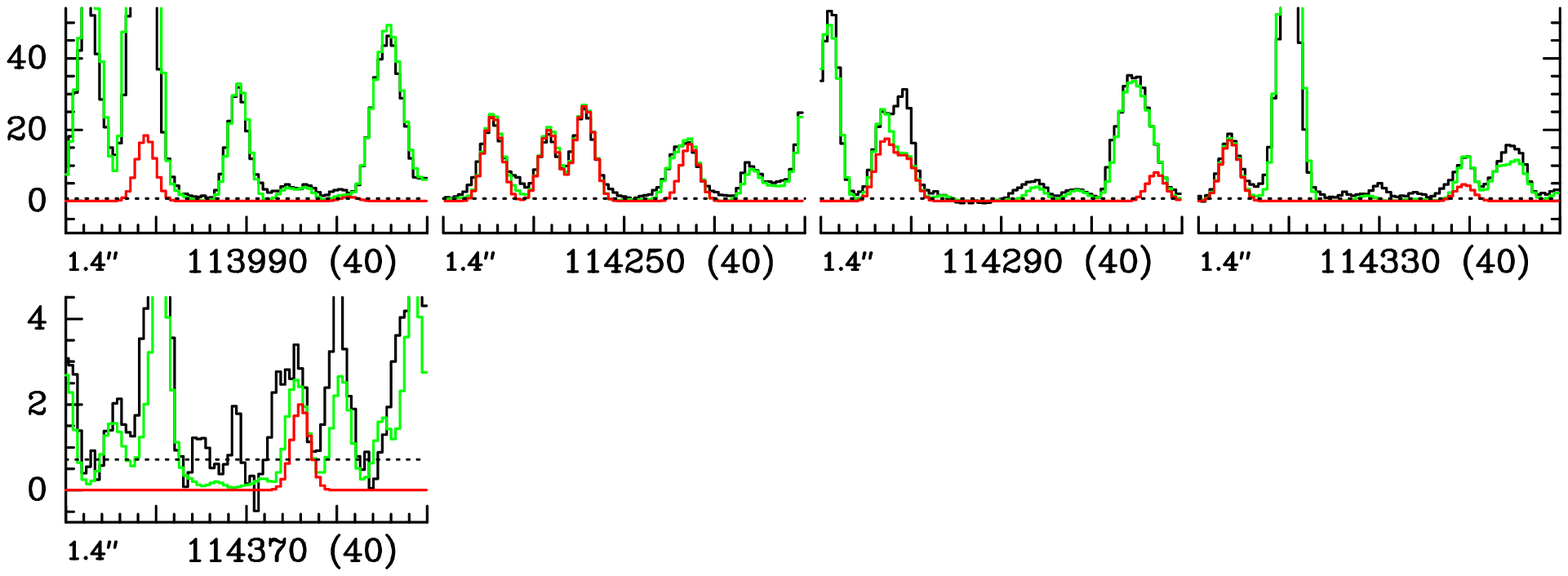}}}
\caption{continued.}
\end{figure*}
}
\addtocounter{figure}{-1}

\onlfig{
\clearpage
\begin{figure*}
%\centerline{\resizebox{0.9\hsize}{!}{\includegraphics[angle=0]{/homes/belloche/SgrB2/ALMA/Analysis/Weeds/R-NC/Figs/c2h3cn_v15e1_p2.eps}}}
\centerline{\resizebox{0.9\hsize}{!}{\includegraphics[angle=0]{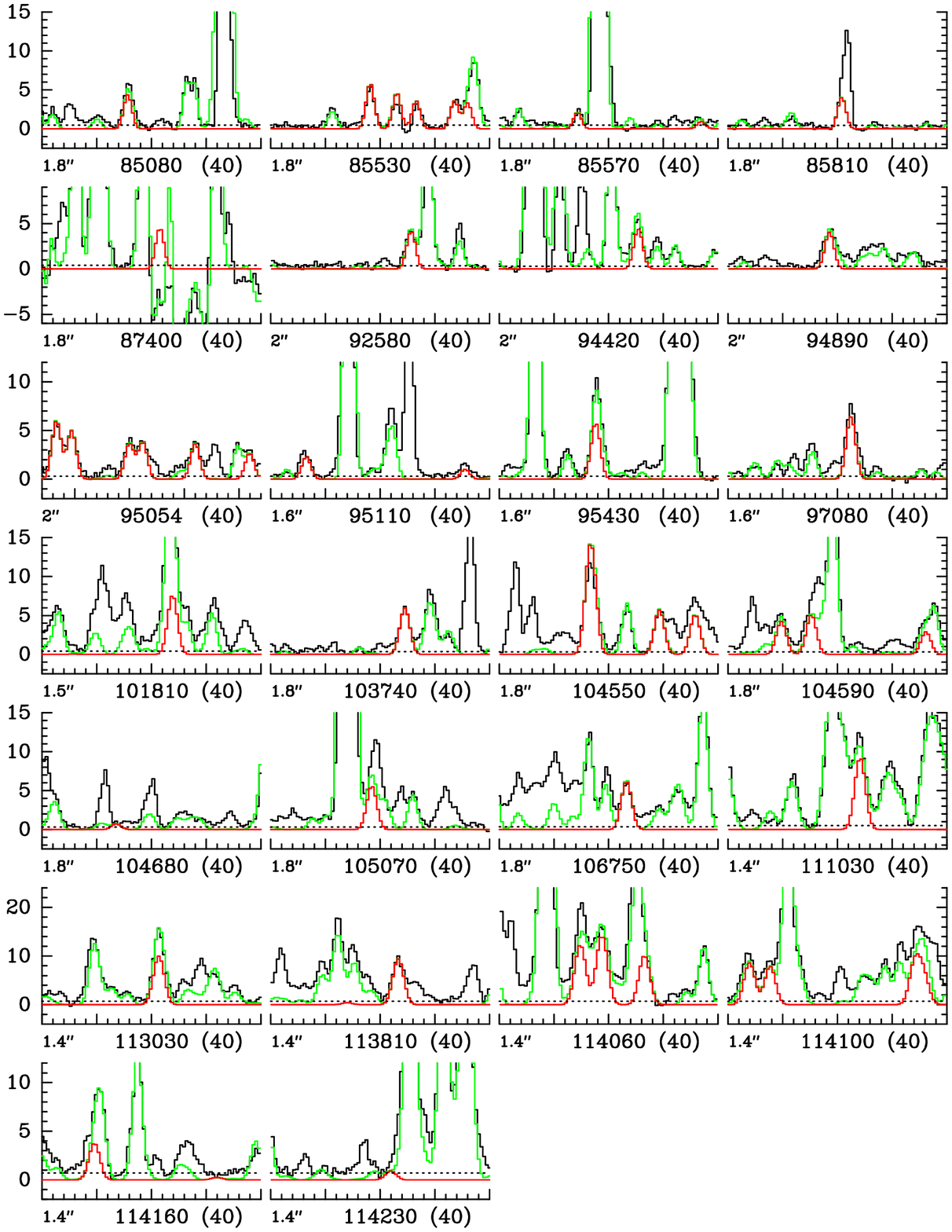}}}
\caption{Same as Fig.~\ref{f:spec_c2h5cn_ve0} for C$_2$H$_3$CN, $\varv_{15}=1$.
}
\label{f:spec_c2h3cn_v15e1}
\end{figure*}
}

\onlfig{
\clearpage
\begin{figure*}
%\centerline{\resizebox{0.9\hsize}{!}{\includegraphics[angle=0]{/homes/belloche/SgrB2/ALMA/Analysis/Weeds/R-NC/Figs/c2h3cn_v11e2_p2.eps}}}
\centerline{\resizebox{0.9\hsize}{!}{\includegraphics[angle=0]{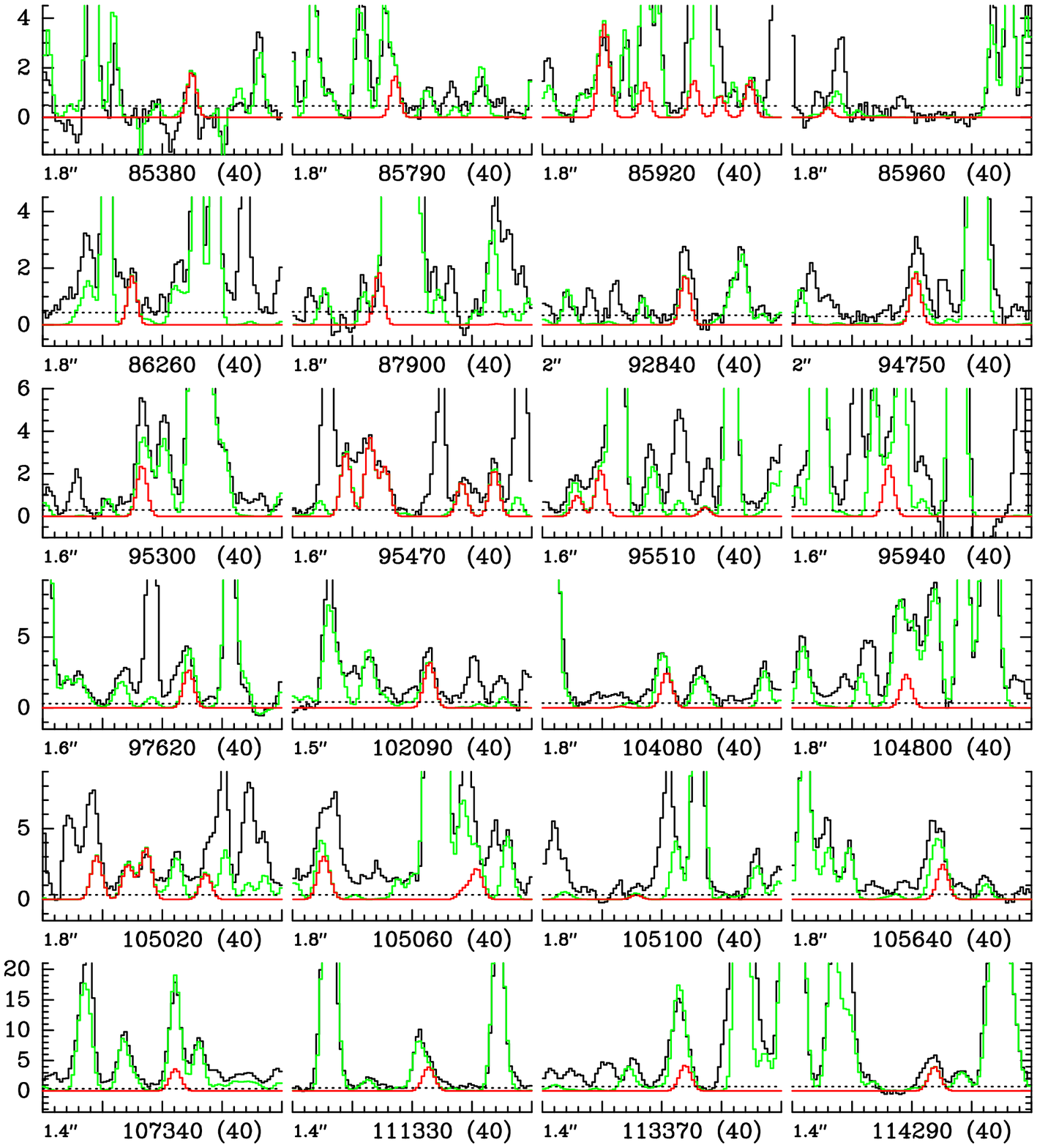}}}
\caption{Same as Fig.~\ref{f:spec_c2h5cn_ve0} for C$_2$H$_3$CN, $\varv_{11}=2$.
}
\label{f:spec_c2h3cn_v11e2}
\end{figure*}
}

\onlfig{
\clearpage
\begin{figure}
%\centerline{\resizebox{1.0\hsize}{!}{\includegraphics[angle=0]{/homes/belloche/SgrB2/ALMA/Analysis/Weeds/R-NC/popdiag_c2h3cn_p2.eps}}}
\centerline{\resizebox{1.0\hsize}{!}{\includegraphics[angle=0]{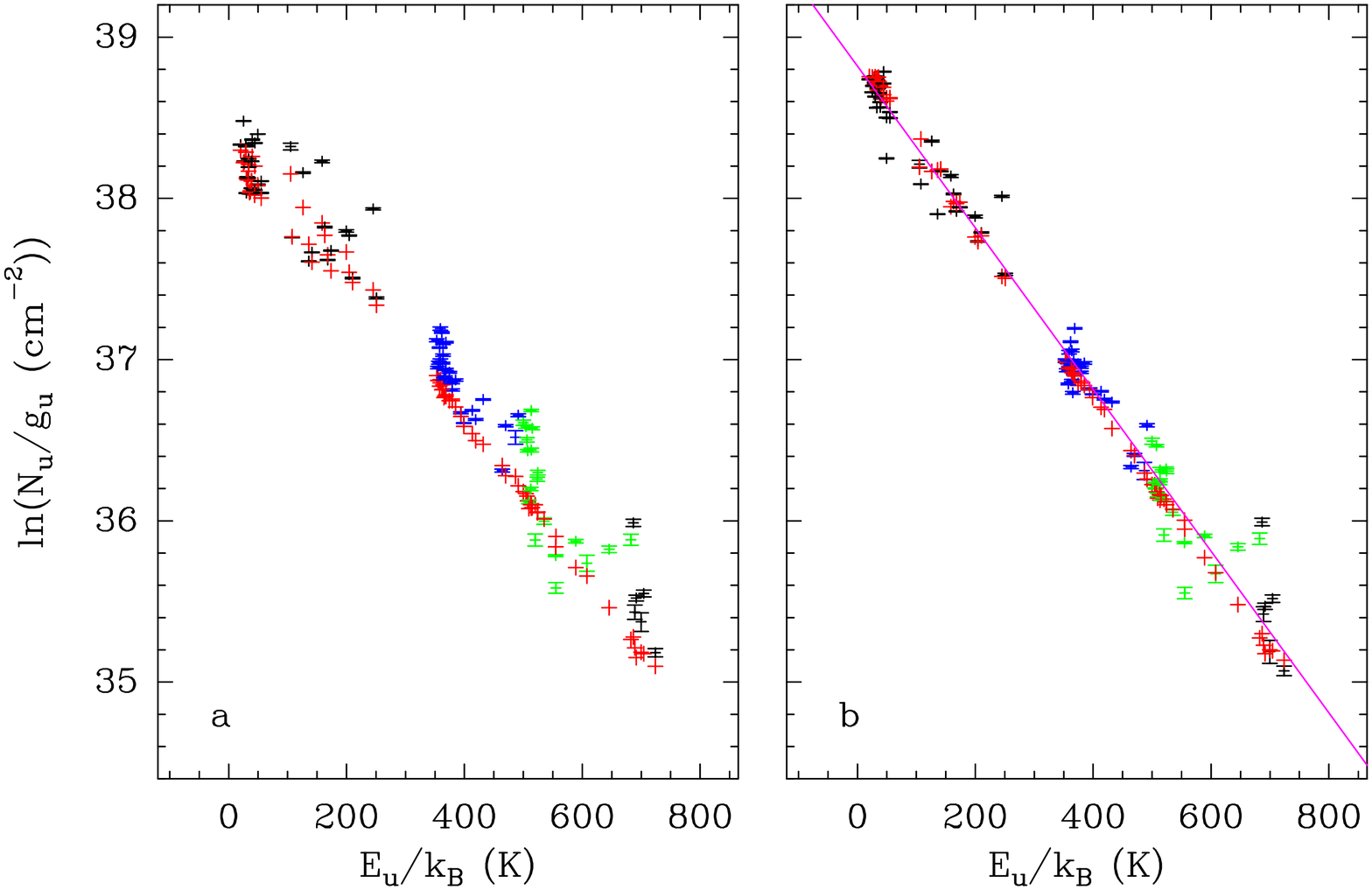}}}
\caption{Same as Fig.~\ref{f:popdiag_c2h5cn} for C$_2$H$_3$CN, $\varv=0$, 
$\varv_{11}=1$, $\varv_{15}=1$, and $\varv_{11}=2$.}
\label{f:popdiag_c2h3cn}
\end{figure}
}

Transitions of all three singly-substituted $^{13}$C isotopologues of vinyl 
cyanide are also clearly detected \citep[Figs.~\ref{f:spec_c2h3cn_13c1_ve0}, 
\ref{f:spec_c2h3cn_13c2_ve0}, and \ref{f:spec_c2h3cn_13c3_ve0}; see also][ for 
a previous single-dish detection]{VyCN-isos_rot_2008}. Only a few
lines are sufficiently free of contamination to allow for a size measurement
in the corresponding integrated intensity maps. The outcome is more uncertain
than for the main isotopologue, but is consistent with the source size adopted
above. Due to the smaller number of detected lines, the population diagrams 
have a higher dispersion than for the main isotopologue and the rotational
temperature is less well constrained but the fits to all three diagrams are 
consistent with a temperature of about 200~K 
(Figs.~\ref{f:popdiag_c2h3cn_13c1}, \ref{f:popdiag_c2h3cn_13c2}, and 
\ref{f:popdiag_c2h3cn_13c3}).

\onlfig{
\clearpage
\begin{figure*}
%\centerline{\resizebox{0.9\hsize}{!}{\includegraphics[angle=0]{/homes/belloche/SgrB2/ALMA/Analysis/Weeds/R-NC/Figs/c2h3cn_13c1_ve0_p2.eps}}}
\centerline{\resizebox{0.9\hsize}{!}{\includegraphics[angle=0]{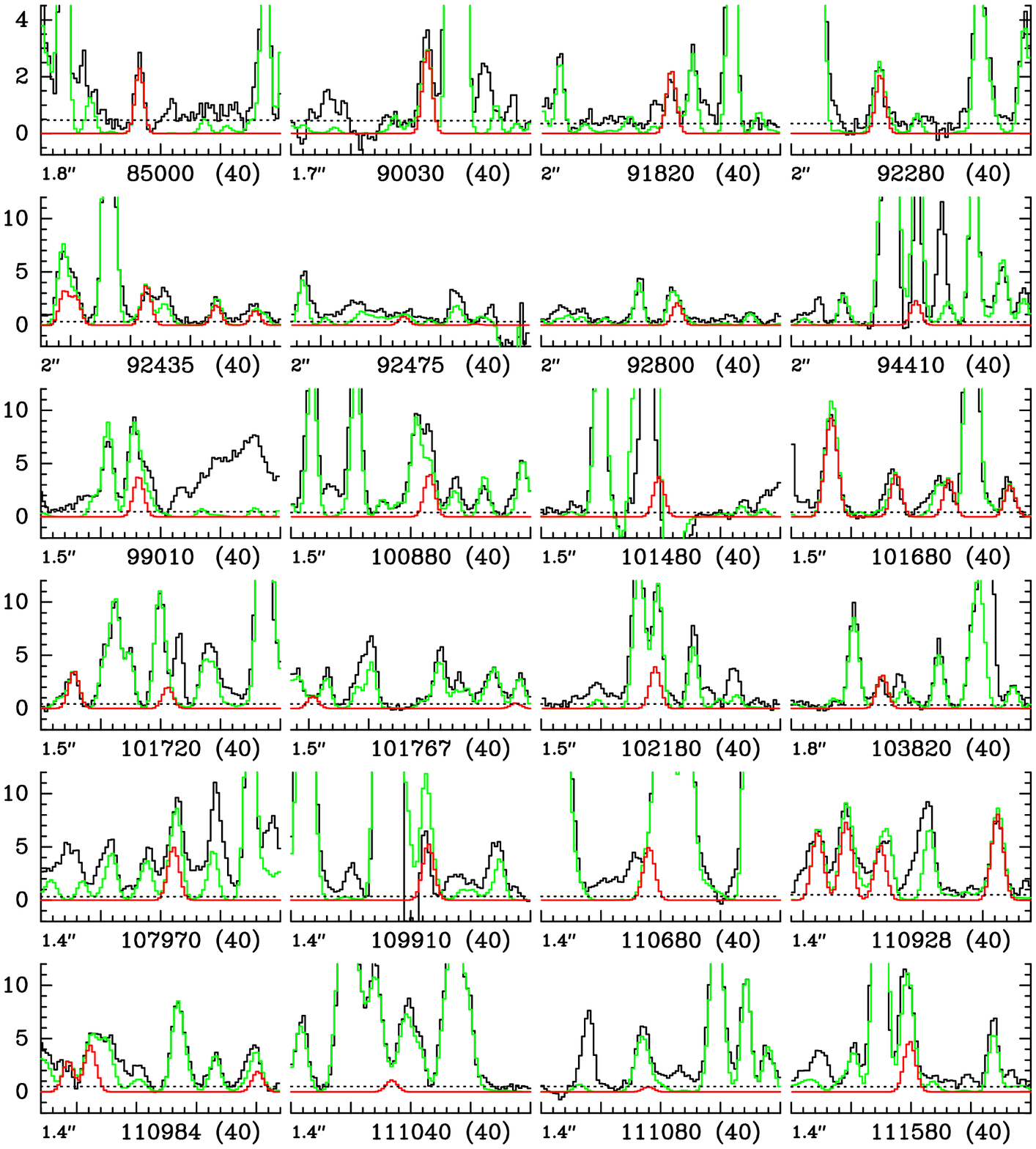}}}
\caption{Same as Fig.~\ref{f:spec_c2h5cn_ve0} for $^{13}$CH$_2$CHCN, $\varv=0$.
}
\label{f:spec_c2h3cn_13c1_ve0}
\end{figure*}
}

\onlfig{
\clearpage
\begin{figure*}
%\centerline{\resizebox{0.9\hsize}{!}{\includegraphics[angle=0]{/homes/belloche/SgrB2/ALMA/Analysis/Weeds/R-NC/Figs/c2h3cn_13c2_ve0_p2.eps}}}
\centerline{\resizebox{0.9\hsize}{!}{\includegraphics[angle=0]{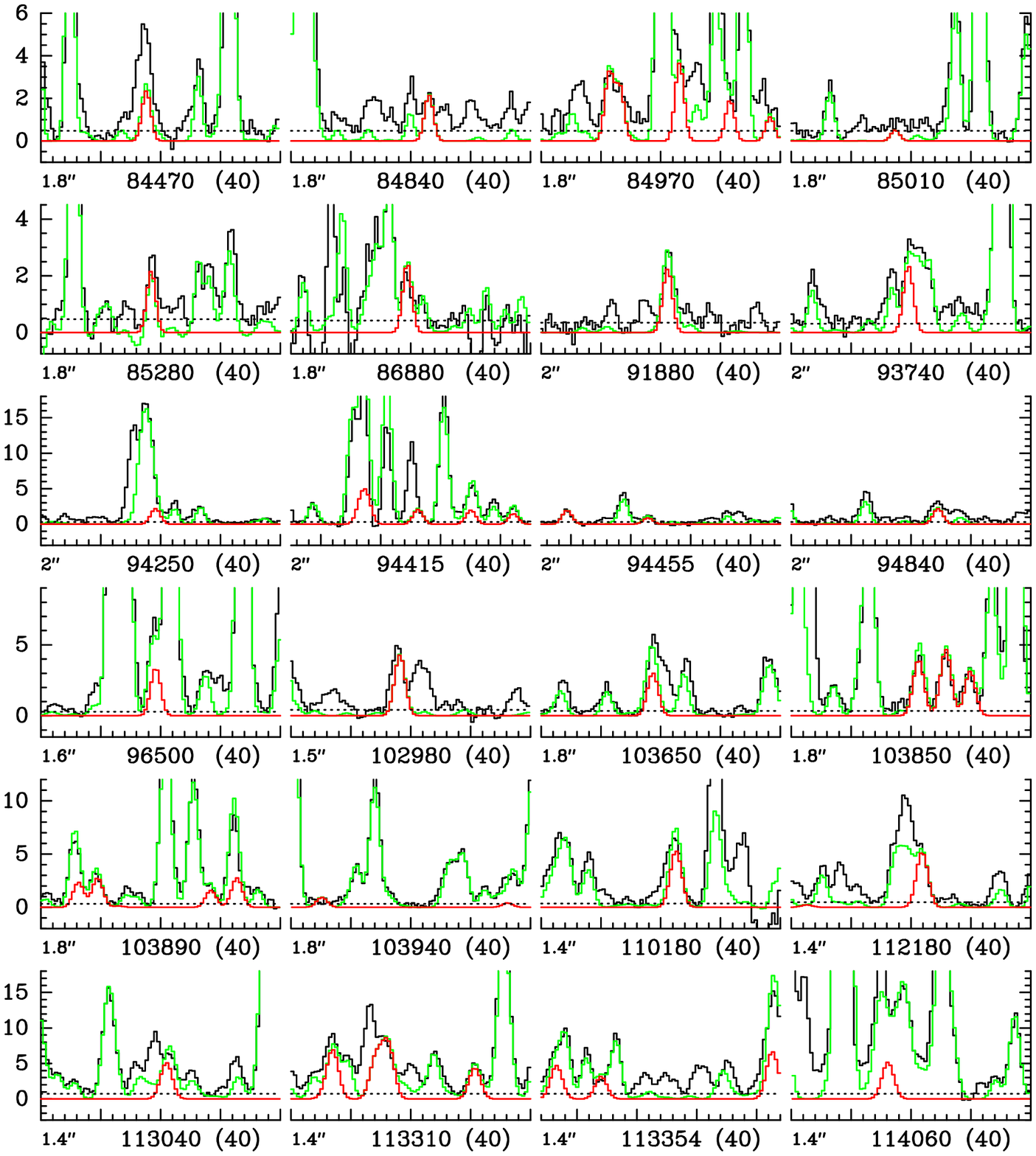}}}
\caption{Same as Fig.~\ref{f:spec_c2h5cn_ve0} for CH$_2$$^{13}$CHCN, $\varv=0$.
}
\label{f:spec_c2h3cn_13c2_ve0}
\end{figure*}
}

\onlfig{
\clearpage
\begin{figure*}
%\centerline{\resizebox{0.9\hsize}{!}{\includegraphics[angle=0]{/homes/belloche/SgrB2/ALMA/Analysis/Weeds/R-NC/Figs/c2h3cn_13c3_ve0_p2.eps}}}
\centerline{\resizebox{0.9\hsize}{!}{\includegraphics[angle=0]{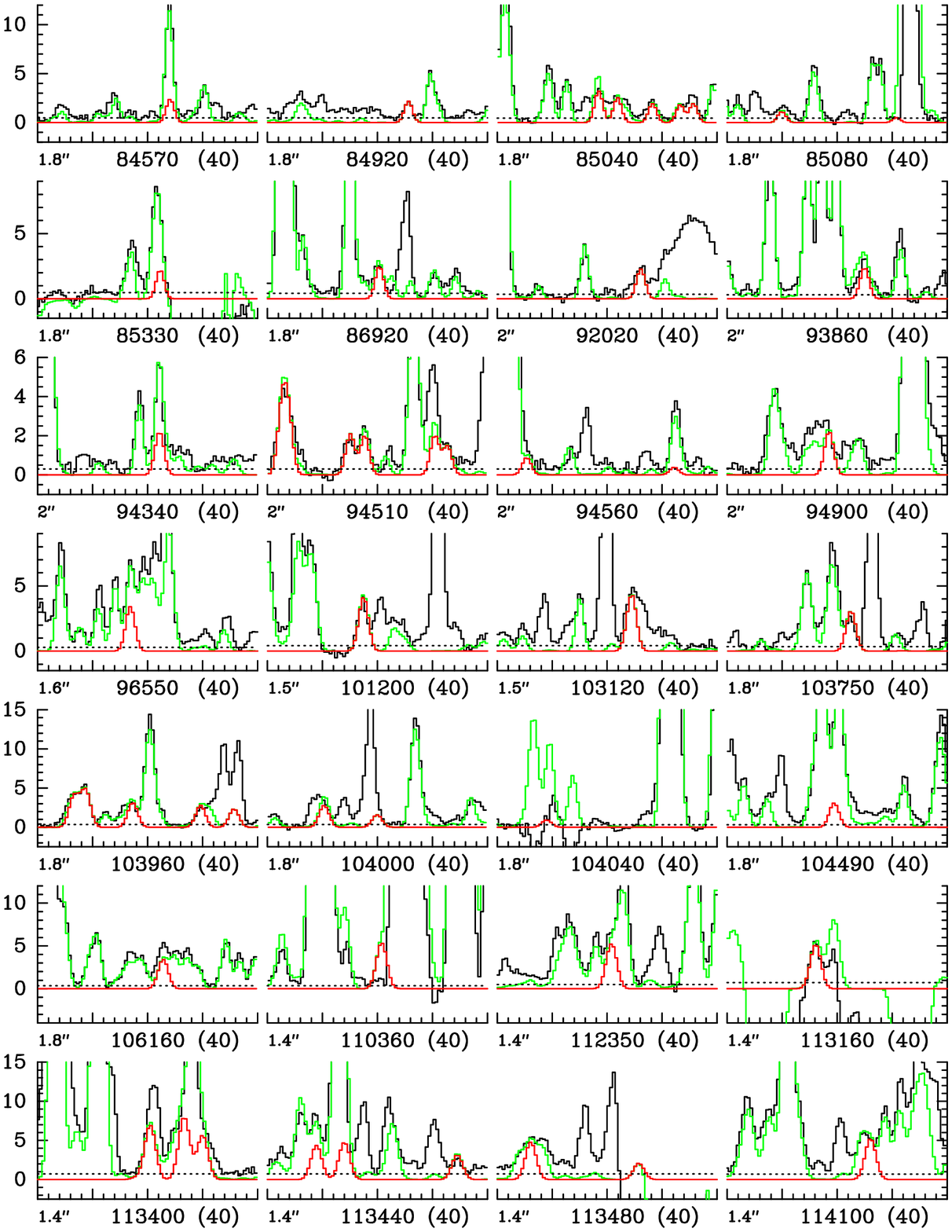}}}
\caption{Same as Fig.~\ref{f:spec_c2h5cn_ve0} for C$_2$H$_3$$^{13}$CN, 
$\varv=0$.
}
\label{f:spec_c2h3cn_13c3_ve0}
\end{figure*}
}

\onlfig{
\clearpage
\begin{figure}
%\centerline{\resizebox{1.0\hsize}{!}{\includegraphics[angle=0]{/homes/belloche/SgrB2/ALMA/Analysis/Weeds/R-NC/popdiag_c2h3cn_13c1_p2.eps}}}
\centerline{\resizebox{1.0\hsize}{!}{\includegraphics[angle=0]{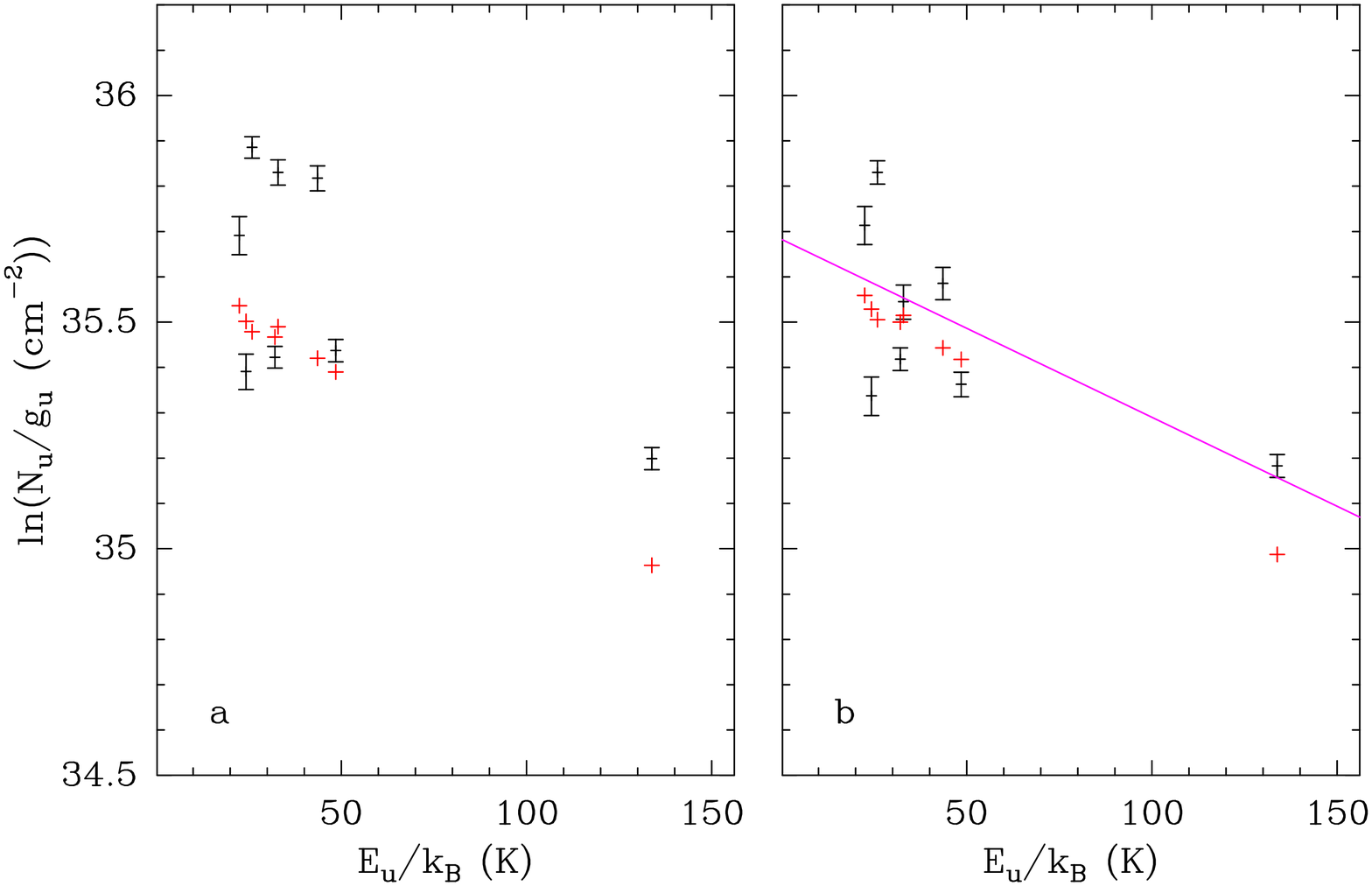}}}
\caption{Same as Fig.~\ref{f:popdiag_c2h5cn} for $^{13}$CH$_2$CHCN, $\varv=0$.}
\label{f:popdiag_c2h3cn_13c1}
\end{figure}
}

\onlfig{
\begin{figure}
%\centerline{\resizebox{1.0\hsize}{!}{\includegraphics[angle=0]{/homes/belloche/SgrB2/ALMA/Analysis/Weeds/R-NC/popdiag_c2h3cn_13c2_p2.eps}}}
\centerline{\resizebox{1.0\hsize}{!}{\includegraphics[angle=0]{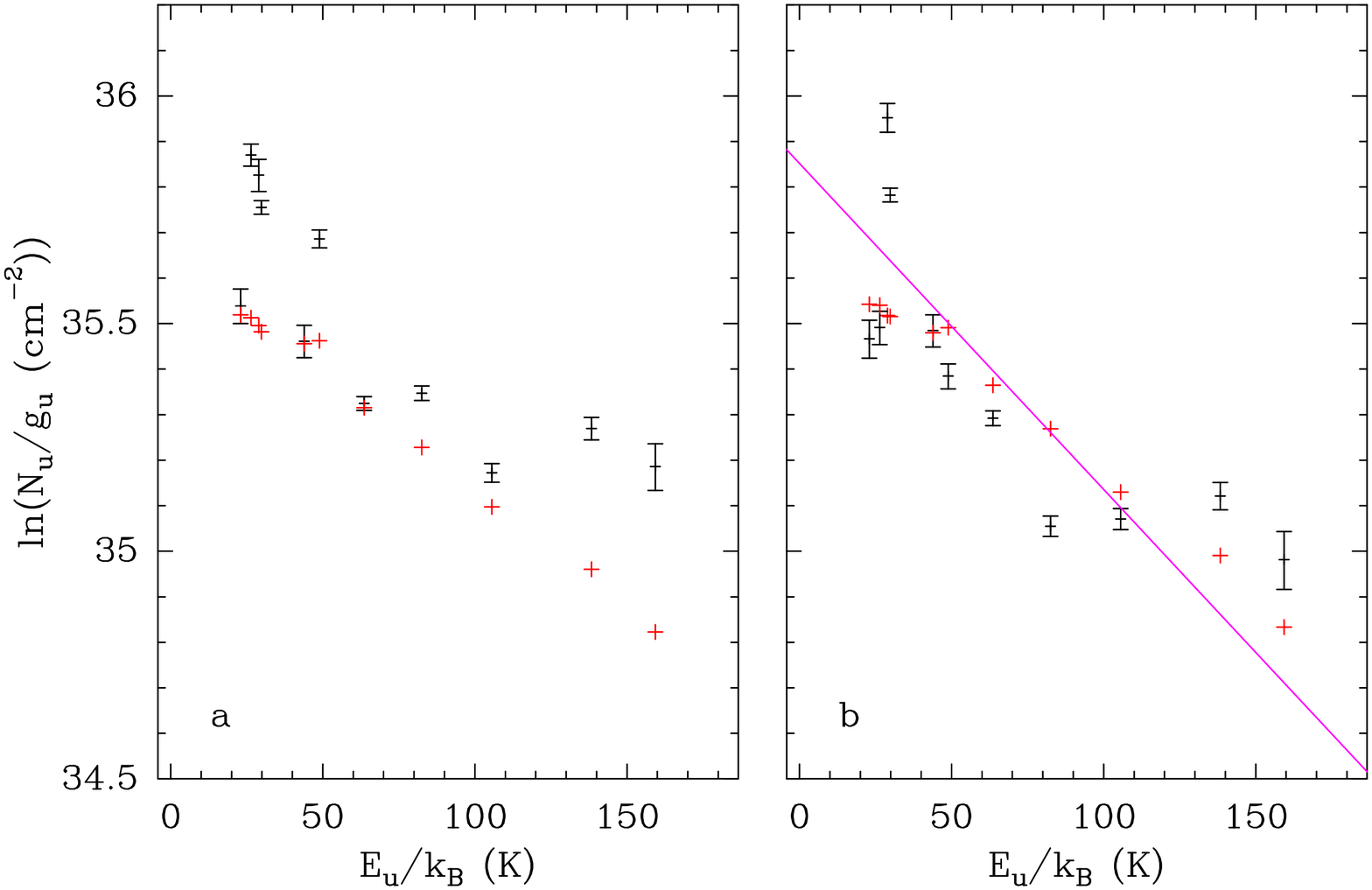}}}
\caption{Same as Fig.~\ref{f:popdiag_c2h5cn} for CH$_2$$^{13}$CHCN, $\varv=0$.}
\label{f:popdiag_c2h3cn_13c2}
\end{figure}
}

\onlfig{
\begin{figure}
%\centerline{\resizebox{1.0\hsize}{!}{\includegraphics[angle=0]{/homes/belloche/SgrB2/ALMA/Analysis/Weeds/R-NC/popdiag_c2h3cn_13c3_p2.eps}}}
\centerline{\resizebox{1.0\hsize}{!}{\includegraphics[angle=0]{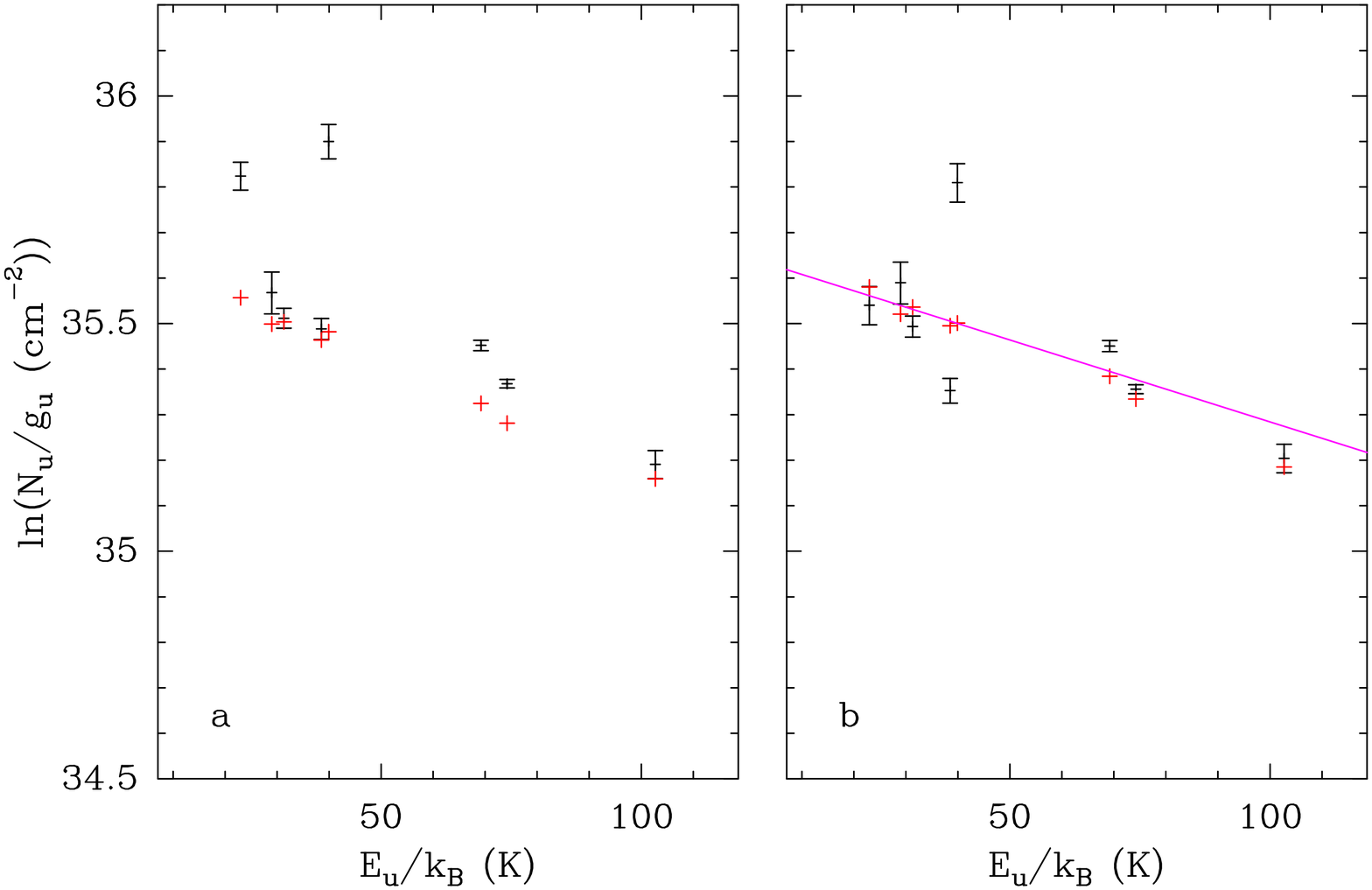}}}
\caption{Same as Fig.~\ref{f:popdiag_c2h5cn} for CH$_2$CH$^{13}$CN, $\varv=0$.}
\label{f:popdiag_c2h3cn_13c3}
\end{figure}
}

As a result of this analysis, we adopt a source size of $1.1\arcsec$ and a 
temperature of 200~K for vinyl cyanide and its isotopologues. With these 
parameters, we do not detect the $^{15}$N isotopologue. We also looked for the 
singly-deuterated species \textit{cis}-CHDCHCN, \textit{trans}-CHDCHCN, and 
CH$_2$CDCN,  but did not detect them. Column density upper limits are reported 
in Table~\ref{t:coldens}.

\subsubsection{Deuterated ethanol CH$_3$CH$_2$OD, CH$_3$CHDOH, and CH$_2$DCH$_2$OH}
\label{sss:c2h5oh}

Ethanol and its $^{13}$C isotopologues are well detected toward 
Sgr~B2(N2). The detected lines and detailed modeling of these species is
presented in a companion paper \citep[][]{Mueller15c}. We report in 
Table~\ref{t:coldens} the parameters derived in that paper for the main 
isotopologue based on the analysis of all isotopologues.

Assuming the same LTE parameters as for the main isotopologue, we searched for 
all singly deuterated isotopologues of ethanol. None is detected. Upper limits 
to their column densities are reported in Table~\ref{t:coldens}.

\subsubsection{Deuterated methyl formate CH$_2$DOCHO}
\label{sss:ch3ocho}

Methyl formate is clearly seen toward Sgr~B2(N2), with dozens of 
transitions detected in both its ground and first torsional states
(Figs.~\ref{f:spec_ch3ocho_ve0} and \ref{f:spec_ch3ocho_ve1}). We derive
a median source size of $1.5''$ from fits to the integrated intensity maps of
its numerous uncontaminated lines. The formal fit to its population diagram 
including both states yields a rotational temperature of $\sim 140$~K 
(Fig.~\ref{f:popdiag_ch3ocho} and Table~\ref{t:popfit}). We used a temperature 
of 150~K in our model, which fits very well the ALMA spectrum, apart from a few
discrepancies that we describe now. The reason
why the synthetic spectrum of the ground state poorly fits the ALMA spectrum
at 100080~MHz is unclear. It may be due to the nearby HC$_3$N 11--10 transition 
at 100076~MHz that is probably affected by self-absorption and/or spatial 
filtering and is by far overestimated by our simple LTE model. The discrepancy 
around 110226~MHz is due to contamination by diffuse cloud absorption in 
$^{13}$CO 1--0 that is not yet included in our full model. Similar 
contamination by \textit{c}-C$_3$H$_2$ absorption features not yet implemented
in our full model likely explains the small discrepancies for the 
$\varv_{\rm t} = 1$ transitions around 85370~MHz.

\onlfig{
\clearpage
\begin{figure*}
%\centerline{\resizebox{0.9\hsize}{!}{\includegraphics[angle=0]{/homes/belloche/SgrB2/ALMA/Analysis/Weeds/R-D/Figs/ch3ocho_ve0_p2_1.eps}}}
\centerline{\resizebox{0.9\hsize}{!}{\includegraphics[angle=0]{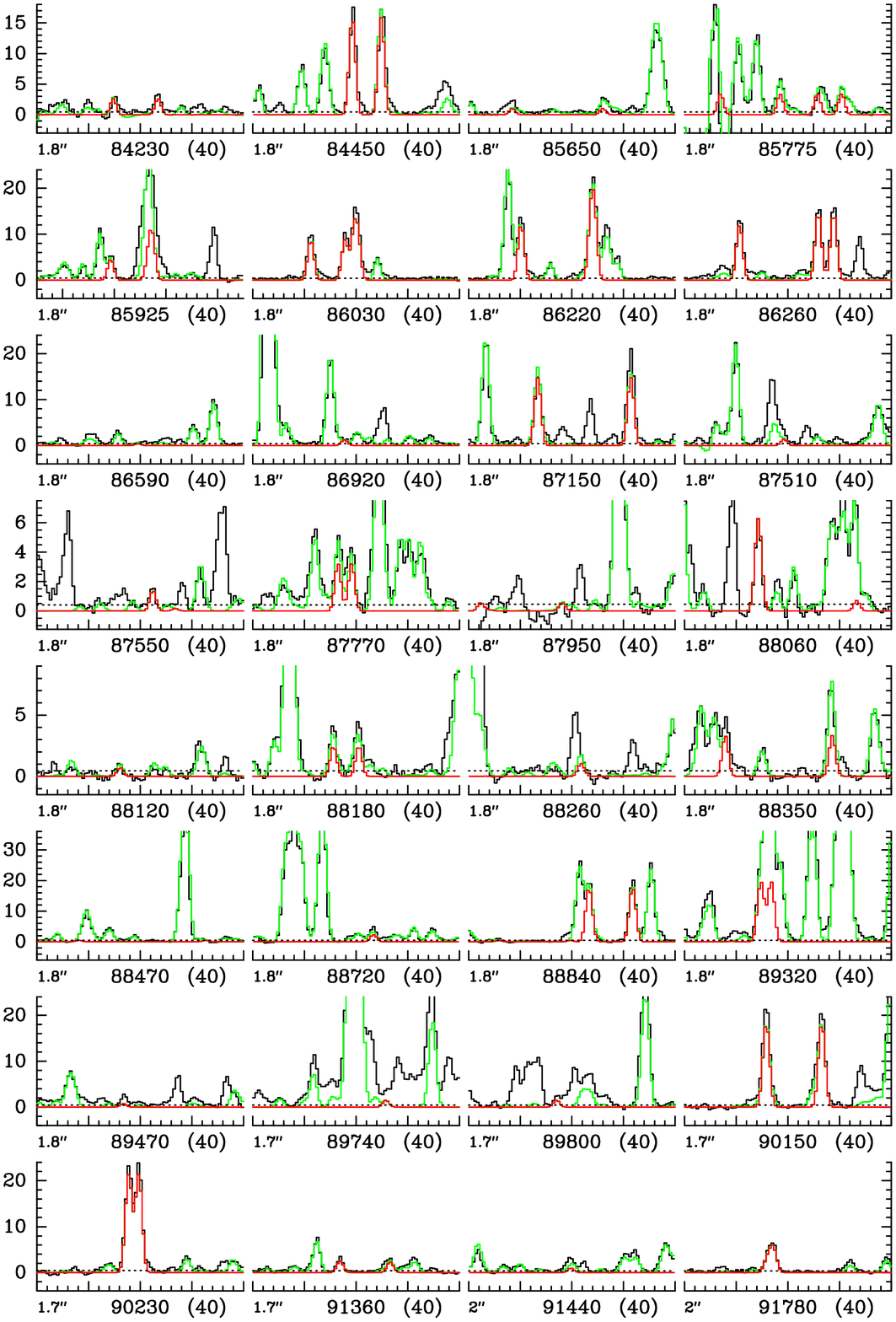}}}
\caption{Same as Fig.~\ref{f:spec_c2h5cn_ve0} for CH$_3$OCHO, $\varv_{\rm t}=0$.
}
\label{f:spec_ch3ocho_ve0}
\end{figure*}
}

\onlfig{
\clearpage
\begin{figure*}
\addtocounter{figure}{-1}
%\centerline{\resizebox{0.9\hsize}{!}{\includegraphics[angle=0]{/homes/belloche/SgrB2/ALMA/Analysis/Weeds/R-D/Figs/ch3ocho_ve0_p2_2.eps}}}
\centerline{\resizebox{0.9\hsize}{!}{\includegraphics[angle=0]{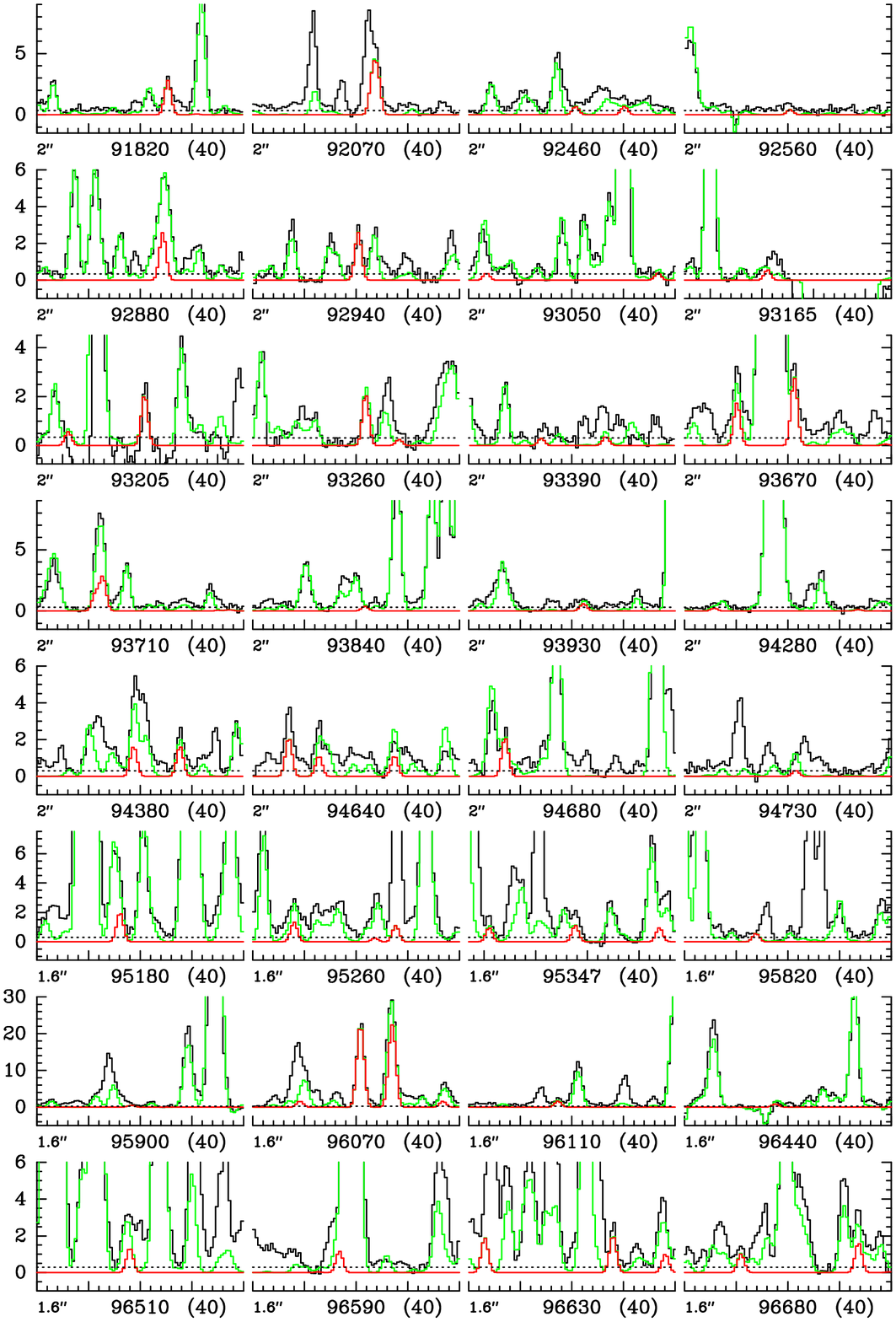}}}
\caption{continued.}
\end{figure*}
}
\addtocounter{figure}{-1}

\onlfig{
\clearpage
\begin{figure*}
\addtocounter{figure}{-1}
%\centerline{\resizebox{0.9\hsize}{!}{\includegraphics[angle=0]{/homes/belloche/SgrB2/ALMA/Analysis/Weeds/R-D/Figs/ch3ocho_ve0_p2_3.eps}}}
\centerline{\resizebox{0.9\hsize}{!}{\includegraphics[angle=0]{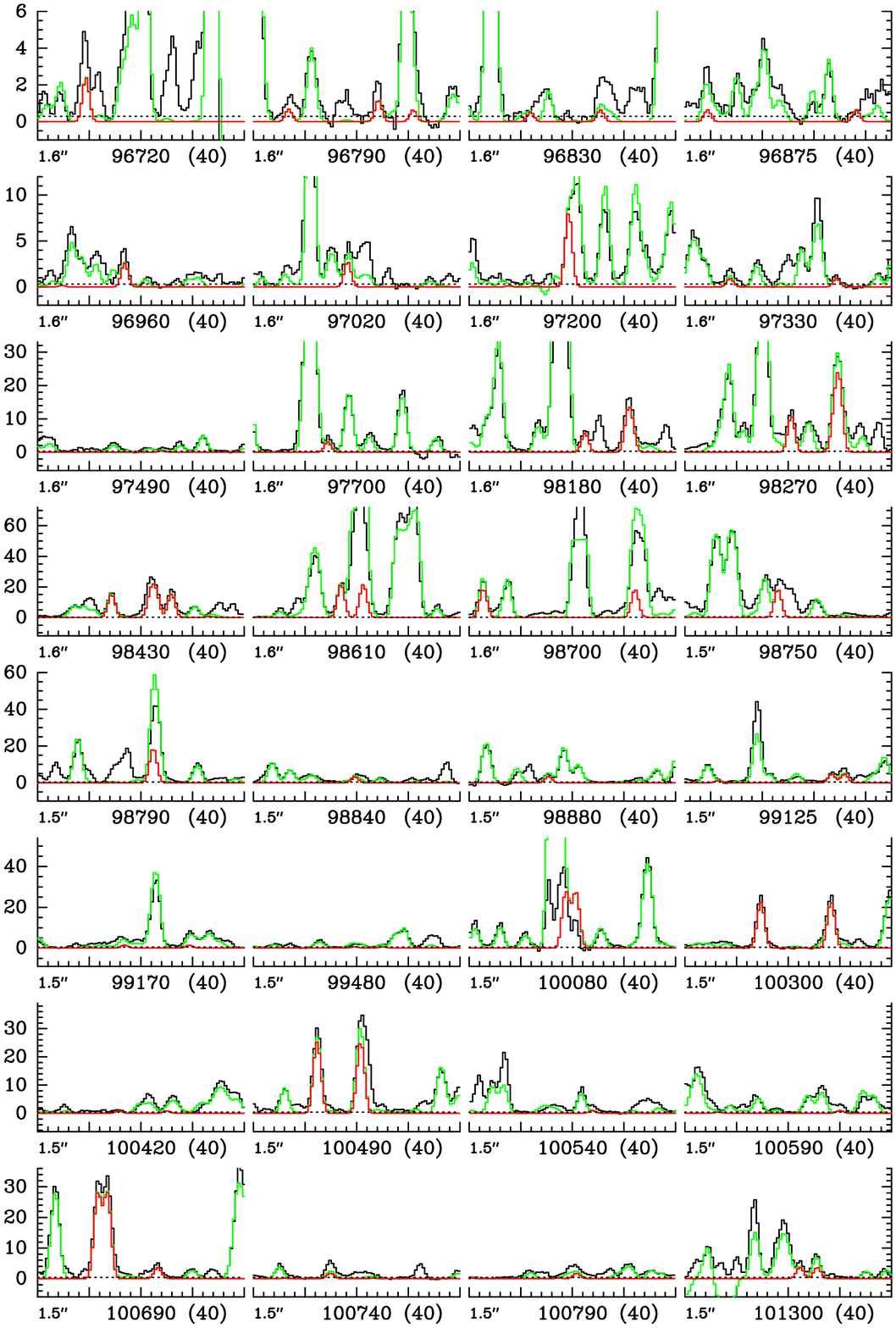}}}
\caption{continued.}
\end{figure*}
}
\addtocounter{figure}{-1}

\onlfig{
\clearpage
\begin{figure*}
\addtocounter{figure}{-1}
%\centerline{\resizebox{0.9\hsize}{!}{\includegraphics[angle=0]{/homes/belloche/SgrB2/ALMA/Analysis/Weeds/R-D/Figs/ch3ocho_ve0_p2_4.eps}}}
\centerline{\resizebox{0.9\hsize}{!}{\includegraphics[angle=0]{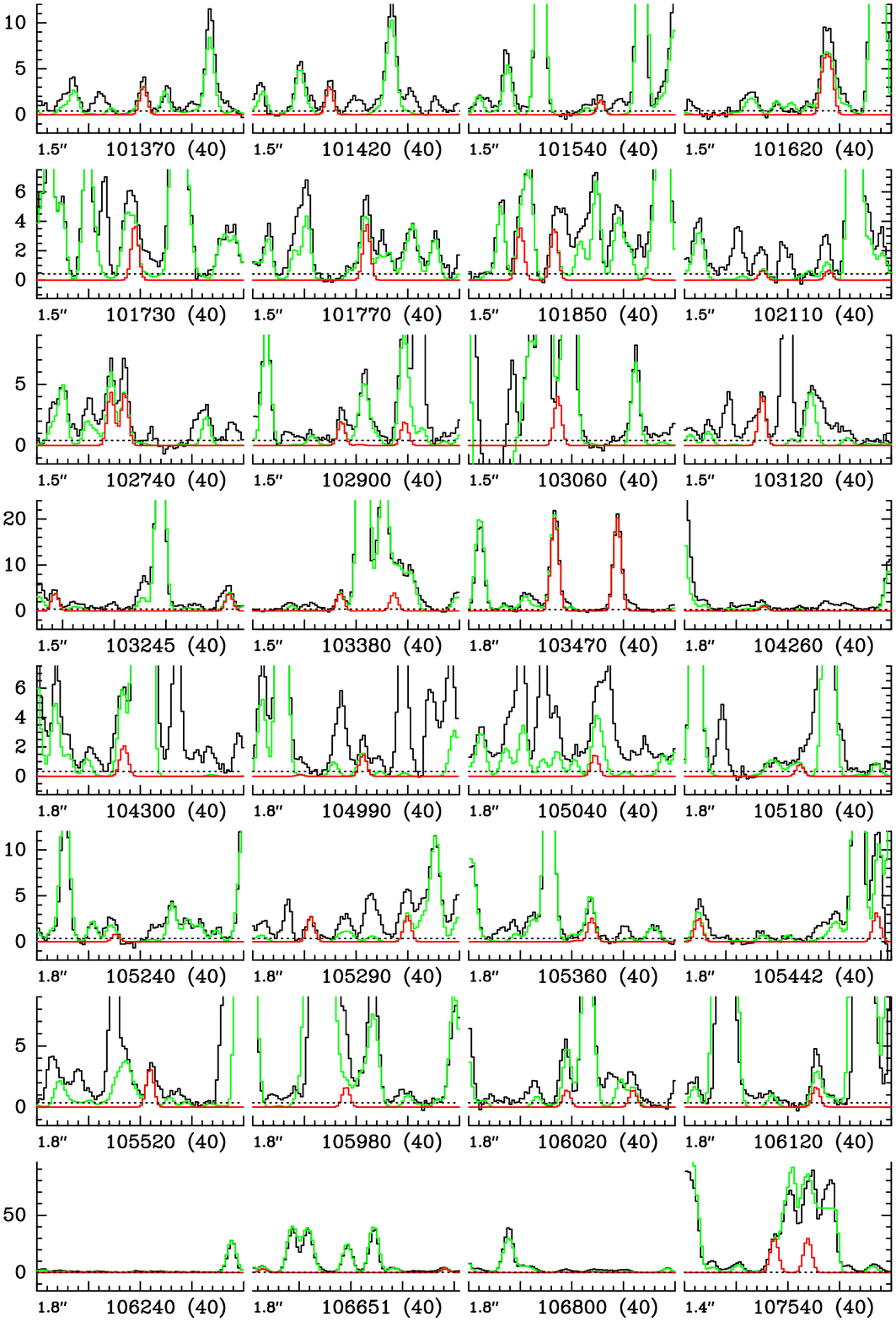}}}
\caption{continued.}
\end{figure*}
}
\addtocounter{figure}{-1}

\onlfig{
\clearpage
\begin{figure*}
\addtocounter{figure}{-1}
%\centerline{\resizebox{0.9\hsize}{!}{\includegraphics[angle=0]{/homes/belloche/SgrB2/ALMA/Analysis/Weeds/R-D/Figs/ch3ocho_ve0_p2_5.eps}}}
\centerline{\resizebox{0.9\hsize}{!}{\includegraphics[angle=0]{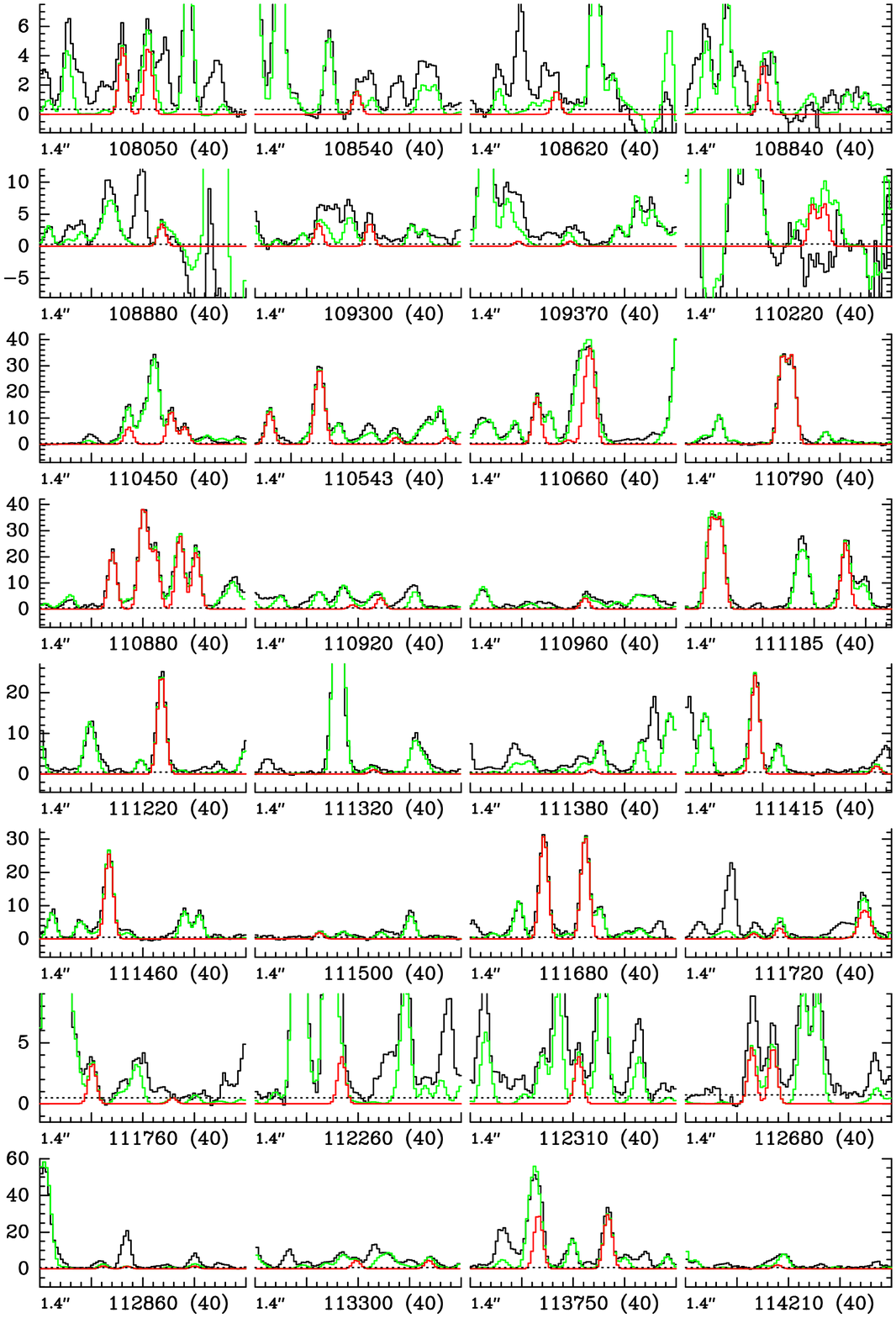}}}
\caption{continued.}
\end{figure*}
}
\addtocounter{figure}{-1}

\onlfig{
\clearpage
\begin{figure*}
\addtocounter{figure}{-1}
%\centerline{\resizebox{0.45\hsize}{!}{\includegraphics[angle=0]{/homes/belloche/SgrB2/ALMA/Analysis/Weeds/R-D/Figs/ch3ocho_ve0_p2_6.eps}}}
\centerline{\resizebox{0.45\hsize}{!}{\includegraphics[angle=0]{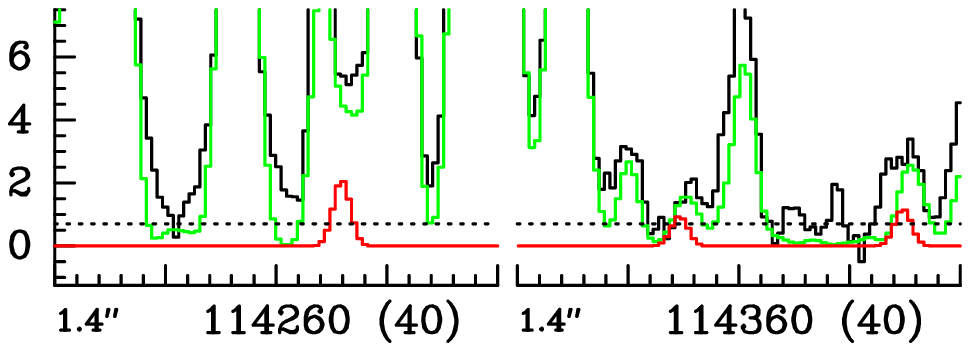}}}
\caption{continued.}
\end{figure*}
}
\addtocounter{figure}{-1}

\onlfig{
\clearpage
\begin{figure*}
%\centerline{\resizebox{0.9\hsize}{!}{\includegraphics[angle=0]{/homes/belloche/SgrB2/ALMA/Analysis/Weeds/R-D/Figs/ch3ocho_ve1_p2_1.eps}}}
\centerline{\resizebox{0.9\hsize}{!}{\includegraphics[angle=0]{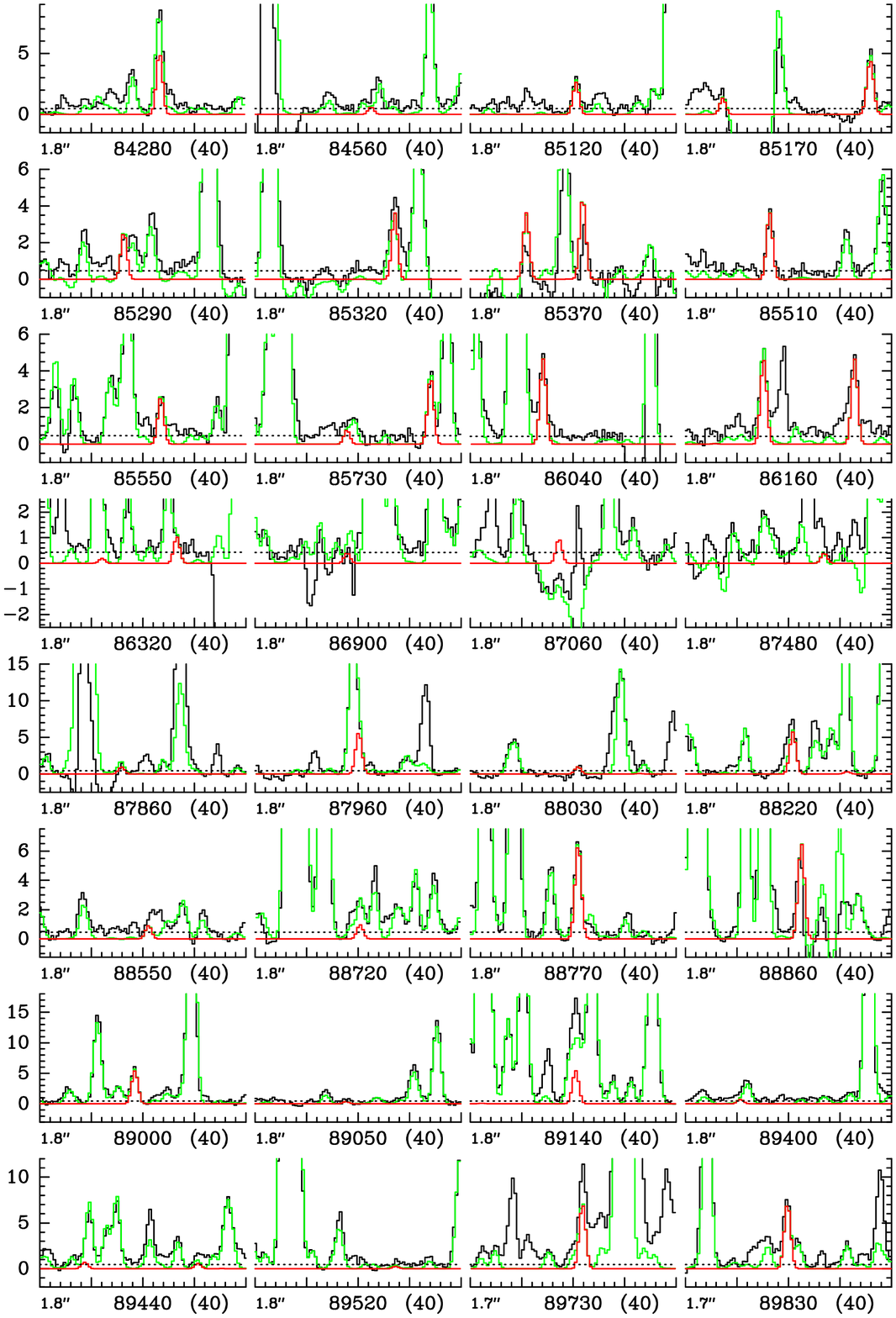}}}
\caption{Same as Fig.~\ref{f:spec_c2h5cn_ve0} for CH$_3$OCHO, $\varv_{\rm t}=1$.
}
\label{f:spec_ch3ocho_ve1}
\end{figure*}
}

\onlfig{
\clearpage
\begin{figure*}
\addtocounter{figure}{-1}
%\centerline{\resizebox{0.9\hsize}{!}{\includegraphics[angle=0]{/homes/belloche/SgrB2/ALMA/Analysis/Weeds/R-D/Figs/ch3ocho_ve1_p2_2.eps}}}
\centerline{\resizebox{0.9\hsize}{!}{\includegraphics[angle=0]{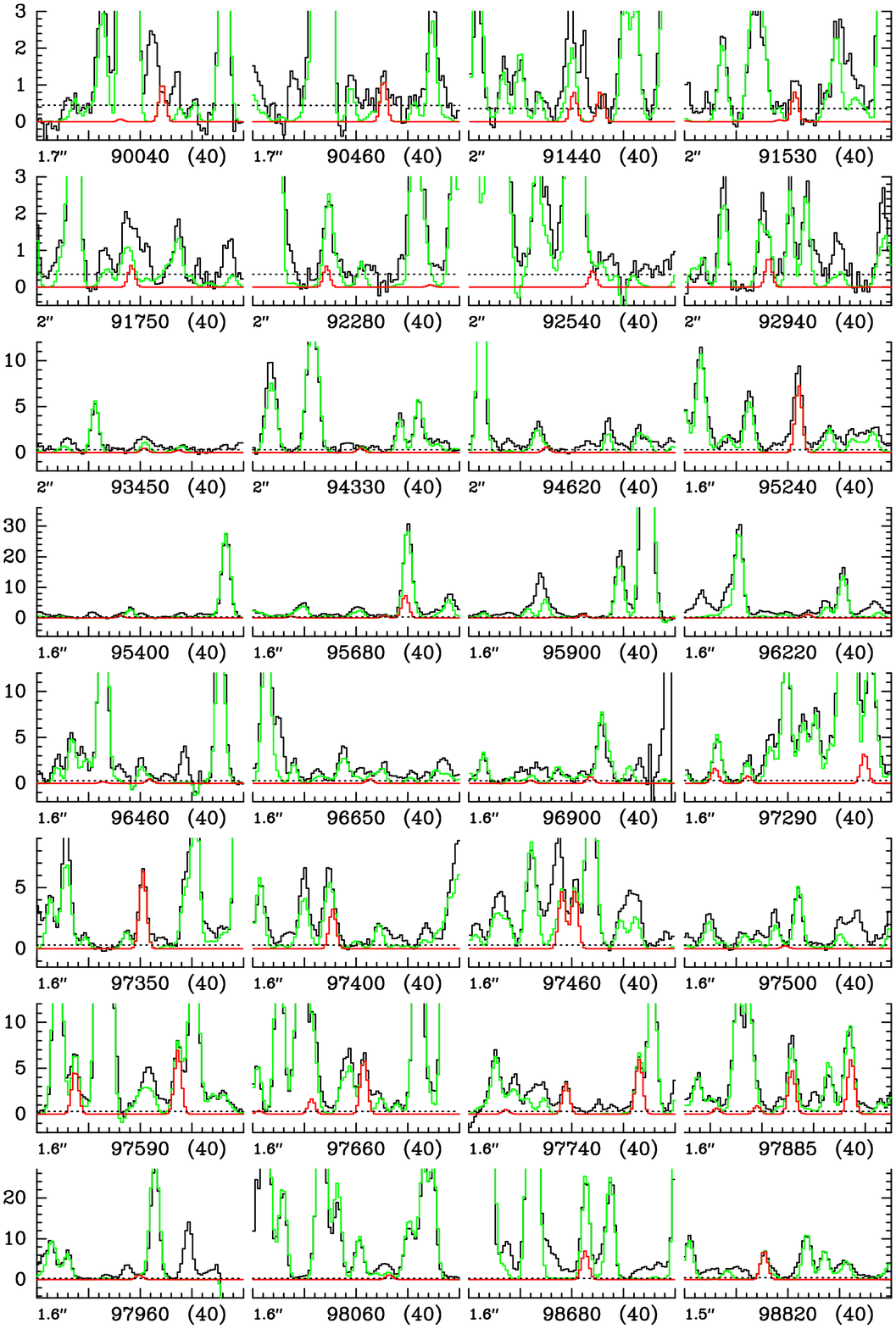}}}
\caption{continued.}
\end{figure*}
}
\addtocounter{figure}{-1}

\onlfig{
\clearpage
\begin{figure*}
\addtocounter{figure}{-1}
%\centerline{\resizebox{0.9\hsize}{!}{\includegraphics[angle=0]{/homes/belloche/SgrB2/ALMA/Analysis/Weeds/R-D/Figs/ch3ocho_ve1_p2_3.eps}}}
\centerline{\resizebox{0.9\hsize}{!}{\includegraphics[angle=0]{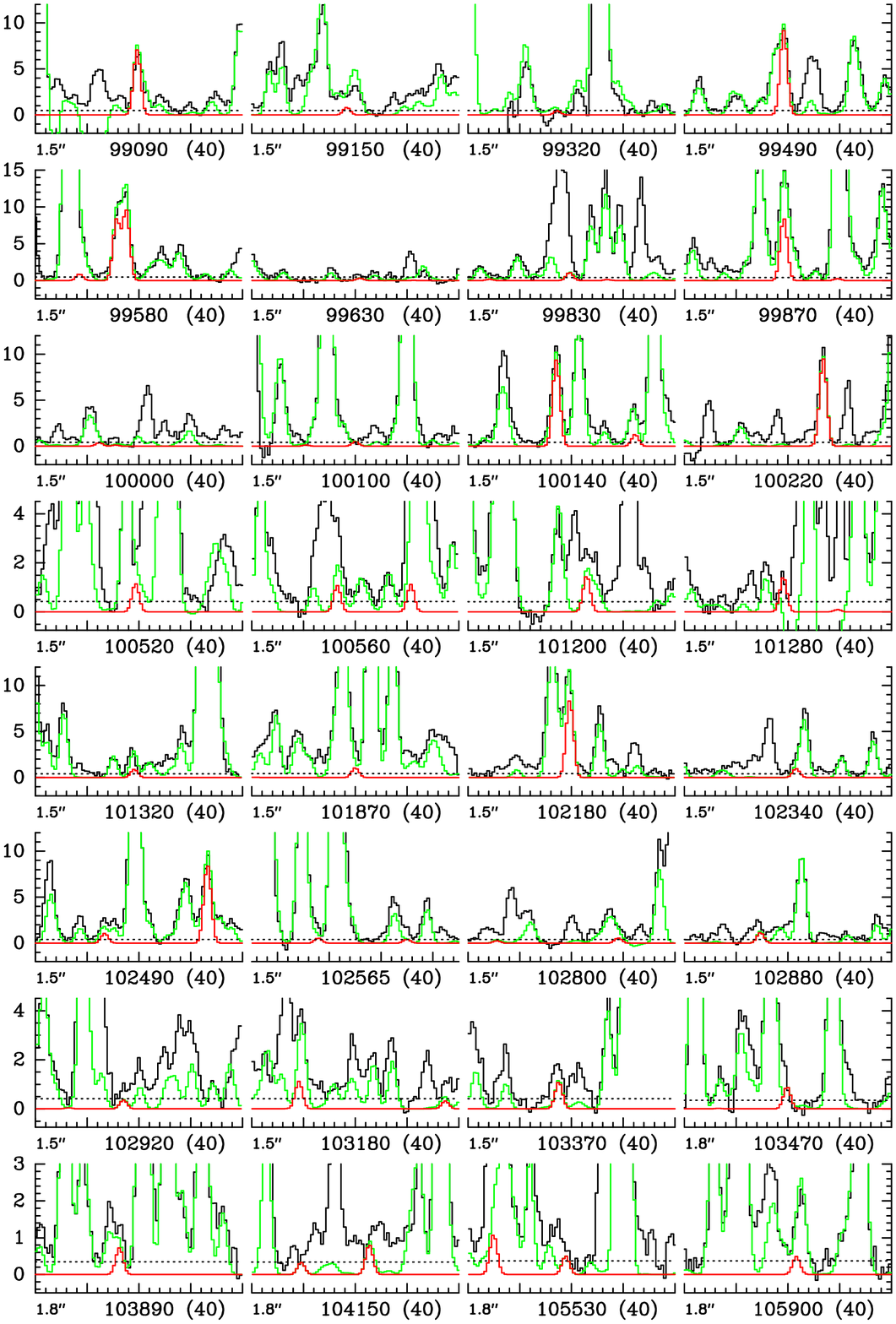}}}
\caption{continued.}
\end{figure*}
}
\addtocounter{figure}{-1}

\onlfig{
\clearpage
\begin{figure*}
\addtocounter{figure}{-1}
%\centerline{\resizebox{0.9\hsize}{!}{\includegraphics[angle=0]{/homes/belloche/SgrB2/ALMA/Analysis/Weeds/R-D/Figs/ch3ocho_ve1_p2_4.eps}}}
\centerline{\resizebox{0.9\hsize}{!}{\includegraphics[angle=0]{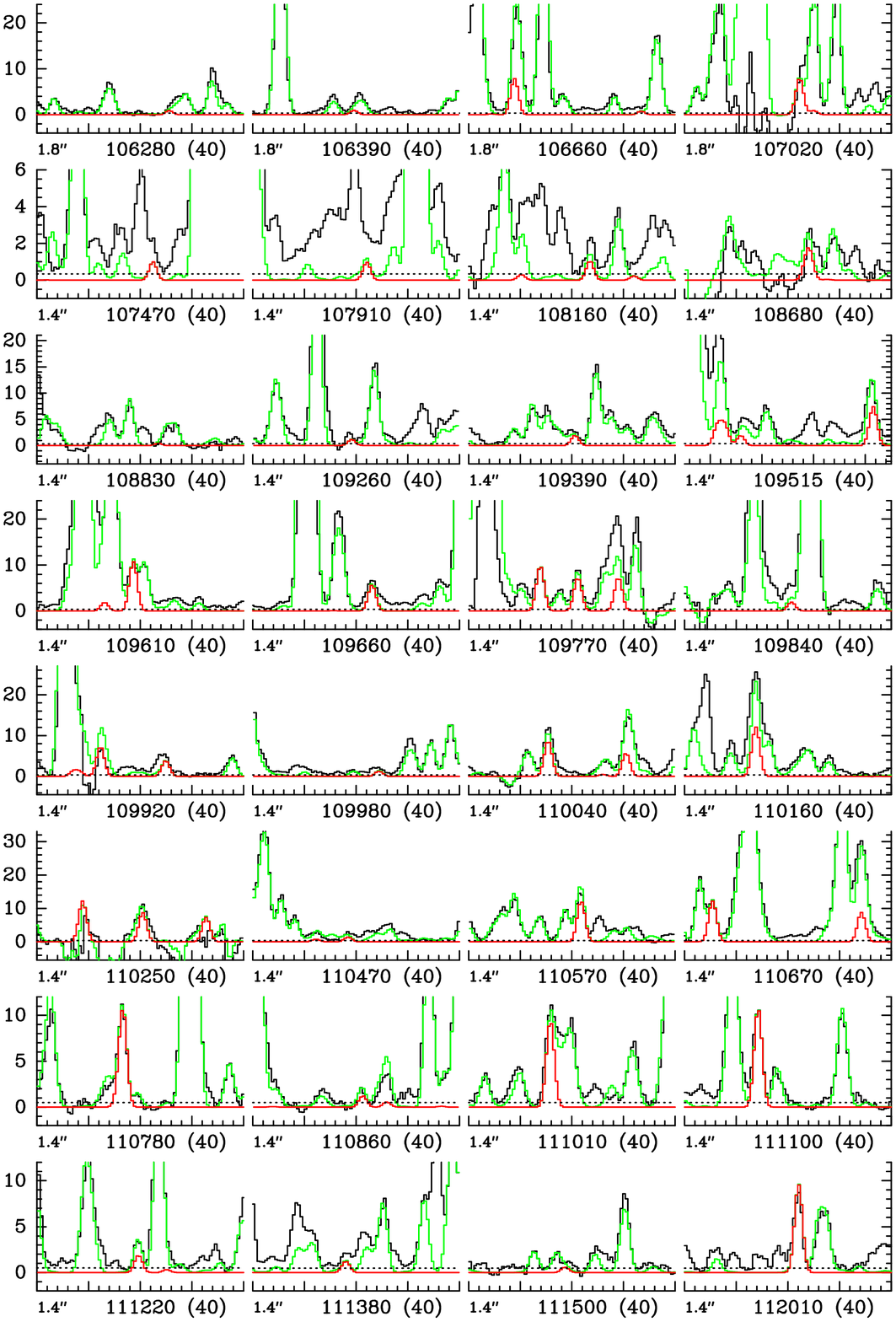}}}
\caption{continued.}
\end{figure*}
}
\addtocounter{figure}{-1}

\onlfig{
\clearpage
\begin{figure}
\addtocounter{figure}{-1}
%\centerline{\resizebox{0.9\hsize}{!}{\includegraphics[angle=0]{/homes/belloche/SgrB2/ALMA/Analysis/Weeds/R-D/Figs/ch3ocho_ve1_p2_5.eps}}}
\centerline{\resizebox{0.9\hsize}{!}{\includegraphics[angle=0]{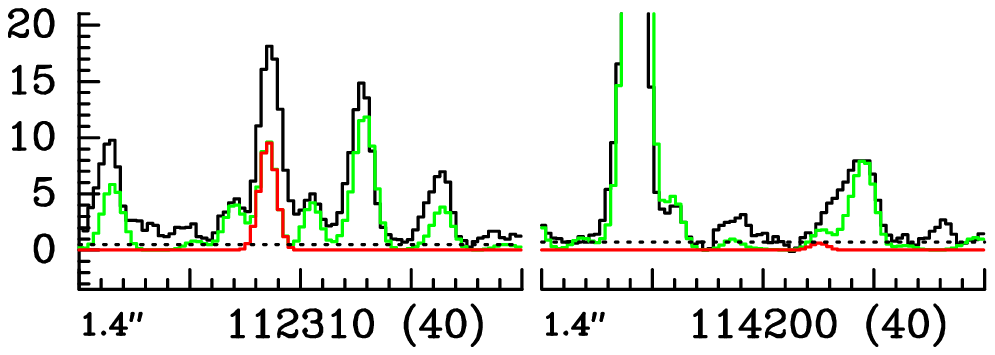}}}
\caption{continued.}
\end{figure}
}
\addtocounter{figure}{-1}

\onlfig{
\begin{figure*}
%\centerline{\resizebox{0.45\hsize}{!}{\includegraphics[angle=0]{/homes/belloche/SgrB2/ALMA/Analysis/Weeds/R-D/popdiag_ch3ocho_p2.eps}}}
\centerline{\resizebox{0.45\hsize}{!}{\includegraphics[angle=0]{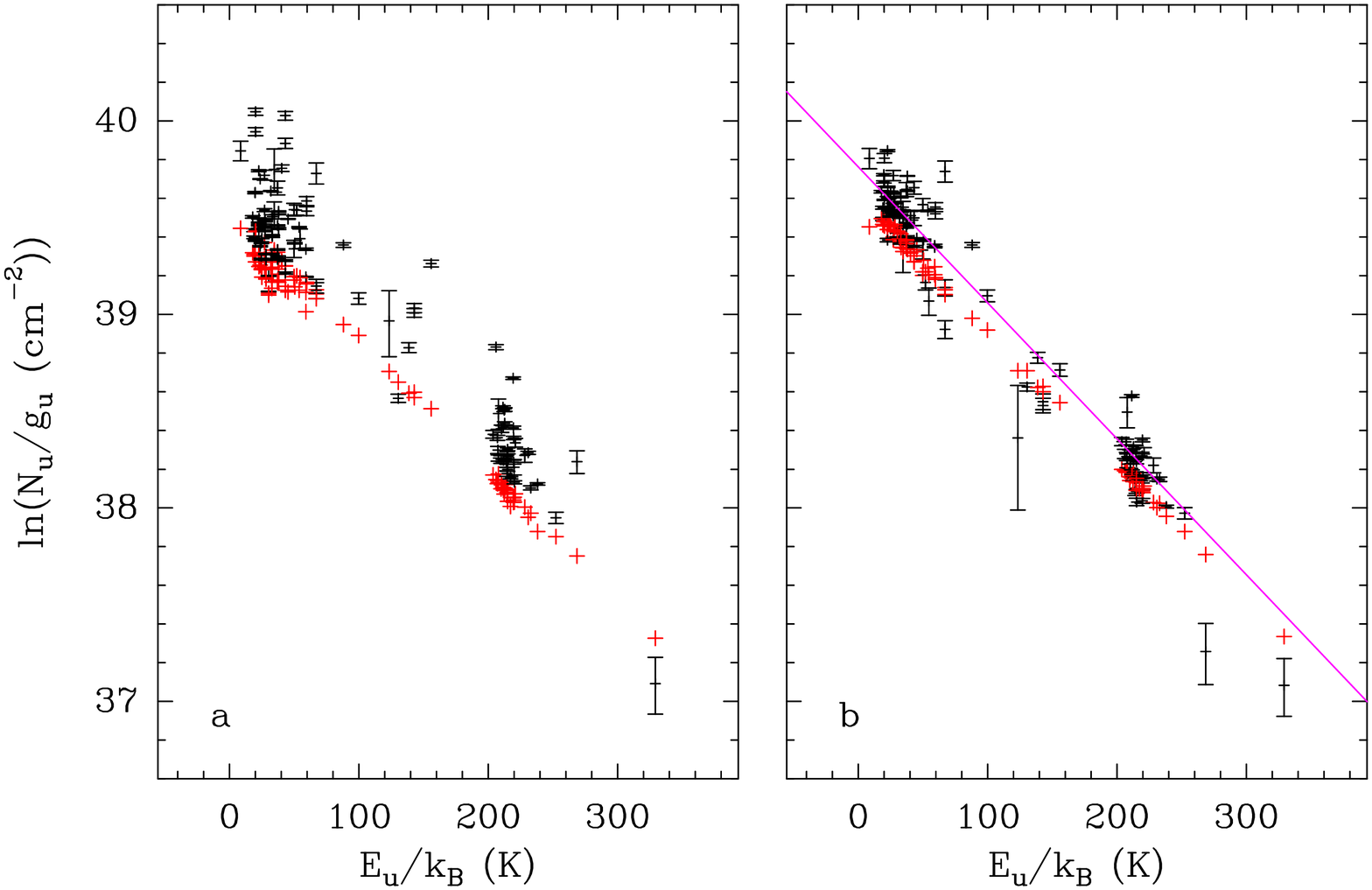}}}
\caption{Same as Fig.~\ref{f:popdiag_c2h5cn} for CH$_3$OCHO, $\varv_{\rm t}=0$ 
and $\varv_{\rm t}=1$.}
\label{f:popdiag_ch3ocho}
\end{figure*}
}

Assuming the same LTE parameters as for the main isotopologue, we searched for
the in-plane and out-of-plane conformers of CH$_2$DOCHO toward Sgr~B2(N2) but
none of them is detected. Upper limits to their individual column densities 
are reported in Table~\ref{t:coldens}.

%\subsubsection{Deuterated dimethyl ether?}
%blah 
%\subsubsection{Deuterated ethenone}

%__________________________________________________________________

\section{Discussion}
\label{s:discussion}

\subsection{Comparison to other observations}

\begin{figure}[!t]
%\centerline{\resizebox{1.0\hsize}{!}{\includegraphics[angle=0]{/homes/belloche/SgrB2/ALMA/Analysis/Weeds/R-D/plot_dfrac.eps}}}
\centerline{\resizebox{1.0\hsize}{!}{\includegraphics[angle=0]{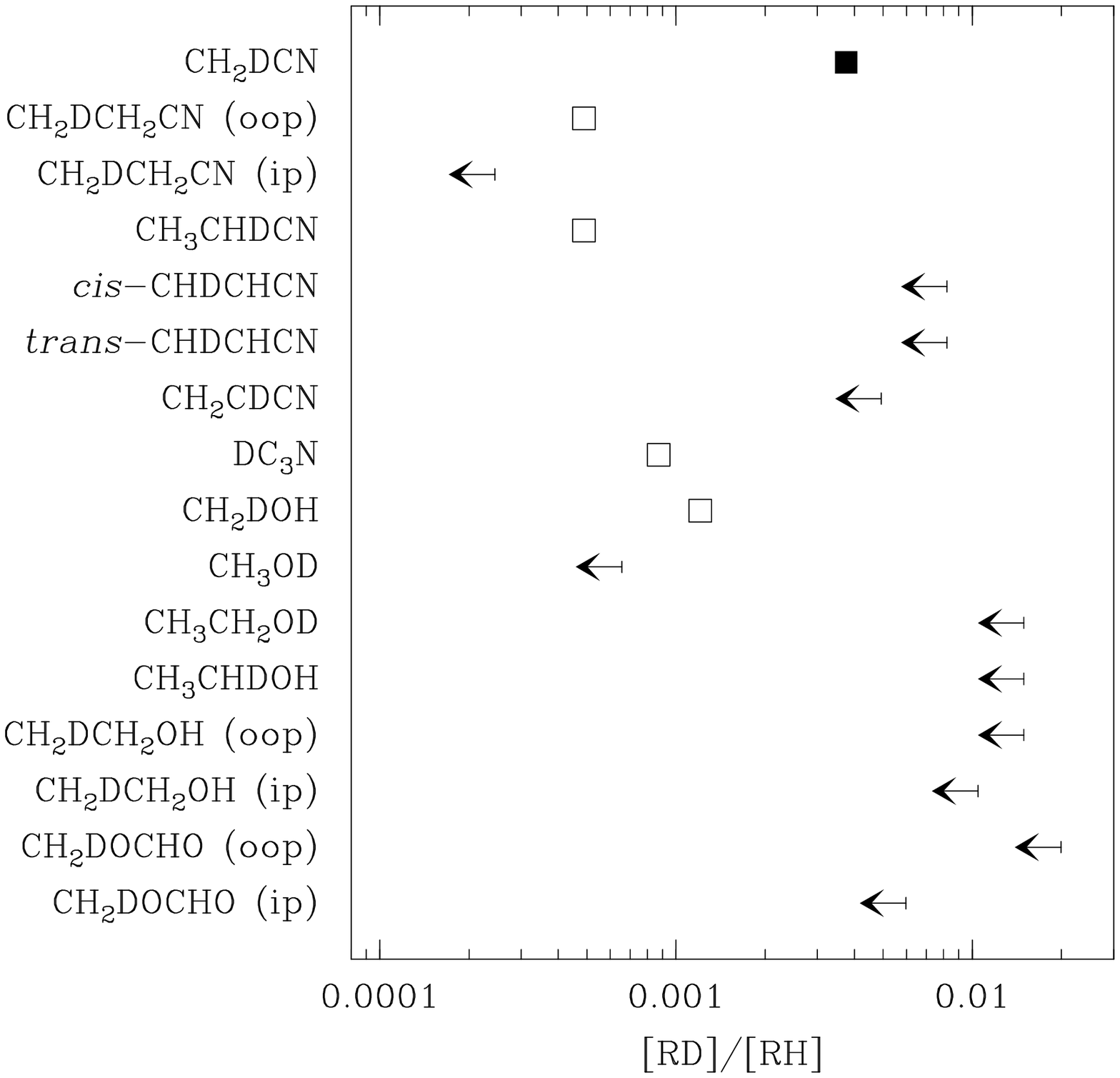}}}
\caption{Deuterium fractionation of (complex) organic molecules toward
Sgr~B2(N2). Secure detections are indicated with a filled square, tentative 
detections with an empty square, and upper limits with an arrow pointing to 
the left. 'oop' and 'ip' describe the position of the deuterium and stand for
'out of plane' and 'in plane', respectively.}
\label{f:dfrac}
\end{figure}

The levels of deuterium fractionation derived in Sect.~\ref{s:results} for 
(complex) organic molecules toward Sgr~B2(N2) are summarized in 
Table~\ref{t:dfrac} and displayed in Fig.~\ref{f:dfrac}.

\begin{table}[!t]
 {\centering
 \caption{
 Deuterium fractionation of selected (complex) organic molecules toward Sgr~B2(N2) compared to predictions of astrochemical models.
}
 \label{t:dfrac}
 \vspace*{0.0ex}
 \begin{tabular}{lcrlcc}
 \hline\hline
 \noalign{\smallskip}
 \multicolumn{1}{c}{Molecule} & \multicolumn{1}{c}{\hspace*{-3.0ex}Status\tablefootmark{a}} & \multicolumn{4}{c}{\hspace*{-2ex}[XD]/[XH]} \\ 
 \cline{3-6}
 \noalign{\smallskip}
  & & \multicolumn{2}{c}{\hspace*{-2ex}N2\tablefootmark{b}} & \multicolumn{1}{c}{\hspace*{-2ex}T14\tablefootmark{c}} & \multicolumn{1}{c}{\hspace*{-2ex}A12\tablefootmark{d}} \\ 
  & & \multicolumn{2}{c}{\hspace*{-2ex}\%} & \multicolumn{1}{c}{\hspace*{-2ex}\%} & \multicolumn{1}{c}{\hspace*{-2ex}\%} \\ 
 \hline
 \noalign{\smallskip}
 CH$_2$DCN &\hspace*{-3.0ex} d &\hspace*{-2ex} & \hspace*{-2.5ex}    0.38 & \hspace*{-2ex} 3.6 -- 0.15  & \hspace*{-2ex} 2.3 -- 6.3 \\ 
 CH$_2$DCH$_2$CN (oop) &\hspace*{-3.0ex} t &\hspace*{-2ex} & \hspace*{-2.5ex}    0.05 & \hspace*{-2ex} ---  & \hspace*{-2ex} --- \\ 
 CH$_2$DCH$_2$CN (ip) &\hspace*{-3.0ex} n &\hspace*{-2ex} $<$ & \hspace*{-2.5ex}   0.024 & \hspace*{-2ex} ---  & \hspace*{-2ex} --- \\ 
 CH$_3$CHDCN &\hspace*{-3.0ex} t &\hspace*{-2ex} & \hspace*{-2.5ex}    0.05 & \hspace*{-2ex} ---  & \hspace*{-2ex} --- \\ 
 \textit{cis-}CHDCHCN &\hspace*{-3.0ex} n &\hspace*{-2ex} $<$ & \hspace*{-2.5ex}     0.8 & \hspace*{-2ex} ---  & \hspace*{-2ex} --- \\ 
 \textit{trans-}CHDCHCN &\hspace*{-3.0ex} n &\hspace*{-2ex} $<$ & \hspace*{-2.5ex}     0.8 & \hspace*{-2ex} ---  & \hspace*{-2ex} --- \\ 
 CH$_2$CDCN &\hspace*{-3.0ex} n &\hspace*{-2ex} $<$ & \hspace*{-2.5ex}     0.5 & \hspace*{-2ex} ---  & \hspace*{-2ex} --- \\ 
 DC$_3$N &\hspace*{-3.0ex} t &\hspace*{-2ex} & \hspace*{-2.5ex}    0.09 & \hspace*{-2ex} ---  & \hspace*{-2ex} 5.4 -- 1.1 \\ 
 CH$_2$DOH &\hspace*{-3.0ex} t &\hspace*{-2ex} & \hspace*{-2.5ex}    0.12 & \hspace*{-2ex} 5.5 -- 0.51  & \hspace*{-2ex} 2.4 -- 2.4\tablefootmark{e} \\ 
 CH$_3$OD &\hspace*{-3.0ex} n &\hspace*{-2ex} $<$ & \hspace*{-2.5ex}    0.07 & \hspace*{-2ex} 3.5 -- 0.3  & \hspace*{-2ex} ---\tablefootmark{e} \\ 
 CH$_3$CH$_2$OD &\hspace*{-3.0ex} n &\hspace*{-2ex} $<$ & \hspace*{-2.5ex}     1.5 & \hspace*{-2ex} ---  & \hspace*{-2ex} --- \\ 
 CH$_3$CHDOH &\hspace*{-3.0ex} n &\hspace*{-2ex} $<$ & \hspace*{-2.5ex}     1.5 & \hspace*{-2ex} ---  & \hspace*{-2ex} --- \\ 
 CH$_2$DCH$_2$OH (oop) &\hspace*{-3.0ex} n &\hspace*{-2ex} $<$ & \hspace*{-2.5ex}     1.5 & \hspace*{-2ex} ---  & \hspace*{-2ex} --- \\ 
 CH$_2$DCH$_2$OH (ip) &\hspace*{-3.0ex} n &\hspace*{-2ex} $<$ & \hspace*{-2.5ex}     1.0 & \hspace*{-2ex} ---  & \hspace*{-2ex} --- \\ 
 CH$_2$DOCHO (oop) &\hspace*{-3.0ex} n &\hspace*{-2ex} $<$ & \hspace*{-2.5ex}     2.0 & \hspace*{-2ex} 14--0.43\tablefootmark{f}  & \hspace*{-2ex} --- \\ 
 CH$_2$DOCHO (ip) &\hspace*{-3.0ex} n &\hspace*{-2ex} $<$ & \hspace*{-2.5ex}     0.6 & \hspace*{-2ex} 7 -- 0.22\tablefootmark{f}  & \hspace*{-2ex} --- \\ 
 \hline
 \end{tabular}
 }\\[1ex] 
 \tablefoot{'oop' and 'ip' describe the position of the deuterium and stand for 'out of plane' and 'in plane', respectively.
 \tablefoottext{a}{d: detection, t: tentative detection, n: non-detection.}
 \tablefoottext{b}{Deuterium fractionation measured toward Sgr~B2(N2).}
 \tablefoottext{c}{Deuterium fractionation predicted by the model of \citet{Taquet14} in the hot corino at the beginning and end of the Class 0 phase.}
 \tablefoottext{d}{Deuterium fractionation predicted by the model of \citet{Aikawa12} in the hot corino at the beginning and end of the Class 0 phase.}
 \tablefoottext{e}{The model of \citet{Aikawa12} was not designed to predict the abundance ratios of deuterated isomers: it assumes statistical branching ratios.}
 \tablefoottext{f}{The model of \citet{Taquet14} does not distinguish between the in-plane and out-of-plane conformers. The values listed here assume a statistical distribution (2:1).}
 }
 \end{table}

\subsubsection{Deuterated methyl cyanide}
\label{sss:ch2dcn}

The detection of CH$_2$DCN toward Sgr~B2(N2) is the most secure among the 
deuterated species reported here (Sect.~\ref{ss:ch3cn} and 
Fig.~\ref{f:spec_ch2dcn}). On the basis of the LTE modeling of methyl
cyanide and its various isotopologues, we derive a deuterium fractionation of 
$0.4\%$ for this molecule. This is a factor 2.6 lower than 
the fractionation reported by \citet{Gerin92} toward Orion~KL ($1\%$). Along 
with this first interstellar detection, these authors also reported a 
tentative detection toward the hot core G34.26+0.15 that, if true, would 
indicate a similar level of deuterium fractionation as toward Orion~KL. The 
difference with the level measured in Sgr~B2(N2) is probably not significant 
because the Orion KL and G34.26+0.15 values may suffer, as mentioned by these 
authors, from a lack of knowledge of the source size and opacity of the lines 
of the main isotopologue. 

A detection of deuterated methyl cyanide toward the Class~0 
(low-mass) protostar IRAS~16293--2422 was also reported in \citet{Taquet14}
based on an unpublished analysis. They quote a deuterium fractionation of 
$1.3\%$, a factor 3.4 higher than the one obtained for Sgr~B2(N2).

\subsubsection{Deuterated ethyl cyanide}
\label{sss:c2h5cn_d}

Both deuterated isotopologues of ethyl cyanide are tentatively detected toward 
Sgr~B2(N2) (Sect.~\ref{ss:c2h5cn} and Figs.~\ref{f:spec_ch3chdcn} and
\ref{f:spec_ch2dch2cn-oop}). We derive a deuterium fractionation of 
$\sim 0.05\%$ for both CH$_2$DCH$_2$CN (in its out-of-plane conformation) and 
the chiral molecule CH$_3$CHDCN. The upper limit obtained for the 
in-plane conformer of 
CH$_2$DCH$_2$CN (Fig.~\ref{f:spec_ch2dch2cn-ip}) is still consistent with the 
expectation that it should be twice less abundant than the out-of-plane one. 
If we assume such a ratio, then the total
deuterium fractionation for CH$_2$DCH$_2$CN would be $\sim 0.075\%$, a 
factor 1.5 times higher than for CH$_3$CHDCN. This would be consistent 
with the statistical expectation because the methyl group at the end of the
carbon chain has three equivalent hydrogen atoms while the middle-chain one has 
only two.

The deuterium fractionation derived for ethyl cyanide toward Sgr~B2(N2) is 
nearly one order of magnitude lower than for methyl cyanide (0.4\%), but 
similar within a factor two to methanol (0.12\%) and cyanoacetylene (0.09\%).

\citet{Margules09} reported a detection of the $^{15}$N isotopologue of ethyl 
cyanide toward Orion KL but obtained only an upper limit for CH$_2$DCH$_2$CN. 
They derived a column density ratio 
[CH$_2$DCH$_2$CN (oop)]/[C$_2$H$_5$C$^{15}$N] $< 0.33$,
which translates into [CH$_2$DCH$_2$CN (oop)]/[C$_2$H$_5$CN] < 0.2\%
using the $^{14}$N/$^{15}$N isotopic ratio of $148 \pm 74$ derived by 
\citet{Daly13}. We note that \citet{Daly13} claimed
tentative detections of both deuterated isotopologues of ethyl cyanide 
with a deuterium fractionation of 2\%, based on the same survey of Orion~KL.
No detected transitions are shown in that study, though, and their Table 3 
actually reports upper limits for the deuterated species. Given the lower 
deuterium fractionation obtained by \citet{Gerin92} for methyl cyanide toward
Orion~KL (1\%) and the order of magnitude difference in deuterium fractionation
between methyl cyanide and ethyl cyanide obtained here toward Sgr~B2(N2), a 
deuterium fractionation of 2\% for ethyl cyanide in Orion~KL sounds unlikely
and questions the tentative detection of \citet{Daly13}.
 
\subsubsection{Deuterated methanol}
\label{sss:ch3oh_d}

CH$_2$DOH is tentatively detected toward Sgr~B2(N2) (Sect.~\ref{ss:ch3oh} and
Fig.~\ref{f:spec_ch2doh}). The deuterium fractionation we derive for this 
isotopologue is 0.12\%, a factor $\sim 3$ lower than the value 
we obtain for methyl cyanide, the deuterated species of which is securely 
identified in our ALMA spectrum of Sgr~B2(N2) (Sect.~\ref{sss:ch2dcn}). 
Therefore, even if the difference in 
deuterium fractionation between methanol and methyl cyanide is a priori 
surprising, the fact that it is lower for the former gives us more confidence 
in the detection of CH$_2$DOH. The deuterium fractionation derived for 
CH$_2$DOH is a factor five lower than toward the Compact Ridge in Orion~KL
\cite[0.58\%,][]{Neill13}, about one order of magnitude lower than toward the
high-mass protostellar objects surveyed by \citet{Fontani15} and the 
intermediate-mass protostar NGC~7129 FIRS~2 
\citep[2\%,][]{Fuente14}, and more than 
two orders of magnitude lower than toward Class 0 protostars which have values 
between 19\% and 33\% \citep[][]{Parise06}\footnote{\label{fn:ch2doh}There was 
an issue with the spectroscopic predictions used in the early studies 
reporting CH$_2$DOH column densities (B. Parise, priv. comm.). We compared the 
$S\mu^2$ values listed in Table 1 of \citet{Parise02}, which were also used in 
\citet{Parise06}, with the current JPL catalog. The new values of the selected 
transitions are a factor $2.1 \pm 0.4$ times higher than the old ones on 
average. The partition function is the same in both cases. As a result, the 
column densities reported for CH$_2$DOH in both articles were overestimated by 
a factor of $\sim 2$.}.

Evidence for the presence of CH$_3$OD is too tenuous in our ALMA spectrum of
Sgr~B2(N2) to claim a detection, even a tentative one (Sect.~\ref{sss:ch3od}). 
Still, we cannot completely exclude that CH$_3$OD is present at the level 
indicated by our upper limit. The synthetic spectrum shown in 
Fig.~\ref{f:spec_ch3od_ve0} indicates that a large fraction ($>50\%$) of the 
flux density detected at 90744, 99964, 110951, and 113352~MHz may well be 
emitted by this molecule. If this is true, the deuterium fractionation of 
methanol would then be $\sim 0.07\%$ for CH$_3$OD. This would be nearly one 
order of magnitude lower than toward the Compact Ridge in Orion~KL 
\citep[0.5\%,][]{Neill13} and about 50 times lower than toward Class~0 
protostars \citep[1.6\%--4.7\%,][]{Parise06}.

A detection of CH$_3$OD toward Sgr~B2 was reported by \citet{Gottlieb79}
with the 36 foot radio telescope of the National Radio Astronomy Observatory
at Kitt Peak ($HPBW \sim 74''$). They detected a line at the frequency 
expected for the pair of partially blended transitions 2$_{-1}$--1$_{-1}$ E 
(90703.6~MHz, $E_{\rm u}/k_{\rm B} = 11.3$~K) and 2$_0$--1$_0$ A (90705.8~MHz, 
$E_{\rm u}/k_{\rm B} = 6.5$~K), but they did not detect the nearby 
2$_1$--1$_1$ E transition (90743.5~MHz, $E_{\rm u}/k_{\rm B} = 15.6$~K) that
is in fact expected to be partially blended with the 10$_1$--9$_2$ A 
transition (90741.7~MHz, $E_{\rm u}/k_{\rm B} = 124$~K). The 
former two transitions are blended with deep HNC absorption features produced 
by diffuse clouds along the line of sight in our ALMA spectrum of Sgr~B2(N2).
This prevents their detection in our spectrum (see 
Fig.~\ref{f:spec_ch3od_ve0}). Our LTE model shows that, for a temperature 
of 160~K, the latter two transitions are expected to be as strong as
the former two. We conclude from this that either the assignment of the 
90704~MHz line to CH$_3$OD in the Kitt Peak spectrum was not correct, or the 
line reported by \citet{Gottlieb79} traces low-excitation emission of CH$_3$OD.
Given that the line detected in emission in the Kitt Peak spectrum dominates 
over the absorption features, opposite to what is seen in the ALMA spectrum, 
this emission line, if real, must come from a region more extended than the 
Sgr~B2 continuum emission that is absorbed by the diffuse clouds along the 
line of sight. Such an extended emission would be filtered out in our ALMA 
spectrum. 

\citet{Gottlieb79} derived a ratio [CH$_3$OD]/[$^{13}$CH$_3$OH]
$\sim 0.18$ for Sgr~B2. This translates into [CH$_3$OD]/[CH$_3$OH] 
$\sim 0.7\%$, assuming a $^{12}$C/$^{13}$C isotopic ratio of 25 as derived for 
methanol toward Sgr~B2(N2) in the companion paper \citet{Mueller15c}. This 
is an order of magnitude higher than our upper limit of 0.07\% 
derived in Sect.~\ref{ss:ch3oh}. This discrepancy seriously questions the 
detection of CH$_3$OD reported by \citet{Gottlieb79} toward Sgr~B2, unless 
deuteration of methanol is more efficient by one order of magnitude on large
scales in the Sgr~B2 cloud compared to the embedded hot cores. We note that
\citet{Comito03} derived an abundance ratio [HDO]/[H$_2$O] $\sim 0.06\%$
toward the Sgr~B2 hot cores and their $T<100$~K envelope, and even lower 
values of $\sim 0.013\%$ and $\sim 0.02\%$ (uncertain within a factor two) 
were obtained for [DCN]/[HCN] and [DCO$^+$]/[HCO$^+$] in the molecular ridge 
close to Sgr~B2(M) \citep[][]{Jacq99}. Deuterium fractionation thus does not 
appear to be generally more efficient on larger scales in Sgr~B2, which again 
questions the detection of CH$_3$OD reported by \citet{Gottlieb79}.

\subsubsection{Deuterated cyanoacetylene}
\label{sss:dc3n}

The detection of DC$_3$N reported toward Sgr~B2(N2) is only tentative 
(Sect.~\ref{ss:hc3n} and Fig.~\ref{f:spec_dc3n}). We obtain a deuterium
fractionation of 0.09\%, similar to the values obtained for methanol and
ethyl cyanide. DC$_3$N was first detected toward TMC~1 with a deuterium
fractionation of 2--8\% \citep[][]{Langer80}, revised to a lower value of
1.5\% by \citet{Turner01}. High values were reported with single dish
telescopes for a number of other cold dense cores \citep[5\%--10\%,][]{Howe94} 
as well as for a protostar in a stage of ``Warm Carbon-Chain Chemistry'' 
\citep[$\sim 3\%$,][]{Sakai09}. A tentative detection toward 
the Compact Ridge 
and the Hot Core of Orion~KL was recently reported with a deuterium 
fractionation of $1.5\% \pm 0.9\%$ \citep[][]{Esplugues13}. 
A tentative detection was also recently reported toward the high-mass 
protostar NGC~2264 CMM3 \citep[1.8\% $\pm$ 1.5\%,][]{Watanabe15}.
The deuterium 
fractionation of cyanoacetylene tentatively derived toward Sgr~B2(N2) is thus
at least one order of magnitude lower than in Orion and NGC~2264 CMM3 
(if confirmed) and even two 
orders of magnitude lower than in cold dense gas.

\subsubsection{Deuterated vinyl cyanide}
\label{sss:c2h3cn_d}

The column density upper limits reported in Sect.~\ref{sss:c2h3cn} yield
deuterium fractionations $<0.8\%$, $<0.8\%$, and $<0.5\%$ for 
\textit{cis-}CHDCHCN, \textit{trans-}CHDCHCN, and CH$_2$CDCN, respectively.
We are not aware of any reliable detection of deuterated vinyl cyanide in the
interstellar medium.

\subsubsection{Deuterated ethanol}
\label{sss:c2h5oh_d}

The column density upper limits reported in Sect.~\ref{sss:c2h5oh} yield
deuterium fractionations $<1.5\%$, $<1.5\%$, $<1.5\%$, and $<1.0\%$ for 
CH$_3$CH$_2$OD, CH$_3$CHDOH, and the out-of-plane and in-plane conformers of 
CH$_2$DCH$_2$OH, respectively. The latter two translate into a total deuterium 
fractionation $<2.5\%$ for CH$_2$DCH$_2$OH.
These upper limits are about one order of 
magnitude higher than the deuterium fractionation measured for methanol 
(Sect.~\ref{sss:ch3oh_d}). They are thus not very constraining. We are not 
aware of any detection of deuterated ethanol in the interstellar medium.

\subsubsection{Deuterated methyl formate}
\label{sss:ch3ocho_d}

The column density upper limits reported in Sect.~\ref{sss:ch3ocho} yield
deuterium fractionations $<2.0\%$ and $<0.6\%$ for the out-of-plane and 
in-plane conformers of CH$_2$DOCHO, respectively. This implies an upper limit
of $2.6\%$ for the molecule as a whole, or even $1.8\%$ if we assume a 
statistical distribution and use the most stringent constraint provided by the
in-plane conformer. A detection of CH$_2$DOCHO toward the Compact Ridge of
Orion~KL was reported by \citet{Coudert13}. They derived a deuterium 
fractionation of $4\% \pm 2\%$. \citet{Fuente14} reported a tentative 
detection of CH$_3$OCDO and both conformers of CH$_2$DOCHO at a deuteration
level of 6\% toward the intermediate-mass protostar NGC~7129 FIRS~2, but the 
value is uncertain due to blends with other species.
Our upper limit for Sgr~B2(N2) lies a factor two to three below these values.

\subsection{Comparison to predictions of chemical models}
\label{ss:comp_chem}

We compare the deuterium fractionation derived for complex organic molecules
toward Sgr~B2(N2) to predictions of astrochemical models that 
investigated the deuteration of such species. The case of 
deuterated methanol is more specifically discussed in 
Sect.~\ref{ss:ch2doh_ch3od_ratio} below. The two models
considered here follow the evolution of molecular abundances in a collapsing 
dense core that is heated up by the central protostar once it is formed 
\citep[][]{Taquet12,Taquet14,Aikawa12}. These models were designed for
Class~0 protostars, i.e. low-mass objects, so they are not optimum for a
comparison to Sgr~B2(N2). However they are the only models dealing with 
deuterated complex molecules that are presently available in the literature. 
No model following the deuteration of complex molecules in a high-mass 
star forming region like Sgr~B2(N2) has been published so far. As we will see 
in Sect.~\ref{ss:deuterium_gc} below, there is a clear need for that.

Both models couple the chemistry in the
gas-phase and on the dust grains. The layering of the icy grain mantles
and the presence of \textit{ortho-} and \textit{para-}H$_2$ are taken into 
account in \citet[][, hereafter T14]{Taquet14} but not in 
\citet[][, hereafter A12]{Aikawa12}. Model A12 uses a 
one-dimensional radiation hydrodynamic model of gravitational collapse, 
preceded by a hydrostatic phase that lasts for one million years.
Model T14 relies on a parametric description of the quasi-static 
formation of the dense core followed by a free-fall collapse phase during which
the protostar quickly appears after one free-fall time. In both models, the 
gas and dust temperatures are kept equal. In model T14, the temperature 
decreases from 17~K at the external radius down to about 7~K in the inner 
parts of the core at the end of the prestellar phase. In model A12, the 
external temperature is about 10 K and drops to about 5~K at the end of the 
prestellar phase. In both models, the temperature increases once the protostar 
appears.

Table~\ref{t:dfrac} lists the deuterium fractionations predicted by both 
astrochemical models in the hot corino central region at the beginning and end 
of the Class 0 phase. Nearly all predicted values are at least a factor four 
higher than the ones obtained toward Sgr~B2(N2). The only exception is
the deuterium fractionation of CH$_3$CN: it is predicted to decline during
the Class 0 phase in model T14 and the measured value lies close to (a 
factor 2.5 above) the lower end of the predicted range. The decrease 
of the deuterium fractionation of complex organic molecules during the hot 
corino phase in model T14 is due to the collapse of the envelope: as time 
goes by, the gas passing through the hot corino is originating from shells 
that were initially at larger radii, i.e. lower densities, and underwent lower 
deuteration, imposed by a lower level of CO depletion and a higher 
[\textit{ortho}-H$_2$]/[\textit{para}-H$_2$] ratio. However, model A12
predicts an increase of the deuterium fractionation of CH$_3$CN with time
during the protostellar phase, with values a factor $>5$ higher than measured 
toward Sgr~B2(N2), i.e. in worse agreement with the observations than model 
T14. The fact that model A12 does not predict a decrease of the deuteration of 
complex organic molecules with time may be due to the model not taking the 
multi-layered structure of ice mantles into account.

The deuterium fractionation predicted by model T14 for methanol is 1.5--3.4 
times higher than the one of methyl cyanide while model A12 finds the opposite,
with methyl cyanide becoming 2.6 times more deuterated than methanol at the 
end of the Class~0 phase. The ratio measured toward Sgr~B2(N2) is more in
agreement with the latter model.

Model T14 predicts higher deuterium fractionations for methyl formate and 
dimethyl ether because these more complex molecules are formed from the 
photodissociation of highly deuterated formaldehyde and methanol at the surface 
of dust grains when they start to warm up. Both molecules are predicted to be 
even more deuterated than methanol. The upper limit we derive for the 
deuteration of methyl formate -- an order of magnitude higher than that of 
methanol -- is not 
stringent enough to test this prediction. The model predictions for these
more complex molecules could let us expect a similar behaviour for ethyl 
cyanide compared to methyl cyanide. The tentative deuteration level we 
measured for ethyl cyanide is, however, not consistent with this naive 
expectation: ethyl cyanide is (at least) about a factor five less deuterated 
than methyl cyanide in Sgr~B2(N2). 

Unlike all other complex organic molecules, the formation of methyl cyanide is 
dominated by gas phase processes in our own models 
\citep[e.g.,][]{Garrod08,Garrod13}. This occurs via the reaction CH$_3^+$ + HCN 
$\rightarrow$ CH$_3$CNH$^+$, followed by dissociative recombination. In these 
models, the CH$_3^+$ ion is formed via  H$_3^+$ + CH$_3$OH $\rightarrow$ 
CH$_3^+$ + H$_2$O + H$_2$ and similar reactions of methanol with He$^+$ and 
C$^+$. If equivalent gas phase reactions dominate the formation of deuterated 
methyl cyanide also, then the apparent discrepancy between the degrees of 
deuteration of methyl cyanide and ethyl cyanide may come from their different 
formation pathways, the former in the gas phase but the latter on the surface 
of dust grains. Numerical simulations should be performed to test this 
hypothesis.

The rotational temperature and source size derived for vinyl cyanide toward 
Sgr~B2(N2) seem to indicate an origin in a somewhat warmer and more compact 
region than for ethyl cyanide. This is a priori consistent with the early 
model of \citet{Caselli93} where the gas-phase abundance of vinyl cyanide is 
predicted to increase with time in the hot core where it is formed via 
protonation of ethyl cyanide followed by dissociative recombination. However, 
this channel has not been measured in the laboratory nor properly calculated. 
It is thus not 
clear whether this path dominates the formation of vinyl cyanide. If it does, 
we might expect a similar deuterium fractionation as for ethyl cyanide, but 
this is a simplistic assumption. The upper limit we derive for the deuteration 
of vinyl cyanide toward Sgr~B2(N2) -- more than one order of magnitude higher 
than the deuteration measured for ethyl cyanide -- is not low enough to test 
this hypothesis.
 
Finally, model A12 predicts a decrease of the deuterium fractionation of 
cyanoacetylene with time, but the value at the end of the Class 0 phase is 
still more than one order of magnitude higher than the one tentatively derived 
toward Sgr~B2(N2).

\subsection{The [CH$_2$DOH]/[CH$_3$OD] abundance ratio}
\label{ss:ch2doh_ch3od_ratio}

If the deuteration of methanol were purely statistical, then one would 
expect CH$_2$DOH to be three times more abundant than CH$_3$OD because the
methyl group contains three equivalent hydrogen atoms while the hydroxyl group
has only one. Early chemical models producing methanol (and its deuterated
species) on the surface of dust grains predicted such a value 
\citep[][]{Charnley97}. However various chemical processes have been shown or
suggested to lead potentially to different values. \citet{Osamura04} found
theoretically that it is difficult to exchange protons and deuterons between 
the methyl and hydroxyl parts of methanol in the gas phase. Therefore, 
CH$_2$DOH and CH$_3$OD are distinct species that can have different chemical
timescales. In particular, protonation of methanol in the gas phase (after 
sublimation) followed by dissociative recombination is thought to lead to a 
more rapid destruction of CH$_3$OD 
compared to CH$_2$DOH \citep[][]{Charnley97}. This process would increase the
[CH$_2$DOH]/[CH$_3$OD] ratio after sublimation of methanol from the grain 
ices, on a timescale of $10^5$~yr for a temperature of 50~K and an H$_2$
density of $10^6$~cm$^{-3}$ \citep[][]{Osamura04}. On the contrary, at high 
temperature (100~K), the gas phase model of \citet{Osamura04} leads to a
decrease of the [CH$_2$DOH]/[CH$_3$OD] ratio because the release of water and 
its deuterated isotopologue into the gas phase opens a new channel for the 
production of CH$_3$OD via H$_2$DO$^+$. However this model assumes a high 
initial [HDO]/[H$_2$O] ratio of 0.1. The same study showed that, with an 
initial ratio of 0.01, [CH$_2$DOH]/[CH$_3$OD] remains roughly constant.

A preferential decrease of the abundance of CH$_3$OD could also occur on the
grains before desorption because hydrogenation of CH$_3$OD by D--H exchange 
with water was experimentally shown to occur (above 120~K) while no equivalent 
process exists for CH$_2$DOH \citep[][]{Ratajczak09}. \citet{Nagaoka05} showed
experimentally that the formation of deuterated methanol in interstellar ice 
analogs proceeds via H--D substitution of methanol with atomic D rather than 
successive addition of H and D on CO. The formation of CH$_3$OD was not found
to occur during this process, which \citeauthor{Nagaoka05} justify by 
invoking the higher barrier of H abstraction from a hydroxyl group compared to 
the H abstraction from a methyl group. Further constraints on the H/D 
addition/abstraction reactions on formaldehyde and methanol were obtained by 
\citet{Hidaka09}. Gas-grain coupled astrochemical models incorporating 
these new substitution and abstraction reactions on the grain surfaces indeed 
predict that an enhancement of the [CH$_2$DOH]/[CH$_3$OD] ratio beyond the 
statistical value can be reached at low temperature on a timescale depending 
on the density \citep[$> 2 \times 10^{4}$~yr for a density of 
$\sim 5 \times 10^6$~cm$^{-3}$,][]{Taquet12}. However, at earlier times and/or 
lower densities, the [CH$_2$DOH]/[CH$_3$OD] ratio is predicted to have a low 
value of $\sim 1.4$. In a more recent version of this model that includes the 
spin of H$_2$ in the chemical network and follows the dynamical evolution and 
warming-up of a collapsing envelope, the [CH$_2$DOH]/[CH$_3$OD] ratio in the 
gas phase after sublimation is predicted to be $\sim 1.5$ \citep[][]{Taquet14}.

Observationally, CH$_3$OD is found to be underabundant in the 
vicinity of low-mass protostars compared to the statistical expectation, with 
typical [CH$_2$DOH]/[CH$_3$OD] ratios in the range 7--13 
\citep[][, see footnote~\ref{fn:ch2doh}]{Parise06}. Similar high ratios (6 
and 13) were reported for two intermediate-mass protostars 
\citep[][]{Ratajczak11}\footnote{\label{fn:ch2doh-2}We have divided the ratios 
by 1.5 because the CH$_2$DOH column densities reported by \citet{Ratajczak11} 
are affected by the same issue as described in footnote~\ref{fn:ch2doh}. We 
use the ratios reported by the authors based on the transition 
$3_{03}$--$2_{02}$ E0 that has a $S\mu^2$ value in their Table A.1 1.5 times 
lower than in the current JPL catalog.}.
However, this ratio was found to be close to unity in Orion KL: 
\citet{Jacq93} reported a ratio of 1.1--1.5 on the basis of selected 
transitions observed with the IRAM 30\,m telescope; \citet{Peng12} obtained a 
ratio of $0.7 \pm 0.3$ in the central region based on a few transitions mapped 
at high-angular resolution with ALMA\footnote{\citet{Neill13} pointed 
out that the use of more recent spectroscopic predictions for CH$_2$DOH would 
lead to a higher ratio of approximately 1.}; \citet{Neill13} derived a ratio of
$1.2 \pm 0.3$ toward the Compact Ridge based on the analysis of a complete 
line survey performed with \textit{Herschel}. The upper limit of 0.6 obtained 
toward W3(H2O) \citep[][, see footnote~\ref{fn:ch2doh-2}]{Ratajczak11} adds 
another piece of evidence 
suggesting that the [CH$_2$DOH]/[CH$_3$OD] ratio is significantly lower than 
the statistical value in high mass star forming regions.

Our analysis of the ALMA spectrum of Sgr~B2(N2) leads to 
[CH$_2$DOH]/[CH$_3$OD] $> 1.8$. This lower limit is consistent with 
the ratio derived for Orion KL by \citet{Neill13} within the uncertainties 
($2\sigma$). If the actual value in Sgr~B2(N2) is close to this lower
limit, then it would be a further indication that this ratio is indeed 
lower than the statistical value in the hot portions of high-mass star forming 
regions, also in the Galactic Center region. However, taken at face value, our 
lower limit does not exclude a [CH$_2$DOH]/[CH$_3$OD] ratio in Sgr~B2(N2) 
closer to (or even higher than) the statistical value.

The low [CH$_2$DOH]/[CH$_3$OD] ratio found in Orion~KL and the lower
limit obtained for Sgr~B2(N2)
are consistent with the value predicted by model T14. In this model, the ratio
lies below the statistical value because of the abstraction reactions included
in the chemical network following the experiments of \citet{Hidaka09}.
These reactions occur only on the methyl group of methanol, and thus reduce the
[CH$_2$DOH]/[CH$_3$OD] ratio.

%The deuterium 
%fractionations predicted by this model for 
%CH$_2$DOH ($\sim 4\%$ at $t \sim 10^4$~yr after the birth of the protostar) 
%and CH$_3$OD ($\sim 2.7\%$) in the gas phase after their sublimation from the
%grain surfaces are much higher than those measured in Orion~KL and Sgr~B2(N2) 
%but they decrease with time ($\sim 0.5\%$ and $\sim 0.3\%$, respectively, at 
%$t \sim 10^5$~yr) because the gas that falls in and populates the hot corino 
%as time goes by originates from further and 
%further shells that encountered lower densities where deuteration is less 
%efficient. 
%Since deuteration is also less efficient with increasing 
%temperature, the higher temperatures expected in the large-scale envelope of 
%Sgr~B2 \citep[30--50~K in dense clumps interferometrically traced with 
%ammonia,][]{Ott14} compared to the temperature assumed  by \citet{Taquet14} 
%-- 17~K in the external layers, dropping to $6$~K in the inner layers just 
%before the formation of the protostar -- may also explain the overall 
%lower deuterium fractionation in Sgr~B2(N2).

\subsection{Deuterium in the Galactic Center region}
\label{ss:deuterium_gc}

With the only exception of methyl cyanide, the levels of deuterium 
fractionation derived for complex organic molecules in the Sgr~B2(N2) hot core
are all at least a factor four lower than those predicted by current gas-grain
chemical models. Since deuteration is sensitive to temperature -- a higher
temperature implies a lower CO depletion and a higher ortho/para ratio of
H$_2$, both limiting the formation of H$_2$D$^+$ -- the overall lower 
deuteration measured in Sgr~B2(N2) may result from the somewhat higher 
temperatures measured in the Galactic Center region: gas kinetic temperatures 
of 30--50~K were reported in dense clumps interferometrically traced with 
ammonia \citep[][]{Ott14} and dust temperatures derived from infrared 
measurements with the \textit{Herschel} Space Observatory lie in the range 
20--28~K toward Sgr~B2 (Guzm\'an et al., submitted)\footnote{We note, however, 
that high density material with lower dust temperatures may exist and be 
masked by warmer, outer layers traced with \textit{Herschel}.}.
The two numerical studies to which we compared our results were focused on 
(low-mass) hot corinos that are embedded in colder regions: models T14 and A12 
assume a dust temperature in the external layers of the protostellar envelope 
of 17~K and 10~K, respectively, and the temperature in the inner layers drops 
to 7~K and 5~K, respectively, just before the formation of the protostar. 
Changing the dust temperature has a significant impact on the deuteration in
model T14: for instance, the degree of deuterium fractionation of 
methanol (CH$_2$DOH) in the ices at the end of the prestellar phase was found 
to be reduced by a factor 2.8 when the external temperature is set to 20~K 
instead of 17~K \citep[][]{Taquet14}. Dedicated numerical simulations are thus 
needed to evaluate the effect of the dust temperature in the prestellar phase 
on the deuterium composition of hot cores in the Galactic Center region.

Another explanation for the low deuteration level of complex organic molecules 
in Sgr~B2(N2) could be the abundance of deuterium itself in the Galactic 
Center region. Several studies 
\citep[][]{Jacq90,Jacq99,Lubowich00,Polehampton02} suggested that
deuterium may be up to $\sim 10$ times less abundant in the Galactic Center 
region than in the local interstellar medium, due to stellar processing.
The levels of COM deuteration found in Sgr~B2(N2) are indeed systematically
lower (by a factor of a few up to a factor $\sim$10) than those reported in the
Orion-KL Hot Core or Compact Ridge. While a chemical or evolutionary origin 
for this difference cannot be excluded without further modeling, the low
deuteration level of complex organic molecules in Sgr~B2(N2) tends to support 
the idea that deuterium is less abundant in the Galactic Center region by 
maybe up to a factor ten.

% Jacq et al. 1990: the HDO/H2O abundance ratio which we derive (0.002) is of
% the same order as in other galactic sources. We conclude that the fractional
% abundance of deuterated molecules in Sgr B2 does not differ greatly from that
% found in the solar neighbourhood. This is somewhat surprising in that
% deuterium is expected to be destroyed by stellar processing. Moreover, the
% gas kinetic temperature in the galactic center region is typically 70 K
% (see e.g. Guesten, 1990) and hence one cannot expect a large degree of 
% fractionation in deuterated species. A priori, this implies a D/H abundance
% ratio in the galactic center region at least as large as 10^-5. We conclude
% that either fresh deuterium rich gas is being accreted into the center region
% or that there are processes which produce as well as destroy deuterium.

% Jacq et al. 1999: we conclude that the ratio of D/H in the Sgr B2 cloud is 
% 5e-6 with a factor 10 uncertainty. This result, while it has large 
% uncertainty, suggests that deuterium is roughly one order of magnitude less
% abundant in the galactic center region than in the solar neighbourhood.

% CH2DCN: see Gerin et al. 1992 for detections in Orion and (tentative) G34.3.

% Lubowich et al. 2000: D/H = 1.7e-6 toward Sgr A.
% Polehampton et al. 2002

%__________________________________________________________________

\section{Conclusions}
\label{s:conclusions}

We searched for deuterated complex organic molecules in a complete 3~mm 
interferometric line survey performed toward the hot core Sgr~B2(N2) with
ALMA. We report the secure detection of CH$_2$DCN and tentative detections of 
CH$_2$DOH, CH$_2$DCH$_2$CN in its out-of-plane conformation, the chiral
molecule CH$_3$CHDCN, and 
DC$_3$N. We also derive column density upper limits for CH$_3$OD, 
CH$_2$DCH$_2$CN in its 
in-plane conformation, the three deuterated isotopologues of vinyl cyanide, 
the four deuterated species of ethanol, and CH$_2$DOCHO in its in-plane 
and out-of-plane conformations. The detections are characterized by compact 
emission consistent with a hot-core origin. We obtain a deuterium 
fractionation level of 0.4\% for CH$_2$DCN and values between 0.05\% and 
0.12\% for the other (tentatively detected) species. Stringent upper limits
are derived for the in-plane conformer of CH$_2$DCH$_2$CN ($<0.024\%$) and
CH$_3$OD ($<0.07\%$). The deuteration upper limits for the other non-detected 
deuterated species lie in the range 0.5--1.8\%.

\vspace*{1ex}
The following conclusions arise from this analysis:
\begin{enumerate}
 \item Ethyl cyanide is less deuterated than methyl cyanide by at least a 
factor five. This may be due to their different formation pathways (grain 
surface versus gas phase).
 \item The relative abundances of the three deuterated species of ethyl cyanide 
are most likely consistent with a statistical distribution.
 \item The [CH$_2$DOH]/[CH$_3$OD] abundance ratio in Sgr~B2(N2) is higher than 
1.8. It may still be consistent with the value obtained in Orion~KL but a 
value closer to (or, like in low-mass Class 0 protostars, even higher than) 
the statistical value cannot be excluded.
 \item Except for methyl cyanide, the levels of deuterium fractionation 
measured toward the hot core Sgr~B2(N2) lie at least a factor four below the 
predictions of current astrochemical models that were designed for lower mass 
objects (hot corinos in the Class 0 phase). 
 \item The levels of deuterium fractionation derived for complex organic 
molecules in Sgr~B2(N2) are lower by a factor of a few up to a factor ten
than in the Orion-KL Hot Core or Compact Ridge.
\end{enumerate}
The discrepancy between the deuteration levels measured in Sgr~B2(N2) and the
predictions of current chemical models for complex organic molecules, and the 
difference between Sgr~B2(N2) and Orion~KL may both be due to the higher 
kinetic temperatures that prevail in the Galactic Center region compared to 
nearby star forming regions. Dedicated numerical simulations are needed to 
test this hypothesis. An alternative explanation for the low deuteration level 
of complex organic molecules in Sgr~B2(N2) could be a lower abundance of 
deuterium itself in the Galactic Center region by up to a factor ten, as was 
also suggested by previous studies based on smaller molecules.

\begin{acknowledgements}
We thank B\'ereng\`ere Parise for enlightening discussions about deuterium 
fractionation.
This work has been supported in part by the Deutsche Forschungsgemeinschaft
through the collaborative research grant SFB 956 ``Conditions and
Impact of Star Formation,'' project area B3.
R.T.G. acknowledges support from the NASA Astrophysics Theory Program through 
grant NNX11AC38G.
This paper makes use of the following ALMA data:
ADS/JAO.ALMA\#2011.0.00017.S, ADS/JAO.ALMA\#2012.1.00012.S. 
ALMA is a partnership of ESO (representing its member states), NSF (USA) and 
NINS (Japan), together with NRC (Canada), NSC and ASIAA (Taiwan), and KASI 
(Republic of Korea), in cooperation with the Republic of Chile. The Joint ALMA 
Observatory is operated by ESO, AUI/NRAO and NAOJ.
The interferometric data are available in the ALMA archive at 
https://almascience.eso.org/aq/.
\end{acknowledgements}


\begin{thebibliography}{}
\bibitem[Aikawa et al.(2012)]{Aikawa12} Aikawa, Y., Wakelam, V., Hersant, F., 
Garrod, R.~T., \& Herbst, E.\ 2012, \apj, 760, 40
\bibitem[Anderson et al.(1988)]{Anderson88} Anderson, T., 
Crownover, R.~L., Herbst, E., \& De Lucia, F.~C.\ 1988, \apjs, 67, 135 
\bibitem[Baskakov et al.(1996)]{VyCN_rot_1996} 
Baskakov, O.~I., Dyubko, S.~F., Ilyushin, V.~V., et al. 
1996, J. Mol. Spectrosc., 179, 94 
\bibitem[Bauer \& Maes(1969)]{MeCN_rot-8_1969} 
Bauer, A. \& Maes, S. 
1969, J. Phys. (Paris), 30,  169
\bibitem[Bauer(1971)]{MeCN_rot-16_1971} 
Bauer, A. 
1971, J. Mol. Spectrosc., 40,  183
\bibitem[Belloche et al.(2008)]{Belloche08} Belloche, A., Menten, K.~M., 
Comito, C., et al.\ 2008, \aap, 482, 179
\bibitem[Belloche et al.(2009)]{Belloche09} Belloche, A., Garrod, R.~T., 
M{\"u}ller, H.~S.~P., et al.\ 2009, \aap, 499, 215 
\bibitem[Belloche et al.(2013)]{Belloche13} Belloche, A., M{\"u}ller, 
H.~S.~P., Menten, K.~M., Schilke, P., \& Comito, C.\ 2013, \aap, 559, A47
\bibitem[Belloche et al.(2014)]{Belloche14} Belloche, A., Garrod, 
R.~T., M{\"u}ller, H.~S.~P., \& Menten, K.~M.\ 2014, Science, 345, 1584 
\bibitem[Brauer et al.(2009)]{EtCN_rot_2009} 
Brauer, C.~S., Pearson, J.~C., Drouin, B.~J., \& Yu, S. 
2009, \apjs, 184, 133 
\bibitem[Caselli et al.(1993)]{Caselli93} Caselli, P., Hasegawa, T.~I., \& 
Herbst, E.\ 1993, \apj, 408, 548 
%\bibitem[Caselli et al.(2002)]{Caselli02} Caselli, P., Stantcheva, T., 
%Shalabiea, O., Shematovich, V.~I., \& Herbst, E.\ 2002, \planss, 50, 1257 
\bibitem[Caselli \& Ceccarelli(2012)]{Caselli12} Caselli, P., \& Ceccarelli, 
C.\ 2012, \aapr, 20, 56
\bibitem[Cazzoli \& Kisiel(1988)]{VyCN_rot-vib_1988} 
Cazzoli, G., \& Kisiel, Z. 
1988, J. Mol. Spectrosc., 130, 303 
\bibitem[Ceccarelli et al.(2014)]{Ceccarelli14} Ceccarelli, C., Caselli, P., 
Bockel{\'e}e-Morvan, D., et al.\ 2014, Protostars and Planets VI, 859 
\bibitem[Charnley et al.(1997)]{Charnley97} Charnley, S.~B., 
Tielens, A.~G.~G.~M., \& Rodgers, S.~D.\ 1997, \apjl, 482, L203 
\bibitem[Colmont et al.(1997)]{VyCN-isos_rot_1997} 
Colmont, J.~M., Wlodarczak, G., Priem, D., et al. 
1997, J. Mol. Spectrosc., 181, 330 
\bibitem[Comito et al.(2003)]{Comito03} Comito, C., Schilke, P., Gerin, M., 
et al.\ 2003, \aap, 402, 635
\bibitem[Coudert et al.(2013)]{Coudert13} Coudert, L.~H., Drouin, 
B.~J., Tercero, B., et al.\ 2013, \apj, 779, 119 
\bibitem[Creswell et al.(1977)]{HC3N-isos_rot_1977} 
Creswell, R.~A., Winnewisser, G., \& Gerry, M.~C.~L. 
1977, J. Mol. Spectrosc., 65, 420 
\bibitem[Daly et al.(2013)]{Daly13} Daly, A.~M., Berm{\'u}dez, 
C., L{\'o}pez, A., et al.\ 2013, \apj, 768, 81 
\bibitem[Demaison et al.(1979)]{MeCN-isos_rot_1979} 
Demaison, J., Dubrulle, A., Boucher, D., Burie, J., \& Typke, V. 
1979, J. Mol. Spectrosc., 76, 1 
\bibitem[Duan et al.(2003)]{Duan03} Duan, Y.-B., Ozier, I., Tsunekawa, S., \& 
Takagi, K.\ 2003, J. Mol. Spectrosc., 218, 95 
\bibitem[Esplugues et al.(2013)]{Esplugues13} Esplugues, G.~B., 
Cernicharo, J., Viti, S., et al.\ 2013, \aap, 559, A51 
\bibitem[Favre et al.(2014)]{Favre14} Favre, C., Carvajal, M., Field, D., et 
al.\ 2014, \apjs, 215, 25 
\bibitem[Fayt et al.(2004)]{HC3N-15_rot-vib_2004} 
Fayt, A., Vigouroux, C., Willaert, F., et al. 
2004, J. Mol. Struct., 695, 295 
\bibitem[Fontani et al.(2015)]{Fontani15} Fontani, F., Busquet, G., Palau, A., 
et al.\ 2015, \aap, 575, A87 
\bibitem[Fuente et al.(2014)]{Fuente14} Fuente, A., Cernicharo, J., 
Caselli, P., et al.\ 2014, \aap, 568, A65
\bibitem[Fukuyama et al.(1996)]{EtCN_rot_1996} 
Fukuyama, Y., Odashima, H., Takagi, K., \& Tsunekawa, S. 
1996, \apjs, 104, 329 
\bibitem[Garrod et al.(2008)]{Garrod08} Garrod, R.~T., Weaver, S.~L.~W., \& 
Herbst, E.\ 2008, \apj, 682, 283 
\bibitem[Garrod(2013)]{Garrod13} Garrod, R.~T.\ 2013, \apj, 765, 60 
%\bibitem[Gensheimer et al.(1996)]{Gensheimer96} Gensheimer, P.~D., 
%Mauersberger, R., \& Wilson, T.~L.\ 1996, \aap, 314, 281
\bibitem[Gerin et al.(1992)]{Gerin92} Gerin, M., Combes, F., Wlodarczak, G., 
et al.\ 1992, \aap, 259, L35 
\bibitem[Goldsmith \& Langer(1999)]{Goldsmith99} Goldsmith, P.~F., \& Langer, 
W.~D.\ 1999, \apj, 517, 209
\bibitem[Gottlieb et al.(1979)]{Gottlieb79} Gottlieb, C.~A., Ball, J.~A., 
Gottlieb, E.~W., \& Dickinson, D.~F.\ 1979, \apj, 227, 422 
\bibitem[Heise et al.(1981)]{EtCN_IR_1981} Heise, H.~M., Winther, F., \& 
Lutz, H. 1981, J. Mol. Spectrosc., 90, 531 
\bibitem[Herbst \& van Dishoeck(2009)]{Herbst09} Herbst, E., \& van Dishoeck,
E.~F.\ 2009, \araa, 47, 427
\bibitem[Hidaka et al.(2009)]{Hidaka09} Hidaka, H., Watanabe, M., Kouchi, A., 
\& Watanabe, N.\ 2009, \apj, 702, 291 
\bibitem[Howe et al.(1994)]{Howe94} Howe, D.~A., Millar, T.~J., Schilke, P., 
\& Walmsley, C.~M.\ 1994, \mnras, 267, 59 
\bibitem[Ilyushin et al.(2009)]{Ilyushin09} Ilyushin, V., Kryvda, A., \& 
Alekseev, E.\ 2009, J. Mol. Spectrosc., 255, 32 
\bibitem[Jacq et al.(1990)]{Jacq90} Jacq, T., Walmsley, C.~M., Henkel, C., et 
al.\ 1990, \aap, 228, 447 
\bibitem[Jacq et al.(1993)]{Jacq93} Jacq, T., Walmsley, C.~M., 
Mauersberger, R., et al.\ 1993, \aap, 271, 276 
\bibitem[Jacq et al.(1999)]{Jacq99} Jacq, T., Baudry, A., Walmsley, C.~M., \& 
Caselli, P.\ 1999, \aap, 347, 957 
\bibitem[Kisiel et al.(2015)]{VyCN_IR_2015} Kisiel, Z., Martin-Drumel, M.~A., 
\& Pirali, O.\ 2015, J. Mol. Spectrosc., 315, 83 
\bibitem[Langer et al.(1980)]{Langer80} Langer, W.~D., Schloerb, F.~P., 
Snell, R.~L., \& Young, J.~S.\ 1980, \apjl, 239, L125 
\bibitem[Lauvergnat et al.(2009)]{CH2DOH_FIR_2009} Lauvergnat, D., Coudert, 
L.~H., Klee, S., \& Smirnov, M.\ 2009, J. Mol. Spectrosc., 256, 204 
\bibitem[Linsky(2003)]{Linsky03} Linsky, J.~L.\ 2003, \ssr, 106, 49
\bibitem[Lubowich et al.(2000)]{Lubowich00} Lubowich, D.~A., Pasachoff, J.~M., 
Balonek, T.~J., et al.\ 2000, \nat, 405, 1025 
\bibitem[Maret et al.(2011)]{Maret11} Maret, S., Hily-Blant, P., Pety, J., 
Bardeau, S., \& Reynier, E.\ 2011, \aap, 526, A47
\bibitem[Margul{\`e}s et al.(2009)]{Margules09} Margul{\`e}s, L., 
Motiyenko, R., Demyk, K., et al.\ 2009, \aap, 493, 565 
\bibitem[M{\"u}ller et al.(2001)]{Mueller01} M{\"u}ller, H.~S.~P., Thorwirth, 
S., Roth, D.~A., \& Winnewisser, G.\ 2001, \aap, 370, L49
\bibitem[M{\"u}ller et al.(2005)]{Mueller05} M{\" u}ller, H.~S.~P.,
Schl{\" o}der, F., Stutzki, J., \& Winnewisser, G.\ 2005, J. Mol. Struct.,
742, 215
\bibitem[M{\"u}ller et al.(2008)]{VyCN-isos_rot_2008} 
M{\"u}ller, H.~S.~P., Belloche, A., Menten, K.~M., Comito, C., \& Schilke, P. 
2008, J. Mol. Spectrosc., 251, 319 
\bibitem[M{\"u}ller et al.(2009)]{MeCN-isos_rot_2009} 
M{\"u}ller, H.~S.~P., Drouin, B.~J., \& Pearson, J.~C. 
2009, \aap, 506, 1487 
\bibitem[M{\"u}ller et al.(2015a)]{MeCN_rot-16_2015} 
M{\"u}ller, H.~S.~P., Brown, L.~R., Drouin, B.~J., et al. 
2015a, J. Mol. Spectrosc., 312, 22 
\bibitem[M{\"u}ller et al.(2015b)]{MeCN_isos-v8_2015}
M{\"u}ller, H.~S.~P., Drouin, B.~J., \& Pearson, J.~C., et al. 2015b, \aap, 
submitted
\bibitem[M\"uller et al.(2015c)]{Mueller15c} M\"uller, H.~S.~P., 
Belloche A., Xu, L.-H., et al.\ 2015c, \aap, submitted
\bibitem[Nagaoka et al.(2005)]{Nagaoka05} Nagaoka, A., Watanabe, N., \& 
Kouchi, A.\ 2005, \apjl, 624, L29
\bibitem[Neill et al.(2013)]{Neill13} Neill, J.~L., Crockett, N.~R., 
Bergin, E.~A., Pearson, J.~C., \& Xu, L.-H.\ 2013, \apj, 777, 85
\bibitem[Nguyen et al.(2013)]{D-MeCN_rot_2013} 
Nguyen, L., Walters, A., Margul{\`e}s, L., et al. 
2013, \aap, 553, A84 
\bibitem[Osamura et al.(2004)]{Osamura04} Osamura, Y., Roberts, H., \& 
Herbst, E.\ 2004, \aap, 421, 1101 
\bibitem[Ott et al.(2014)]{Ott14} Ott, J., Wei{\ss}, A., 
Staveley-Smith, L., Henkel, C., \& Meier, D.~S.\ 2014, \apj, 785, 55
\bibitem[Parise et al.(2002)]{Parise02} Parise, B., Ceccarelli, C., 
Tielens, A.~G.~G.~M., et al.\ 2002, \aap, 393, L49
\bibitem[Parise et al.(2004)]{Parise04} Parise, B., Castets, A., Herbst, E., 
et al.\ 2004, \aap, 416, 159
\bibitem[Parise et al.(2006)]{Parise06} Parise, B., Ceccarelli, C., 
Tielens, A.~G.~G.~M., et al.\ 2006, \aap, 453, 949 
\bibitem[Parise et al.(2009)]{Parise09} Parise, B., Leurini, S., Schilke, P., 
et al.\ 2009, \aap, 508, 737 
\bibitem[Pearson et al.(2012)]{CH2DOH_rot_2012} 
Pearson, J.~C., Yu, S., \& Drouin, B.~J. 
2012, J. Mol. Spectrosc., 280, 119 
%\bibitem[Peng et al.(1993)]{Peng93} Peng, Y., Vogel, S.~N., 
%\& Carlstrom, J.~E.\ 1993, \apj, 418, 255
\bibitem[Peng et al.(2012)]{Peng12} Peng, T.-C., Despois, D., Brouillet, N., 
Parise, B., \& Baudry, A.\ 2012, \aap, 543, A152
\bibitem[Pickett et al.(1998)]{Pickett98} Pickett, H.~M., Poynter, R.~L., 
Cohen, E.~A., Delitsky, M.~L., Pearson, J.~C., \& M{\"u}ller, H.~S.~P.\ 1998, 
\jqsrt, 60, 883
\bibitem[Pillai et al.(2007)]{Pillai07} Pillai, T., Wyrowski, F., 
Hatchell, J., Gibb, A.~G., \& Thompson, M.~A.\ 2007, \aap, 467, 207 
\bibitem[Polehampton et al.(2002)]{Polehampton02} Polehampton, E.~T., 
Baluteau, J.-P., Ceccarelli, C., Swinyard, B.~M., \& Caux, E.\ 2002, \aap, 
388, L44 
\bibitem[Qin et al.(2011)]{Qin11} Qin, S.-L., Schilke, P., Rolffs, R., et al.\ 
2011, \aap, 530, L9 
\bibitem[Ratajczak et al.(2009)]{Ratajczak09} Ratajczak, A., Quirico, E., 
Faure, A., Schmitt, B., \& Ceccarelli, C.\ 2009, \aap, 496, L21
\bibitem[Ratajczak et al.(2011)]{Ratajczak11} Ratajczak, A., Taquet, V., 
Kahane, C., et al.\ 2011, \aap, 528, L13
\bibitem[Reid et al.(2014)]{Reid14} Reid, M.~J., Menten, K.~M., Brunthaler, 
A., et al.\ 2014, \apj, 783, 130 
\bibitem[Remijan et al.(2015)]{Remijan15} Remijan, A., et al.\ 2015, ALMA
Cycle 3 Technical Handbook Version 1.0, ALMA, ISBN 978-3-923524-66-2
\bibitem[Richard et al.(2012)]{13C-EtCN_rot_2012} 
Richard, C., Margul{\`e}s, L., Motiyenko, R.~A., \& Guillemin, J.-C. 
2012, \aap, 543, A135 
\bibitem[Rolffs et al.(2011)]{Rolffs11} Rolffs, R., Schilke, P., Wyrowski, F., 
et al.\ 2011, \aap, 527, A68
\bibitem[Sakai et al.(2009)]{Sakai09} Sakai, N., Sakai, T., Hirota, T., \& 
Yamamoto, S.\ 2009, \apj, 702, 1025 
\bibitem[Spahn et al.(2008)]{DC3N_rot_2008} 
Spahn, H., M{\"u}ller, H.~S.~P., Giesen, T.~F., et al. 
2008, Chem. Phys., 346, 132 
%\bibitem[Stantcheva \& Herbst(2003)]{Stantcheva03} Stantcheva, T., \& 
%Herbst, E.\ 2003, \mnras, 340, 983
\bibitem[Taquet et al.(2012)]{Taquet12} Taquet, V., Ceccarelli, C., \& 
Kahane, C.\ 2012, \apjl, 748, L3 
\bibitem[Taquet et al.(2014)]{Taquet14} Taquet, V., Charnley, S.~B., \& 
Sipil{\"a}, O.\ 2014, \apj, 791, 1
\bibitem[Thorwirth et al.(2000)]{HC3N_rot-vib_2000} 
Thorwirth, S., M{\"u}ller, H.~S.~P., \& Winnewisser, G. 
2000, J. Mol. Spectrosc., 204, 133 
\bibitem[Thorwirth et al.(2001)]{HC3N-isos_rot-vib_2001} 
Thorwirth, S., M{\"u}ller, H.~S.~P., \& Winnewisser, G. 
2001, Phys. Chem. Chem. Phys., 3, 1236 
\bibitem[Turner(2001)]{Turner01} Turner, B.~E.\ 2001, \apjs, 136, 579 
\bibitem[Uyemura et al.(1982)]{HC3N_DC3N_IR_1982} Uyemura, M., Deguchi, S., 
Nakada, Y., \& Onaka, T.\ 1982, Bull. Chem. Soc. Jpn., 55 384
\bibitem[Walters et al.(2015)]{D-EtOH_rot_2015}
Walters, A., Sch{\"a}fer, M., Ordu, M.~H., et al.
2015, J. Mol. Spectrosc., 314, 6
\bibitem[Watanabe et al.(2015)]{Watanabe15} Watanabe, Y., Sakai, N., 
L{\'o}pez-Sepulcre, A., et al.\ 2015, \apj, 809, 162
\bibitem[Yamada \& Creswell(1986)]{HC3N_rot-vib_1986} 
Yamada, K.~M.~T., \& Creswell, R.~A. 
1986, J. Mol. Spectrosc., 116, 384 
\end{thebibliography}

%%%%%%%%%%%%%%%%%%%%%%%%%%%%%%%%%%%%%%%%%%%%%%%%%%%%%%%%%%%%%%
%Example below of non-structurated natbib references  
%To use the v8.3 macros with this form of composition of bibliography, 
%the option "bibyear" should be added to the command line 
%"\documentclass[bibyear]{aa}".
%%%%%%%%%%%%%%%%%%%%%%%%%%%%%%%%%%%%%%%%%%%%%%%%%%%%%%%%%%%%%%

\end{document}